\documentclass[aps,rmp,twocolumn,superscriptaddress]{revtex4}
\usepackage{graphicx}
\usepackage{amsthm} 
\usepackage{amsbsy} 
\usepackage{dcolumn}   
\usepackage{verbatim}   
\usepackage{color}     	 
\usepackage{amsmath,amsfonts,amssymb}
\usepackage{natbib}

\theoremstyle{remark}

\theoremstyle{definition}

\newcommand{\eps}{\epsilon}

\newcommand{\ie}{{\it i.e.}}

\definecolor{nblue}{rgb}{0.2,0.2,0.9}
\definecolor{nblack}{rgb}{0,0,0}
\definecolor{nred}{rgb}{1,0.0,0.0}

\newcommand{\tr}{\mathrm{tr}\,} 
\newcommand{\bra}[1]{\left\langle#1\right|} 
\newcommand{\ket}[1]{\left|#1\right\rangle} 

\newcommand{\expectation}[1]{\left\langle #1\right\rangle}

\newcommand{\identity}{\openone}
\newcommand{\ba}{\begin{eqnarray}}
\newcommand{\ea}{\end{eqnarray}}
\newcommand{\be}{\begin{equation}}
\newcommand{\ee}{\end{equation}}
\newcommand{\si}{\sigma}
\newcommand{\mb}{\mathbf}

\newcommand{\id}{\openone}

\begin{document}
\title{Bell nonlocality}

\author{Nicolas Brunner} 
\affiliation{D\'epartement de Physique Th\'eorique, Universit\'e de Gen\`eve, 1211 Gen\`eve, Switzerland}
\affiliation{H.H. Wills Physics Laboratory, University of Bristol, Tyndall Avenue, Bristol, BS8 1TL, United Kingdom}
\author{Daniel Cavalcanti} 
\affiliation{Centre for Quantum Technologies, National University of Singapore, 2 Science Drive 3, Singapore 117543}
\affiliation{ICFO-Institut de Ci\`encies Foto\`niques, Mediterranean Technology Park, 08860 Castelldefels (Barcelona), Spain}
\author{Stefano Pironio} 
\affiliation{Laboratoire d'Information Quantique, Universit\'e Libre de Bruxelles (ULB), Belgium}
\author{Valerio Scarani}
\affiliation{Centre for Quantum Technologies, National University of Singapore, 2 Science Drive 3, Singapore 117543}
\affiliation{Department of Physics, National University of Singapore, 3 Science Drive 2, Singapore 117542}
\author{Stephanie Wehner} 
\affiliation{Centre for Quantum Technologies, National University of Singapore, 2 Science Drive 3, Singapore 117543}
\affiliation{School of Computing, National University of Singapore, 13 Computing Drive, Singapore 117417}

\begin{abstract}
Bell's 1964 theorem, which states that the predictions of quantum theory cannot be accounted for by any local theory, represents one of the most profound developments in the foundations of physics. In the last two decades, Bell's theorem has been a central theme of research from a variety of perspectives, mainly motivated by quantum information science,  where the nonlocality of quantum theory underpins many of the advantages afforded by a quantum processing of information. The focus of this review is to a large extent oriented by these later developments. We review the main concepts and tools which have been developed to describe and study the nonlocality of quantum theory, and which have raised this topic to the status of a full sub-field of quantum information science.
\end{abstract}

\maketitle
\tableofcontents

\section{Introduction}
\label{Introduction}

In 1964, Bell proved that the predictions of quantum theory are incompatible with those of any physical theory satisfying a natural notion of locality\footnote{To avoid any misunderstanding from the start, by ``locality" we do not mean the notion used within quantum mechanics and quantum field theory that operators defined in spacelike separated regions commute. Bell's notion of locality is different and is clarified below.} \cite{Bell64}. Bell's theorem has deeply influenced our perception and understanding of physics, and arguably ranks among the most profound scientific discoveries ever made. 
With the advent of quantum information science, a considerable interest has been devoted to Bell's theorem. 
In particular, a wide range of concepts and technical tools have been developed for describing and studying the nonlocality of quantum theory. These represent the main focus of the present review.
Hence we will not discuss, at least not directly, the extensive literature dealing with the conceptual implications of Bell's theorem from a traditional foundational perspective. Skipping shamelessly many important contributions before and after Bell's ground-breaking discovery, the most notable one being the famous Einstein-Podolosky-Rosen paper \cite{EPR35}, we start straightaway with the mathematical formulation of a locality constraint in the context of certain experiments involving separate systems, and its violation by the predictions of quantum theory. 

\subsection{Non-locality in a nutshell}\label{nutshell}
\begin{figure}[t]
\includegraphics[width = 0.4\textwidth]{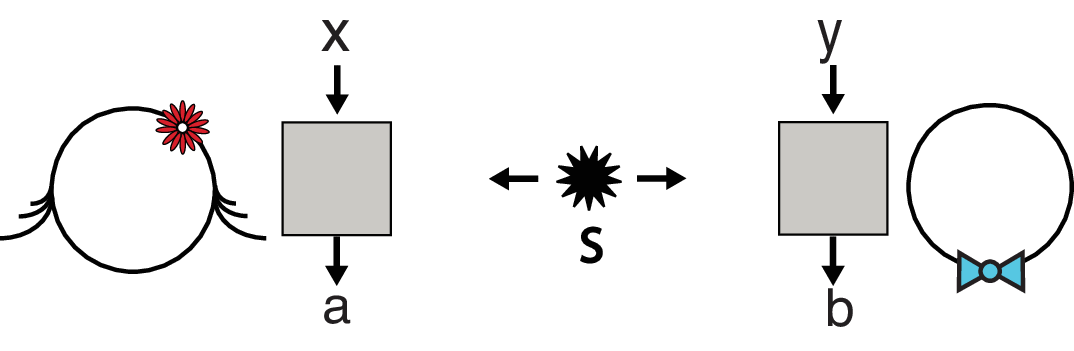}
    \caption{Sketch of a Bell experiment. A source (S) distributes two physical systems to distant observers, Alice and Bob. Upon receiving their systems, each observer performs a measurement on it. The measurement chosen by Alice is labeled $x$ and its outcome $a$. Similarly, Bob chooses measurement $y$ and gets outcome $b$. The experiment is characterized by the joint probability distribution $p(ab|xy)$ of obtaining outcomes $a$ and $b$ when Alice and Bob choose measurements $x$ and $y$.}
\label{bellexp}
\end{figure}

In a typical ``Bell experiment", two systems which may have previously interacted -- for instance they may have been produced by a common source -- are now spatially separated and are each measured by one of two distant observers, Alice and Bob (see Fig.~\ref{bellexp}). Alice may choose one out of several possible measurements to perform on her system and we let $x$ denote her measurement choice. For instance, $x$ may refer to the position of a knob on her measurement apparatus. Similarly, we let $y$ denote Bob's measurement choice. Once the measurements are performed, they yield outcomes $a$ and $b$ on the two systems. The actual values assigned to the measurement choices $x,y$ and outcomes $a,b$ are purely conventional; they are mere macroscopic labels distinguishing the different possibilities. 

From one run of the experiment to the other, the outcomes $a$ and $b$ that are obtained may vary, even when the same choices of measurements $x$ and $y$ are made. These outcomes are thus in general governed by a probability distribution $p(ab|xy)$, which can of course depend on the particular experiment being performed. By repeating the experiment a sufficient number of times and collecting the observed data, one can get a fair estimate of such probabilities. 

When such an experiment is actually performed -- say, by generating pairs of spin-$\frac{1}{2}$ particles and measuring the spin of each particle in different directions -- it will in general be found that
\be 
p(ab|xy)\neq p(a|x)\,p(b|y)\,,
\ee
implying that the outcomes on both sides are not statistically independent from each other. Even though the two systems may be separated by a large distance -- and may even be space-like separated -- the existence of such correlations is nothing mysterious. In particular, it does not necessarily imply some kind of direct influence of one system on the other, for these correlations may simply reveal some dependence relation between the two systems which was established when they interacted in the past. This is at least what one would expect in a local theory. 

Let us formalize the idea of a \emph{local} theory more precisely. The assumption of locality implies that we should be able to identify a set of past factors, described by some variables $\lambda$, having a joint causal influence on both outcomes, and which fully account for the dependence between $a$ and $b$. Once all such factors have been taken into account, the residual indeterminacies about the outcomes must now be decoupled, that is, the probabilities for $a$ and $b$ should factorize: 
\be
p(ab|xy,\lambda)=p(a|x,\lambda)p(b|y,\lambda)\,.
\ee
This factorability condition simply expresses that we have found an explanation according to which the probability for $a$ only depends on the past variables $\lambda$ and on the \emph{local} measurement $x$, but \emph{not on the distant} measurement and outcome, and analogously for the probability to obtain $b$. The variable $\lambda$ will not necessarily be constant for all runs of the experiment, even if the procedure which prepares the particles to be measured is held fixed, because $\lambda$ may involve physical quantities that are not fully controllable. The different values of $\lambda$ across the runs should thus be characterized by a probability distribution $q(\lambda)$. Combined with the above factorability condition, we can thus write
\be\label{intro.loc}
p(ab|xy)=\int_\Lambda\!\mathrm{d}\lambda\,q(\lambda)p(a|x,\lambda)p(b|y,\lambda)\,,
\ee
where we also implicitly assumed that the measurements $x$ and $y$ can be freely chosen in a way that is independent of $\lambda$, i.e., that $q(\lambda|x,y)=q(\lambda)$. This decomposition now represents a precise condition for locality in the context of Bell experiments\footnote{Bell also used the term \emph{local causality} instead of locality. \emph{Local hidden-variable} or \emph{local realistic} models are also frequently used to refer to the existence of a decomposition of the form (\ref{intro.loc}); see \cite{Norsen09,Goldstein2011} for a critical discussion of these terminologies.}. Note that no assumptions of determinism or of a ``classical behaviour" are being involved in the condition (\ref{intro.loc}): we assumed that $a$ (and similarly $b$) is only \emph{probabilistically} determined by the measurement $x$ and the variable $\lambda$, with no restrictions on the physical laws governing this causal relation. Locality is the crucial assumption behind (\ref{intro.loc}). In relativistic terms, it is the requirement that events in one region of space-time should not influence events in space-like separated regions.

It is now a straightforward mathematical theorem\footnote{It is relatively frequent to see on the arXiv a paper claiming to ``disprove"  Bell's  theorem or that a mistake in the derivation of Bell inequalities has been found. However, once one accepts the definition (\ref{intro.loc}), it is a quite trivial \emph{mathematical theorem} that this definition is incompatible with certain quantum predictions. Such papers are thus either using (possibly unawarely) a different definition of locality or they are erroneous. Quantum Randi challenges have been proposed to confront Bell-deniers in a pedagogical way \cite{vongehr,Gill12}.} that the predictions of quantum theory for certain experiments involving entangled particles do not admit a decomposition of the form (\ref{intro.loc}). To establish this result, let us consider for simplicity an experiment where there are only two measurement choices  per observer $x,y \in \{0,1\}$ and where the possible outcomes take also two values labelled $a,b \in \{-1,+1\}$. Let $\langle a_xb_y\rangle = \sum_{a,b}ab\,p(ab|xy)$ be the expectation value of the product $ab$ for given measurement choices $(x,y)$ and consider the following expression
$S=\langle a_0b_0\rangle+\langle a_0b_1\rangle+\langle a_1b_0\rangle-\langle a_1b_1\rangle$, 
which is a function of the probabilities $p(ab|xy)$. If these probabilities satisfy the locality decomposition (\ref{intro.loc}), we necessarily have that
\ba \label{intro.chsh}
S=\langle a_0b_0\rangle+\langle a_0b_1\rangle+\langle a_1b_0\rangle-\langle a_1b_1\rangle\leq 2\,,
\ea
which is known as the Clauser-Horne-Shimony-Holt (CHSH) inequality \cite{CHSH69}. To derive this inequality,  we can use (\ref{intro.loc}) in the definition of $\langle a_xb_y\rangle$, which allows us to express this expectation value as an average $\langle a_xb_y\rangle=\int\!\mathrm{d}\lambda\, q(\lambda) \langle a_x\rangle_\lambda\langle b_y\rangle_\lambda$ of a product of local expectations $\langle a_x\rangle_\lambda=\sum_a a p(a|x,\lambda)$ and $\langle b_y\rangle_\lambda=\sum_b b p(b|y,\lambda)$ taking values in $[-1,1]$. Inserting this expression in (\ref{intro.chsh}), we can write $S=\int\!\mathrm{d}\lambda\, q(\lambda) S_\lambda$, with $S_\lambda = \langle a_0\rangle_\lambda\langle b_0\rangle_\lambda+
\langle a_0\rangle_\lambda\langle b_1\rangle_\lambda+
\langle a_1\rangle_\lambda\langle b_0\rangle_\lambda-
\langle a_1\rangle_\lambda\langle b_1\rangle_\lambda
$. Since $\langle a_0\rangle_\lambda,\langle a_1\rangle_\lambda\in[-1,1]$, this last expression is smaller than $S_\lambda\leq |\langle b_0\rangle_\lambda+\langle b_1\rangle_\lambda|+|\langle b_0\rangle_\lambda-\langle b_1\rangle_\lambda|$. Without loss of generality, we can assume that $\langle b_0\rangle_\lambda\geq\langle b_1\rangle_\lambda\geq 0$ which yields $S_\lambda=2\langle b_0\rangle_\lambda\leq 2$, and thus $S=\int\!\mathrm{d}\lambda\, q(\lambda) S_\lambda\leq 2$. 

Consider now the quantum predictions for an experiment in which the two systems measured by Alice and Bob are two qubits in the singlet state 
$\ket{\Psi^-}=\frac{1}{\sqrt{2}}\left(\ket{01}-\ket{10}\right)$,
where we have used the shortcut notation $\ket{ab}\equiv\ket{a}\otimes\ket{b}$, and where $\ket{0}$ and $\ket{1}$ are conventionally the eigenstates of $\sigma_z$ for the eigenvalues $+1$ and $-1$ respectively. Let the measurement choices $x$ and $y$ be associated with vectors $\vec{x}$ and $\vec{y}$ corresponding to measurements of $\vec{x}\cdot\vec{\sigma}$ on the first qubit and of $\vec{y}\cdot\vec{\sigma}$ on the second qubit, where $\vec{\sigma}=\left(\sigma_1,\sigma_2,\sigma_3\right)$ denotes the Pauli vector. According to quantum theory, we then have the expectations $
\expectation{a_xb_y}=-\vec{x}\cdot\vec{y}$.
Let the two settings $x\in \{0,1\}$ correspond to measurements in the orthogonal directions $\hat e_1$ and $\hat e_2$ respectively and the settings $y\in \{0,1\}$ to measurements in the directions $-(\hat e_1+\hat e_2)/\sqrt{2}$ and $(-\hat e_1+\hat e_2)/\sqrt{2}$. We then have  $\expectation{a_0b_0}=\expectation{a_0b_1}=\expectation{a_1b_0}=1/\sqrt{2}$ and $\expectation{a_1b_1}=-1/\sqrt{2}$, whence \ba\label{intro.chsh.viol}
S\,=\,2\sqrt{2}>2\,,\ea
in contradiction with (\ref{intro.chsh}) and thus with the locality constraint (\ref{intro.loc}). 
This is the content of Bell's theorem, establishing the non-local character of quantum theory and of any model reproducing its predictions.

The CHSH inequality (\ref{intro.chsh}) is an example of a \emph{Bell inequality}, a linear inequality for the probabilities $p(ab|xy)$ that is necessarily verified by any model satisfying the locality condition (\ref{intro.loc}), but which can be \emph{violated} by suitable measurements on a pair of quantum particles in an entangled state. The violation of these inequalities and the predictions of quantum theory were first confirmed experimentally by \cite{FC72}, then more convincingly by Aspect \emph{et al.} \cite{AGR82}, and in many other experiment since.

Before outlining in more detail the content of the present article, let us first reconsider Bell's locality condition from a more operational perspective, which illustrates the spirit underlying this review. 

\subsection{The limitations of non-communicating PhD students}\label{phd}
Consider a quantum apparatus which can perform a measurement on a quantum system in a state $\rho_A$. If measurement $x$ is chosen, an output $a$ is obtained. Quantum theory predicts the statistics $p(a|x)$ for the outcomes given the measurements. Suppose that a PhD student, who cannot realize such a quantum experiment, is instead provided with unlimited classical computational power and a source of random numbers. If the student is competent, he can simulate the \textit{same} statistics as in the quantum experiment based only on the \textit{description} of the state $\rho_A$ and of the measurement $x$ to be performed on it. This is not a particularly deep remark: it is the daily job of physicists all over the world and an obvious consequence of the fact that the theory allows predicting the results.

Now, consider two quantum devices in two distant locations $A$ and $B$ performing measurements $x$ and $y$ on two systems in a joint state $\rho_{AB}$. Quantum theory allows computing the joint probabilities $p(ab|xy)$, so certainly the above student can simulate the experiment if he is given all the relevant information. However, in the quantum experiment the two locations can be sufficiently separated so that no information on $y$ is available at the location $A$ before a result is obtained, and similarly no information on $x$ is available at $B$. Can \emph{two} students, one at $A$ and the other at $B$, simulate the quantum statistics in the same circumstances? As before, the students cannot manipulate any quantum systems, but they have unlimited computational power, access to a source of random numbers, and a perfect description of the joint state $\rho_{AB}$. Though they cannot communicate once the measurements are specified, they may have set-up in advance a common strategy and have shared some common classical data $\lambda$, which may vary across different simulation runs according to a probability distribution $q(\lambda)$. In full generality, the output of the first student will thus be characterized by a probability distribution $p(a|x,\lambda)$, which is fixed by their common strategy and the joint state $\rho_{AB}$, but which may depend on the specific measurement $x$ chosen and of the data $\lambda$ shared with the second student.  Similarly the output of the second student is given by a probabilistic function $p(b|y,\lambda)$. The joint statistics simulated by the two students are thus characterized by the probabilities
\be \label{intro.loc2}
p(ab|xy)=\int d\lambda q(\lambda) p(a|x,\lambda) p(b|y,\lambda)\,,
\ee
which is nothing but the locality condition (\ref{intro.loc}). This condition thus admits a very simple and operational interpretation: it characterizes the correlations that can be reproduced with classical resources by our two non-communicating students. The fact that certain experiments  involving entangled quantum states violate Bell inequalities then imply that the two students cannot simulate such experiments. The violations of Bell inequalities can thus be interpreted as establishing a gap between what non-communicating observers having access to classical or to quantum entangled resources can achieve. Note that locality, i.e., the constraint that the two observers cannot communicate, is the important limitation here. As we said above, if all the information about $x$ and $y$ is available to one of the students, it is always possible to reproduce the quantum statistics using only classical resources. 

There is another point worth stressing here. The fact that entangled quantum systems are able to do things completely differently than classical systems is well known. Indeed, for more than one century physicists have discovered that classical physics does not explain everything. However, if given only the statistics of a real quantum experiment and of a classical simulation of it, there is no way to tell the difference. The brute measurement data produced, for instance, by a Stern-Gerlach experiment can be simulated classically; its ``non-classicality"  becomes evident only when one takes into account that a magnetic moment is being measured and that the measurements are associated to the direction of the gradient of a magnetic field. In the case of Bell nonlocality, however, the real quantum experiment and its (attempted) simulation can be distinguished solely from the measurement data, without having to specify which physical degree of freedom is involved or which measurements are performed. This property is referred to as \emph{device-independence}. Interpreted in this way, the violation of Bell inequalities can be seen as a detector of entanglement that is robust to any experimental imperfection: as long as a violation is observed, we have the guarantee, independently of any implementation details, that the two systems are entangled. This remark is important: since entanglement is at the basis of many protocols in quantum information, and in particular quantum cryptographic protocols, it opens the way to device-independent tests of their performance. 
 
\subsection{Scope of this review}

We have given here only a very succinct and intuitive presentation of the locality condition from which Bell's theorem follows. This naturally raises a first series of questions: What are the precise physical assumptions on which this condition is based? Can we rigorously justify, in particular on relativistic grounds, the notion of locality captured by this condition? To what extent does nonlocality, i.e. the violation of (\ref{intro.loc}), conflict with relativity? What do the various interpretations of quantum theory have to say about this issue? We do not address here such questions, that have been the subject of extensive analysis and discussions by both physicists and philosophers of science since Bell's discovery. A recent concise review has been written from this perspective \cite{Goldstein2011}. Bell's collection of papers on the subject \cite{Bell04} is a must read, in which he explains and develops his main result from a variety of perspectives. In particular, the two articles \cite{Bell75} and \cite{Bell90} introduce the principle of \emph{local causality} -- a precise formulation of the notion of relativistic locality  -- from which the condition (\ref{intro.loc}) can be derived, see also \cite{Norsen07}. For a discussion of the implications of non-locality for relativity we refer to \cite{Maudlin02}.

The present review has a more technical flavour: How can one show that the measurement statistics of a given experiment do not satisfy the condition (\ref{intro.loc})? How can one derive Bell inequalities in a systematic way? Which entangled states violate these inequalities, which ones do not? Can quantum nonlocality be exploited for information processing, and if yes how? How should one design the best experimental test of quantum nonlocality? Etc. Though they may have foundational motivations or implications, the works discussed here have an original technical component. Many of them also follow a recent trend in which non-locality is considered from an operational perspective and where its relations with other topics in quantum information science, such as the theory of entanglement or cryptography, are investigated. Finally, we focus on progress reported in the last fifteen years or so. For works on Bell non-locality before this period or for aspects not covered here, we refer to the following reviews \cite{CS78,HS91,Genovese05,WW01b,BCMW10,KT92,Tsirelson93,Mermin93,Zeilinger99} and references therein.

\subsection{Outline}
A succinct outline of this review is as follows. Section~\ref{secgeom} is devoted to setting-up some general definitions and presenting a mathematical characterization of nonlocal correlations. In particular, we study the general properties of correlations that can arise between local, quantum and no-signalling systems. We address the problem of deriving Bell inequalities from the locality condition (\ref{intro.loc}) and of determining their maximal quantum violations.
Section~\ref{Nonlocality and quantum theory} addresses nonlocality in quantum theory. The main question is to understand how quantum nonlocality relates to certain properties of quantum resources, such as entanglement and Hilbert space dimension.  
The relation between nonlocality and information is discussed in sections~\ref{appl_nl} and \ref{Information}. We first present in section \ref{appl_nl} various applications of quantum nonlocality, such as communication complexity, quantum cryptography, and device-independent quantum information processing. Section \ref{Information} provides an information-theoretic perspective on nonlocality, in which nonlocal correlations are viewed as a fundamental resource. Notably, these ideas stimulated a series of work trying to recover the structure of quantum correlations (and more generally of quantum theory itself) from information-theoretic principles. 
Section~\ref{Multipartite} is devoted to the nonlocality of multipartite systems. The notions of genuine multipartite nonlocality and monogamy of correlations are discussed, as well as their relevance for quantum multipartite systems.
In Section~\ref{Experimental aspects}, we succinctly review the impressive experimental work that has been achieved on quantum nonlocality, where Bell inequality violations have been demonstrated using a variety of different physical systems and experimental configurations. We also discuss the loopholes that may affect Bell experiments, and report recent progress made towards a loophole-free Bell experiment.
Finally, section~\ref{Related concepts} deals with variations around Bell's theorem, in which different notions of non-locality -- stronger or weaker than Bell's one -- are considered. Finally, the appendix provides a guide referencing Bell inequalities for a wide range of Bell scenarios.

\section{Mathematical characterization of nonlocal correlations}
\label{secgeom}

This section presents the main concepts and tools for characterizing nonlocal correlations. The notations introduced here will be used throughout this review. For the sake of clarity, the discussion focuses  mainly on the case of two observers, generalizations to the multipartite case being usually straightforward (see also sections~\ref{multi} and \ref{Multipartite} for results specific to the multipartite case).

\subsection{General definitions}
As in the Introduction, we consider two distant observers, Alice and Bob, performing measurements on a shared physical system, for instance a pair of entangled particles.
Each observer has a choice of $m$ different measurements to perform on his system. Each measurement can yield $\Delta$ possible outcomes. Abstractly we may describe the situation by saying that Alice and Bob have access to a ``black box''. Each party selects locally an input (a measurement setting) and the box produces an output (a measurement outcome). We refer to this scenario as a \emph{Bell scenario}. 

We label the inputs of Alice and Bob $x,y\in\{1,\ldots,m\}$ and their outputs $a,b\in\{1,\ldots, \Delta\}$, respectively. The labels attributed to the inputs and outputs are purely conventional, and all the results presented here are independent of this choice. Some parts of this review might use other notations for convenience. In particular when the outputs are binary, it is customary to write $a,b\in\{-1,1\}$ or $a,b\in \{0,1\}$.

Let $p(ab|xy)$ denote the joint probability to obtain the output pair $(a,b)$ given the input pair $(x,y)$. A Bell scenario is then completely characterized by $\Delta^2m^2$ such joint probabilities, one for each possible pairs of inputs and outputs. Following the terminology introduced in \cite{Tsirelson93}, we will refer to the set $\mb{p}=\{p(ab|xy)\}$ of all these probabilities as a \emph{behavior}. Informally, we will simply refer to them as the \emph{correlations} characterizing the black box shared by Alice and Bob.
A behavior can be viewed as a point $\mb{p}\in\mathbb{R}^{\Delta^2m^2}$ belonging to the probability space $\mathcal{P}\subset \mathbb{R}^{\Delta^2m^2}$ defined by the positivity constraints $p(ab|xy)\geq 0$ and the normalization constraints $\sum_{a,b=1}^{\Delta} p(ab|xy)=1$. Due to the normalization constraints $\mathcal{P}$ is a subspace of $\mathbb{R}^{\Delta^2m^2}$ of dimension $\dim \mathcal{P}=(\Delta^2-1)m^2$.

The existence of a given physical model behind the correlations obtained in a Bell scenario translates into additional constraints on the behaviors $\mb{p}$. Three main types of correlations can be distinguished.

\subsubsection{No-signaling correlations}
A first natural limitation on behaviors $\textbf{p}$ are the \emph{no-signaling} constraints~\cite{Cirelson80,PR94}, formally expressed as
\begin{eqnarray}\label{sec2.nosig}
\sum_{b=1}^\Delta p(ab|xy)=\sum_{b=1}^\Delta p(ab|xy')& &\text{for all } a,x,y,y'\nonumber\\
\sum_{a=1}^\Delta p(ab|xy)=\sum_{a=1}^\Delta p(ab|x'y)& &\text{for all } b,y,x,x'\,.
\end{eqnarray}
These constraints have a clear physical interpretation: they imply that the the local marginal probabilities of Alice $p(a|x)\equiv p(a|xy)=\sum_{b=1}^\Delta p(ab|xy)$ are independent of Bob's measurement setting $y$, and thus Bob cannot signal to Alice by his choice of input (and the other way around). In particular, if Alice and Bob are space-like separated, the no-signaling constraints (\ref{sec2.nosig}) guarantee that Alice and Bob cannot use their black box for instantaneous signaling, preventing from a direct conflict with relativity.

Let $\mathcal{NS}$ denote the set of behaviors satisfying the no-signaling constraints (\ref{sec2.nosig}). It is not difficult to see that $\mathcal{NS}$ is an affine subspace of $\mathbb{R}^{\Delta^2m^2}$ of dimension 
\begin{align}\label{eq:tDef}
	\dim \mathcal{NS} =2(\Delta-1)m+(\Delta-1)^2m^2 =: t\ , 
\end{align}
see e.g. \cite{Pironio05}. One can thus parametrize points in $\mathcal{NS}$ using $t$ numbers rather than the $\Delta^2m^2$ numbers (or $(\Delta^2-1)m^2$ taking into account normalization) necessary to specify a point in the general probability space $\mathcal{P}$. A possible parametrization is given by the set of probabilities: $\{p(a|x),p(b|y),p(ab|xy)\}$ where $a,b=1,\ldots,\Delta-1$ and $x,y=1,\ldots,m$. 
There are indeed $t$ such probabilities and their knowledge is sufficient to reconstruct the full list of $p(ab|xy)$ for any $a,b,x,y$.
Seen as a subset of $\mathbb{R}^t$, the no-signaling set is thus uniquely constrained by the $\Delta^2m^2$ positivity constraints $p(ab|xy)\geq 0$ (which have to be re-expressed in term of the chosen parametrization).

In the case of binary outcome ($\Delta=2$), an alternative parametrization is provided by the $2m+m^2$ \emph{correlators} $\{\langle A_x\rangle,\langle B_y\rangle,\langle A_{x}B_y\rangle\}$, where
\be
\langle A_x\rangle = \sum_{a \in \{\pm 1\}} a\ p(a|x)\quad  \, , \,\, \langle B_y\rangle = \sum_{b\in \{\pm 1\}} b\ p(b|y)\,,
\ee
\be \label{sec2.correlator}
\langle A_xB_y\rangle = \sum_{a,b \in \{\pm 1\}} ab\ p(ab|xy)\,,
\ee
and we have assumed $a,b\in\{-1,1\}$.
Joint probabilities and correlators are related as follows: $p(ab|xy)=\left[1+a\langle A_x\rangle+b\langle B_y\rangle+ab\langle A_xB_y\rangle\right]/4$. Thus an arbitrary no-signaling behavior must satisfy $1+a\langle A_x\rangle+b\langle B_y\rangle+ab\langle A_xB_y\rangle\geq 0$ for all $a,b,x,y$.
See \cite{BGP10b} for a more general definition of correlators for the $\Delta>2$ case.

A particular subset of interest of $\mathcal{NS}$ in the $\Delta=2$ case is the one for which $\langle A_x\rangle =\langle B_y\rangle=0$. We will refer to this set as the \emph{correlation space} $\mathcal{C}$. In term of the $m^2$ correlators (\ref{sec2.correlator}), an arbitrary point in $\mathcal{C}$ is only constrained by the inequalities $-1\leq \langle A_xB_y\rangle\leq 1$. 
Bell inequalities that involve only the quantities $\langle A_xB_y\rangle$, like the CHSH inequality, are called correlation inequalities.

\subsubsection{Local correlations}
A more restrictive constraint than the no-signaling condition is the locality condition discussed in the Introduction. Formally, the set $\mathcal{L}$ of \emph{local behaviors} is defined by the elements of $\mathcal{P}$ that can be written in the form
\be \label{sec2.loc}
p(ab|xy)=\int_\Lambda \!\mathrm{d}\lambda \,q(\lambda) p(a|x,\lambda)p(b|y,\lambda)
\ee
where the (hidden) variables $\lambda$ are arbitrary variables taking value in a space $\Lambda$ and distributed according to the probability density $q(\lambda)$ and where $p(a|x,\lambda)$ and $p(b|y,\lambda)$ are local probability response functions for Alice and Bob, respectively. Operationally, one can also think about $\lambda$ as shared randomness, that is some shared classical random bits, where Alice will choose an outcome $a$ depending both on her measurement setting $x$ as well as $\lambda$ and similarly for Bob.

Whereas any local behavior satisfies the no-signaling constraint, the converse does not hold. There exist no-signaling behaviors which do not satisfy the locality conditions. Hence the set of local correlations is strictly smaller than the set of no-signaling correlations, that is $\mathcal{L}\subset \mathcal{NS}$. 

Correlations that cannot be written in the above form are said to be \emph{nonlocal}. Note that this can happen only if $\Delta\geq 2$ and $m\geq 2$ (otherwise it is always possible to build a local model for any behavior in $\mathcal{P}$). In the following, we thus always assume $\Delta\geq 2$, $m\geq 2$.

\subsubsection{Quantum correlations}
Finally, we consider the set of behaviors achievable in quantum mechanics. Formally, the set $\mathcal{Q}$ of \emph{quantum behaviors} corresponds to the elements of $\mathcal{P}$ that can be written as
\be \label{sec2.quantum}
p(ab|xy)=\tr\left(\rho_{AB}\, M_{a|x}\otimes M_{b|y}\right)
\ee
where $\rho_{AB}$ is a quantum state in a joint Hilbert space $\mathcal{H}_A\otimes \mathcal{H}_B$ of arbitrary dimension, $M_{a|x}$ are measurement operators (POVM elements) on $\mathcal{H}_A$ characterizing Alice's measurements (thus $M_{a|x}\geq 0$ and $\sum_{a=1}^\Delta M_{a|x}=\identity$), and similarly $M_{b|y}$ are operators on $\mathcal{H}_B$ characterizing Bob's measurements.

Note that without loss of generality, we can always assume the state to be pure and the measurement operators to be orthogonal projectors, if necessary by increasing the dimension of the Hilbert space. That is, we can equivalently write a quantum behavior as
\be \label{sec2.quantum2}
p(ab|xy)=\langle \psi|M_{a|x}\otimes M_{b|y}|\psi\rangle
\ee
where $M_{a|x}M_{a'|x}=\delta_{aa'}M_{a|x}$, $\sum_{a}M_{a|x}=\identity_A$ and similarly for the operators $M_{b|y}$.

A different definition of quantum behaviors is also possible, where instead of imposing a tensor product structure between Alice's and Bob's systems, we merely require that their local operators commute \cite{Tsirelson93}. We call the corresponding set $\mathcal{Q}'$, i.e., a behavior $\mb{p}$ belongs to $\mathcal{Q}'$ if
\be \label{sec2.quantum2c}
p(ab|xy)=\langle \psi|M_{a|x} M_{b|y}|\psi\rangle
\ee
where $|\psi\rangle$ is a state in a Hilbert space $H$, and $M_{a|x}$ and $M_{b|y}$ are orthogonal projectors on $H$ defining proper measurements and satisfying $[M_{a|x},M_{b|y}]=0$.
The former definition (\ref{sec2.quantum2}) is standard in non-relativistic quantum theory, while the second one (\ref{sec2.quantum2c}) is natural in relativistic quantum field theory. Since $[M_{a|x}\otimes \identity_B,\identity_A\otimes M_{b|y}]=0$ it is immediate that $\mathcal{Q}\subseteq \mathcal{Q}'$. It is an open question, known as Tsirelson's problem, whether the inclusion is strict, i.e., $\mathcal{Q}\neq \mathcal{Q}'$ \cite{Tsirelson93,SW08,JNP+10,Fritz10}. In the case where the Hilbert spaces $\mathcal{H},\mathcal{H}_A,\mathcal{H}_B$ are finite, it is known that the definitions (\ref{sec2.quantum2}) and (\ref{sec2.quantum2c}) coincide (see e.g. \cite{Tsirelson93,DLTW08, NCP+11}). 
It is also known that $\mathcal{Q}=\mathcal{Q}'$ if Alice has a binary choice of inputs with two outputs each, independently of Bob's number of inputs and outputs \cite{NCP+11}. More precisely, in this case any element of $\mathcal{Q}'$ can be approximated arbitrarily well by an element of $\mathcal{Q}$.
For many applications and results, it does not matter whether we consider the quantum sets $\mathcal{Q}$ or $\mathcal{Q}'$.  In the following, we will drop the distinction and use the notation $\mathcal{Q}$ to refer to both sets, except when results are specific to only one definition. 

It can easily be shown that any local behavior admits a description of the form \eqref{sec2.quantum} and thus belongs to $\mathcal{Q}$ (see e.g. \cite{Pitowsky86}). Moreover, any quantum behavior satisfies the no-signaling constraints. 
However, there are quantum correlations that do not belong to the local set (this follows from the violation of Bell inequalities) and, as we will see, there are no-signaling correlations that do not belong to the quantum set \cite{KT85,Rastall85,PR94}. In general, we thus have the strict inclusions $\mathcal{L}\subset \mathcal{Q}\subset\mathcal{NS}$ (see Fig.~\ref{sec2.geometry}). Furthermore, it can be shown that $\dim \mathcal{L}=\dim\mathcal{Q}=\dim\mathcal{NS}=t$ \cite{Pironio05} where $t$ is defined in~\eqref{eq:tDef}.

In the following sections, we discuss the properties of $\mathcal{L}$, $\mathcal{Q}$, and $\mathcal{NS}$ in more details. In particular, we will see how it is possible to decide if a given behavior belongs or not to one of these sets. We will show how each set can be characterized in terms of Bell-type inequalities and discuss how to compute bounds for Bell-type expression for behaviors in $\mathcal{L}$, $\mathcal{Q}$, and $\mathcal{NS}$.

\subsection{Bell inequalities}\label{seclocal}
\begin{figure}[t]
\centering
\includegraphics[width = 0.4\textwidth]{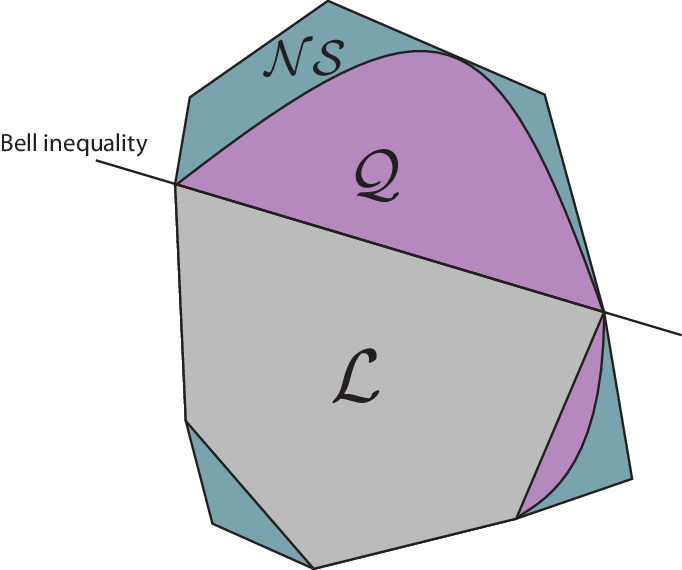}
\caption{Sketch of the no-signaling ($\mathcal{NS}$), quantum ($\mathcal{Q}$), and local ($\mathcal{L}$) sets. Notice the strict inclusions $\mathcal{L}\subset \mathcal{Q}\subset\mathcal{NS}$. Moreover, $\mathcal{NS}$ and $\mathcal{L}$ are polytopes, \ie, they can be defined as the convex combination of a finite number of extremal points. The set $\mathcal{Q}$ is convex, but not a polytope. The hyperplanes delimiting the set $\mathcal{L}$ correspond to Bell inequalities.}
\label{sec2.geometry}
\end{figure}

The sets $\mathcal{L}$, $\mathcal{Q}$, and $\mathcal{NS}$ are closed, bounded, and convex. That is, if $\mb{p}_1$ and $\mb{p}_2$ belong to one of these sets, then the mixture $\mu \mb{p}_1 +(1-\mu) \mb{p}_2$ with $0\leq\mu\leq 1$ also belongs to this set. The convexity of $\mathcal{Q}$ can be established for instance by following the argument in \cite{Pitowsky86}. By the hyperplane separation theorem, it follows that for each behavior $\hat{\mb{p}}\in \mathbb{R}^t$ that does not belong to one of the sets $\mathcal{K}=\mathcal{L}$, $\mathcal{Q}$, or $\mathcal{NS}$ there exists a hyperplane that separates this $\hat{\mb{p}}$ from the corresponding set (see Figure~\ref{sec2.geometry}). That is if $\hat{\mb{p}}\notin \mathcal{K}$, then there exists an inequality of the form
\begin{align}
	\mb{s}\cdot\mb{p}=\sum_{abxy}s^{ab}_{xy}\ p(ab|xy)\leq S_k 
\end{align}
that is satisfied by all $\mb{p}\in \mathcal{K}$ but which is violated by $\hat{\mb{p}}$: $\mb{s}\cdot \hat{\mb{p}}>S_k$. In the case of the local set $\mathcal{L}$, such inequalities are simply \emph{Bell inequalities}. Thus any nonlocal behavior violates a Bell inequality.
An example of such an inequality is the CHSH inequality \eqref{intro.chsh} that we have introduced in section~\ref{nutshell}. 
The inequalities associated to the quantum set, which characterize the limits of $\mathcal{Q}$, are often called \emph{quantum Bell} inequalities or \emph{Tsirelson} inequalities.

In the following, we will refer to an arbitrary $\mb{s}\in\mathbb{R}^t$ as a \emph{Bell expression} and to the \emph{mininal} value $S_l$ such that $\mb{s}\cdot \mb{p}\leq S_l$ holds for all $\mb{p}\in\mathcal{L}$ as the \emph{local bound} of this Bell expression. Similarly, we define the \emph{quantum bound} $S_q$ and the \emph{no-signaling bound} $S_{ns}$ as the analogue quantities for the sets $\mathcal{Q}$ and $\mathcal{NS}$.
If $S_{q} > S_{l}$ we also say that quantum mechanics \emph{violates} the Bell inequality $\mb{s}\cdot \mb{p} \leq S_l$. When such a behavior is observed one speaks of a \emph{Bell inequality violation}.

\subsubsection{The local polytope}
Let us now investigate how Bell inequalities, i.e., the hyperplanes characterizing the set $\mathcal{L}$, can be found.
To this end, it is useful to note that we can express local correlations in a simpler form.
A first step is to realize that local correlations can, equivalently to the definition (\ref{sec2.loc}), 
be defined in terms of \emph{deterministic} local hidden-variable models. In a deterministic model, the local response functions $p(a|x,\lambda)$ and $p(b|y,\lambda)$ only take the values $0$ or $1$, that is, the hidden variable $\lambda$ fully specifies the outcome that is obtained for each measurement. No such requirement is imposed on the general stochastic model (\ref{sec2.loc}).
That both definitions are equivalent follows from the fact that any local randomness present in the response function $p(a|x,\lambda)$ and $p(b|y,\lambda)$ can always be incorporated in the shared random variable $\lambda$. To see this, introduce two parameters $\mu_1,\mu_2\in[0,1]$ in order to define a new hidden-variable $\lambda'=(\lambda,\mu_1,\mu_2)$. Let
\begin{equation}
\label{sec2.detlocalice}
p'(a|x,\lambda')=\begin{cases}
1 &\text{if } F(a-1|x,\lambda)\leq \mu_1<F(a|x,\lambda)\\
0 &\text{otherwise},
\end{cases}
\end{equation}
where $F(a|x,\lambda)=\sum_{\widetilde{a}\leq a}p(\widetilde{a}|x,\lambda)$, be a new response function for Alice and define a similar one for Bob. If we choose $q'(\lambda')=q'(\lambda,\mu_1,\mu_2)=q(\lambda)$ for the new hidden variable distribution, that is if we uniformly randomize over $\mu_1$ and $\mu_2$, we clearly recover the predictions of the general, stochastic model (\ref{sec2.loc}). The newly defined model, however, is deterministic. This equivalence between the two models was first noted in \cite{Fine82}.

We can further simplify the definition by noting that we only need to consider a finite number of hidden variables. Indeed, in a deterministic model, each hidden variable $\lambda$ defines an assignment of one of the possible outputs to each input. The model as a whole is a probabilistic mixture of these deterministic assignments of outputs to inputs, with the hidden variable specifying which particular assignment is chosen in each run of the experiment. Since the total number of inputs and outputs is finite, there can only be a finite number of such assignments, and hence a finite number of hidden variables. 

More precisely, we can rephrase the local model (\ref{sec2.loc}) as follows.
Let $\lambda=(a_1,\ldots,a_{m}\,;\,b_0,\ldots,b_{m})$ define an assignment of outputs $a_{x}$ and $b_{y}$ for each of the inputs ${x}=1,\ldots,m$ and $y=1,\ldots,m$. Let $\mathbf{d}_\lambda\in \mathcal{L}$ denote the corresponding deterministic behavior:
\begin{equation}\label{sec2.detpoint}
d_\lambda(ab|xy)=\left\{\begin{array}{ll} 1 &\text{if } a=a_x \text{ and } b=b_y\\ 0&\mbox{otherwise.}\end{array}\right.
\end{equation}
There are $\Delta^{2m}$ possible output assignments and therefore $\Delta^{2m}$ such local deterministic behaviors. A behavior $\mb{p}$ is local if and only if it can be written as a convex combination of these deterministic points, that is if
\begin{equation}\label{sec2.locpoly}
\mb{p}=\sum_\lambda q_\lambda \mb{d}_\lambda,\text{ with }
 q_\lambda \geq 0,\, \sum_\lambda q_\lambda=1\,.
\end{equation}

This last representation is particularly useful as it provides an algorithm for determining if a given behavior $\mb{p}$ is local \cite{ZKBL99, KGZMZ01}.
Indeed, determining whether there exist weights $q_\lambda$ satisfying the linear constraints in Eq.~(\ref{sec2.locpoly}) is a typical instance of a \emph{linear programming} problem~\cite{boyd:book} for which there exist algorithms that run in time that is polynomial in the number of variables. 
Note, however, that since there are $\Delta^{2m}$ possible $\lambda$ the size of this particular linear program is extremely large and hence the algorithm is not efficient by itself.
Every linear program comes in a primal and dual form. The dual form of the linear program associated to (\ref{sec2.locpoly}) has an interesting physical interpretation. Indeed, it can be formulated as
\begin{equation}\label{lploc}
\begin{array}{ll}
\max_{(\mb{s},S_l)} &\mb{s}\cdot \mb{p}-S_l\\
\text{s.t.} &\mb{s}\cdot \mb{d}_\lambda-S_l\leq 0,\qquad\lambda=1,\ldots,\Delta^{2m},\\
&\mb{s}\cdot \mb{p}-S_l\leq 1\,.
\end{array}
\end{equation}
If $\mb{p}$ is local, the maximum $S$ of the above program is $S\leq 0$. If $\mb{p}$ is non-local, the maximum is $S=1$, i.e., the program returns an inequality $\mb{s}\cdot \mb{d}_\lambda\leq S_l$ satisfied by all deterministic points (and hence, by convexity, by all local points), but violated by $\mb{p}$: $\mb{s}\cdot \mb{p} = S_l+1>S_l$. 
That is the program (\ref{lploc}) provides a procedure for finding, for any $\mb{p}$, a Bell inequality that detects its non-locality.

Since the set $\mathcal{L}$ is the convex hull of a finite number of points, it is a \emph{polytope}. The local deterministic behaviors $\mb{d}_\lambda$ correspond to the vertices, or extreme points, of the polytope.
It is a basic result in polyhedral theory, known as Minkowski's theorem, that a polytope can, equivalently to the representation (\ref{sec2.locpoly}) as the convex hull of its vertices, be represented as the intersection of finitely many half-spaces. Hence, we have that $\mb{p}\in\mathcal L$ if and only if
\be
\mb{s}^i\cdot \mb{p}\leq S^i_l\quad \forall\, i\in I\label{sec2.locineq},
\ee
where $I$ indexes a finite set of linear inequalities. If on the other hand $\mb{p}$ is nonlocal, it necessarily violates one of the inequalities in (\ref{sec2.locineq}). Thus the local set $\mathcal{L}$ can be characterized by a finite set of Bell inequalities.

\subsubsection{Facet Bell inequalities}
If $\mb{s}\cdot \mb{p}\leq S_l$ is a valid inequality for the polytope $\mathcal{L}$, then $F=\{\mb{p}\in \mathcal{L}\mid \mb{s}\cdot \mb{p}=S_l\}$ is called a face of $\mathcal{L}$. Faces of dimension $\dim F=\dim\mathcal{L}-1=t-1$ are called facets of $\mathcal{L}$ and the corresponding inequalities are called facet Bell inequalities. The terminology ``tight Bell inequalities" is also used\footnote{Note however that in polytope theory a tight inequality refers merely to an inequality that ``touches" the polytope, i.e., such that $F\neq \emptyset$.}.
Facet inequalities are important because they provide a \emph{minimal} representation of the set $\mathcal{L}$ in the form (\ref{sec2.locineq}): minimal as they are necessarily required in the description (\ref{sec2.locineq}), and since any other Bell inequality can be written as a non-negative combination of the facet inequalities. These notions are easily understood and visualised in two or three dimensions (note, however, that our low-dimensional intuition is often unreliable in higher dimensions). A more general discussion of polytope theory (not applied to Bell inequalities) can be found in, e.g., \cite{Schrijver89} or \cite{Ziegler95}. The connection between optimal Bell inequalities and polytope theory was realized early by \cite{Froissard81} and later by different authors \cite{GM84,Pitowsky89,Peres99,WW01}.

Facet Bell inequalities provide a practical description of the local polytope $\mathcal{L}$. Usually, however, we start from the vertices of $\mathcal{L}$, which are the local deterministic behaviors $\mb{d}_\lambda$. The task of determining the facets of a polytope, given its vertices, is known as the facet enumeration or convex hull problem. For sufficiently simple cases, it is possible to obtain all the facets with the help of computer codes, such as \texttt{cdd} \cite{cdd} or \texttt{porta} \cite{porta}, which are specifically designed for convex hull computations. However, such programs become prohibitively time consuming as the number of parties, inputs or outputs grow.
Note also that the simpler problem of determining whether a behavior is local using the linear program associated to (\ref{lploc}) becomes also rapidly impractical for large number of inputs $m$ since the number of deterministic points scale exponentially with $m$.  Results in computer science tell us that this problem is in general extremely hard \cite{fortnow}. Evidence in this direction was first given in \cite{Pitowsky89}. Then, it was proven that deciding whether a behavior is local for the class of Bell scenarios with binary outputs ($\Delta=2$) and $m$ inputs is NP-complete \cite{aiis04}. 
It is therefore highly unlikely that the problem of characterizing the local polytope admits a simple solution in full-generality.

In the following, we list some facet Bell inequalities of interest. Note that the positivity conditions (corresponding to $p(ab|xy)\geq 0)$ are always facets of the local polytope, but obviously are never violated by any physical theory. All other facet inequalities are violated by some no-signaling behaviors, and possibly by some quantum behaviors. It is in fact an open question whether there exist facet inequalities in the bipartite case (different than the positivity ones) that are not violated by any quantum behaviors (such inequalities are known in Bell scenarios with more parties \cite{ABBA+10}).
Note also that if an inequality defines a facet of the local polytope then it is obviously also the case for all the inequalities obtained from it by relabeling the outputs, inputs, or parties. What we mean thus in the following by ``an inequality" is the whole class of inequalities obtained by such operations.
Finally, it was shown in \cite{Pironio05} that there exists a hierarchical structure in the facial structure of local polytopes, in the sense that a facet Bell inequality of a given polytope with $\Delta$ outputs and $m$ inputs can always be  extended (or lifted) to any polytope with $\Delta'\geq \Delta$ and/or $m'\geq m$ (and also to polytopes corresponding to more parties) in such a way as to define a facet of the new polytope.

\subsubsection{Examples}\label{examples}
The simplest non-trivial Bell scenario corresponds to the case $\Delta=2$, $m=2$. The corresponding local polytope was completely characterized in \cite{Froissard81} and independently in \cite{Fine82}. In this case, there is only one (non-trivial) facet inequality: the CHSH inequality introduced in Eq.~(\ref{intro.chsh}). It is shown in \cite{Pironio04} that the CHSH inequality is also the only facet inequality for all polytopes with 2 inputs and 2 outputs for Alice and an arbitrary number of inputs and outputs for Bob.

The case $\Delta=2,m=3$ was computationally solved in \cite{Froissard81} who found that, together with the CHSH inequality, the following inequality
\begin{eqnarray}\label{froissard}
&p^A_1&+p^B_1-p_{11}-p_{12}-p_{13}\nonumber\\
&&-p_{21}-p_{31}-p_{22}+p_{23}+p_{32}\geq -1
\end{eqnarray}
is facet defining, where $p^A_x=p(a=1|x)$, $p^B_y=p(b=1|y)$, and $p_{xy}=p(a=1,b=1|xy)$. This result was later on rederived in \cite{Sliwa03,CG04}. The Froissard inequality is also refereed as the $I_{3322}$ inequality, following the terminology of \cite{CG04}. Note that this inequality could be generalized for the case of an arbitrary number of measurements $m$ with binary outcomes, a family known as the $I_{mm22}$ inequalities \cite{CG04}, proven to be facets in \cite{AI07}.

For $\Delta$ arbitrary and $m=2$, \cite{CGL+02} introduced the following inequality (we use the notation of \cite{AGG05}
\begin{equation}\label{cglmp} [a_1-b_1]+[b_1-a_2]  +[a_2-b_2]+ [b_2-a_1-1] \geq d-1 \end{equation}
where $[a_x-b_y]=\sum_{j=0}^{\Delta-1} j p(a-b=j\text{ mod } \Delta|xy)$ and similarly for the other terms. Note that for convenience the measurement outcomes are now denoted $a,b\in\{0,1,...,\Delta-1\}$. This inequality is known as the CGLMP inequality. For $\Delta=2$, it reduces to the CHSH inequality. It has been shown to be facet-defining for all $\Delta$ in \cite{Masanes03}.

The above inequality can be extended to an arbitrary number of inputs $m$ in the following way \cite{BKP07}
\ba \label{chained}
&& [a_1-b_1]+[b_1-a_2]  +[a_2-b_2]+\ldots \nonumber\\
&&\ldots+ [a_m-b_m]+ [b_m-a_1-1] \geq d-1
\ea
Though this Bell inequality is not a facet inequality, it is useful in several contexts. In the case $\Delta=2$, it reduces to the chained inequality introduced in \cite{Pearle70,BC90}.

Beyond these simple cases, a huge zoology of Bell inequalities has been derived and it would be impossible to discuss them all here in detail, in particular given the increase of complexity with larger values of $\Delta$ and $m$. For instance, in the case $\Delta=2$, there is only one (non-trivial) facet Bell inequality for $m=2$, two inequalities for $m=3$, but already for $m=4$ their number is not known \cite{BG08}. For $m=10$, there are at least $44,368,793$ inequalities \cite{aiis04} (and this figure is probably a gross underestimate)! To complete the simple examples given above, let us mention some recent papers where new Bell inequalities have been derived. In \cite{CG04,BG08}, several facet Bell inequalities have been obtained by solving numerically the convex hull problem for small values of $\Delta$ and $m$. In \cite{aiis04,AIIS05,AI07} large families of Bell inequalities are obtained by establishing a relation between the local polytope for $\Delta=2$ and a high-dimensional convex polytope called the cut polytope in polyhedral combinatorics. In \cite{VP08,Vertesi08,PV09} new algorithms have been proposed to construct families of facet and non-facet Bell inequalities in the $\Delta=2$ case. Methods exploiting symmetries to generate Bell inequalities for arbitrary $\Delta$ and $m$ (and an arbitrary number $n$ of parties) are investigated in \cite{BGP10b,BBBG+12}. Finally, while we focused here on the case where $\Delta$ and $m$ are finite, it is also possible to define Bell inequalities taking a continuous set of values for the outputs \cite{CFRD07,SCA+10} or the inputs \cite{KZ99,AMR+12}.

Finally, note that nonlinear Bell inequalities have also been considered. Quadratic inequalities were discussed in \cite{Uffink02}, while \cite{CFRD07,SCA+10} considered Bell inequalities based on moments of the probability distribution. Another approach, based on entropic quantities, was introduced in \cite{BC88} and further developed in \cite{CA97} and \cite{CF12}.

\subsubsection{Nonlocal games}\label{sec:games}

Bell inequalities are also referred to as \emph{nonlocal games} or sometimes simply as \emph{games}. Looking at Bell inequalities through the lens of games often provides an intuitive understanding of their meaning. Such games enjoy a long history
in computer science where they are known as \emph{interactive proof systems}; see~\cite{condon:oldSurvey} for an early survey. 
More recently, they have also been studied in the quantum setting, under the name of \emph{interactive proof systems with entanglement}~\cite{CHT+04}. In order to make such literature accessible, let us
see how the two concepts can be translated into each other\footnote{For the purpose of illustration, we will thereby restrict ourselves to the case of only two parties, Alice and Bob. However, the relation holds for an arbitrary amount of parties.}.

When talking about a game, we imagine that there is an outside party, the \emph{referee} that plays the game against Alice and Bob. 
In this context, parties or systems are referred to as \emph{players}.
Papers dealing with interactive proof systems
also refer to the referee as the \emph{verifier}, and to the players as \emph{provers}.
The referee chooses a question $x \in X$ for Alice and $y \in Y$ for Bob according to some probability distribution $\pi: X \times Y \rightarrow [0,1]$ from some set of possible
questions $X$ and $Y$. Upon receiving $x$ from the referee, Alice (Bob) returns an answer $a \in R_A$ ($b \in R_B$) from some set of possible answers $R_A$ ($R_B$).
The referee then decides whether these answers are winning answers for the questions he posed according to the rules of the game. These rules are typically expressed
by means of a predicate $V: R_A \times R_B \times X \times Y \rightarrow \{0,1\}$ where $V(a,b,x,y) = 1$ if and only if Alice and Bob win against the referee by giving answers
$a,b$ for questions $x$ and $y$. To emphasize the idea that the correct answers depend on the questions given, one often writes the predicate as $V(a,b|x,y)$.

Alice and Bob are fully aware of the rules, that is, they know the predicate $V$ and the distribution $\pi$. Before the game starts, they can agree on any strategy that may help them thwart the referee. However,
once the game started they can no longer communicate. In particular, this means that Alice does not know which question is given to Bob and vice versa. In the classical setting, such a strategy consists of shared randomness, which is the computer science name for local hidden variables. In the quantum case, Alice and Bob's strategy consists of a choice of shared quantum state and measurements.

The relation between games and Bell inequalities becomes apparent by noting that the questions are simply labels for measurement settings. 
That is, using our earlier notation we can take $X = Y = \{1,\ldots,m\}$. Note that we can without loss of generality assume that the number of settings $|X|$ and $|Y|$ are the same - otherwise, we can simply extend the number of settings for each party but never employ them.
Similarly, the answers correspond to measurement outcomes. That is, we can take $R_A = R_B = \{1,\ldots,\Delta \}$.

Any strategy leads to some particular probabilities $p(a,b|x,y)$ that Alice and Bob give answers $a,b$ for questions $x,y$ respectively.
In the language of Bell inequalities, this is simply the probability that Alice and Bob obtain measurement outcomes $a$ and $b$ when performing the measurements
labelled $x$ and $y$.
The probability that Alice and Bob win against the referee for some particular strategy can thus be written as
\begin{align}
	p_{\rm win}
	&= \sum_{x,y} \pi(x,y) \sum_{a,b} V(a,b|x,y) p(a,b|x,y)\ .
\end{align}
In the classical or quantum setting, one can consider the maximum winning probability that Alice and Bob can achieve. For instance, considering classical resources, we have that
\begin{align}
	\max p_{\rm win} = S_l \, , 
\end{align}
where the maximization is taken over all deterministic strategies of Alice and Bob. Note that this leads to the familiar form of a Bell inequality
\begin{align}
	p_{\rm win} = \mb{s}\cdot \mb{p} \leq S_l \, 
\end{align}
where the coefficients are given by
\begin{align}
	s_{x,y}^{a,b} = \pi(x,y) V(a,b|x,y)\ .
\end{align}
Hence games form a subset of general Bell inequalities. In complexity theory, the winning probability is also often referred to as the \emph{value} of the game.

\paragraph{XOR games.}

A class of games that is very well understood are so-called XOR-games~\cite{CHT+04}.
In an XOR game, each player only has two possible answers $a,b \in \{0,1\}$. To decide whether Alice and Bob win, 
the referee computes the XOR $c = a \oplus b := a + b \mod 2$ and then bases his decision solely on $c$.
For such games the predicate is generally written as $V(c|x,y) := \sum_{a} V(a,b=c\oplus a|x,y)$.
We will see later that it is easy to find the optimal quantum strategy for XOR games, 
and indeed the structure of their optimal measurements is entirely understood. Also, multi-player XOR games are reasonably well-understood and bounds relating the classical and quantum winning probabilities are known~\cite{thomas:xorGame}.

Let us simply note for the moment that XOR games are equivalent to \emph{correlation Bell inequalities} with binary outcomes. Indeed, from Eqs. (\ref{sec2.correlator}) it follows that we can write $p(a \oplus b = 0|x,y)= \frac{1}{2}\left(1+\langle A_xB_y\rangle\right)$ and $p(a \oplus b = 1|x,y)= \frac{1}{2}\left(1-\langle A_xB_y\rangle\right)$. 
The winning probability for an XOR game can thus be written as
\ba \label{eq:XORwinning2}
	p_{\rm win} &=& \frac{1}{2}\sum_{x,y}\pi(x,y) \times \\
	& & \sum_{c \in \{0,1\}}V(c|x,y)\left(1 + (-1)^c \langle A_x B_y\rangle\right)\,,\nonumber
\ea
which is the general form of a correlation Bell inequalities. XOR games can thus be recast as correlation inequalities and vice versa. 

\paragraph{An example: CHSH as a game.}

An illustrative example of how correlation Bell inequalities transform into games and vice versa is provided by the CHSH inequality. 
For convenience, we will here take $X = Y = \{0,1\}$ (instead of $\{1,2\}$), as well as $R_A = R_B = \{0,1\}$.
Viewing CHSH as a game, the rules state that Alice and Bob win
if and only if $x \cdot y = a \oplus b$. Plugging this into~\eqref{eq:XORwinning2} one obtains
\begin{align}
	p_{\rm win}^{\rm CHSH} = \frac{1}{2}\left(1 + \frac{ S}{4}\right)\ ,
\end{align}
where $S$ is the CHSH expression as given in \eqref{intro.chsh}. 
Indeed one has that $S\leq 2$ for any classical strategy. Hence, the probability for Alice and Bob to win the game using classical resources is at most $3/4$. Using quantum resources, the winning probability is at most $(1+ 1/\sqrt{2})/2 \approx 0.85$, as given by Tsirelson's bound $S\leq 2\sqrt{2}$.

\paragraph{Projection and unique games.}\label{sec:projection}

A \emph{projection game} is a game in which for 
every pair of questions $x$ and $y$ to Alice and Bob, and for every answer $a$ of Alice, there exists a \emph{unique} winning answer for Bob. 
In the quantum information literature, these are also often simply called \emph{unique games}. However, in the classical computer science
literature and also some of the quantum information literature the term \emph{unique game} may also refer to a game for which for
any pair of questions $(x,y)$ there exists a permutation $\pi_{x,y}$ over the set $\{1,\ldots, \Delta\}$ of possible answers such 
that Alice and Bob win if and only if their answers obey $a = \pi_{x,y}(b)$. In terms of the predicate this means that
$V(a,b|x,y) = 1$ if and only if $b = \pi_{x,y}(a)$. Note that in this language, every unique game is a projection game because
there is only one correct answer for Bob for each $x$,$y$, and $a$. However, not every projection game forms a unique game. 

A more general notion which imposes a limit on the number of winning answers are \emph{k-to-k'} games~\cite{KRT07}. More precisely, a game is \emph{k-to-k'}
if for all questions $x$ and $y$ the following two conditions hold: for all answers $a$ of Alice there exist at most $k$ winning answers for Bob, and 
for all answers $b$ of Bob there exist at most $k'$ winning answers for Alice. A projection game is thus a k-to-k' game for $k = k' = 1$.

\paragraph{Other special classes of games.}

Several other special classes of games have been studied on occasion. A {linear game} is a game for which one can associate the set of possible answers $\{1,\ldots,\Delta\}$
with an Abelian group $G$ of size $\Delta$ and find a function $W: \{1,\ldots,m\}^{\times 2} \rightarrow G$ such that $V(a,b|x,y) = 1$ if and only if $a-b = W(x,y)$. 
Any linear game is a unique game, and has been shown to have the special property to be a \emph{uniform game}, that is, a game in which there exists an optimal quantum 
strategy such that
the marginal distributions $p(a|x)$ and $p(b|y)$ are the uniform distributions over $R_A$ and $R_B$ respectively~\cite{KRT07}.
Furthermore, a game may be called \emph{free} if the questions
are drawn from a product distribution, that is, $\pi(x,y) = \pi_A(x) \times \pi_B(y)$ for some distributions $\pi_A$ and $\pi_B$~\cite{kempeVidick:parallel}. 
A game is called \emph{symmetric} if for all questions $x$,$y$ and all answers $a$,$b$, we have $V(a,b|x,y) = V(b,a|y,x)$~\cite{kempeVidick:parallel,dinur:parallel}.
An example of a game that is both free and symmetric is given by the CHSH game above.
Another class of games that has drawn attention in the computer science literature is characterized merely by the
fact that there exist a quantum strategy that wins the game with probability $p_{\rm win} = 1$. Such games are sometimes also called Kochen-Specker games (or pseudo-telepathy games or Greenberger-Horne-Zeilinger games) due to the fact that the optimal quantum strategy yields a so-called Kochen-Specker set~\cite{renatoStefan:KSset}, a concept in contextuality which is outside the scope of this survey (see~\cite{broadbent:survey} for a survey on such games, see also the related discussion in section~\ref{withoutBI}). 

\subsection{Bell inequality violations}

In the above discussion, we saw that it is in principle possible to decide (albeit very inefficiently) whether a given behavior is local and to compute the local bound $S_l$ of an arbitrary Bell expression. In this section, we look at the analogous problem in the quantum and no-signaling cases. We review the existing methods for computing the quantum and no-signaling bounds, \ie, the maximal quantum and no-signaling violations, of an arbitrary Bell expression $\mb{s}$. Such methods can also be used to determine if a given behavior admits a  quantum or no-signaling representation and thus this section is more generally concerned with the problem of practical characterizations of the quantum and no-signaling sets beyond the formal definitions (\ref{sec2.quantum}) and (\ref{sec2.nosig}).

\subsubsection{Quantum bounds}

\paragraph{Properties of quantum correlations.}\label{geometry quantum}\label{sec:bellOpCHSH}
Before discussing in more detail how one can compute the quantum bound $S_q$ of a Bell expression, let us briefly discuss the general structure of the quantum set $\mathcal{Q}$. 
Recall that a behavior $\mb{p}$ is quantum if, as defined in Eq.~(\ref{sec2.quantum2c}) 
it can be written as $p(ab|xy)=\langle \psi|M_{a|x} M_{b|y}|\psi\rangle$
where $|\psi\rangle$ is a state in a Hilbert space $\mathcal{H}$, and $M_{a|x}$ and $M_{b|y}$ are orthogonal projectors on $\mathcal{H}$ defining proper measurements and satisfying $[M_{a|x},M_{b|y}]=0$. (For characterizing the quantum set it is convenient to assume we impose commutation relations rather than a tensor product structure and we will follow this approach in the remainder of this section.)

As we already mentioned, the local set $\mathcal{L}$ is strictly contained in the quantum set $\mathcal{Q}$, i.e., there are quantum behaviors that are nonlocal, and thus in general $S_q>S_l$.  There are two basic requirements that any quantum behavior must satisfy to be nonlocal. First Alice's different measurements must be non-commuting as well as Bob's ones \cite{Fine82}. Second the state $\rho$ must be entangled. Without surprise, quantum nonlocality can thus be traced back to the two features usually seen as distinguishing quantum from classical physics: non-commutativity and entanglement.

Contrarily to the local set, the set $\mathcal{Q}$ of quantum correlations is generally not a polytope. It can therefore not be described by a finite number of extreme points or a finite number of linear inequalities. It is not difficult to see though that all extremal points of $\mathcal{L}$, i.e the local deterministic behaviors, are also extremal points of $\mathcal{Q}$. Furthermore certain faces of $\mathcal{L}$ are also faces of $\mathcal{Q}$. An example is provided by the $\Delta-1$-dimensional face associated to the hyperplanes $p(ab|xy)=0$ (note, however, that the corresponding Bell inequalities, $p(ab|xy)\geq 0$, cannot be violated by any physical behavior). Thus while $Q$ is not a polytope, its boundary contains some flat regions. In \cite{LPSW07}, it is shown that the local and quantum set have common faces which correspond to Bell inequalities that can be violated by certain no-signaling behaviors. As we mentioned earlier, it is an open question whether there exist such examples of maximal dimension, i.e. whether there exist facets of $\mathcal{L}$ corresponding to Bell inequalities that are not violated by $\mathcal{Q}$ but which can be violated by $\mathcal{NS}$ (such examples are known in the tripartite case \cite{ABBA+10}).

The boundary of the \emph{nonlocal} part of $\mathcal{Q}$ may also contain flat regions, i.e., the maximal violation of a Bell inequality may sometimes be realized with two or more different non-local quantum behaviors. The question of when an extremal quantum behavior can be realized by a unique quantum representation (up to unitary equivalence) has been considered in \cite{FFW11}, where it has in particular been shown that in the correlation space $\mathcal{C}$ all non-local extremal behaviors are uniquely realizable in the cases $m=2$ and $m=3$. Examples of non-correlation Bell inequalities that are maximally violated by a unique quantum behaviors have been given in \cite{AMP12}. These inequalities are maximally violated by partially entangled states, thus showing that these state are necessary to characterize the boundary of the quantum region, see also \cite{LVB11,VW11}. Note, however, that the so-called embezzling state~\cite{patrick:embezzle} is universal in the sense that any two-party Bell inequality can be maximally violated using an embezzling state~\cite{embezzle:games} up to a small error term.

Let us now focus more specifically on the problem of computing the quantum bound of a Bell expression. Recall that $\mathcal{Q}$ as any convex compact set can be described by an infinite system of linear inequalities of the form $\mb{s}\cdot \mb{p} \leq S_q$, here the quantum Bell inequalities. Given an arbitrary Bell expression $\mb{s}$, its corresponding quantum bound is given by
\be \label{sec2.qbound}
S_q=\max_{\mb{p}\in \mathcal{Q}} \mb{s}\cdot \mb{p} = \max_{\mathcal{S}} || \mathcal{S}||
\ee
where 
\begin{align}\label{eq:bellPoly}
	\mathcal{S}=\sum_{abxy} s^{ab}_{xy}\, M_{a|x}M_{b|y}
\end{align}
is the \emph{Bell operator} associated to $\mb{s}$, $|| \mathcal{S}||$ denotes the spectral norm (largest eigenvalue) of $\mathcal{S}$, and the above optimisation is performed over all possible Bell operators $\mathcal{S}$ associated to $\mb{s}$. That is, over all possible measurements $\{M_{a|x}\}_a$ and $\{M_{b|x}\}_b$ where the coefficients $s^{ab}_{xy}$ are given by the choice of $\mb{s}$.
In the case of the CHSH expression, the Bell operator takes the form $\mathcal{S}=\hat A_1(\hat B_1+\hat B_2)+\hat A_2(\hat B_1-\hat B_2)$, where $\hat A_x,\hat B_y$ are arbitrary $\pm 1$-eigenvalued observables. Following \cite{Landau87}, we can derive the quantum  bound of the CHSH inequality \cite{Cirelson80} by noting that $\mathcal{S}^2=4+[\hat A_1,\hat A_2][\hat B_1,\hat B_2]$, from which it follows that $||\mathcal{S}^2||\leq 8$ and hence $||\mathcal{S}|| \leq 2\sqrt{2}$. Computing the quantum bound of other Bell expessions is a more complicated business. It has been shown to be an NP-hard problem in the tripartite case~\cite{KKV07}.

\paragraph{Correlation inequalities.}

The case of quantum correlations inequalities defined in the correlation space $\mathcal{C}$ is particularly well understood thanks to Tsirelson \cite{Cirelson80,Tsirelson87,Tsirelson93}. Recall that in the correlation space, a behavior is defined by the $m^2$ correlators $\langle A_x B_y\rangle$. It is easy to see that such a behavior is quantum if we can write $\langle A_x B_y\rangle=\langle\psi|\hat A_x\otimes \hat B_y|\psi\rangle$ for some quantum state $|\psi\rangle$ in $\mathcal{H}_A\otimes \mathcal{H}_B$ and some $\pm 1$-eigenvalued quantum observables $A_x$ on $\mathcal{H}_A$ and $B_y$ on $\mathcal{H}_B$. Tsirelson showed that it is sufficient to consider $\dim \mathcal{H}_A=\dim \mathcal{H}_B=2^{m}$ if $m$ is even and $\dim \mathcal{H}_A=\dim \mathcal{H}_B=2^{m+1}$ if $m$ is odd, and $|\psi\rangle$ to be a maximally entangled state in $\mathcal{H}_A\otimes \mathcal{H}_B$. Furthermore, he showed that
the $m^2$ correlators $\langle A_x B_y\rangle$ are quantum if and only if there exit $2m$ unit vectors $\hat{v}_x$ and $\hat{w}_y$ in $\mathbb{R}^{2m}$ such that
\be\label{2sec.corrvec}
\langle A_x B_y\rangle= \hat{v}_x\cdot \hat{w}_y
\ee
for all $x,y\in \{1,\ldots,m\}$. This last representation is particularly useful, as deciding if a behavior can be written in the form (\ref{2sec.corrvec}) can be cast as a \emph{semidefinite program} (SDP) for which efficient algorithms are available \cite{CHT+04,Wehner06}. This means that the problem of computing the winning probability of the game, $p_{\rm win}$, is in the complexity class EXP (exponential time, since SDPs can be solved in polynomial time but the input is of exponential size).
However, combining~\cite{qipPspace} and \cite{wehner:simulate}  one now knows that the problem of computing $p_{\rm win}$ for XOR games lies in the complexity
class PSPACE. 

This technique can be used to compute tight bounds for 2-outcome correlation inequalities, i.e., XOR-games. In particular, the quantum bounds for the CHSH inequality and the chained inequalities (inequality (\ref{chained}) in the case $\Delta=2$) can easily be obtained in this way \cite{Wehner06}. It should be noted that this SDP technique can be seen as a special case of the general SDP method discussed in Section~\ref{hierarchy}.

In the $\Delta=2$, $m=2$, this SDP approach can be used to yield a complete description of $\mathcal{Q}\cap\mathcal{C}$ (i.e. the quantum part of the correlation space $\mathcal{C})$ in terms of a finite set of non-linear inequalities: a behavior is quantum if and only if satisfies
\be \label{asin}
|\text{asin}\langle A_1B_1\rangle +\text{asin} \langle A_1B_2\rangle+\text{asin} \langle A_2B_1\rangle-\text{asin} \langle A_2B_2\rangle|\leq \pi
\ee
together with the inequalities obtained by permuting the $\langle A_xB_y\rangle$ in the above expression \cite{Cirelson80, Tsirelson87,Landau88,Masanes03}. For further results and a more detailed discussion of the characterization of $\mathcal{Q}$ in the correlation space $\mathcal{C}$, we refer to \cite{Tsirelson87,Tsirelson93,AMO09}.

It is interesting to note that it is much harder to determine the optimal local bound $S_l$ for a correlation Bell inequality than it is to compute the quantum one unless P=NP~\cite{CHT+04}. That is, the quantum problem is actually easier than the classical one. 

\paragraph{State and measurement dependent bounds.}
Let us now go back to the general case of quantum correlations in the probability space $\mathcal{P}$. To compute the quantum bound (\ref{sec2.qbound}) of a Bell expression, a first simple
approach is to introduce an explicit parametrization of a family of Bell operators $\mathcal{S}$ in an Hilbert space $H=H_A\otimes H_B$ of fixed dimension $\dim H=d_H$, and to maximize $||\mathcal{S}||$ over all operators in this family. In general, however, we have no a priori guarantee that the optimal quantum bound can be realized using a Bell operator from this particular family. Furthermore, most optimization methods cannot guarantee convergence to the global extremum. This approaches therefore typically only yields lower-bounds on $S_q$.  It is nevertheless very useful when looking for an explicit quantum violation of a Bell inequality $\mb{s}\cdot \mb{p}\leq S_l$ (though we will have no guarantee that this is the optimal quantum violation).

Rather than directly trying to obtain a state-independent bound by maximizing the norm of the Bell operator, it is often easier to compute the quantum bound for a fixed quantum state $|\psi\rangle$, i.e. maximize $\langle \psi|\mathcal{S}|\psi\rangle$ over all Bell operators $\mathcal{S}$. This optimization can be dealt with as above by introducing an explicit parametrization of a family of Bell operators. An other possibility, introduced in \cite{LD07}, is to exploit the fact that, for a given quantum state, a Bell expression is bilinear in the measurement operators, that is, it is linear in the operators $\{M_{a|x}\}$ for fixed $\{M_{b|y}\}$ and linear in the $\{M_{b|y}\}$ for fixed $\{M_{a|x}\}$. When the measurements on one system are fixed, the problem of finding the optimal measurements for the other system can therefore be cast as a SDP. This SDP can then be used as the basis for an iterative algorithm: fix Bob's measurements and find Alice's optimal ones; with these optimized measurements for Alice now fixed, find the optimal ones for Bob; then optimize again over Alice's measurements and so on, until the quantum value converges within the desired numerical precision. A similar  iterative algorithm was introduced in \cite{WW01b} for correlation inequalities. In this case, once the measurements  for one party are fixed, optimization of the other party's measurements can be carried out explicitly. This turns out
to be true not only for correlation inequalities but for any Bell expression with binary outcomes \cite{LD07}. Finally, let us note that a method for finding an optimal Bell operator for a fixed quantum state, can again be used in an iterative algorithm to find a state-independent bound \cite{PV09}: starting with an initial quantum state (e.g. a maximally entangled state), find the corresponding optimal Bell operator; find then the optimal quantum state associated to this Bell operator (i.e. the eigenvector associated to the largest eigenvalue); repeat these steps starting from this new state.

\paragraph{General bounds.}\label{hierarchy}
The techniques that we just described provide lower-bounds on $S_q$. Looking at~\eqref{sec2.qbound} it becomes clear that finding $S_q$ can be understood as an instance of polynomial optimization. More specifically, we wish to optimize~\eqref{eq:bellPoly} over non-commutative variables $M_{a|x}$, $M_{b|y}$ subject to certain constraints, namely that such variables form quantum measurements and Alice measurement operators commute with those of Bob. It is known that in principle any polynomial optimization problem in \emph{commutative} variables can be solved using a hierarchy of SDPs---two general methods that are dual to each other have been introduced by~\cite{lasserre} and~\cite{parillo:sdp} respectively. 

\begin{figure}
\begin{center}
\includegraphics[width = 0.5\textwidth]{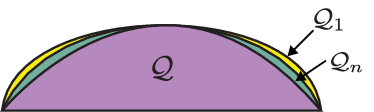}
    \caption{Hierarchy of sets $\mathcal{Q}_\nu$ generated by the hierarchy of SDPs defined in \cite{NPA07} (see text). Each set in the hierarchy approximates better the set of quantum correlations $\mathcal{Q}$. In the CHSH scenario, the set $\mathcal{Q}_1$ already achieves the maximum quantum value of the CHSH inequality, \ie, Tsirelson's bound.}
\label{NPA}
\end{center}
\end{figure}

It turns out that these techniques can be extended to the quantum setting~\cite{NPA07,NPA08,DLTW08}, yielding a powerful approach to obtaining upper-bounds on $S_q$, i.e., of deriving constraints satisfied by the entire quantum set. This method was originally introduced in \cite{NPA07}, which follows the ideas of~\cite{lasserre}. The idea is basically the following. Let $|\psi\rangle$ and $\{M_{a|x}\}$, $\{M_{b|y}\}$ define a quantum realization of a behavior $\mb{p}\in\mathcal{Q}'$, i.e. $p(ab|xy)=\langle\psi|M_{a|x}M_{b|y}|\psi\rangle$. Let $\mathcal{O}$ be a set of $k$ operators consisting of all the operators $M_{a|x}$ and $M_{b|y}$ together with some finite products of them. For instance $\mathcal{O}$ may consist of all operators of the form $M_{a|x}$, $M_{b|y}$, $M_{a|x}M_{a'|x'}$, $M_{a|x}M_{b|y}$, $M_{b|y}M_{b'|y'}$. Denote by $O_i$ ($i=1,\ldots,k$) the elements of $\mathcal{O}$ and introduce the  $k\times k$ matrix $\Gamma$ with entries $\Gamma_{ij}=\langle\psi| O^\dagger_iO_j|\psi\rangle$, called the \emph{moment matrix} associated to $\mathcal{O}$. Then the following properties are easily established (independently of the particular quantum realization considered): $i)$ $\Gamma\succeq 0$ is semidefinite positive, $ii)$ the entries of $\Gamma$ satisfy a series of linear inequalities, $iii)$ the probabilities $p(ab|xy)$ defining the behavior $\mb{p}$ correspond to a subset of the entries of $\Gamma$.
A necessary condition for a behavior $\mb{p}$ to be quantum is therefore that there exist a moment matrix $\Gamma$ with the above properties, a problem that can be determined using SDP. For any $\mathcal{O}$,  the set of behaviors $\mb{p}\in\mathcal{P}$ for which there exist such a moment matrix thus define a set $\mathcal{Q}_\mathcal{O}$ that contains the quantum set $\mathcal{Q}'$ (and thus also $\mathcal{Q}$). Optimizing a Bell expression (which is linear in $\mb{p}$) over this set $\mathcal{Q}_\mathcal{O}$ is also a SDP and yields an upper-bound on $S_q$.
Consider in particular the case where $\mathcal{O}$ is the set of all operators consisting of product of at most $\nu$ of the operators $M_{a|x}$ and $M_{b|y}$ and denote the corresponding set of behaviors $Q_{\nu}$. Then the associated SDP defines for $\nu=1,2,\ldots$ a hierarchy  $\mathcal{Q}_1\supseteq\mathcal{Q}_2\supseteq\ldots \supseteq\mathcal{Q}$ of relaxations approximating better and better the quantum region from the outside (See Fig. \ref{NPA}). Or, equivalently, they define a decreasing series of upper-bounds on the quantum bound $S_q$ of any Bell expression. 

Subsequently, following the ideas of \cite{parillo:sdp}, ~\cite{DLTW08} constructed the SDP hierarchy that is dual to~\cite{NPA07}. It relies on the fact that 
for any Bell operator $\mathcal{S}$ we have $\xi = \hat{b}_q \id - \mathcal{S} \geq 0$ (i.e., $\|\mathcal{S}\| \leq \hat{b}_q$) if and only if the polynomial $\xi$ can be written as a (weighted) sum of squares of other polynomials. 
We can thus think of minimizing $\hat{b}_q$ such that $\xi$ is a sum of squares of polynomials in order to find $\|B\|$. 
If we limit the degree of these polynomials, the problem can be cast as an SDP. 
Very roughly, at level $\ell$ of the SDP hierarchy we then limit the degree to be at most $2\ell$, leading to better and better bounds for
increasing values of $\ell$. 

In \cite{NPA08,DLTW08}, it is shown that this hierarchy of SDP relaxations converges in the asymptotic limit to the set $\mathcal{Q}'$ (see also \cite{PNA10} for a more general approach not limited to quantum correlations). It is also possible to certify that a behavior $\mb{p}$ belongs to the quantum set $\mathcal{Q}$ or to obtain the optimal bound $S_q$ of a Bell expression at a finite step in the hierarchy (see e.g.~\cite{DLTW08,NPA08} for a number of examples). A criterion has been introduced in \cite{NPA08} to determine when this happens and to reconstruct from the moment matrix $\Gamma$ a quantum realization of this optimal solution in term of an explicit state $|\psi\rangle$ and operators $M_{a|x}$ and $M_{b|y}$. Optimality at a finite step in the hierarchy can also be determined by comparing the SDP upper-bound with a lower-bound obtained by searching over explicit families of quantum Bell operators. In \cite{PV09}, for instance, the optimal quantum value $S_q$ of 221 Bell expressions has been determined in this way at the 3rd step of the hierarchy. Even if they do not always provide an optimal bound, numerical examples show that low-order steps of the hierarchy usually already approximate very well the quantum bound. In \cite{KRT07}, it is proven that for a certain particular family of Bell scenarios, known as \emph{unique games} (see section \ref{sec:projection}), the first step of the hierarchy always provides a good approximation of the quantum set. Let us also note that for correlation inequalities, the first step of the hierarchy always provides the optimal solution as it equivalent to the SDP approaches based on Tsirelson results mentioned earlier.

In the $\Delta=2$, $m=2$ case, the set $\mathcal{Q}_1$ corresponding to the first step of the hierarchy has been analytically characterized in \cite{NPA07}. A behavior $\mb{p}$ belongs to $\mathcal{Q}_1$ if and only if $\langle A_i \rangle^2=1$ or $\langle B_j \rangle^2=1$ for some $i,j=1,2$ or if it satisfies the inequality
\be \label{NLineq}
|\text{asin}\langle D_{11}\rangle +\text{asin} \langle D_{12}\rangle+\text{asin} \langle D_{21}\rangle-\text{asin} \langle D_{22}\rangle|\leq \pi
\ee
together with the inequalities obtained from this one by permuting the $D_{ij}$, where 
\ba D_{ij}=\frac{ \langle A_{i}B_j\rangle-\langle A_i\rangle\langle B_j\rangle}{\sqrt{(1-\langle A_i\rangle^2)(1-\langle B_j\rangle^2}}. \ea 
The non-linear inequalities \eqref{NLineq} thus form a necessary condition for a behavior to be quantum. They strengthen the inequalities (\ref{asin}) to which they reduce when $\langle A_i\rangle=\langle B_j\rangle=0$.

\subsubsection{No-signaling bounds}\label{NScorrs}
Let us now consider the problem of computing bounds on Bell expressions for no-signaling correlations. Contrary to the case of local and quantum correlations, this turns out to be a rather easy task. 
To understand why note that, as already mentioned, once the no-signaling constraints~(\ref{sec2.nosig}) are taken into account, e.g. by introducing a parametrization of the relevant affine subspace $\mathbb{R}^t$, the set $\mathcal{NS}$ of no-signaling behaviors is uniquely determined by the set of $\Delta^2m^2$ positivity inequalities $p(ab|xy)\geq 0$. Deciding whether a behavior belongs to $\mathcal{NS}$ can thus easily be verified by checking that all positivity inequalities are satisfied. Since there are $\Delta^2m^2$ such inequalities, this is a problem whose complexity scales polynomially with the number of inputs and outputs. More generally, linear programming can be used to determine efficiently the no-signaling bound $S_{ns}$ of an arbitrary Bell expression $\mb{s}$, as used, e.g. in \cite{Toner09}. Especially in the case of multipartite correlations it is sometimes convenient to compute $S_{ns}$ to obtain a (crude) bound for $S_q$.

Finally, let us remark that since $\mathcal{NS}$ is defined by a finite number of linear inequalities, it is, as the local set, a polytope and can also be described as the convex hull of a finite set of vertices. These can be obtained from the list of facets (the inequalities $p(ab|xy)\geq 0$) using the same polytope algorithms that allow one to list the facets of $\mathcal{L}$ given its vertices. 
The vertices of $\mathcal{L}$, the local deterministic points $\mb{d}_\lambda$, are clearly also vertices of $\mathcal{NS}$ (since they cannot be written as a convex combination of any other behavior). All other vertices of $\mathcal{NS}$ are non-local.

\begin{figure}[t]
\begin{center}
\includegraphics[width = 0.45\textwidth]{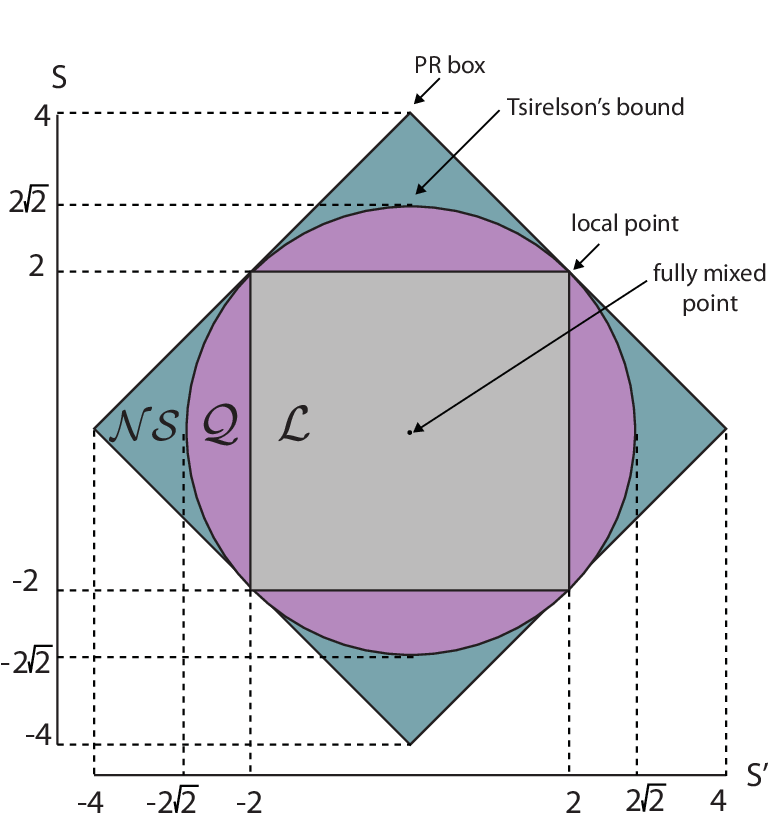}
    \caption{A two-dimensional section of the no-signaling polytope in the CHSH scenario ($m=\Delta=2$). The vertical axis represents the CHSH value $S$, while the horizontal axis represents the value of a symmetry of the CHSH expression $S'$ (where inputs have been relabeled). Local correlations satisfy $|S|\leq 2$ and $|S'|\leq 2$. The PR box is the no-signaling behavior achieving the maximum CHSH value S=4. Tsirelson's bound corresponds to the point where $S=2\sqrt{2}$, \ie, the maximum CHSH value that a quantum behavior can achieve.}
\label{sec2.fig2}
\end{center}
\end{figure}

The geometry of the no-signaling set and its relation to $\mathcal{L}$ is particularly simple for the $\Delta=2,m=2$ Bell scenario. In this case, the no-signaling behaviors form an $8$-dimensional subspace of the full probability space $\mathcal{P}$. The local polytope consists of $16$ vertices, the local deterministic points, and $24$ facets. $16$ of these facets are positivity inequalities and $8$ are different versions, up to relabeling of the inputs and outputs, of the CHSH inequality. The no-signaling polytope, on the other hand, consists of $16$ facets, the positivity inequalities, and $24$ vertices. $16$ of these vertices are the local deterministic ones and $8$ are non-local vertices, all equivalent up to relabeling of inputs and outputs to the behavior
\be\label{sec2.prbox}
p(ab|xy)=\left\{\begin{array}{ll}
1/2&\text{if }a\oplus b=xy\\
0&\text{otherwise}
\end{array}\right.
\ee
which is usually referred to as a \emph{PR-box}.
It is easily verified that the PR-box violates the CHSH inequality (\ref{intro.chsh}) up to the value $\mb{s}\cdot\mb{p}=4>2$, the algebraic maximum. In the language
of games, this means that the CHSH game can be won with probability $p_{\rm win}^{CHSH} = 1$.
There exists a one-to-one correspondence between each version of the PR-box and of the CHSH inequality, in the sense that each PR-box violates only one of the CHSH inequalities. The PR-box was introduced in \cite{KT85,Rastall85,PR94}. Since the maximal quantum violation of the CHSH inequality is $2\sqrt{2}$, it provides an example of a no-signaling behavior that is not quantum, implying that in general $Q\neq NS$.
 The relation between $\mathcal{L}$, $\mathcal{Q}$, and $\mathcal{NS}$ in the $\Delta=2,m=2$ case is represented in Figure~\ref{sec2.fig2}.
 
The complete list of all no-signaling vertices is also known in the case of two inputs ($m=2$) and an arbitrary number of outputs \cite{BLMP+05} and in the case of two outputs ($\Delta=2$) and an arbitrary number of inputs \cite{JM05,BP05}. In both cases, the corresponding non-local vertices can be seen as as straightforward generalizations of the PR-box.

\subsection{Multipartite correlations} \label{multi}

Though we focused for simplicity in the preceding sections on Bell scenarios involving $n=2$ systems, most of the above definitions and basic results extend straightforwardly to the case of an arbitrary number $n>2$ of systems. 
For instance, in the tripartite case a behavior $p(abc|xyz)$ is no-signaling when
\ba \sum_c p (abc|xyz) = \sum_{c'}p(abc'|xyz') \,\, \forall \, a,b,x,y,z,z' \ea
and similar relations obtained from permutations of the parties; a behavior is local if it can be written as a convex combination of a finite number of deterministic behaviors $d_\lambda(abc|xyz)$; Bell inequalities correspond to faces of the corresponding polytope; and so on. Below, we discuss a few notable results obtained in the multipartite case. Note that many references cited in the previous subsections also contain results for more than 2 parties.

As in the bipartite case, one can consider Bell inequalities that involve only full correlators in the case where all measurements have binary outcomes. In the $n=3$ case, for instance, such an inequality would involve only terms of the form $\langle A_xB_yC_z\rangle=\sum_{a,b,c=\pm1} abc\, p(abc|xyz)$, and similarly for more parties. All correlation Bell inequalities with $m=2$ inputs have been derived in \cite{WW01} and \cite{ZB02} for an arbitrary number $n$ of parties. There are $2^{2^n}$ such inequalities (with redundancies under relabelling) which can be summarized in a single, but non-linear inequality. Notable inequalities that are part of this family are the Mermin inequalities introduced in \cite{Mermin90} and further developed in \cite{BK93,Ardehali92}. In the case $n=3$, the Mermin inequality takes the form
\be\label{sec2.mermin}
| \langle A_1B_2C_2\rangle + \langle A_2B_1C_2\rangle +\langle A_2B_2C_1\rangle -\langle A_1B_1C_1\rangle |\leq 2\,.
\ee
It is associated to the GHZ paradox (see section \ref{withoutBI}) in the sense that correlations that exhibit the GHZ paradox violate it up to the algebraic bound of 4.
In \cite{WW01} the structure of the quantum region in the full correlation space is also investigated. In particular, it is shown that the quantum bound of all inequalities introduced in \cite{WW01,ZB02} is achieved by the $n$-partite GHZ state $(|00\ldots 0\rangle+|11\ldots 1\rangle)/\sqrt{2}$.

In \cite{Sliwa03}, all facet Bell inequalities (in the full probability space) have been derived for $3$ parties in the case $\Delta=2,m=2$. There are 46 inequivalent such inequalities. All of these are violated in quantum mechanics, except for the inequality considered in \cite{ABBA+10}. In \cite{PBS11}, all vertices of the no-signaling polytope corresponding to the same Bell scenario have been listed. Interestingly, they are also 46 inequivalent classes of no-signaling vertices. In  \cite{Fritz12}, it is shown that there actually exists a bijection between facet Bell inequalities and no-signaling vertices for every Bell scenario with two inputs and outputs, independently of the number of parties.

Evidently, the structure of non-local correlations is much richer (and less understood) in the multipartite case than in the bipartite one. In particular, there exist different definitions of non-locality that refine the straightforward extension of the bipartite definition. This question and other ones that are more specific to the multipartite scenario are discussed in section~\ref{sec:multipartite}.

\subsection{Nonlocality without ``inequalities"}\label{withoutBI}
To demonstrate that some quantum correlations $\mb{p}$ are non-local it is sufficient, as discussed in the previous subsections, to exhibit a Bell inequality that is violated by $\mb{p}$. In certain cases, however, it is  possible to show directly that quantum predictions are incompatible with those of any local model via a simple logical contradiction that does not involve any inequality (though such arguments can obviously always be rephrased as the violation of a Bell inequality). Here, we present two examples of such ``Bell's theorem without inequalities", namely the Greenberger-Horne-Zeilinger paradox and a construction due to Hardy.

The situation considered by \cite{GHZ89} (see also \cite{GHSZ90,Mermin90pt}) involves three players Alice, Bob and Charlie. Each player receives a binary input, denoted by $A_i$, $B_i$, and $C_i$, with $i=1,2$.
For each input, players should provide a binary output: $\pm1$. With a slight abuse of notation, let us denote by $A_i=\pm 1$ the answer to the question ``$A_i$'' and so on. Suppose that the players share a state of the form $\ket{GHZ}=\frac{1}{\sqrt{2}}(\ket{000}+\ket{111})$, and upon receiving input `1' (`2') they perform a local Pauli measurement $\sigma_x$ ($\sigma_y$).

It is not difficult to see that their measurement outcomes will always satisfy the following relations
\ba\label{GHZ} \nonumber
A_1 B_1 C_1 &=& +1\ , \\ 
A_1 B_2 C_2 &=& -1\ , \\\nonumber
A_2 B_1 C_2 &=& -1\ , \\\nonumber
A_2 B_2 C_1 &=& -1 \,.
\ea
Let us contrast these quantum predictions with those of a local model, where the answer of each party only depends on the question he receives and on some shared random data $\lambda$. Since the correlations in (\ref{GHZ}) are perfect (i.e. exactly $+1$ or $-1$), each answer must clearly be a \emph{deterministic} function of the local question and of $\lambda$. For fixed $\lambda$, a local model thus amounts to assigning a definite value $\pm 1$ to all of the variables $A_i$, $B_i$, and $C_i$. But then this is in direct contradiction with the conditions (\ref{GHZ}). To see this, consider the product of all four left-hand-side terms. Since $A_i^2=B_i^2=C_i^2=1$, this product is necessary equal to $1$, but the product of the right-hand side is $-1$.  This argument demonstrates in a simple way the incompatibility between the predictions of quantum theory and those of any local model.

Note that the above GHZ paradox can be recasted as the violation of Mermin's inequality, given in Eq. \eqref{sec2.mermin}, \ie, the GHZ correlations (\ref{GHZ}) violate the inequality \eqref{sec2.mermin} up to its algebraic maximum $4$. In the language of non-local games, it provides an example of a game for which there exists a quantum strategy that wins it with probability $p_\text{win}=1$ (see section \ref{sec:games}). GHZ paradoxes of the above types are also known as ``pseudo-telepathy games" \cite{broadbent:survey} or ``Kochen-Specker games" \cite{Mermin93,renatoStefan:KSset}. 
Other multipartite GHZ-type paradoxes, as well as a more detailed discussion of the nonlocal correlations of GHZ states can be found in section~\ref{multiGHZ}. Notable examples of ``nonlocality proofs without inequalities" of the GHZ-type but in the bipartite case are presented in \cite{Aravind02,Cabello01,AGA+12}.

Another interesting demonstration of quantum nonlocality without inequalities was given in \cite{Hardy93}. Consider a bipartite Bell test, in which each observer chooses between two dichotomic measurements. Hardy considered a situation in which the joint probability distribution satisfies the following relations:

\ba p(+1,+1|A_1,B_1) &=& 0, \nonumber \\  \label{hardy}  p(+1,-1|A_2,B_1) &=& 0, \\ \nonumber p(-1,+1|A_1,B_2) &=& 0. \ea 
For any distribution that is local, it then follows that 
\ba p_{\text{Hardy}} \equiv p(+1,+1|A_2,B_2) =0. \ea
Hardy realized that this logical implication does not hold in quantum mechanics. 

Consider an entangled state of two qubits of the form
\ba \ket{\psi}=\alpha (\ket{01}+\ket{10})+\beta \ket{00}, \ea
where $2|\alpha|^2+|\beta|^2=1$. Both parties perform the same measurements. The first measurement is in the computational basis, with result +1 for state $\ket{0}$, and -1 for state $\ket{1}$. For the second measurement, the result +1 corresponds to a projection on the qubit state $\ket{\phi}=\cos{\theta}\ket{0}+\sin{\theta}\ket{1}$, while the result -1 is associated to the orthogonal projector. Setting $\alpha = \beta \tan{\theta}$, one obtains Hardy's paradox: all three relations \eqref{hardy} are satisfied, but we obtain $p_{\text{Hardy}}= 2 \beta \sin^2{\theta}>0$ if $ 0 < |\beta | <1$. An interesting aspect of this construction is that the paradox occurs for any entangled state of two qubits, with the notable exception of the maximally entangled state ($\beta=0$). This represents one of the first hints that entanglement and nonlocality are not monotonically related (see section~\ref{MoreNL}).

The strongest demonstration of Hardy's paradox gives $p_{\text{Hardy}} = (5 \sqrt{5}-11)/2 \approx 9 \%$ \cite{Hardy93}, which turns out to be the maximal possible value in quantum theory \cite{RZS12}. That is, Hardy's paradox cannot be strengthened by using higher dimensional quantum entangled states. For interesting extensions of Hardy's paradox see \cite{Fritz11} and references therein.

\subsection{Quantifying nonlocality}\label{measures}
So far, we have mainly discussed the problem of detecting nonlocal correlations, i.e. determining whether given correlations $\mb{p}$ belong to $\mathcal{L}$ or not. Another relevant question is how to \emph{quantify} nonlocality. 

A common choice for quantifying nonlocality is through the amount of violation of a Bell inequality, i.e., $\mb{p}$ is more non-local than $\mb{q}$ if $\mb{s}\cdot \mb{p}>\mb{s}\cdot \mb{q}$ for some Bell expression $\mb{s}$. The problem with this approach is that there can be another Bell expression $\mb{s}'$ such that $\mb{s}'\cdot \mb{q}>\mb{s}'\cdot \mb{p}$. Another problem is that a given Bell inequality can be written in many equivalent ways using the normalization conditions $\mb{1}\cdot \mb{p}=\sum_{abxy}p(ab|xy)=m^2$ (we recall that $m$ denotes the number of possible inputs $x$ and $y$). Let for instance $\mb{s}$ be the CHSH expression (\ref{intro.chsh}), for which $S_l=2$ and $S_q=2\sqrt{2}$. Consider the Bell expression $\mb{s}_\alpha=\alpha\, \mb{s} +(1-\alpha)/2\, \mb{1}$ obtained from the CHSH expression through irrelevant rescaling and addition of an offset. For any $\mb{p}$, we thus have $\mb{s}_\alpha\cdot p=\alpha\, \mb{s}\cdot \mb{p}-2\alpha+2$, which implies that the local bound $S_l^\alpha=2$ of the new Bell expression is identical to the one of the original CHSH expression, but now its maximal quantum violation $S_q^\alpha=2+2(\sqrt(2)-1)\alpha$ can (artificially) be made arbitrarily large by increasing $\alpha$. 

If the amount of violation of Bell inequalities is used to quantify non-locality, this amount of violation must thus first be normalized in some proper way. If this normalization is well-chosen, one can then often relate the amount of violation of Bell inequalities to an operational measure of non-locality. 

Possible operational measures of non-locality are simply given by the tolerance of non-local correlations to the addition of noise, such as white noise \cite{KGZMZ01,ADGL02}, local noise \cite{PWPV+08,JPPV+10}, or detection inefficiencies \cite{Massar02,MPRG02}. In particular, it is shown in \cite{PWPV+08} (see also section \ref{grothendieck}) that the tolerance of $\mb{p}$ to any local noise, defined as the minimal value of $r$ such that $r\mb{p}+(1-r) \mb{q}\notin \mathcal{L}$ for all $\mb{q}\in\mathcal{L}$, is given by $r=2/(\nu+1)$ where $\nu$ is the maximal possible violation by $\mb{p}$ of a Bell inequality, defined in the following way
\begin{equation}
\nu=\max_{\mb{s}}\frac{|\mb{s}\cdot \mb{p}|}{\max_{\mb{q}\in\mathcal{L}}|\mb{s}\cdot\mb{q}|}\,.
\end{equation}
Note that taking the ratio and the absolute value is crucial for a
meaningful definition of this amount of violation. If instead of the ratio, one takes for instance the difference, a change of scale $\mb{s}\rightarrow \gamma \mb{s}$ would lead to arbitrary violations. If one removes instead the absolute value, the same happens via an offset, as in the example discussed above. 

Another operational measure of the non-locality of correlations $\mb{p}$ is given by the amount of classical communication between the two wings of the Bell experiment by which a local model has to be supplemented for reproducing these correlations. This approach was adopted in \cite{Maudlin92,BCT99,Steiner00,BT03,TB03} (see also discussion in section \ref{simulation}). In \cite{Pironio03}, it is shown that any Bell expression $\mb{s}$ can be rewritten in a normalized form $\mb{s}^*$ -- through an appropriate change of scale and an offset -- such that the minimal average amount of classical communication $C(\mb{p})$ necessary to reproduce $\mb{p}$ is given by $C(\mb{p}) = \max_{\mb{s}^*} \mb{s}^*\cdot \mb{p}$. Techniques for estimating the communication complexity of arbitrary no-signaling correlations and their relation to Bell violations were further developed in \cite{DKLR11}.

Finally, a third proposed approach to measure nonlocality is through its ``statistical strength'' \cite{DGG05}: that is, the amount of confidence that the measurement outcomes of $n$ independent Bell experiments governed by a behavior $\mb{p}$ could not have been reproduced by a local behavior. Indeed, statistical fluctuations on a finite  sample allow for the possibility of apparent Bell inequality violations even by a local model (this issue for the interpretation of experimental results of a Bell test is specifically discussed in section~\ref{finitestat}). In an experiment, the goal is to test in a finite number of trials whether the system obeys a Bell local model (hypothesis LOC) or whether it is governed by some quantum model that is non-local (hypothesis QM). The statistical tool that quantifies the asymptotic average amount of support in favor of QM against LOC per independent trial is the Kullback-Leibler (or relative entropy) divergence \cite{DGG05}. This quantity can be see as a distance $D(\mb{p})$ between a given behavior $\mb{p}$ and the set of local behaviors. 

The statistical strength of the most common nonlocality tests have been estimated in \cite{DGG05,AGG05} and are summarized here in Table \ref{table_KL}. It is worth to note that the CHSH scenario is the strongest test among bipartite Bell tests involving qubits \cite{DGG05}. When, considering higher dimensional systems, optimal tests \cite{AGG05} involve partially entangled states (rather than maximally entangled ones), illustrating the astonishing relation between entanglement and nonlocality (see section \ref{MoreNL}). Finally, the Mermin-GHZ test (see section \ref{multi}), involving three qubits appears to be much stronger than the considered bipartite Bell tests \cite{DGG05}. 

    \begin{table}[h!] 
    \caption{Kullback-Leibler divergence for the most common quantum Bell tests. ME stands for maximally entangled.}\label{table_KL}
    \centering
    \begin{tabular}{ c | c | c  }
    \hline \hline
    Bell inequality  & Quantum state & KL divergence (bits) \\
    \hline  \\
     CHSH & ME 2-qubit & 0.046 \\
     CGMLP & ME 2-qutrit  & 0.058 \\
     CGMLP & Optimal 2-qutrit & 0.077\\
     Mermin-GHZ & GHZ 3-qubit &  0.208 \\
    \hline \hline
    \end{tabular}
    \end{table}

\subsection{Multiple-rounds and parallel repetition}\label{repetition}

So far, we have characterized the predictions $\mb{p}=\{p(a,b|x,y)\}$ of local, quantum, or no-signaling systems in \emph{single-round} Bell experiments where a single choice of input pair $(x,y)$ is made and the two devices produce a single output pair $(a,b)$. More generally, we can also consider \emph{multiple-round} Bell experiments in which a sequence $(x_1,y_1),\ldots,(x_n,y_n)$ of input pairs is used in the two devices, resulting in a sequence $(a_1,b_1),\ldots (a_n,b_n)$ of output pairs.  A physical model for such an experiment will thus be characterized by the joint probabilities $\mb{p}_n=\{p(a_1b_1\ldots a_nb_n|x_1y_1\ldots x_ny_n)\}$. The motivation for considering such multiple-round Bell scenarios is clear: it corresponds to the situation of real experimental tests of Bell inequalities in which the two quantum devices are probed many times to gather sufficient measurement statistics.

Three general multiple-round scenarios can be distinguished \cite{BCHK+02}. First, the $n$ output pairs can be obtained by measuring $n$ independent\footnote{Note that we allow though some correlations between the different systems through some global shared randomness $\lambda$, see definition of $\mathcal{L}_n^I$ further in the text. The $n$ systems are only independent with respect to sharing the inputs and outputs.} systems. Effectively, this means that the measurement settings are applied sequentially, \ie, the next input pair is introduced in the two devices after outcomes have been produced for the previous round, and furthermore the devices have no memory of the previous round. In this scenario, we say that $\mb{p}_n$ is local, which we denote $\mb{p}_n\in\mathcal{L}_n^I$, if $p(a_1b_1\ldots a_nb_n|x_1y_1\ldots x_ny_n)=\sum_\lambda q_\lambda p_1(a_1b_1|x_1y_1,\lambda)\times\ldots\times p_n(a_nb_n|x_ny_n,\lambda)$ (similar definitions apply to the quantum and no-signaling case). 

In the second scenario, the measurement settings are also applied sequentially, but the devices' behavior in a given round can depend on the previous measurement settings and outputs, \ie, the devices have memory of the previous rounds\footnote{Formally, we consider here  two-sided memory models, where each device has a memory of every previous input and output, including those of the other device. One can also consider one-sided memory models, where each device has only a memory of the inputs and outputs relative to his side of the Bell experiment but not the other one \cite{BCHK+02}).}. In this case, we say that $\mb{p}_n$ is local, which we denote $\mb{p}_n\in\mathcal{L}_n^M$, if we can write $p(a_1b_1\ldots a_nb_n|x_1y_1\ldots x_ny_n)=\sum_\lambda q_\lambda p_1(a_1b_1|x_1y_1,\lambda)\times p_2(a_2b_2|x_2y_2,w_1,\lambda)\times\ldots\times p_n(a_nb_n|x_ny_n,w_{n-1},\lambda)$, where $w_{i}=(a_1b_1\ldots a_ib_i,x_1y_1\ldots x_ny_n)$ denotes all inputs and outputs up to round $i$. This situation is the most general one that characterizes usual experimental tests of Bell inequality.

Finally, we may also consider a third scenario in which Alice and Bob apply all their $n$ inputs at the same time, and then, at a later time, the device produces all $n$ outputs. We then say that $\mb{p}_n$ is local, which we denote $\mb{p}_n\in\mathcal{L}_n^S$, if $p(a_1b_1\ldots a_nb_n|x_1y_1\ldots x_ny_n)=\sum_\lambda q_\lambda p(a_1b_1\ldots a_nb_n|x_1y_1\ldots x_ny_n,\lambda)$. In this case, the devices can exhibit a collective behavior where the outputs $a_i$ of Alice's device at round $i$ depends on the values of inputs and outputs of her device at any other round, and similarly for Bob's device. This multiple-round model  is formally equivalent to a single-round model with ``big" inputs $x=x_1\ldots x_n$ and $y=y_1\ldots y_n$ and outputs $a=a_1\ldots a_n$ and $b=b_1\ldots b_n$.

The memory model $\mathcal{L}_n^M$ and the simultaneous model $\mathcal{L}_n^S$ are strictly more powerful than the independent model $\mathcal{L}_n^I$. It is easy to see though that local strategies exploiting such memory or collective effects cannot reproduce non-local correlations \cite{BCHK+02}, which necessarily require some genuine non-local resource, such as an entangled quantum state. 

Another potential problem though is that in experimental tests, the correlations $\mb{p}_n$, which characterize the \emph{probabilities} with different set of events can be realized, are not directly observable. Instead one observes a finite number of events, representing only one particular realization of the set of possibilities encoded in $\mb{p}_n$. If the local models $\mathcal{L}_n^I$, $\mathcal{L}_n^M$, or $\mathcal{L}_n^S$ cannot reproduce non-local correlations \emph{on average}, it could nevertheless be possible that clever choices of such multiple-round stragegies, in particular exploiting memory or collective effects, could increase the chance of a statistical fluctuation resulting in an apparent violation of a Bell inequality. In the case of the independent and memory models, which are the most relevant to experimental tests and to applications of quantum non-locality, such statistical fluctuations are harmless and can easily be controlled \cite{BCHK+02,Gill03}. See section~\ref{finitestat} for a more detailed discussion.  This is due to the fact that at any given round $i$, independent and memory local models are constrained to satisfy the Bell inequalities, even when conditioning on events up to round $i-1$. That is if $\mb{p}_{i|w_{i-1}}$ denote the correlations at round $i$ conditioned on the past variables $w_{i-1}$, we necessarily have $\mb{s}\cdot \mb{p}_{i|w_{i-1}}\leq S_l$ for every $i$, $w_{i-1}$, and Bell expression $\mb{s}$. 

This last property can nicely be rephrased in the language of non-local games. It implies that to win $n$ instances of a game, there is no better strategy than using each time the strategy that is optimal for a single-round. This is not the case in the simultaneous model, where all inputs are given at the same time and all output produced at the same time. In this case, which is known as a \emph{parallel repetition} of the game in computer science, there may exist collective strategies to win $n$ instances of the games that are better than using each time the optimal single-round strategy. It is in fact known that for example the CHSH game can be played better locally over many rounds (see \cite{BCHK+02} for an explicit example in the case $n=2$), that is, when playing the CHSH game many times in parallel the gap between the local and quantum bound \emph{shrinks}. 

The question of whether there exists a better strategy for parallel repetition of the game is particularly interesting from the perspective of computer science, see e.g. \cite{cleve:xorParallel}. However, it also tells us something about the strength of correlations between physical systems when Alice and Bob hold many particles to be measured simultaneously. 

Note that if there exists a strategy that lets the players win with probability $p_{\rm win}$ in a single round, then they can win with probability $p_{\rm win}^n$ when playing the game $n$ times. The question is then whether there exists a strategy that beats this value. We speak of (strong) parallel repetition if there exists a non-trivial $q$ such that the winning probability when playing the game $n$ times is always upper bounded by $q^n$. The term \emph{perfect} parallel repetition refers to the case where $q = p_{\rm win}$. It is known that for classical strategies, i.e., local models, parallel repetition holds~\cite{raz:parallel}. More precisely, if $p_{\rm win} = 1-\eps$, then for all games $p_{\rm win}^n = (1-\eps^c)^{\Omega{(n/s)}}$ for some $c \geq 2$, where $s$ is the length of the answers~\cite{raz:parallel,holenstein}. A strong parallel repetition theorem has $c = 1$. It is furthermore known that for unique games, $p_{\rm win}^n = (1-\eps^2)^{\Omega(n)}$~\cite{rao:parallel}. However, for the so-called odd cyle game we require $c \geq 2$, and thus strong parallel repetition does not always hold~\cite{raz:counterexample}.

For no-signaling strategies, it is known~\cite{holenstein} that parallel repetition also holds. 
As quantum and classical theory obey the no-signaling principle this also gives a bound for quantum and classical correlations. Yet, since for many games (e.g. unique games) 
we have $p_{\rm win} = 1$ in the no-signaling case, this bound is not always insightful. For quantum correlations, it is known that for XOR-games ($2$-outcome correlation Bell inequalities)
perfect parallel repetition holds~\cite{cleve:xorParallel}. 
Again, this also gives a bound for classical correlations, but already for the CHSH game it is not known how tight this bound actually is. Parallel repetition in the quantum setting also holds for unique games~\cite{KRT07}. 
A more general result is known for quantum correlations~\cite{vidick:parallel}, however, requires the game to be modified slightly to include ``check'' rounds. A similar construction can be made for local correlations~\cite{feige:parallel}.

\section{Nonlocality and quantum theory}
\label{Nonlocality and quantum theory}

In this section, we analyze the quantum resources---in terms of entanglement or Hilbert space dimension---that are necessary to produce nonlocal correlations by performing local measurements on quantum states \footnote{Another resource that can be considered is the time required to achieve a certain Bell inequality violation, given the range of energy available during the measurements \cite{DW:njp}.}. Here we focus on the case of bipartite states, whereas the nonlocal correlations of multipartite quantum states will be discussed in Section \ref{sec:multipartite}.

\subsection{Nonlocality vs entanglement}
In order to obtain nonlocal correlations from measurements on a quantum state, it is necessary that the latter is entangled. That is, the state 
cannot be written in the separable form 
\be\label{sep state}
\rho_{AB}=\sum_\lambda p_\lambda \rho_A^\lambda\otimes\rho_B^\lambda.
\ee
Indeed, if a state is of the above form, the correlations obtained by performing local measurements on it are given by
\ba
p(ab|xy)&=&\tr\left[\sum_\lambda p_\lambda \left(\rho_A^\lambda\otimes\rho_B^\lambda\right) M_{a|x}\otimes M_{b|y}\right]\nonumber\\
&=&\sum_\lambda p_\lambda \tr(\rho_A^\lambda M_{a|x})\tr(\rho_B^\lambda  M_{b|y})\nonumber\\
&=&\sum_\lambda p_\lambda p(a|x,\lambda)p(b|y,\lambda),
\ea
which is of the local form (\ref{sec2.loc}). Hence the observation of nonlocal correlations implies the presence of entanglement.

It is interesting to investigate whether this link can be reversed. That is, do all entangled state lead to nonlocality? 
In the case of pure states, the answer is positive. Specifically, for any entangled pure state, it is possible to find local measurements such that the resulting correlations violate a Bell inequality\footnote{Note that this result also holds for all multipartite pure entangled states \cite{PR92}, as discussed in more detail in section~\ref{sec:multipartite}.}, in particular the CHSH inequality. This was shown for the case of two-qubit states in \cite{CFS73} and for bipartite states of arbitrary Hilbert space dimension in \cite{HS91,Gisin91}\footnote{Note that this result was also stated, without giving an explicit construction, in \cite{Werner89}.}. Therefore, all pure entangled states are nonlocal. The only pure states that do not violate Bell inequalities are the product states $|\Psi\rangle=|\psi\rangle_A\otimes |\phi\rangle_B$.

For mixed states, it turns out that the relation between entanglement and nonlocality is much more subtle, and in fact not fully understood yet. First, \cite{Werner89} discovered a class of mixed entangled states which admit a local model (i.e. of the form (\ref{sec2.loc})) for any possible local measurements. Hence the resulting correlations cannot violate any Bell inequality. While Werner considered only projective measurements, his results where later extend to the case of general measurements (POVMs) in \cite{Barrett02}. 

\begin{figure*}[tbp]
\includegraphics[width = 1\textwidth]{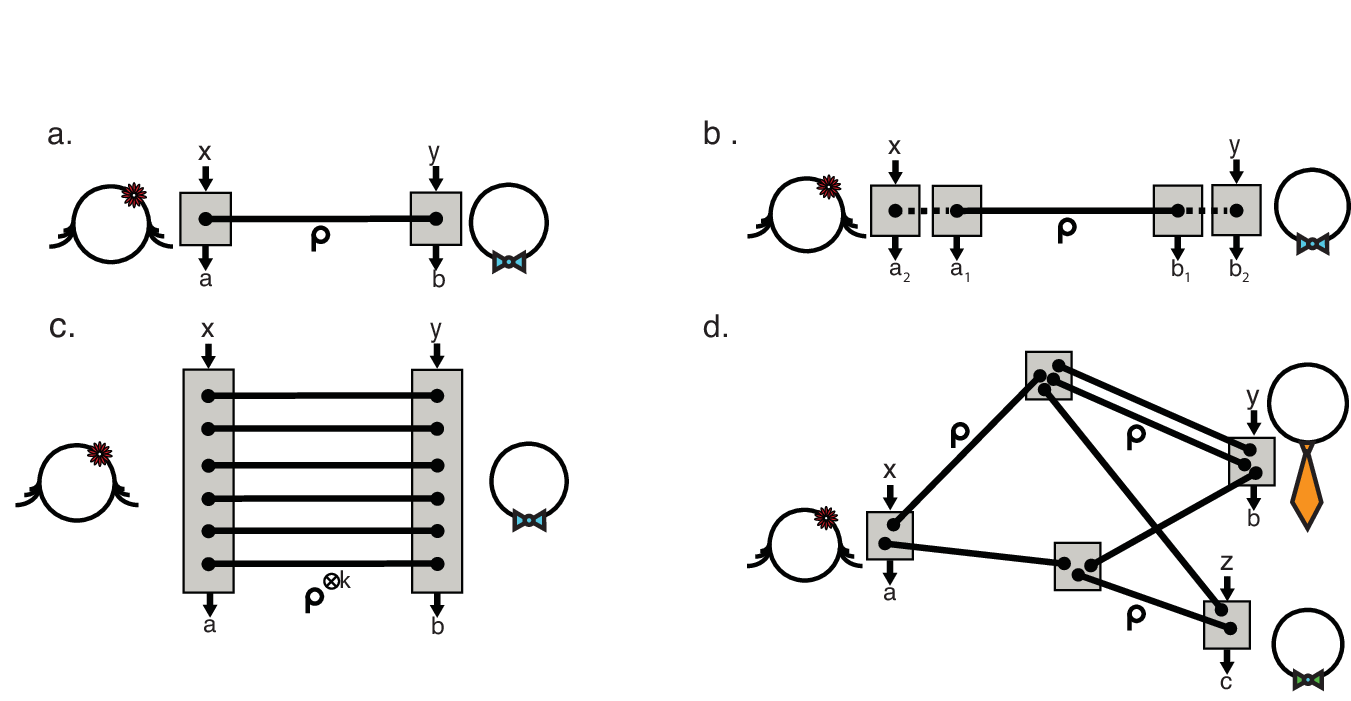}
\caption{The nonlocality of a quantum state $\rho$ can be revealed in different scenarios. a) The simplest scenario: Alice and Bob directly perform local measurements on a single copy of $\rho$. b) The hidden nonlocality scenario: Alice and Bob first apply a filtering to the state; upon successful operation of the filter, they perform the local measurements for the Bell test. c) Many-copy scenario: Alice and Bob measure collectively many copies of the state $\rho$. d) Network scenario: several copies of $\rho$ are distributed in a quantum network, where each observers performs local measurements.}
\label{scheme}
\end{figure*}

The situation is complicated by the fact that directly performing measurements on a mixed state $\rho$ is not always the best way to reveal its non-locality. For instance, it may be necessary to perform joint measurements on several copies of the state, that is considering the state $\rho\otimes\rho\otimes
\ldots\otimes \rho$ \cite{Palazuelos12}. Alternatively one may need to apply a judicious pre-processing to $\rho$, for instance a filtering, before performing the measurements \cite{popescu95}. Therefore, there exist different possible scenarios for revealing the nonlocality of mixed entangled states, some examples of which are represented in Fig.~\ref{scheme} and are discussed in more details below. Importantly a state may lead to nonlocal correlations in a given scenario but not in others. It is also worth mentioning that when many copies of a state can be jointly pre-processed before the measurements, the problem becomes closely related to entanglement distillation. Indeed, any state from which pure bipartite entanglement can be distilled will lead to nonlocality. For undistillable (or bound) bipartite entangled states, it is not yet known whether Bell inequality violations can be obtained, or whether these states admit a local model, as conjectured in \cite{Peres99}. Nevertheless, recent results suggest that nonlocality might in fact be generic for all entangled states \cite{MLD08}.

\subsubsection{Single-copy nonlocality}\label{nonlocal quantum}
The simplest possibility to reveal nonlocality of an entangled state $\rho$ is to find suitable local measurements such that the resulting correlations violate a Bell inequality. In the case of pure states this is always sufficient to reveal nonlocality. In particular, as mentioned above, all pure entangled states violate the CHSH inequality \cite{HS91,Gisin91}. For mixed states, a necessary and sufficient condition for any two-qubit state to violate the CHSH inequality was given in \cite{HHH95}. This criterion works as follows. Associate to any two-qubit state $\rho$ a correlation matrix $T_\rho$ with entries
$t_{ij}=\tr[\rho(\sigma_i\otimes\sigma_j)]$ for $i,j=1,2,3,$ where $\sigma_i$ are the Pauli matrices. The maximum CHSH value $S$ for $\rho$ (considering the most general measurements) is then simply given by
\be\label{horodecki criterion}
S_\rho=2\sqrt{m_{11}^2+m_{22}^2},
\ee
where $m^2_{11}$ and $m^2_{22}$ are the two largest eigenvalues of the matrix $T_\rho T_\rho^T$ ($T_\rho^T$ denotes the transpose of $T_\rho$).
Using the above criterion, it is possible to relate the entanglement of $\rho$, as measured by its concurrence, to its maximal violation of CHSH \cite{VW02}.

From the above criterion it is straightforward to see that not every entangled two-qubit mixed state violates the CHSH inequality. However, contrary to the case of pure states, it is here not enough to focus on the CHSH inequality. In particular, there exist two-qubit states which do not violate CHSH, but violate a more sophisticated Bell inequality ($I_{3322}$, see eq. \eqref{froissard}) involving three measurement settings per party \cite{CG04}. Another example is the two-qubit Werner state, given by a mixture of a maximally entangled state $\ket{\phi_+}=(\ket{00}+\ket{11})/\sqrt{2}$ and the maximally mixed state, \ie:
\be\label{two-qubit werner}
\rho_W=p\ket{\phi_+}\bra{\phi_+}+(1-p)\frac{\openone}{4}.
\ee
This state is separable for $p\leq1/3$ (and thus does not violate any Bell inequality) and entangled otherwise. Using the criterion \eqref{horodecki criterion} one finds that $S=p2\sqrt{2}$, which leads to violation of CHSH for $p>1/\sqrt{2}\approx 0.707$. However, it was shown in \cite{Vertesi08} that the state \eqref{two-qubit werner} violates a Bell inequality involving 465 settings per party for $p \gtrsim 0.7056$. 

If explicit Bell inequality violations yield upper-bounds on the critical value of $p$ necessary to reveal the nonlocality of the state (\ref{two-qubit werner}), it is also possible to obtain lower-bounds by constructing explicit local models. In a seminal paper, \cite{Werner89} showed that the correlations resulting from projective measurements on the state \eqref{two-qubit werner} admit a local model if $p\leq1/2$, even though the state is entangled for $p>1/3$. Entangled states admitting a local model are usually termed \emph{local states}.
Here we describe Werner's model, following the presentation of \cite{Popescu94}. Note first that it is sufficient to construct a local model for $p=1/2$, since then the model can be extended for any $p<1/2$ by mixing with completely uncorrelated and random data. Let Alice and Bob measure the spin polarization of their particles in the $\hat{n}_A$ and $\hat{n}_B$ directions respectively, where vectors describe the measurements on the Bloch sphere. The probability that both Alice and Bob get the outcome `0' is given by
\be\label{qm prob}
p(00|\hat{n}_A,\hat{n}_B)=\frac{1}{4}(1-\frac{1}{2}\cos\alpha),
\ee
 where $\alpha$ is the angle between $\hat{n}_A$ and $\hat{n}_B$. Now, we give a local hidden variable model that gives the same statistics. Here the hidden variable, shared by Alice and Bob, is a vector on the Bloch sphere $\hat{\lambda}=(\sin\theta\hat{x}+\sin{\theta}\sin{\phi}\hat{y}+\cos{\theta}\hat{z})$. In each run of the experiment a different $\hat{\lambda}$ is sent, chosen according to the uniform distribution $dq(\hat{\lambda})=\sin{\theta}d\theta d\phi/4\pi$. After receiving $\hat{\lambda}$, Alice gives the outcome `0' with probability
\be
p_A(0,\hat{n}_A,\hat{\lambda})=\cos^2(\alpha_A/2),
\ee
where $\alpha_A$ is the angle between $\hat{n}_A$ and $\hat{\lambda}$. At the same time, Bob gives the outcome `0' with probability
\be
p_B(0|\hat{n}_B,\hat{\lambda})=
\left\{
    \begin{array}{ll}
        1  & \mbox{if } 2\cos^2(\alpha_B/2) < 1, \\
        0 &  \mbox{if } 2\cos^2(\alpha_B/2) > 1,
    \end{array}
\right.
\ee
where $\alpha_B$ is the angle between $\hat{n}_B$ and $\hat{\lambda}$. Now one can check that the joint probability distribution obtained by Alice and Bob using this local model is given by
\be\label{lhv prob}
p_{LHV}(00|\hat{n}_A,\hat{n}_B)=\int dq(\hat{\lambda})p_A(0,\hat{n}_A,\hat{\lambda})p_B(0,\hat{n}_B,\hat{\lambda}),
\ee
which is indeed equal to \eqref{qm prob}. It is straightforward to check that the above model reproduces the desired correlations for all measurements outcomes. 

Later on, it was proven that two-qubit Werner states are local for $p\lesssim 0.66 $ in \cite{AGT06}, using a connection to Grothendieck constant (see Section~\ref{grothendieck}). Furthermore, \cite{Barrett02} extended the result of Werner to the most general (non-sequential) quantum measurements (so-called POVMs), where a local model is given for $p\leq5/12$. For the interval $0.66 \lesssim p \lesssim 0.7056$ (or $5/12 < p \lesssim 0.7056$ if considering POVMs) it is not known whether the nonlocality of the state (\ref{two-qubit werner}) can be revealed by performing measurements on a single copy of the state at a time.

Werner and Barrett also derived a local model for a family of states generalizing the two-qubit state (\ref{two-qubit werner}) to arbitrary Hilbert space dimension $d$. These are called Werner states, given by
\be\label{werner}
\rho_W = p \frac{2 P_{\text{anti}}}{d(d-1)} + (1-p) \frac{\openone}{d^2}
\ee
where $\openone$ is a $d\times d$ identity matrix, and $P_{\text{anti}}$ denotes the projector on the antisymmetric subspace.
These states have a particular symmetry, being invariant under unitary operations of the form $U\otimes U$.  
The values of $\alpha$ for which $\rho_W$ is entangled and admits a local model (for projective or general measurements) are summarized in Table \ref{table local bounds}.
\begin{table*}
\begin{tabular}{|c||c|c|}
\hline
&    Werner state \eqref{werner}    &      Isotropic state \eqref{isotropic}   \\
\hline \hline
Separable &  $p \leq \frac{1}{d-1} $ & $p\leq \frac{1}{d+1}$\\
\hline
Local for general measurements & $p  \leq\frac{(d-1)^{(d-1)} (3d-1)}{(d+1)d^d}$  & $p\leq \frac{(d-1)^{(d-1)}(3d-1)}{(d+1)d^d} $\\
\hline
Local for projective measurements &$p \leq\frac{d-1}{d}$& $p\leq \frac{(-1+\sum_{k=1}^d 1/k)}{d-1} $ \\
\hline
\end{tabular}
\caption{Separability and locality bounds for Werner states \eqref{werner} and for isotropic states \eqref{isotropic}. For Werner states, bounds for projective measurements were derived in \cite{Werner89} and in \cite{Barrett02} for POVMs. For isotropic states, bounds were derived in \cite{APBT+07}, as well as in \cite{WJD07} for projective measurements.}\label{table local bounds}
\end{table*}

\begin{figure}[tbp]
\includegraphics[width = 0.48\textwidth]{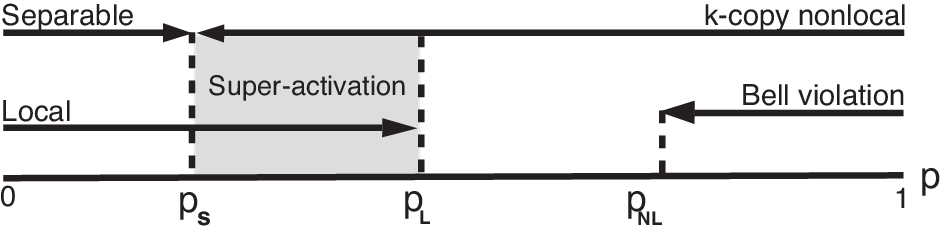}
\caption{Nonlocal properties of the isotropic state \eqref{isotropic}. The state is separable for $p\leq p_s=1/(d+1)$, admits a local model for $p\leq p_L$ \cite{APBT+07}, and violates a Bell inequality for $p_L<p_{NL}<p$ \cite{CGL+02}. In the interval $p_L<p<p_{NL}$, it is not known whether the state admits a local model or, on the contrary, violates a Bell inequality. Finally, when several copies of the isotropic state can be measured jointly, nonlocality is obtained whenever a single copy of the state is entangled, that is, if $p>p_s$ \cite{CABV12}. In the grey region, superactivation of quantum nonlocality occurs \cite{Palazuelos12}.}
\label{bounds iso}
\end{figure}

The local models discussed above were further extended in \cite{APBT+07} and \cite{WJD07} to another family of states generalizing the two-qubit state (\ref{two-qubit werner}), namely the isotropic states 
\be\label{isotropic}
\rho_{iso}=p\ket{\Phi_+}\bra{\Phi_+}+(1-p)\frac{\openone}{d^2},
\ee
where $\ket{\Phi_+}=(1/\sqrt{d})\sum_{i=0}^{d-1} \ket{ii}$ is a maximally entangled state of local dimension $d$. Again, for such states there exist a range of the parameter $p$ for which $\rho_{iso}$ is entangled but admits a local model (see Table \ref{table local bounds}). Note also that $\rho_{iso}$ violates the CGLMP inequality (see eq. \eqref{cglmp}) when $p$ is above a critical value, $p_{NL}$, that decreases with the local dimension $d$. In particular $p_{NL}\rightarrow0.67$ when $d\rightarrow\infty$ \cite{CGL+02} (see Fig.\ref{bounds iso}).

More generally, the approach of \cite{APBT+07} allows one to construct a local model for general states, of the form $\rho = p \ket{\psi}\bra{\psi} + (1-p) \openone / d^2$, where $\ket{\psi}$ is an arbitrary entangled pure state in $\mathbb{C}^d\otimes\mathbb{C}^d$. It is found that for $p\leq \Theta(\frac{\log(d)}{d^2}) $ the state $\rho$ admits a local model for projective measurements. 
Interestingly there is a $\log(d)$ gap in the asymptotic limit between the above bound and the separability limit, which is given by $p\leq \Theta(1/d^2)$. 
An upper bound on $p$ follows from the result of \cite{ADGL02}, where it is shown that a state of the form $\rho=p\ket{\phi_+}\bra{\phi_+}+(1-p)\frac{\openone}{d^2}$ (where $\ket{\phi_+}$ denotes a two-qubit maximally entangled state) violates the CHSH inequality for $p\geq \Theta(\frac{4}{(\sqrt{2}-1)d})$, which tends to zero when $d\rightarrow\infty$. This shows that there exist entangled states embedded in $\mathbb{C}^d\otimes\mathbb{C}^d$ which are highly robust against white noise.

Finally, it is worth pointing out the connection~\cite{Werner89} between the fact that a quantum state admits a local model and the existence of a 
symmetric extension~\cite{doherty1} for this state. 
A bipartite state $\rho_{AB}$ has a $k$-symmetric extension (with respect to part B) if there exists a quantum state of $k+1$ parties, $\rho'_{AB_1...B_{k}}$, such that $\rho'_{AB_i}=\rho_{AB}$ for every $i=1,...,k$, where $\rho'_{AB_i}$ denotes the reduced state of subsystems $A$ and $B_i$. 
In \cite{TDS03} it is shown that if Alice and Bob share a state $\rho_{AB}$ that has a $k$-symmetric extension, every experiment where Bob uses at most $k$ measurement settings (independently of the number of outputs) can be simulated by a local model. 
Note that there is no restriction on the number of measurement settings for Alice. This result can be understood as follows: consider a Bell scenario where Alice chooses among $m$ measurements, represented by operators $M_{a|x}$ with $x=1,..,m$ 
and Bob among $k$ measurements, given by $M_{b|y}$ with $y=1,...,k$. 
Since $\rho_{AB}$ has a $k$-symmetric extension, for each measurement $x$ of Alice, the joint probability distribution $p(a,b_1\ldots b_k|x,y_1=1\ldots y_k=k)$ is well defined via the Born rule\footnote{Note that the same argument holds if, instead of a $k$-symmetric extension, $\rho_{AB}$ has a $k$-symmetric quasi-extension, where, instead of a state of $k+1$ parties $\rho'_{AB_1...B_{k}}$, one has an entanglement witness of $k+1$ parties, $W_{AB_1...B_k}$, with unit trace and such that the reduced states satisfy $W_{AB_i}=\rho_{AB}$ for all $i$.} 
\begin{multline}
 p(a,b_1\ldots b_k|x,y_1=1\ldots y_k=k)\\
=\tr[\rho'_{AB_1...B_k}(M_{a|x}\otimes M_{b_1|1}\otimes...\otimes M_{b_k|k})].
\end{multline}

From these distributions one can then define a joint probability distribution for all possible measurements as
\begin{multline}
p(a_1\ldots a_m, b_1\ldots b_k|x=1\ldots x=m,y_1=1\ldots y_k=k)\\
=\frac{\prod_{i=1}^{m} p(a_ib_1\ldots b_k|x_iy_1=1\ldots y_k=k)}{p(b_1\ldots b_k|y_1=1\ldots y_k=k)^{m-1}}.
\end{multline}
This joint probability distribution provides the local model \footnote{Indeed, it is easy to see that the existence of a joint distribution for all possible measurements that Alice and Bob can make is equivalence to the existence of a local model \cite{Fine82}. Simply think of $\lambda=(a_1\ldots a_m,b_1\ldots,b_k)$ as the hidden variable instructing which outcome every party must output for any measurements that they perform and the joint probability $p(a_1\ldots a_m, b_1\ldots b_k|x=1\ldots x=m,y_1=1\ldots y_k=k)$ as the distribution $q(\lambda)$s over hidden variables.}. Notice that if a state has a $\infty-$symmetric extension it is separable \cite{DPS04}.

\subsubsection{Hidden nonlocality}\label{hiddenNL}

In \cite{popescu95} a more general way of obtaining nonlocal correlations from an entangled quantum state was proposed. Instead of performing a single measurement (in each run of the test), each observer now performs a sequence of measurements. For instance, the observers may first perform a local filtering to their systems before performing a standard Bell test, as in Fig. \ref{scheme}b. That is, they each apply some physical operation (e.g. a measurement) to their system and proceed with the standard Bell test only if that physical operation yields a desired outcome. If one (or both) local operations does not yield the desired outcome, the parties discard this run of the test. Popescu demonstrated the power of this sequential scenario by showing explicitly that certain entangled states admitting a local model can display nonlocality if a judicious local filtering is performed. Hence, the filtering reveals the `hidden nonlocality' of the state. In particular, Popescu showed that this occurs for Werner states (see eq. \eqref{werner}) of local dimension $d \geq 5$. 

One can intuitively understand hidden nonlocality in the following way. Alice and Bob share a mixed entangled state $\rho$. Importantly, even if $\rho$ is local, it may be viewed as a statistical mixture involving one (or more) nonlocal state. In order to extract nonlocality from $\rho$, Alice and Bob first apply a local measurement for which a given outcome can occur only (or most likely) for a nonlocal state in the mixture. Hence, by post-selecting those events in which this particular measurement outcome occurs, Alice and Bob can filter out the nonlocal state. Finally, by performing appropriate local measurements, they can violate a Bell inequality. 

In order to exclude the existence of a local model reproducing this sequential measurement scenario, it is essential that Alice and Bob choose the measurement basis of the final measurement after a successful operation of the filter. If this is not the case, a local strategy could fake Bell inequality violation by adapting the outcome of the first measurement based on the knowledge of which basis has been chosen for the second measurement. A formal account of this argument can be found in \cite{TBD+97,ZHHH98}. A general framework for Bell tests with sequential measurements is discussed in \cite{TBD+97,GWCA+13}. 

A question left open in the work of \cite{popescu95}, is whether hidden nonlocality can also be demonstrated for an entangled state admitting a local model for the most general non-sequential measurements. Note that \cite{popescu95} considered Werner states, which admit a local model for projective measurements, but are not known to be local when POVMs are considered. This question was answered recently in \cite{HQBB13}, where it is shown that there exist entangled states featuring `genuine hidden nonlocality'. That is, states which admit a local model for non-sequential POVMs, but violate a Bell inequality using judicious filtering.  

Other examples of hidden nonlocality were reported. In \cite{Gisin96}, it is shown that there exist two-qubit states which do not violate the CHSH inequality, but do violate CHSH after a judicious local filtering is applied.
In \cite{Peres96} it is demonstrated that five copies of a two-qubit state Werner state \eqref{two-qubit werner} admitting a local model for projective measurements display hidden nonlocality. It is worth noting that these works on hidden nonlocality eventually led to the concept of distillation of entanglement, a central notion in quantum information theory. 

Finally, an important question in this area is whether all entangled states feature nonlocality when local filtering is considered. Although this question is still to be answered, important progress was recently achieved. 
It was show in \cite{MLD08} that for every entangled state $\rho$, there exists another state $\sigma$ which does not violate the CHSH inequality, such that $\rho\otimes\sigma$ violate CHSH after local filtering (see also \cite{LMR12}). 
In particular, if one  chooses $\rho$ such that it does not violate CHSH, a phenomenon of ``activation" of CHSH nonlocality occurs. 

\subsubsection{Multi-copy nonlocality}\label{activation}

Another relevant scenario consists in allowing the parties to perform measurements on several copies of the state $\rho$ in each run of the Bell test. However, here no initial filtering is allowed, contrary to the scenario of hidden nonlocality. 
In the multi-copy scenario, represented in Fig. \ref{scheme}c, Alice and Bob can perform measurements on $k$ copies of the state $\rho$, that is, they measure effectively a state of the form $\rho^{\otimes k}= \rho \otimes \rho \otimes ... \otimes \rho$ ($k$ times). The key point here is that the parties can now perform joint measurements on their $k$ subsystems, that is measurements featuring eigenstates which are entangled. 
Remarkably, the maximal violation of the CHSH inequality for certain states can be increased if several copies of the state are jointly measured \cite{LD06}. In fact, there exist states $\rho$ which do not violate the CHSH inequality, but $\rho^{\otimes 2}$ does \cite{NV11}. 

In more general terms, the possibility of performing measurements on several copies of a state leads to a phenomenon of \emph{activation of nonlocality}. Notably, it was recently demonstrated that quantum nonlocality can be superactivated \cite{Palazuelos12}, that is, the combination of a number of local quantum states can become nonlocal. This demonstrates that nonlocality is not an additive quantity. Specifically, it is shown in \cite{Palazuelos12} that by performing joint measurements on many copies of a local isotropic state $\rho_{iso}$ (see eq. \eqref{isotropic}) of local dimension $d=8$, it is possible to violate a Bell inequality, without involving any pre-processing. This is remarkable given that the initial state $\rho_{iso}$ admits a local model for the most general measurements (i.e. including POVMs). 

More recently, it was shown that for every state $\rho \in \mathbb{C}^d \otimes \mathbb{C}^d$ with singlet fidelity\footnote{The singlet fidelity (or equivalently entanglement fraction) of a state $\rho$ is defined as the maximal fidelity $SF$ of $\rho$ with a maximally entangled state ($MES$), \ie $SF(\rho)=\max_{\ket{\psi} \in MES}\bra{\psi}\rho\ket{\psi}$.} larger than $1/d$, there exist a number of copies $k$ of $\rho$ such that $\rho^{\otimes k}$ is nonlocal \cite{CABV12}. This result implies that every entangled isotropic state \eqref{isotropic} is a nonlocal resource, and establishes a direct connection between the usefulness of a state in quantum teleportation and its nonlocality (see Sec. \ref{nl and tel}). Whether superactivation of nonlocality is possible for any entangled state admitting a local model is an interesting open question.

\subsubsection{More general scenarios}

It is also relevant to investigate the case in which several copies of a bipartite entangled state $\rho$ are distributed in a network of $n$ observers, as sketched in Fig. \ref{scheme}d. It turns out that here a phenomenon of activation of nonlocality can also occur. That is, by judiciously placing several copies of a state $\rho$ admitting a local model, nonlocal correlations among the $n$ observers can be obtained. The state $\rho$ is then termed a ``nonlocal resource''. Again, activation of nonlocality is possible here due to the fact that one (or more) observer can perform a joint measurement on several subsystems (see Section~\ref{activation}).

Examples of activation of nonlocality in networks were reported. First, by concatenating many copies of a state which does not violate the CHSH inequality in an entanglement swapping scenario one obtains a state which violates CHSH \cite{SSB+05,KLMG12}. Second, it was shown that many copies of a two-qubit Werner states \eqref{two-qubit werner} distributed in a star network violate a Bell inequality for $p\gtrsim 0.64$, hence for states which admit a local model for projective measurements \cite{SSB+05,CASA11}. The case of isotropic states, as well as other examples of activation of nonlocality, were in discussed in \cite{CRS12}. 

\subsubsection{Entanglement distillation and nonlocality}\label{peres}

As already hinted in Section~\ref{hiddenNL}, the notion of hidden nonlocality is intimately related to entanglement distillation. For instance, in \cite{Peres96}, the local filtering that is applied on several copies of a state can be used to distill entanglement. Hence the protocol of \cite{Peres96} can be decomposed as entanglement distillation followed by a standard (single-copy) Bell test. In this sense, every entangled state that is distillable can be used to obtain nonlocal correlations.

An interesting question then arises concerning bound entangled states, \ie, states from which no entanglement can be distilled \cite{HHH98}. In fact, a long standing open conjecture---referred to as the Peres conjecture---is that every state with a positive partial transposition (PPT), hence undistillable, admits a local model \cite{Peres99}. More generally, the goal is to understand the link between distillability and nonlocality. Notably, several works established a partial link between both concepts, showing that important classes of Bell inequalities cannot be violated by any PPT state \cite{WW00, SCA08}. For instance, the violation of the CHSH (and more generally of all Mermin inequalities) certifies that the state can be distilled \cite{Acin01,ASW03,Masanes06}. More recently, a method for upper-bounding the possible violation of a given Bell inequality for PPT states (in arbitrary Hilbert space dimension) was presented in \cite{MBLH+13}, from which it can be shown that many bipartite Bell inequalities cannot be violated by PPT states. Finally, note that in the case of more parties, it is proven that nonlocality does not to imply distillability of entanglement \cite{VB12}, hence disproving the Peres conjecture in the multipartite case.

\subsubsection{Nonlocality and teleportation}\label{nl and tel}

Quantum teleportation \cite{BBC+93} is another ``non-local phenomenon'' based on quantum entanglement. As it is the case with nonlocality, it turns out that not every entangled state is useful for teleportation, in the sense of outperforming classical strategies \cite{HHH99}. It is then natural to ask if the fact that a state is useful for teleportation is related to its nonlocality.

This question was first raised by \cite{Popescu94}, who noticed that certain two-qubit entangled Werner state admitting a local model can nevertheless be useful for teleportation. This led to the conclusion that usefulness in teleportation and nonlocality are unrelated. Interestingly, this difference vanishes when considering more general scenarios for revealing nonlocality. In particular, it was recently shown that in the multi-copy scenario, where several copies of the state can be jointly measured, any state that is useful for teleportation is a nonlocal resource \cite{CABV12}. Hence, this work establishes a direct link between teleportation and nonlocality. 

Note also, that a more qualitative relation between the amount of CHSH violation and usefulness for teleportation was derived in \cite{HHH96}. Specifically, the maximal violation $S_\rho$ of the CHSH inequality of a two-qubit state $\rho$ is shown to lower bound its average fidelity for teleportation as follows
\be
F_{telep}\geq\frac{1}{2}(1+\frac{S_\rho^2}{12}).
\ee
Notice that the optimal classical strategy achieves $F_{telep}=2/3$ in the qubit case.
For a device-independent version, see \cite{ho}.

\subsubsection{More nonlocality with less entanglement}
\label{MoreNL}

As discussed above, the relation between entanglement and nonlocality is subtle. Another interesting question is to see whether a quantitative link can be established between both concepts. Astonishingly, it turns out that in certain cases, and depending of which measure of nonlocality is adopted, less entanglement can lead to more nonlocality.
 
An example is provided by certain Bell inequalities, whose maximal violation can only be achieved with partially entangled states \cite{ADGL02} (considering states of a  given Hilbert space dimension). More importantly, there exist simple Bell inequalities, the maximal violation of which cannot be obtained from maximally entangled states of any dimension, but requires partially entangled states \cite{LVB11,VW11}. Also, it is known that there exist Bell inequalities for which partially entangled states give violations which are arbitrarily larger compared to maximally entangled states \cite{JP11,Regev12} (see Section~\ref{grothendieck}).

Interestingly it turns out that this phenomenon, sometimes referred to as an \emph{anomaly of nonlocality} (see \cite{MS07} for a short review), occurs for other measures of nonlocality as well. Notably, this effect was discovered in 1993 by Eberhard \cite{Eberhard93}, who showed that weakly entangled two-qubit states are more resistant to the detection loophole compared to maximally entangled states (see Sec. \ref{the eff}). Moreover, the anomaly of nonlocality was also shown to occur when considering the statistical strength of Bell tests \cite{AGG05}, and the simulation of quantum correlations with nonlocal resources \cite{BGS05}.

\subsection{Nonlocality vs Hilbert space dimension}
In this subsection, we consider the link between nonlocality and another property of quantum systems: the dimension of the Hilbert space in which the quantum state and measurements are defined. Indeed, the Hilbert space dimension generally represents a resource, in the sense that higher-dimensional Hilbert spaces contain more complex quantum states.

Formally, we say that the correlations $p(ab|xy)$ have a
$d$-dimensional representation if there exist a state $\rho_{AB}$ in $\mathbb{C}^d\otimes\mathbb{C}^d$, and measurement operators $M_{a|x}$ and $M_{b|y}$ acting on $\mathbb{C}^d$, such that
\begin{equation}\label{quditpr}
p(ab|xy)=\text{tr}(\rho_{AB}\, M_{a|x}\otimes M_{b|y})\,.
\end{equation}
It some cases, it is also admitted that $p(ab|xy)$ has a $d$-dimensional representation if $p(ab|xy)$ can be written as a convex combination of correlations of the form \eqref{quditpr}.

In the following, we discuss two natural questions. First, what is the minimal dimension $d$ necessary to reproduce a given set of correlations $p(ab|xy)$? This question is closely related to the concept of ``dimension witnesses''. Second, how much nonlocality can correlations of the form \eqref{quditpr} contain as a function of $d$?

\subsubsection{Minimal Hilbert space dimension and dimension witnesses}

Here the general question is to determine what quantum resources, in terms of Hilbert space dimension are necessary to reproduce certain quantum correlations. For instance, if we consider a Bell scenario with a given number of inputs and outputs, what is the minimal dimension $d$ such that all quantum correlations (i.e. all correlations attainable in quantum mechanics) can be reproduced? 
This is in general a very difficult problem. In the case of binary inputs and outputs, we know that qubits ($d=2$) are sufficient, if convex combinations are taken into account \cite{Cirelson80}. However, beyond this simple case, very little is known. In fact, we do not even know if a finite $d$ is sufficient for a scenario involving a finite number of measurements and outcomes. Actually, recent work suggests that this might not be the case \cite{VP10}, giving evidence that the maximal violation of the $I_{3322}$ Bell inequality (see Section~\ref{examples}) can only be attained using a quantum state of infinite dimension. 

A related question is the following. Given some correlations originating from measurements on a quantum system, can we place a lower bound on the Hilbert space dimension of the state and measurements necessary to reproduce them? That is, can we show that certain correlations are impossible to obtain with arbitrary quantum states and measurements of a given dimension. The concept of a dimension witness allows one to address this question. Specifically, a dimension witness for quantum systems of dimension $d$ is a linear function of the probabilities $p(ab|xy)$ described by a vector $\mb{w}$ of real coefficients $w_{abxy}$, such that
\begin{equation}\label{dimwitn}
    W \equiv \sum_{a,b,x,y} w_{abxy}p(ab|xy)\leq  w_d
\end{equation}
for all probabilities of the form \eqref{quditpr} with $\rho_{AB}$ in
$\mathbb{C}^d\otimes\mathbb{C}^d$, and such that there exist quantum
correlations for which $W>w_d$ ~\cite{BPAG+08}. When some
correlations violate \eqref{dimwitn}, they can thus only be
established by measuring systems of local dimension strictly larger than $d$.
The simplest examples of dimension witnesses involve Bell inequalities featuring measurements with ternary outcomes, the maximal violation of which cannot be reached with qubits, but requires qutrits \cite{BPAG+08}. Other examples will be discussed in the next subsection. 

It is also possible to devise entropic dimension witnesses \cite{WCD08}, which were discussed in the context of information-theoretic tasks. Finally, the dimension of a single system can be witnessed in a prepare and measure scenario \cite{GBHA10}. Note however that, since this approach is not based on nonlocal correlations, it is not possible to separate quantum and classical behavior in general; indeed, any quantum behavior can be simulated classically by using systems of high enough dimension.

\subsubsection{Grothendieck's constant and Bell inequalities with unbounded violation}\label{grothendieck}
As mentioned in Section~\ref{measures}, there exist several possible measures of nonlocality. A natural option consists in quantifying the strength of a given nonlocal correlations $\mb{q}$ through the following quantity
\ba \label{nu}  \nu(\mb{q}) \equiv \text{sup}_{\mb{s}} \frac{| \langle \mb{s},\mb{q} \rangle  |}{\text{sup}_{\mb{p} \in \mathcal{L}}| \langle \mb{s},\mb{p} \rangle  |}.  \ea
This represents the ratio between the maximal quantum value for a Bell expression $\mb{s}$ (i.e. $| \langle \mb{s},\mb{q} \rangle  |$) and its local bound (i.e. $\text{sup}_{\mb{p} \in \mathcal{L}}| \langle \mb{s},\mb{p} \rangle  |$), maximized over all possible Bell expressions $\mb{s}$. Note that the absolute value is important here, otherwise the quantity could be ill-defined. This quantity quantifies how much \emph{local} noise (considering any possible local noise) must be added to $\mb{q}$ such that the global distribution becomes local \cite{PWPV+08,JPPV+10}. 
An interesting feature of this quantity is that it provides a unified measure of nonlocality, allowing one to compare the violations of different Bell inequalities.

In \cite{Tsirelson87}, a connection between Grothendieck's inequality, which arose in the study of tensor norms, and the quantum violation of certain Bell inequalities was pointed out. Tsirelson showed that $\nu(\mb{q})$ is upper bounded by Grothendieck's constant $K_G$ for any 2-outcome correlation Bell inequality (i.e. XOR games). Although the exact value of the latter is not known, it is proven that $1.6769 \leq K_G \leq \frac{\pi}{2\log (1+\sqrt{2})}\approx 1.7822$. Importantly, this bound holds for quantum systems of arbitrary dimension.

Moreover, Tsirelson showed that, when restricting to qubits, one has that $ \nu(Q)\leq K_3$, where $K_3$ is Grothendieck's constant of order 3. 
Since it is known that $K_3<K_G$, it follows that there exist 2-outcome correlation Bell inequalities which are dimension witnesses for qubits \cite{BPAG+08}. Explicit examples have been constructed in \cite{VP08}. Moreover, it was proven that dimension witnesses for any Hilbert space dimension $d$ can be obtained from XOR games \cite{BBT11,VP09}.

Tsirelson also raised the question of whether it would be possible to have unbounded
violations of Bell inequalities. That is, does there exist a family of Bell scenarios for which the quantity $\nu(Q)$ is unbounded.

The first result in this direction is due to \cite{Mermin90}, who considered a multipartite scenario. Specifically, he introduced a family of Bell inequalities for an arbitrary number of parties $n$ (now referred to as the Mermin inequalities, see section \ref{multi}), and showed that by performing measurements on an $n$-party GHZ state one obtains a violation of these inequalities that grows exponentially with $n$, while the local bound remains constant.

A natural question is then whether unbounded Bell violations can also occur in the case of a fixed number of parties. This is however a very hard problem, mainly because of the difficulty of finding Bell inequalities and to estimate their quantum violations. It was discovered recently that the abstract concepts of operator space theory and tensor norms provide a useful framework for the study of violations of Bell inequalities in quantum mechanics (see \cite{JPPV+10} for an introduction). This line of research started in \cite{PWPV+08}, where the existence of tripartite correlation Bell inequalities with unbounded quantum
violations is proven. Later the same authors showed that similar results
hold for (non-correlation) bipartite Bell inequalities
\cite{JPPV+10}. More formally, these studies focus on the quantity $\nu(Q)$ (see Eq. \eqref{nu}), i.e. the maximal quantum violation of any Bell inequality
$\mb{s}$, as a function of the number of measurement settings,
outcomes, and Hilbert space dimension. Remarkably they show that $
\nu(Q)$ can be upper and lower bounded by ratios of different
norms of the Bell expression $\mb{s}$ (here viewed as a functional), which have been studied in operator
space theory.

While the works mentioned above give non-constructive proofs of
the existence of Bell inequalities with unbounded violations,
explicit examples have also been found. The strongest result is
due to \cite{JP11} who constructed explicitly
(up to random choices of signs) a bipartite Bell
inequality featuring a violation of order of
$\frac{\sqrt{k}}{\log{k}}$, where each party has $k$ possible
measurements with $k$ outcomes, considering quantum systems of dimension
$d=k$. Notably, this constructions appears to be close to optimal, as a separation between this violation and known
upper bounds is only quadratic in $k$ \cite{JP11,JPPV+10}. Recently, an explicit and simplified
presentation of these Bell inequalities was given in
\cite{Regev12}, based on standard quantum information techniques.
Other explicit examples of Bell inequalities with unbounded
violations have been presented \cite{BRSW11}. Note that, while the construction of \cite{BRSW11} uses maximally entangled states, the works of \cite{JP11} and \cite{Regev12} consider entangled states with low entanglement. Finally, an upper
bound on the maximum Bell violation (for any possible Bell inequality) of a given quantum state was derived in \cite{Palazuelos12b}.

\subsection{Simulation of quantum correlations}
\label{simulation}

So far we have examined which quantum resources are necessary to produce nonlocal correlations, in term of entanglement or Hilbert space dimension of quantum states. Here, we discuss the converse question. How can we use nonlocal resources to characterize and quantify the nonlocality of entangled quantum states? If a state violates a Bell inequality, we know that its measurement statistics cannot be reproduced by a local model. However, we can simulate its correlations if we have access to a nonlocal resource, such as classical communication or nonlocal resources such as the PR-box. The minimal amount of nonlocal resources required can then be considered as a measure of the nonlocality of the state. Here we give a brief review of progress in this direction. 

\subsubsection{Simulating the singlet state}

A classical simulation protocol of a given quantum state $\ket{\psi}$ aims
at reproducing the correlations obtained from local measurements
on $\ket{\psi}$, using only shared randomness and classical
communication. For definiteness, we shall focus here on the
singlet state of two qubits, i.e. $\ket{\psi}=
(\ket{01}-\ket{10})/\sqrt{2}$, which is also the most studied
case.

Alice and Bob first receive as input a unit vector on the Bloch
sphere, i.e. $\hat{n}_A,\hat{n}_B \in \mathbb{R}^3$, representing
projective measurements $\hat{n}_A\cdot \vec{\si}$ and $\hat{n}_B\cdot
\vec{\si}$, where $\vec{\si}$ is the vector of Pauli matrices. Then, they
are allowed to exchange classical communication. Finally, they
must produce binary outcomes $a,b=\pm1$ reproducing the expected
statistics, i.e. $p(ab| \hat{n}_A \hat{n}_B)= 1/4 (1- ab \hat{n}_A \cdot \hat{n}_B)$.

It is then interesting to look for the model using the least
classical communication, since the smallest number of bits
required to simulate $\ket{\psi}$ can be considered as a measure
of the nonlocality of $\ket{\psi}$. This approach was proposed
independently by several authors \cite{Maudlin92,BCT99,Steiner00}.
These first partial results were superseded in \cite{TB03}, where it is shown that a single bit of communication is
sufficient to simulate exactly the correlations of local
projective measurements on a singlet state. Note that in this
model Alice and Bob use infinite shared randomness, which is
proven to be necessary for models with finite communication
\cite{MBCC01}.

It is also interesting to investigate simulation models using only
non-signaling resources, such as the PR box (see Section~\ref{NScorrs}). Remarkably, a single
PR box is enough to simulate the singlet correlations
\cite{CGMP05}. The latter model is even more economical than the
model of \cite{TB03}, since a PR box is a strictly weaker
nonlocal resource; indeed, while it is possible to get a PR box
from one bit of communication, the opposite is impossible since
the PR box is non-signaling.

Finally, it is also possible to devise a simulation model of the
singlet state in which post-selection is allowed \cite{GG99}, that
is, the parties are not required to provide an output in all runs
of the protocol. Indeed post-selection should be considered as a
nonlocal resource, giving rise to the detection loophole (see Sec.
\ref{det loop}).

An elegant unified presentation of all the above models can be found in \cite{DLR05}.

\subsubsection{Other quantum states}

The simulation of quantum correlations of arbitrary bipartite entangled quantum states has also been investigated. Notably, \cite{RT07} showed that the correlations obtained from local measurements with binary outputs on any $\rho_{AB} \in \mathbb{C}^d \otimes \mathbb{C}^d$ can be simulated with only two bits of communication, which is proven to be necessary in general \cite{VB09}. Note however that this model focuses on the correlations between the outcomes of Alice and Bob, and does not, in general, reproduce the expected quantum marginals.

A case of particular interest is that of partially entangled qubit states, i.e. $\ket{\psi_\theta}= \cos{\theta}\ket{00}+\sin{\theta}\ket{11}$. While it is shown that its correlations (including marginals) can be perfectly simulated with two bits of communication for any $\theta$ \cite{TB03}, it is not known whether a single bit of communication would suffice. It is however proven that a simulation model using a single PR box does not exist \cite{BGS05} for weakly entangled states ($\theta\leq \pi/7.8$). Thus it appears that less entangled states require more nonlocal resources to be simulated compared to maximally entangled ones, illustrating the subtle relation between entanglement and nonlocality (see Section \ref{MoreNL}). A simulation model for states $\ket{\psi_\theta}$ using only non-signaling resources has also been presented \cite{BGPS08}.

Moreover, \cite{BCT99} established that the simulation of measurements with $d$ outcomes on a maximally entangled state in $\mathbb{C}^d \otimes \mathbb{C}^d$ requires classical communication of order $d$ bits. Therefore, there exist families of quantum nonlocal correlations requiring an arbitrarily large amount of classical communication for being simulated. Much less is known on the simulation of multipartite entangled states. \cite{BG11} presented a simulation model for equatorial measurements on the 3-qubit GHZ state which requires 3 bits of communication, or 8 PR boxes. The simulation of the protocol of entanglement swapping, which combines entangled states and entangled measurements, was also discussed \cite{BBBC+12}.

More generally the problem of simulating quantum nonlocal correlations is intimately related to the field of communication complexity. Thus, many results on communication complexity are relevant in the context of nonlocality. For more details on communication complexity and on the procedure for converting communication complexity problems into nonlocal tasks, we refer the reader to a recent review \cite{BCMW10}.

\subsubsection{Elitzur-Popescu-Rohrlich decomposition}\label{secEPR2}

A different perspective on simulating quantum correlations was presented in \cite{EPR92}---often referred to as the EPR2 approach. These authors proposed to decompose a quantum probability distribution $p_q(ab|xy) $ into a local and nonlocal part. Formally, that means to write $p_q$ as a convex combination of a local distribution ($p_l$) and a nonlocal one ($p_{ns}$):
\be\label{eqEPR2}
p_q(ab|xy)= w p_l(ab|xy)+(1-w) p_{ns}(ab|xy).
\ee
Note that, since $p_q$ and $p_l$ are no-signaling distributions, $p_{ns}$ is also no-signaling (hence the subscript $ns$). 
Clearly, any distribution can be written in this way (take for instance $w=0$ and $p_q=p_{ns}$). 
To find the EPR2 decomposition, one then finds the maximum of $w$ among all possible decompositions of the form \eqref{eqEPR2}. This quantity, denoted $w_{max}$, defines the local content of the distribution $p_q$. 
The EPR2 decomposition can be understood as a simulation of the distribution $p_q$ where, with probability $w_{max}$ a local distribution is used, and with probability $1-w_{max}$ a nonlocal (no-signaling) distribution is used. Note that $q_{max}$ can also be considered as a measure of the nonlocality of the distribution $p_q$: if $w_{max}=1$, $p_q$ is local; if $w_{max}<1$, $p_q $ is nonlocal; if $w_{max}=0$, $p_q $ is fully nonlocal.

One can bound $w_{max}$, for a given distribution $p_q$, trough the violation of a Bell
inequality $\mb{s}\cdot \mb{p} \leq S_l$ \cite{BKP07}. Denote $Q$, the Bell value of distribution $p_q$.
It is straightforward to see that
\be w_{max}\leq\frac{S_{ns}-Q}{S_{ns}- S_l}, \ee 
where $S_{ns}$ is the maximal value of the Bell expression $\mb{s}$ for any no-signaling distribution. 
Notice that, if $p_q$ reaches the maximal value allowed by no-signaling (i.e. $Q=S_{ns}$), then $w_{max}=0$. This
means that the quantum distribution is maximally nonlocal according to
the EPR2 decomposition, hence no local part can be extracted.

It is also possible to define the local content of a quantum state. To do this, consider all possible measurements that can be applied
to a quantum state and then derive the local content for the distribution obtained from these measurements. 
Originally, \cite{EPR92} showed that the maximally entangled state of two qubits has zero local content, i.e. it is fully nonlocal.
This result was then generalized to any bipartite maximally entangled state \cite{BKP07}, via a generalization of the chained Bell inequality (see Section \ref{examples}), showing that such states can provide maximally nonlocal and monogamous correlations, which is relevant for instance in quantum cryptography. 

The local content of other quantum states has also been discussed. In particular, for the case of two-bit entangled pure states
$\ket{\psi_{\theta}}=\cos{\theta} \ket{00}+\sin{\theta}\ket{11}$, with $\theta \in
[0,\pi/4]$, it was proven that $q_{max}=1-\cos(2\theta)$
\cite{PBG12}. The EPR2 decomposition of pure entangled two-qutrit states was
also sketched in \cite{Scarani08}.

Finally, note that these ideas were generalized to the multipartite case in \cite{ACSA10}.

\section{Applications of quantum nonlocality}\label{appl_nl}
When considering nonlocality as a potential resource for information processing, two intuitive ideas immediately come to mind.  First, since the existence of nonlocal correlations between the two wings of a Bell experiment seems to imply some connection between these two distant wings, one could hope to exploit this connection to communicate, and in particular to communicate faster-than-light. Second, since a local model for a Bell experiment is equivalent, as we saw in Section~\ref{seclocal}, to a deterministic model in which a definite outcome $a(x)$ and $b(y)$ is assigned in advance to every measurement $x$ and to every measurement $y$, nonlocality then suggests, in contrast, that these measurement outcomes are fundamentally undetermined and thus that they could be used to establish cryptographic keys. Both ideas are partly true and partly misleading. In both cases, the no-signaling principle plays a fundamental role.

\subsection{Communication complexity}\label{sec:commcompl}
In the first example discussed above, no-signaling acts as a limitation: we have already seen that the no-signaling conditions (\ref{sec2.nosig}), which are satisfied by any set of correlations arising from measurement on quantum systems, imply that Bob's outcome does not reveal any information about Alice's input $x$ and the other way around. Thus, no-signaling prevents the use of non-local correlations as a substitute for \emph{direct} communication between Alice and Bob. It may then come as a surprise that non-locality can nevertheless be exploited to reduce the amount of communication in certain distributed computing tasks, both in information-theory and the study of communication complexity. In the setting of communication complexity, Alice receives an $n$-bit string $x$ and Bob receives an $n$-bit string $y$ and the aim is for Bob to compute some function $f(x,y)$ with as little communication between Alice and Bob. This can always be achieved if Alice sends her $n$-bit string $x$ to Bob, but for certain functions less communication is sufficient. The minimum number of bits that must be exchanged between Alice and Bob for Bob to determine $f(x,y)$ is known as the \emph{communication complexity} of $f$. It was realized in \cite{CB97} that if Alice and Bob can share systems exhibiting non-local correlations, then they can compute certain functions with \emph{less communication} than would be required without such non-local systems. This phenomenon does not violate the no-signaling principle because the knowledge that Bob obtains about Alice's input through $f(x,y)$ is no greater than what is already conveyed by Alice's communication. The field of communication complexity is an active field of research in computer science, in which strong connections with non-locality have been discovered since \cite{CB97}. For more details, we refer the reader to a recent review on the subject \cite{BCMW10}.

\subsection{Information theory}
Nonlocal correlations can also enhance communication power in the context of information theory. 
Consider two parties, Alice and Bob, communicating via a noisy communication channel.
If we only care about the rate at which information is transmitted from Alice to Bob such that the error rate goes to zero in the large block length limit, then this transmission rate \emph{cannot} be increased using entanglement~\cite{noIncreaseEntangled} or even no-signaling correlations~\cite{noIncreaseNSig}. However, the situation changes when we care about the rate at which information can be sent without any error at all~\cite{noIncreaseNSig}. The maximum such rate is known as the zero-error capacity of a noisy-communication channel. For example, it is known that if Alice and Bob share certain no-signaling correlations, zero-error transmission becomes possible through a noisy-channel, even if that channel's zero-error capacity is zero without the ability to use such correlations~\cite{zeroError1}. 
Notably, even certain quantum correlations are useful in this context, in particular those achieving a unit wining probability in a pseudo-telepathy game (see Section \ref{sec:games}).

\subsection{Quantum cryptography}\label{sec:QKD}
In our second intuition discussed above, that the violation of Bell inequalities guarantees the presence of randomness, the no-signaling principle is no longer a limitation, but a prerequisite.  Indeed, in the same way that every local model can be seen to be equivalent to a local deterministic model, every nonlocal model is equivalent to a nonlocal deterministic model where, for each run of the Bell test, definite outputs $a(x,y)$ and $b(x,y)$ are assigned to every pair of inputs $(x,y)$. In full generality, the violation of Bell inequalities does not therefore guarantee by itself any indeterminacy in the outcomes $a$ and $b$ (as we have already stressed in the introduction, non-locality is -- as its name indicates -- about the violation of locality, not about the violation of determinism). However, every nonlocal deterministic model is necessarily signaling: if $a(x,y)$ depends non trivially on both $x$ and $y$, then Alice can recover some information about Bob's input $y$ from the knowledge of the output $a$ and her choice $x$. In a model that reproduces nonlocal correlations and which is intrinsically no-signaling, the measurement outcomes cannot therefore be fully determined in each run of the Bell test and they must necessarily exhibit some randomness. This intuition is at the basis of \emph{device-independent cryptography} in which the violation of a Bell inequality, which can be asserted without any detailed physical assumptions on the working of the devices, guarantees the production of cryptographic keys that are genuinely random and secure to any adversary limited by quantum theory or, more generally, by the no-signaling principle.

\subsubsection{Initial developments}
One of the earliest connections between nonlocality and cryptography is due to \cite{Herbert75}, who interpreted the $`0'$ and $`1'$ outcomes produced by two distant quantum devices as correlated binary random messages. By considering the error rates in such messages, he presented an elementary derivation of Bell's theorem, but he did not go as far as deducing that quantum nonlocality could be exploited for a secure cryptographic scheme.

The practical application of Bell non-locality to cryptography was first realized by Ekert in his celebrated paper \cite{Ekert91}, which represents more generally one of the founding articles of quantum cryptography. The problem of establishing a secure, encrypted communication between two parties can be reduced to the problem of generating a secure, cryptographic key, i.e., a sufficiently long strings of random bits that are shared between Alice and Bob, but unknown to any potential eavesdropper Eve. Ekert presented a protocol for this key distribution problem which is based on the CHSH inequality and uses a source of two-qubit maximally entangled states $|\phi_+\rangle$---here we present a slight variation of this protocol introduced in \cite{AMP06}. Each party repeatedly receives one qubit from the source and performs a measurement on it. In each run, Alice chooses among three possible measurements $x=0,1,2$ and obtains an outcome bit $a$; Bob chooses among two possible measurements $y=0,1$ and obtains an outcome bit $b$. Once all states have been measured, Alice and Bob announce publicly the settings they have chosen for each particular measurement and divide their results in two groups. The subset of the results corresponding to the measurements $x=1,2$ and $y=0,1$ is used to evaluate the CHSH inequality violation. Hence these measurements are chosen such as to maximize this violation (see Section \ref{nutshell}): for instance, Alice measures in the direction $0,\pi/2$ in the $x-z$ plane of the Bloch sphere for $x=1,2$ and Bob in the directions $-\pi/4,\pi/4$ for $y=0,1$ (also in the $x-z$ plane). The subset of results corresponding to the choices $x=0$ and $y=0$ are used to generate the shared key. Hence the measurement $x=0$ is chosen in the same direction $-\pi/4$ as Bob's measurement $y=0$, in such a way  that the key bits $a$ and $b$ are perfectly correlated.

The CHSH violation guarantees, as discussed earlier, that the key bits are undetermined and in particular that Eve could not have fixed them in advance. More generally, Eve could attempt to obtain information about the values of $a$ and $b$ by performing delayed measurements (after the public disclosure of Alice and Bob's settings) on a system of her own correlated with Alice and Bob's systems. As remarked by Ekert, the protocol is also secure against such attacks as a maximal CHSH violation guarantees that the state shared by Alice and Bob is (essentially equivalent) to a pure entangled $|\phi_+\rangle$ state, which cannot be correlated to any system in Eve's possession. In a realistic implementation, Alice and Bob's key bits will not be perfectly correlated and the CHSH violation will not be maximal, implying that Eve can obtain some finite information on these key bits. But provided that these imperfections are not too important, it should be possible to distill a shared secret key from the raw data of Alice and Bob by applying error-correction and privacy amplification protocols.

The intuition for security in the Ekert protocol is based on the violation of a Bell inequality which can be assessed independently of the protocol's implementation, but this aspect was not fully recognized at the time. When assuming that Alice and Bob's devices perform measurements on qubits in complementary bases, Ekert's protocol was found to be equivalent to an entanglement-based version of the Bennett-Brassard 1984 protocol \cite{BBM92}. This was important in establishing entanglement as a central concept for QKD, but it also implied that the subsequent security proofs used ``qubits" and ``complementary bases" as implicit assumptions\footnote{Quantitative relations between security bounds and the violation of Bell inequalities were pointed out \cite{SG01}; but this link turned out to be an artifact of the assumption called ``individual attacks" and did not survive in stricter security proofs, which were rather derived from the notion of entanglement distillation.}. One crucial point was (understandably) missed in those early days: that the implicit ``qubits" and ``complementary bases" assumptions require a very good control of, and ultimately some trust about, the physical implementation\footnote{It is interesting to notice, though, the note added in print to the \cite{BBM92} paper, which come close to recognizing explicitly the device-independent aspect of Ekert scheme.}. 

However, as the reader knows by now, the violation of Bell inequalities can be established without such knowledge. Therefore, a cryptography protocol based on nonlocality requires fewer assumptions: the devices can in principle be tested and the security of the protocol certified without any detailed characterization of the devices; to some extent, the devices could even be malicious and been prepared by the eavesdropper. This is called a \textit{device-independent} (DI) assessment.

The idea of DI quantum cryptography was first made explicit by Mayers and Yao, who called it \textit{self-testing} \cite{MY98,MY04}. Although their analysis is not directly based on Bell inequalities, it obviously exploits correlations that are nonlocal. The breakthrough that pushed the recent development of DIQKD came with Barrett, Hardy, and Kent who
 introduced a QKD protocol based on the chained Bell inequality (see Section \ref{examples}) and proved it to be secure against ``super-quantum" eavesdroppers that may violate the law of quantum physics but which are constrained by the no-signaling principle \cite{BHK05}. A practical protocol based on the CHSH inequality was then introduced in \cite{AGM06} (though it was proved secure only against a restricted family of attacks), where it was also noticed that proving security assuming only the no-signaling principle implies in particular that one can do away with the ``device-dependent" assumptions of standard QKD. The DI potential of such QKD scheme based on Bell inequalities was then fully perceived in \cite{ABGM+07}, which introduced a DI security proof for collective attacks of the variation of Ekert's protocol presented above against an eavesdropper constrained by the entire quantum formalism and not only the no-signaling principle. 

Finally, the ideas from DIQKD have been adapted to the simpler task of DI randomness generation (DIRNG) in \cite{Colbeck07,PAMB+10} and to distrustful quantum cryptography in \cite{SCA+11}, where a scheme for the device-independent implementation of (imperfect) bit commitment and coin tossing was introduced.

In the following we discuss in more details the status of current security proofs for DIQKD and DIRNG, the assumptions on the devices that underlie them, and the prospects for experimental implementations. We first briefly discuss the quantitative aspects of the relation between randomness and non-locality since it is at the basis of many  security proofs and protocols. 
Note that the development of DIQKD and the recent attacks on standard QKD protocols such as \cite{LWW+11} have led to a series of feasible proposals for QKD that are intermediate between device-dependent and device-independent schemes, see for instance \cite{BCW+12,BP12,LWW+11,LCQ12,TFK+12}. We do not review this work here, as it does not directly rely on nonlocality as a resource.

\subsubsection{Randomness vs non-locality}

\paragraph{Quantitative measures of randomness.}
Imagine Alice holds of a measurement device that produces outcomes $a$ when performing a measurement, where we let $R_A$ denote the random variable of the outcome. 
When can we say that $a$ is random? One way to think about randomness is by means of introducing an observer, Eve, who tries to guess Alice's measurement outcome $a$ - the better the guess the less random $a$ is. 
In order to guess $a$, Eve may perform an arbitrary measurement on a system $E$, which is possibly correlated with the one of Alice. We will use $z$ to label her measurement setting and $e$ to label her measurement outcome. For any given $z$, Eve's best guess for $a$ corresponds to the most probable outcome, the one maximizing $p(a|ez)$.  
The \emph{guessing probability} of Eve is then defined as her average probability to correctly guess $a$, maximized over all her possible measurements
\be 
p_\text{guess}(R_A|E):=\max_z \sum_e p(e|z) \max_a p(a|e,z)\ .
\ee
This guessing probability can also be expressed as the
\emph{min-entropy} $H_\text{min}(R_A|E) = -\log_2 p_\text{guess}(R_A|E)$~\cite{krs:operational}.
It takes on values between $0$ and $\log |R_A|$, corresponding to the cases where Eve can guess perfectly, and where Eve's probability of guessing is no better than for the uniform 
output distribution $1/|R_A|$ respectively. 

The min-entropy is a good measure of how random Alice's measurements outputs are because it tells us exactly how many uniform classical random bits $\ell$ can in principle
be obtained from a classical string $a \in R_A$ by applying some function $f_r: R_A \rightarrow \{0,1\}^\ell$ . It is easy to see that if we only have a guarantee about the min-entropy
of the so-called source $R_A$, then no randomness can be obtained using just one deterministic function $f$. However, if we are willing to invest some perfect randomness labeled $R=r$ from an initial seed, and choose a function $f_r$ depending on it, then we can obtain randomness. This process is known as \emph{randomness extraction} and enjoys a long history in computer 
science (see ~\cite{vadhan:survey} for a survey). Formally, a (strong) extractor produces an output $\rho_{F(R_A)E}$ that is close to uniform and uncorrelated from Eve $ \|\rho_{F(R_A)ER} - \frac{\id_{2^{\ell}}}{2^\ell} \otimes \rho_{ER}\|_1 \leq \epsilon$ for some small $\epsilon$, even if Eve later learns which function $f_r$ we applied. In the context of cryptography, this is also called privacy amplification. If Eve only holds classical side-information about Alice's system it is known that randomness extraction is possible, where the maximum output size obeys $\ell \approx H_{\rm min}(R_A|E)$ \cite{impagliazzo}. This is also true if Eve holds quantum side-information \cite{renner:diss,DPV+09,Tashma09}. More generally the full quantum min-entropy \cite{krs:operational} has been shown to characterize exactly how much randomness can be obtained by making  measurements on $A$ in \cite{bfw:random}.  However, no such general result is known if Eve holds arbitrary no-signaling (\ie, supra-quantum) side-information \cite{HR10}, see section~\ref{sec:diqkd} for a more detailed discussion.

\paragraph{Randomness and Bell violations}
In order to discuss quantitative links between randomness and the violations of Bell inequalities, it is useful, as in the above discussion, to introduce an additional observer and thus consider nonlocal correlations shared between Alice, Bob and Eve.
In such a tripartite setting, the correlations are characterized by the probabilities $p(abe|xyz)$.
If Eve measures $z$ and obtains $e$, then Eve's characterization of Alice's device is 
now given by the conditional probability distributions $p(a|xez)$. If Eve learns $x$, then 
for any given $z$ her best guess for $a$ corresponds to the most probable outcome maximizing $p(a|xez)$. Maximizing over $z$ thus means that Eve can guess $a$ with
probability $p_{\rm guess}(R_A|EX=x)$.
In the case where $a$ can take on two values and Alice and Bob's devices are characterized by a CHSH expectation value $S$ it is shown in~\cite{PAMB+10} (see 
also~\cite{MPA11} for an alternative derivation), that independently of the devices' behaviours and Eve's strategy
\be \label{qrand}
p_\text{guess}(R_A|EX=x)\leq \frac{1}{2}\left(1+\sqrt{2-S^2/4}\right)\,.
\ee
In particular, when $S=2\sqrt{2}$, we get as expected $P_\text{guess}(A|EX=x)\leq 1/2$ corresponding to 1 bit of min-entropy $H_{\rm min}(R_A|E X=x)$, 
implying that Alice's output is fully random. While when the CHSH expectation achieves the local bound $S=2$, we get the trivial bound $p_\text{guess}(A|EX=x)\leq 1$.
Using the SDP hierarchy \cite{NPA07,NPA08} (see Section~\ref{hierarchy}) it is possible to derive analogous bounds for arbitrary Bell inequalities, see Supplementary information of \cite{PAMB+10} for details. 
One can also compute upper-bounds on the guessing probability not only for the local randomness (corresponding to the output $a$ alone), but also of the global randomness (corresponding to the pair of outcomes $a$ and $b$). At the point of maximal CHSH violation, for instance, one finds $p_\text{guess}(R_A R_B|EX=xY=y)\leq 1/4+\sqrt{2}/8\simeq 0.427$ corresponding to 1.23 bits of min-entropy \cite{PAMB+10,AMP12}, where the random variable  $R_B$ corresponds to Bob's outcome. 

The above bounds on the guessing probability are obtained by assuming that the devices and the eavesdropper obey quantum theory. Similar bounds can be obtained assuming only the no-signaling principle. In this case, one obtains the following tight bound for the CHSH inequality \cite{BKP07,MRC+09,PAMB+10} 
\be\label{nsrand}
p_\text{guess}(R_A|EX=x) \leq \frac{3}{2}-\frac{S}{4}\,.
\ee
At the point $S=2\sqrt{2}$ of maximal quantum violation, one finds $p_\text{guess}(A|EX=x)\leq 0.79$ which is, as expected, less constraining than the quantum bound (\ref{qrand}). Maximal randomness, $p_\text{guess}(A|EX=x)\leq 1/2$ is obtained now only at the maximal no-signaling violation $S=4$ of the CHSH inequality, corresponding to a $PR$-box. The above bound has also been generalized for the $\Delta$-output $m$-input chained inequality (see Section~\ref{examples}) in \cite{BKP07}:
\be 
p_\text{guess}(R_A|EX=x) = \frac{1}{d}+\frac{d}{4}S^{(\Delta,m)}_\text{chained}
\ee
For $m\to\infty$ the maximal quantum violation of the chained inequality tends to the maximal no-signaling violation $S_\text{chained}=0$ -- note that the Bell inequality is written as $S^{(\Delta,m)}\geq \Delta-1$, thus a ``high" violation means a lower value for $S^{(\Delta,m)}_\text{chain}$. In this limit, one thus gets $p_\text{guess}(R_A|EX=x)\leq 1/\Delta$, i.e. the outcome can be certified to be fully random even assuming no-signaling alone. This property is central to the security of the QKD protocol introduced in \cite{BHK05} and was further developed in \cite{CR08,CR11} to show that some extensions of quantum theory cannot have improved predictive power. 

Naively one would expect that less nonlocality in a Bell-type experiment implies less randomness. In the quantum setting, this intuition is not always correct. In the case of two-output Bell scenarios, the maximal amount of local randomness (characterizing the single outcome $a$) corresponds to 1 bit of min-entropy and the maximal amount of global randomness (characterizing the joint outcome pair $a,b$) corresponds to 2 bits. It is shown in \cite{AMP12}, through a family of (non-facet) two-input two-output Bell inequalities, that such values can be attained with nonlocal correlations that are arbitrarily close to the local region or which arises from states with arbitrarily little entanglement. 
This work suggests that while non-locality is necessary to certify the presence of randomness, its quantitative aspects are related to the extremality of non-local correlations. Extremality was already identified in \cite{FFW11} as a key property for characterizing the behaviours which are independent of any measurement
results of an eavesdropper. This work also presents different tools to certify and to find extremal behaviours for particular Bell scenarios. Finally, in \cite{DPA12}, it is shown that maximal global randomness can be obtained in a variety of scenarios (including multipartite ones) from the violation of certain Bell inequalities.

\subsubsection{Device-independent randomness generation}\label{DIrandomness}
The above relations between non-locality and randomness immediately suggest to use Bell-violating devices  to certify the generation of random numbers in a DI manner \cite{Colbeck07}. This idea was further developed in \cite{PAMB+10}, where a practical protocol for randomness generation was introduced, the first quantitative bounds on the randomness produced where shown, and a proof-of-principle experimental demonstration was performed. 

The bounds that we have presented above only relate the randomness and expected Bell violation of a pair of quantum devices for a single use of the devices. In an actual protocol for DIRNG, however, the devices are used $n$ times in succession. A typical protocol consists of three main steps \cite{Colbeck07,PAMB+10,CK11}. A measurement step, where the successive pairs of inputs $(x_1,y_1),\ldots,(x_n,y_n)$ are used in the devices, yielding a sequence of outputs $(a_1,b_1),\ldots,(a_n,b_n)$. An estimation step, where the raw data is used to estimate a Bell parameter (if this parameter is too low, the protocol may abort). A randomness extraction step, where the raw output string is processed to  obtain a smaller final string $r=r_1,\ldots,r_m$ which is uniformly random and private with respect to any potential adversary. In addition to the Bell violating devices, the protocol may also consume some initial random seed for choosing the inputs in the measurement step and for processing the raw data in the randomness extraction step. If more randomness is generated than is initially consumed, one has then achieved DI \emph{randomness expansion}.

In \cite{PAMB+10}, a generic family of protocols based on arbitrary Bell inequalities and achieving quadratic expansion are introduced. These protocols are robust to noise and generate randomness for any amount of violation (up to statistical errors). The analysis of the randomness that is produced is based on an extension of the single-copy bounds of the form (\ref{qrand}) and (\ref{nsrand}) to the $n$-copy case. 
A proof-of-principle implementation using two entangled atoms separated by about 1 meter was also reported (see Section~\ref{Experimental aspects}). The  security of these protocols has been proven against quantum or no-signaling adversaries with \emph{classical-side} information. The technical tools for proving security were already introduced in \cite{PAMB+10}, but this was rigorously established only in \cite{FGS11,PM11}. In these later works, it was further shown how to achieve superpolynomial  randomness expansion by repeatedly using the randomness of a pair of devices as input for another pair. 
A scheme based on the CHSH inequality secure against adversaries with \emph{quantum-side} information and achieving superpolynomial expansion with a single pair of quantum devices was obtained in \cite{VV12}. This scheme, though, requires a high violation of the CHSH inequality and is not noise-tolerant. 

The security of the above protocols relies on a series of minimal assumptions. First, the devices and the eavesdropper are constrained by quantum theory, or at least by the no-signaling principle. Second, the initial randomness seed is independent and uncorrelated from the devices' behavior. Third, the two quantum devices are non-interacting during each successive measurements~\footnote{Note that this does not necessary imply that the measurements should be \emph{space-like} separated in the relativistic sense. This space-like separation is required to close the locality loophole in fundamental tests of Bell inequalities, where the aim is to rule out alternative models of Nature that can go beyond present-day physics. In the context of 
DIRNG, we assume however from the beginning the validity of quantum theory and use Bell inequalities as a tool to quantify in a DI way the randomness of quantum theory.  Once we assume quantum theory, they are many ways to ensure that the two systems are not interacting other than placing them in space-like intervals, e.g. by shielding the devices. See Supplementary Information C of \cite{PAMB+10} for a more extensive discussion}.  Fourth, it is also implicit of course that the devices can be secured, in the sense that they do not leak directly unwanted information to the adversary. Apart from these basic requirements, the devices are mostly uncharacterized. In particular, no assumptions are made on the specific  measurements that they implement, on the quantum state that is being measured or on the Hilbert space dimension, etc. 

The level of  confidence in the realization of the above assumptions in an actual implementation or the measures that must be taken to enforce them may vary depending on the adversary model that one is considering. For instance, it may depend on whether the devices are considered to be outright malicious and programmed by a dishonest provider (i.e. the adversary itself) or whether the manufacturer of the device is assumed to be honest and the concept of DI is merely used to account for limited control of the apparatus or unintentional flaws in the devices \cite{PM11}. In the later case, in particular, a weak source of randomness, such as a pseudo-random generator, may be sufficient for all practical purposes to generate the initial seed (in which case the protocol, which produces strong cryptographically secure randomness, is best viewed as a randomness generation protocol than an expansion one). Note in addition that in the honest-provider scenario, the adversary may be considered to be disentangled from the quantum devices, implying that proving security against classical-side information as in \cite{FGS11,PM11} is already sufficient.

Recently, protocols and security analysis have also been introduced where some of the above assumptions are relaxed. In \cite{SPM12}, the separation assumption is relaxed and a small amount of cross-talk between the devices is allowed. This opens up the possibility of using existent experimental systems with high data rates, such as Josephson phase qubits on the same chip. 

In \cite{CR12}, the problem of \emph{randomness amplification} is introduced, which aims at extracting perfect (or arbitrarily close to perfect) randomness from an initial source that is partly correlated with the devices and the adversary. It is shown that if one is given access to certain so-called Santha-Vazirani (SV) sources, then randomness amplification against an adversary limited only by the no-signaling principle is possible for certain parameters of the source. Improving on this first result, it is shown in \cite{GMTD+12} that an arbitrarily SV source can be amplified using certain multipartite quantum correlations. Finally, less stringent models of a compromised random seed than SV have been considered \cite{Hall11} and the conditions for Bell-based randomness expansion against an i.i.d. adversary has been studied in \cite{KHS+12}.

\subsubsection{Device-independent quantum key distribution}\label{sec:diqkd}
The protocols, the underlying assumptions, and the security proofs for DIQKD are similar in spirit to DIRNG with the added complication that DIQKD involves two remote parties that must communicate over a public channel to establish the shared secret key. 
A typical DIQKD protocol consists of the following steps. A measurement step, where Alice and Bob measure a series of entangled quantum systems. An estimation step, in which Alice and Bob publicly announce a fraction of their measurement results to estimate the violation of a Bell inequality and the error rate in their raw data. An error-correction step, in which these errors are corrected using a classical protocol that involves public communication. Finally, a privacy-amplification step in which a shorter, secure key is distilled from the raw key based on a bound on the eavesdropper's information deduced from the Bell violation estimation.

The first DIQKD protocol proven secure against general attacks by a no-signaling eavesdropper was introduced in \cite{BHK05}. The protocol is based on the chained Bell inequality (\ref{chained}) and produces a single secure key bit. It represents mostly a proof-of-principle result as the protocol is inefficient and unable to tolerate reasonable levels of noise. In \cite{BHK05} security is proven assuming that each of the $n$ entangled pair measured in the protocol is isolated from the other pairs. The protocol thus require that Alice and Bob have $n$ separate pairs of devices, rather than a single pair of devices that they use repeatedly $n$ times. The no-signaling conditions are required to hold between each of the $2n$ systems of Alice and Bob. This assumption is removed in \cite{BCK12}, where security is proven in the situation where Alice and Bob have only one device each, which they repeatedly use. Instead of full no-signaling correlations among the $2n$ systems of Alice and Bob, the security is thus based on \emph{time-ordered} no-signaling conditions, where no-signaling is only required from future inputs to previous inputs, but where later outputs can depend arbitrarily on previous inputs.

Efficient and noise-tolerant protocols have been introduced in \cite{AGM06,SGB+06} (see also \cite{AMP06}) where however the security analysis was restricted to individual attacks against no-signaling eavesdroppers. General security against no-signaling eavesdroppers was later proven in \cite{MRC+09,Masanes09,HRW10} under the assumption, as in \cite{BHK05}, that Alice and Bob use $n$ separated pairs of devices constrained by full no-signaling conditions. 
 The question of whether it is possible to prove the security of an efficient and noise-tolerant protocol in the case where Alice and Bob repeatedly use a single pair of devices constrained by time-ordered no-signaling conditions is still open. One of the difficulties in obtaining such a result is related to the possibility of performing privacy amplification against a no-signaling eavesdropper.  It was shown in \cite{HRW09} that if no-signaling is imposed only between Alice's device and Bob's one, but signaling within each device is allowed (so that the output of a device can depend on the inputs of other devices used later in the protocol), then privacy amplification is not possible for protocols based on the CHSH inequality. This result was further extended in \cite{AHT12} for a set of more general conditions, but still less restrictive than the desired time-ordered no-signaling conditions. Recently, it was shown in \cite{RT12} that super-polynomial privacy amplification for protocols based on the chained inequality is impossible under the assumption of time-order no-signaling conditions. This work still leaves open the question of exponential privacy amplification for protocols based on a different Bell inequality or whether linear privacy amplification is possible. 

Another line of results, concerned with security against eavesdroppers that are constrained by the entire quantum formalism and not only the no-signaling principle, was initiated in \cite{ABGM+07} . The advantage in this case is that better key rates and noise-resistance can be expected (as illustrated by the difference between the randomness bounds (\ref{qrand}) and (\ref{nsrand}) and that privacy amplification is possible and well studied. The works \cite{ABGM+07,PABG+09} proved security of the CHSH-based protocol introduced in \cite{AMP06} against collective attacks by a quantum eavesdropper. This proof was extended to a slightly more general setting in  \cite{McKague10}. General security proofs of protocols based on arbitrary Bell inequalities  under the assumption that the devices of Alice and Bob are memoryless (or equivalently that they use $n$ non-interacting pairs of devices instead of a single one) were introduced in \cite{HR10,MPA11}. The memory assumption on the device was removed in \cite{PMA+12}, but security was only proven against quantum adversaries with classical-side information, a condition that is satisfied if the eavesdropper has only access to short-term quantum memories. The keyrates in \cite{HR10, MPA11, PMA+12} are simple expression expressed in term of single-copy bounds on the randomness of the form (\ref{qrand}). The general security of a CHSH-based protocol with no memory assumptions on the devices or the eavesdropper was reported in \cite{RUV12,RUV12l}, albeit it is polynomially inefficient and does not tolerate noisy devices. The security is obtained as a corollary of a more general strong testing result that allows the shared quantum state and operators of the two untrusted devices to be completely characterized. 
Finally, a complete DI proof of security of QKD that tolerates a constant noise rate and guarantees the generation of a linear amount of key was given in \cite{VV12b} for a protocol that is a slight variant of Ekert's  protocol. It is an open question whether this approach can lead to trade-offs between noise rate and key rate as good as the ones that have been shown to be achievable under additional memory assumptions on the devices or the eavesdropper.

The general assumptions that underly the above proofs are similar to the ones for DIRNG: the validity of quantum theory or the no-signaling principle, access to a random seed independent of the devices and the eavesdropper, a separation assumption on the behaviour of the devices, and the implicit assumption that the devices do not leak out directly unwanted information to the eavesdropper. Apart from that, the devices are mostly uncharacterized and no assumptions are made on the Hilbert space dimension, the specific measurements that are implemented, and so on. 

Note that in the dishonest-provider scenario where the devices are outright malicious and assumed to have been prepared by the eavesdropper, repeated implementations of a protocol using the same devices can render an earlier generated key insecure due to device-memory-based attacks \cite{BCK12p}. In such attacks, untrusted devices may record their inputs and outputs and reveal information about them via publicly announced outputs during later implementations of the protocol. See \cite{BCK12p} for a thorough discussion of the general scope of such attacks, including the possibilities of countering them by refined protocols. A countermeasure relying on an  encryption scheme which allows Alice and Bob to exchange data without the devices leaking information about previously generated keys to Eve is presented in \cite{MS12}. 

Finally, let us say a few words about experimental perspectives for DIQKD. The implementation of a DIQKD protocol requires a genuine Bell violation over large distances. Genuine, here, means with the detection loophole closed (at least if one is considering complete DI with no further assumptions on the devices), see subsection~\ref{det loop}. Transmission losses in optical fibres, however, represent a fundamental limitation for the realization of a detection-loophole free Bell test on any distance relevant for QKD. Approaches to circumvent the problem of transmission losses have been proposed based on heralded qubit amplifiers \cite{GPS10,PMW+11} and standard quantum relays based on entanglement swapping with linear optics \cite{CM11}, but an experimental demonstration still represents a great challenge. Quantum repeaters may also provide a possible solution. More recently, another approach based on spin-photon interactions in cavities was also  discussed \cite{BYHR13,MBA13}. Improved data post-processing has also been proposed to increase the tolerance to lost photons \cite{ML12}.

\subsection{Other device-independent protocols}

In a quantum experiment, the violation of a Bell inequality reveals the presence of entanglement in a device-independent way. In fact, in some cases a much stronger statement can be made. Certain quantum correlations can only be reproduced by performing specific local measurements on a specific entangled state. Hence the observation of such correlations allows one to characterize an unknown source of quantum states, as well as the measurement devices, in a device-independent manner.
For instance, the observation of maximal violation of the CHSH inequality implies that the underlying quantum state is necessarily equivalent to a two-qubit singlet state \cite{Cirelson80}. Moreover, the measurement settings of both Alice and Bob must anticommute \cite{PR92,BMR92}. Another method, developed in \cite{MY04}, allows one to reach the same conclusion. Such procedures are termed \emph{self-testing} of the singlet state.

More formally, these works show the following. Consider an experiment involving a state $\ket{\psi}$ and measurement operators $M_A^i$ and $M_B^j$, with $i,j=1,2$. If a CHSH value of $S =2\sqrt{2}$ is achieved, then the state is equivalent (up to local isometries) to a singlet state $\ket{\psi_-}$ and the measurement are to anti-commuting Pauli operators $\sigma_A^i$ for Alice with $\{\sigma_A^i,\sigma_A^k\}=2\delta_{ik}\openone$ (and similarly for Bob $\sigma_B^j$), in the sense that 
\ba  \Phi( \ket{\psi}) &=& \ket{junk} \otimes \ket{\psi_-}  \\  \Phi(M_A^i M_B^j \ket{\psi}) &=& \ket{junk} \otimes  \sigma_A^i \sigma_B^j \ket{\psi_-}\ea
where $\Phi = \Phi_A \otimes \Phi_B$ is a local isometry, and $\ket{junk}$ a state shared by Alice and Bob.

For a self-testing protocol to be practical it should be robust to small deviations from the ideal case, due for instance to experimental imperfections.
A first proof of the robustness of the Mayers-Yao scheme was derived in \cite{MMMO06}, later considerably simplified in \cite{MM11}. In \cite{MYS12}, a framework for studying the robust self-testing of the singlet state was presented, which can be used to certify device-independently the entanglement fraction of a source \cite{BLM+09}.
More generally, it was shown in the ground-breaking work \cite{RUV12,RUV12l} that self-testing can be achieved in the CHSH scenario even if the devices feature a quantum memory. Loosely speaking, this means that the only way to achieve a violation of the CHSH inequality close to $2\sqrt{2}$, is if the measured bipartite states are close to the tensor product of singlet states, and the measurements are the optimal CHSH measurements.

Self-testing of other quantum states was also discussed. In particular, the case of partially entangled bipartite states was addressed in \cite{YN12}. In the multipartite setting, the case of graph states was discussed \cite{McKague10b}, while \cite{MS12b} considered self-testing in XOR games. Also, the device-independent certification of ``entangled measurements" was investigated \cite{RHCB+11,VN11}.

An interesting development of these ideas is the possibility of self-testing a quantum computation. This consists in self-testing a quantum state and a sequence of operations applied to this state. This approach was introduced in \cite{MMMO06}. A full analysis of such a protocol, with a reduced set of assumptions compared to \cite{MMMO06}, has been recently given in \cite{RUV12}.

Moving away from self-testing, an interesting development is the device-independent assessment of multipartite quantum entanglement. Notably, techniques for devising device-independent witnesses of genuine multipartite entanglement \cite{BGLP11} were developed. Moreover, \cite{BSV12} discussed how the structure of multipartite entangled states can be characterized using Bell inequalities, that is, how different classes of multipartite entangled states can be distinguished from each other from their nonlocal correlations.

\section{Information-theoretic perspective on nonlocality}
\label{Information}

As we have seen in the previous section, nonlocality can be seen as a resource for information processing and communication tasks and the no-signaling principle plays a fundamental role in this respect. We have also seen in Section~\ref{secgeom} that there exist no-signaling correlations that are more nonlocal than those of quantum theory, as pointed out in \cite{PR94}. If Alice and Bob had access to such PR-boxes they could implement many of the protocols discussed earlier, from communication complexity to cryptography, often with much higher efficiency than what quantum correlations allow \cite{vanDam05}. No-signaling non-local correlations can thus be viewed as information-theoretic resources and investigated as such \cite{BLMP+05}. This new perspective raises two general questions: Can we develop a resource theory of non-locality, similar to the resource theory of entanglement? What distinguishes quantum correlations from more general no-signaling correlations in this information-theoretic context? To answer them it is first useful to identify the physical properties which are generic to all no-signaling non-local theories. 

\subsection{Properties of no-signaling correlations}\label{properties of ns}

Remarkably, it turns out that many features of quantum mechanics, usually thought as counter-intuitive and genuinely quantum, are in fact general features of any no-signaling theory featuring nonlocality \cite{MAG06,Barrett07}. These include a no-cloning theorem, the monogamy of correlations, a disturbance versus information gain trade-off in measurements, the inherent randomness of measurement outcomes, the complementarity of measurements and uncertainty relations. These physical properties are clearly relevant from an information-theoretic point of view, think for instance about the role that the no-cloning theorem or the monogamy of entanglement play in quantum information science. The fact that such properties are generic to all no-signaling nonlocal theories thus already suggests that such theories offers interesting possibilities for information processing.

We have already given the intuition in subsection~\ref{sec:QKD} of why measurement outcomes must be random in any non-local no-signaling theory. Let us illustrate some of the other above properties with simple examples based on Popescu-Rohrlich type correlations. Consider that Alice and Bob share a PR-box, i.e. correlations of the form
\begin{equation}\label{PR}
p(ab|xy) =
\begin{cases}
\frac{1}{2} & \text{$ a \oplus b = xy$} \\
0 & \text{otherwise}
\end{cases}
\end{equation} where $\oplus$ is addition modulo 2, and $x,y \in \{0,1\}$ denote the inputs and $a,b \in \{0,1\}$ the outputs.
The impossibility of having a perfect cloning machine is here easily derived by contradiction. Assume such a machine exists. Then Bob could apply it to its subsystem, resulting in a tripartite probability distribution $p(ab_1b_2|xy_1y_2)$ satisfying the relations
\ba\label{properties} a \oplus b_1=xy_1  \quad , \quad a \oplus b_2=xy_2 \ea
with $a,b_1,b_2$ locally uniformly distributed.
Combining the above relations leads to
\ba b_1 \oplus b_2 = x (y_1 \oplus y_2) \ea
showing that Bob's marginal probability distribution directly depends on $x$, the input of Alice, when Bob uses inputs such that $y_1\oplus y_2=1$. Thus, Alice can signal to Bob, which contradicts our basic hypothesis that the theory is non-signaling. Therefore, we conclude that a perfect cloning machine cannot exist in a theory featuring PR-box correlations. General and rigorous proofs can be found in \cite{MAG06,Barrett07}. The impossibility of broadcasting no-signaling nonlocal correlations has been discussed in \cite{BBLW07,JGHH+13}.

The above simple example also indicates that no-signaling correlations are constraint by monogamy relations (see Section~\ref{sec:monogamy}). In particular, a PR-box being an extremal point of the no-signaling set must be decoupled from any other system~\cite{BLMP+05,MAG06}. 

As a last example, let us illustrate the existence of a notion of complementarity of measurements in generalized non-signaling theories \cite{MAG06}. Considering again PR-box correlations, the two possible measurements on Bob's side (corresponding to $y=0$ and $y=1$) cannot be compatible, that is, there cannot be a single joint measurement $Y$ returning outcomes $b_0$ and $b_1$ corresponding respectively to $y=0$ and $y=1$. Indeed, this would imply the existence of a distribution $P(ab_0b_1|xY)$ satisfying $b_0\oplus b_1=x$ (since $a \oplus b_0=0$ and $a\oplus b_1=x$), thus violating no-signaling as in the above example. 

\subsection{Nonlocality measures, interconversion and distillation}
If nonlocal boxes can be viewed as a information-theoretic resource, can we define a theoretical framework, analogous e.g. to the framework that has been developed for the study of entanglement, that would allow us to answer unambiguously questions such as, can two given sets of nonlocal correlations be considered equivalent resources, or, what is a good measure of nonlocality?

A prerequisite for addressing these issues is to understand
interconversion between nonlocal boxes, that is, the simulation of
a given nonlocal box using a supply of other nonlocal boxes. In
this context, separated parties are allowed to perform local
operations on their boxes. They can relabel the inputs and
outputs, and also `wire' several boxes, using for instance the
output of one box as an input for another box. Importantly,
classical communication is not allowed, as it represents a
nonlocal resource, which allows trivially for the simulation of any nonlocal box. 

The inter-conversion of bipartite boxes has been studied in
\cite{BLMP+05,JM05,FW11} and is by now relatively well understood.
The main conclusion to be drawn from these works is that the PR
box represents a good unit of bipartite nonlocality---much like
the singlet in the case of entanglement--- in the sense that any
bipartite no-signaling box can be simulated to an arbitrary
precision using a supply of PR boxes \cite{FW11}. In the
multipartite case, the situation is more complicated. On the one
hand several classes of extremal nonlocal boxes can be simulated
exactly using PR boxes \cite{BLMP+05,BP05}. On the other hand
there exist non-signaling boxes which can provably not be
approximated using an arbitrarily large supply of PR boxes
\cite{BP05,PBS11}. In particular, there exist quantum non-local correlations with this property \cite{BP05}. 
 It is still an open question whether there
exists a unit of multipartite nonlocality; in fact even proposing
a good candidate is challenging given the complexity of the set of
multipartite non-signaling correlations (see Section~\ref{multi}).

Another relevant issue is whether nonlocality can be distilled. That is, from a supply of weakly nonlocal boxes is it possible to obtain via local operations (i.e. relabelings and wirings) one copy of a box featuring more nonlocality, in the sense that it violates more a given Bell inequality than the original boxes.  Interestingly, nonlocality distillation is possible for certain classes of nonlocal boxes \cite{FWW09}. Moreover, maximally nonlocal PR box correlations can be distilled out of certain boxes with arbitrarily weak nonlocality \cite{BS09}, i.e. violating a Bell inequality by an arbitrarily small amount. The existence of such distillation protocols has important consequences from an information-theoretic point of view. For instance, if a certain class of boxes can be distilled to a PR box, then all boxes in this class inherit the information-theoretic power of the PR box. Note also a series of negative results, concerning in particular the impossibility of distilling isotropic nonlocal correlations. Such correlations---mixtures of PR box and white noise---are of particular importance, since any nonlocal box can be `depolarized' to an isotropic without decreasing its nonlocality \cite{MAG06}. Partial no-go theorems have been derived \cite{Short09,DW08,Forster11}, but a full proof is still missing.

These developments have opened novel possibilities for defining natural measures of nonlocality, such as the `distillable nonlocality' \cite{FWW09,BCSS11} of a nonlocal box, the maximal amount of nonlocality that can be extracted from an arbitrarily large supply of such boxes.
First steps towards establishing a more general resource theory of nonlocality have recently been taken \cite{BCSS11,GWAN11b}.

Finally, it is interesting to look for sets of correlations which are invariant under local operations. A set is said to be \emph{closed under wirings} if, by combining correlations of this set via local operations, it is impossible to generate correlations outside the set. The study of such sets was initiated in \cite{ABLP+09}. Clearly the sets of local, quantum, and no-signaling correlations are all closed under wirings. Finding other closed sets appears to be a nontrivial problem. An interesting open problem is whether there exist, in the CHSH scenario, a strict subset of the no-signaling polytope, that is closed under wirings and features more nonlocality than quantum mechanics (i.e. violating Tsirelson's bound).

\subsection{Consequences of superstrong non-locality}

The existence of no-signaling correlations stronger than quantum mechanical ones raises fundamental questions.
Why is nonlocality limited in quantum theory? Would there be unlikely consequences from a physical or information-theoretical point of view if supra-quantum correlations were available? Can we identify reasonable principles that allow us to characterize the boundary that separate quantum from supra-quantum correlations? 
The works discussed below address such questions. The first sections below deal with information-theoretic consequences of supra-quantum nonlocality and the second one is based on more physical concepts.

\subsubsection{Information-theoretic consequences}

\paragraph{Communication complexity and nonlocal computation.} 
A first result showing a sharp difference between quantum and super-quantum correlations in their capability of performing information-theoretic tasks was given in \cite{vanDam05} in the context of communication complexity. 
As discussed in subsection~\ref{sec:commcompl}, communication complexity deals with the problem of determining the number of bits that Alice and Bob need to exchange to compute the value $f(x,y)$ of a function whose inputs $x$ and $y$ are distributed among Alice and Bob. The amount of communication that is required depends on the particular function $f$ and the resources that are available to Alice and Bob. Consider binary (or boolean) functions $f(x,y): \{0,1\}^n\times\{0,1\}^n\rightarrow\{0,1\}$ taking $n$-bit strings $x=x_1\ldots x_n$ and $y=y_1\ldots y_n$ as inputs. It is proven that some of these functions have high communication complexity, basically Alice must send her entire bit string $x$ to Bob, even if Alice and Bob are allowed to share unlimited prior entanglement. An example of such a function is the inner product function $f(x,y)=x\cdot y = \sum_i x_iy_i$ \cite{CDNT98}. In contrast, if unlimited PR-boxes were available to Alice and Bob, then a \emph{single} bit of classical communication from Alice to Bob is sufficient for Bob to evaluate \emph{any} binary function, that is, communication complexity collapses. 

Consider again the inner product function. Suppose that Alice and Bob share $n$ PR boxes and receive inputs $x$ and $y$. They input $x_i$ and $y_i$ in box $i$, and then get outcomes $a_i$ and $b_i$ satisfying $a_i \oplus b_i = x_i y_i$. The inner product function can be expressed as
$$f(x,y)=\sum_i x_i y_i=\sum_i a_i\oplus b_i=\underbrace{\sum_i a_i }_{Alice's~side}\oplus \underbrace{\sum_i b_i}_{Bob's~side}.$$
Thus Alice can compute locally $c=\sum_i a_i$, and send the single
bit $c$ to Bob who then outputs $c \oplus b$, where $b=\sum_i
b_i$, which is indeed the inner product. The inner
product function is of particular importance, since any binary
function $f$ can be decomposed into inner products, from which the
result of van Dam follows.

This idea was later generalized to the context of probabilistic
communication complexity where Alice and Bob must compute $f$ with
a minimum probability of success \cite{BBLM+06}. It was shown that
certain noisy PR boxes, with CHSH value $S>4 \sqrt{2/3} \approx 3.266$ make communication complexity trivial in
this scenario. Finally, using nonlocality distillation, it can be shown that (non-quantum) boxes with an arbitrarily small amount of nonlocality
can nevertheless collapse communication complexity \cite{BS09}.

In \cite{LPSW07}, a task closely related to communication complexity, termed \emph{nonlocal computation}, was introduced. The binary function $f$ that Alice and Bob must compute has the special form $f(x,y)=g(x\oplus y)=g(z)$ where $g(z)$ is a boolean function taking as input an $n$-bit string $z$ (with $z_i= x_i \oplus y_i$ and $x_i$ uniform for $i=1,..,n$). Thus each party has locally no information about the function's input $z$. Alice and Bob are asked to output one bit, respectively $a$ and $b$, such that $a\oplus b =f(x,y)=g(z)$. The figure of merit is then the average success probability of Alice and Bob. While strategies based on quantum correlations offer no advantage over classical ones for the nonlocal computation of an arbitrary function, it turns out that certain super-quantum correlations provide an advantage. Remarkably, if one consider as a function the nonlocal AND of two bits $g(z_1,z_2)=z_1z_2$, then the limit at which noisy PR-boxes stop providing an advantage over classical and quantum correlations corresponds exactly to Tsirelson's bound. Note, however, that when the distribution of inputs is not perfectly uniform, i.e. when Alice and Bob have partial knowledge (even arbitrarily small) about the function's input $z$, quantum correlations provide an advantage over classical ones \cite{ABL09}.

\paragraph{Information-causality.}
Suppose Alice sends an $m$-bit message to Bob. How much information is potentially available to Bob? A natural guess is that the amount of information potentially available to Bob is equal to what he receives, that is, $m$ bits. This is in essence the principle of information causality: the amount of information potentially available to Bob about Alice's data is not higher than the amount of information Alice sends to him \cite{PPKS+09}.  While information causality is satisfied in both classical and quantum physics, this is in not the case in general, if supra-quantum correlations were available. Hence information causality can be viewed as a strengthening of the no-signaling principle.

To see how super-quantum correlations can violate information-causality, suppose that Alice is given two classical bits $x_0$ and $x_1$, uniformly distributed. Bob is interested in learning one of these two bits, but Alice does not know which one. To make the task nontrivial, Alice is allowed to send only one bit to Bob. Can they devise a protocol such that Bob can always retrieve the desired bit? In a scenario where Alice and Bob share only classical or quantum correlations, the answer is no. However, if Alice and Bob share a PR box, the task becomes possible \cite{WW05}. Alice first inputs $x_0\oplus x_1$ in her end of the PR box, and gets outcome $a$. She then sends the 1-bit message to Bob: $m=a\oplus x_0$. Bob, who is interested in bit $x_k$ of Alice, inputs $k$ on his end of the PR box, and gets outcome $b$. Upon receiving Alice's message $m$, Bob makes his guess $G=b\oplus m=x_k$. Hence, Bob's guess is always correct.

The principle of information causality allows to recover part of the boundary between quantum and super-quantum correlations \cite{PPKS+09,ABPS09}. Notably, any theory that allows for the violation of Tsirelson's bound violates information causality.

Finally, note that an extension of information causality was recently formulated for quantum information \cite{Pitula13}.

\paragraph{Limitations on multipartite correlations.}
The principles discussed above focus on bipartite correlations. A non-local game termed \emph{guess your neighbour's input} was introduced in \cite{ABBA+10}, which reveals an intriguing separation between quantum and super-quantum correlations in a multipartite context. Consider $n$ distant parties placed on a ring. Each party $i$ is given an input bit $x_i$ according to a joint prior probability distribution $p(x_1\ldots x_n)$. As the name of the game suggests, each party is then asked to give a guess $a_i$ of his right neighbor's input, \ie such that $a_i=x_{i+1}$ for all $i=1...n$. Since a high probability of success at this game would lead to signaling, it is not surprising that quantum resources provide no advantage over classical ones, for any distribution of the inputs. However, it turns out that certain no-signaling super-quantum correlations outperform classical and quantum strategies for certain distributions of the inputs. Remarkably, some of these games correspond to facet Bell inequalities. Hence `Guess your neighbour's input' identifies a portion of the boundary of the quantum set which is of maximal dimension. Moreover, this quite innocuous game has several rather surprising applications, related to generalizations of Gleason's 
theorem \cite{AACC+10,bbb+09} and to unextendible product basis \cite{ASHK+11}.

The motivation for many of the results discussed above is to identify general properties or a set of principles that potentially single out quantum correlations. In \cite{GWAN11}, it was shown that any such principles must be genuinely multipartite. More specifically, there exist tripartite super-quantum correlations which are local among every possible bipartition (even if many copies of them are available and wirings are performed) \cite{GWAN11,YCAT+12}. Thus, no bipartite principle can ever rule out these correlations. Such super-quantum correlations can nevertheless be ruled out by a novel principle termed `local orthogonality' \cite{FSAB+12}, inspired from the game of `Guess your neighbour's input'.

\subsubsection{Physical consequences}
\paragraph{Macroscopic locality.}
Loosely speaking, macroscopic locality is a principle requiring that nonlocal correlations admit a classical limit. More specifically, in a Bell test involving a large number of pairs of particles, the statistics of coarse-grained measurements (not resolving discrete particles) should admit an explanation in terms of a local model, \ie should not violate any Bell inequality \cite{NW09}. This is the case in quantum mechanics \cite{NW09,BBBG+08}, but not in general no-signaling theories. Notably, the set of correlations satisfying macroscopic locality can be completely characterized. It corresponds to the set $Q_1$, the first approximation to the set of quantum correlations in the hierarchy of semi-definite programs  \cite{NPA07} discussed in subsection \ref{hierarchy}. This set  is however strictly larger than the quantum set. Thus, there are super-quantum correlations that still satisfy macroscopic locality. It was shown in \cite{YNSS11} that analytical quantum Bell inequalities can be derived from macroscopic locality. Finally, note that there exist correlations satisfying macroscopic locality which nevertheless violate information causality \cite{CSS10}.

\paragraph{Uncertainty and information.}
	In~\cite{WCD08} it was shown that one can reformulate any Bell inequality in the language of information, which for projection nonlocal games (see Section~\ref{sec:projection}) works as follows. For every question $x$ and answer $a$ of Alice, one can write down a string $s_{x,a} = (s_{x,a}^{(1)},\ldots,s_{x,a}^{(m)})$ where $s_{x,a}^{(y)} = b$ is the answer that Bob must return for question $y$ in order for them to win the game. Written in this way, one may think of the state of Bob's system conditioned on Alice measuring $x$ and obtaining outcome $a$ as an encoding of the string $s_{x,a}$ from which Bob must retrieve entry $s_{x,a}^{(y)}$ correctly. In~\cite{OW10} it was furthermore shown that for any physical theory uncertainty relations can be understood as imposing limits on how well we can retrieve information from an encoding. This information-theoretic perspective is the essential idea behind the relation between nonlocality and uncertainty found in~\cite{OW10}, which holds for \emph{any} physical theory. It should be noted that the aim of~\cite{OW10} is not to derive limits on nonlocality by appealing to intuitive notions on how we expect information to behave, but rather to link it to another concept already existing within quantum mechanics.

\paragraph{Local quantum mechanics.}
In~\cite{bbb+09,AACC+10} it was shown that the correlations of bipartite systems that can be described locally by quantum mechanics, cannot be stronger than quantum correlations. More precisely, if the no-signaling principle holds, and Alice and Bob are locally quantum, then all possible correlations between them admit a quantum mechanical description. However, the situation is different in the multipartite case. There exist tripartite correlations which are locally quantum, which are nevertheless stronger than any quantum correlations~\cite{AACC+10}.

\subsection{Nonlocality in generalized probabilistic theories}
The idea of investigating the information-theoretic power of non-local correlations more general than quantum ones led, following work of \cite{Hardy01} and \cite{Barrett07}, to a very active line of research in which information processing has been considered in the broader framework of ``general probabilistic theories" (GPT) or ``convex-operational" formalism. This framework allows one to define full-fledged theories (\ie \,that include notions of states, evolution, measurements, and not only ``correlations") in which classical and quantum theories are merely two special cases. Given such a formalism, one can compare and contrast quantum theory with other alternative theoretical models. The hope is to better understand quantum theory and identify in what ways it is special. To date, much work has focused on information processing in GPT, investigating for instance cloning, broadcasting, teleportation, or entanglement swapping. Even if these works connect and partly overlap with many of the issues mentioned above, we do not review this very fruitful work here as it does not directly take nonlocality as a starting point. We refer instead to \cite{BW12} for a short review. In what follows, we only mention work that explicitly consider Bell nonlocality in the context of GPT. 

In \cite{SW09}, it was shown that super-strong random access encodings exist in certain theories that violate the CHSH inequality beyond Tsirelson's bound. A quantum random access code is an encoding of an $n$-bit string $x=x_1,\ldots,x_n \in \{0,1\}^n$ into a quantum state $\rho_x \in \mathcal{B}(\mathcal{H})$ such that each bit $x_j$ can be retrieved from $\rho_x$ with some probability $p_j$. In~\cite{nayak99} it was shown that if the state has dimension at most $\dim(\mathcal{H}) = d$, then the success probabilities are bounded
	as $\sum_{x} (1 - h(p_j)) \leq \log d$, where $h(p) = - p \log_2 p - (1-p) \log_2 (1-p)$ is the binary entropy. In~\cite{SW09} it was shown that this inequality can be violated for some theories that allow stronger than quantum correlations, i.e., super-strong random access encodings exist in such theories. In particular, there exists generalized ''states'' in a Hilbert space of dimension $d$ which effectively contain more than $d$ bits of information.

In \cite{JGBB11} it is shown that there is a connection between the strength of nonlocal correlations in a physical theory and the structure of the state spaces of individual systems. In particular, a class of GPTs is presented that allows to study the transition between classical, quantum and super-quantum correlations by varying only the local state space. It is shown that the strength of nonlocal correlations depends strongly on the geometry. As the amount of uncertainty in a theory bounds the geometry of the state space, this provides a very nice insight into~\cite{OW10}.
An intriguing consequence of these results is the existence of models that are locally almost indistinguishable from quantum mechanics, but can nevertheless generate maximally nonlocal correlations \cite{JGBB11}.

\section{Multipartite nonlocality}\label{sec:multipartite}
\label{Multipartite}

In the multipartite case, nonlocality displays a much richer and more complex structure compared to the case of two parties. This makes the study and the characterization of multipartite nonlocal correlations an interesting, but challenging problem. It comes thus to no surprise that our understanding of nonlocality in the multipartite setting is much less advanced than in the bipartite case.

The study of multipartite nonlocality was initiated by the ground-breaking work of \cite{Svetlichny87}. In this paper, the author introduced the concept of genuine multipartite nonlocality, derived a Bell-type inequality for testing it, and showed that that this strong form of nonlocality occurs in quantum mechanics. Later, in particular with the advent of quantum information science, the concepts and tools introduced by Svetlichny were further developed.

In this section, we will start by defining various notions of multipartite nonlocality (with a particular focus on genuine multipartite nonlocality) and discuss the detection of multipartite nonlocality. Next, we discuss the notion of monogamy of nonlocality, which limits nonlocality between different subsets of parties. Finally, we discuss the nonlocality of multipartite quantum systems.

\subsection{Defining multipartite nonlocality}\label{Defining multipartite nonlocality}
The notion of Bell nonlocality that we have introduced in Sections~\ref{Introduction} and \ref{secgeom}) in the case of two separated observers readily extends to three or more observers. For simplicity, we will consider in this Section the case of three separated observers, Alice, Bob and Charlie. Their  measurement settings are denoted $x,y,z$ and their outputs by $a,b,c$, respectively. The experiment is thus characterized by the joint probability distribution $p(abc|xyz)$. We say that these correlation are local if they can be written in the form
\ba \label{locality}  p(abc|xyz) = \int d \lambda q(\lambda)p_\lambda(a|x)p_\lambda(b|y)p_\lambda(c|z) \ea
where $\lambda$ is a shared local random variable and $ \int d \lambda q(\lambda)=1$; and that they are non-local otherwise. This represents the natural generalization of Bell's locality condition \eqref{intro.loc} to the multipartite case. The set of correlations that can be written in the form \eqref{locality} is denoted $\mathcal{L}$.

However, in the multipartite case, there exist several possible refinements of this notion of nonlocality. For instance consider a joint distribution of the form $p(abc|xyz)=p(ab|xy)\times p(c|z)$, i.e., Charles is uncorrelated to Alice and Bob. These correlations can clearly violate the locality condition \eqref{locality} if $p(ab|xy)$ is nonlocal, though no nonlocality at all is exhibited between Alice-Bob and Charles. In other words, such correlations only exibibit bipartite nonlocality. In contrast, one can consider situation where all three parties are nonlocally correlated. This is referred to as \emph{genuine} multipartite nonlocality, which represents the strongest form of multipartite nonlocality. The main purpose of this section is to discuss the problem of defining formally, in the spirit of Bell's definition, this concept of genuine multipartite nonlocality.

\subsubsection{Genuine multipartite nonlocality \`a la Svetlichny}
The first definition of genuine multipartite nonlocality was proposed in \cite{Svetlichny87}. To describe it suppose that $p(abc|xyz)$ can be written in the form
\ba\label{svet}  p(abc|xyz) &=&  \int d \lambda q(\lambda)p_{\lambda}(ab|xy)p_\lambda(c|z)  \nonumber \\
&+&  \int d \mu q(\mu)p_{\mu}(bc|yz)p_\mu (a|x) \\ \nonumber
&+&  \int d \nu q(\nu)p_{\nu}(ac|xz)p_\nu(b|y)  \ea
where  $\int d \lambda 	q(\lambda)+\int d \mu q(\mu)+ \int d \nu q(\nu)=1$. This represents a convex combination of three terms, where in each term at most two of the parties are nonlocally correlated. For instance, the term $\int d \lambda q(\lambda)p_{\lambda}(ab|xy)p_\lambda(c|z)$ represents correlations where Charles is locally correlated (through the hidden variable $\lambda$) with the joint system of Alice and Bob. The correlations between Alice and Bob, however, are arbitrary, and in particular can be nonlocal. Operationally, we can think of such correlations as describing a situation where Alice and Bob are free to share arbitrary nonlocal resources between themselves or are able to communicate freely, while they are prevented to do so with Charles. The convex combination (\ref{svet}) thus represents a situation where only two parties share a nonlocal resource or communicate  in any measurement run. We say that they are $2$-way nonlocal. 
On the other hand, if $p(abc|xyz)$ cannot be written in the above form, then necessarily the three parties Alice, Bob, and Charles must share some common nonlocal resource. We then say that they are $3$-way nonlocal or genuinely tripartite nonlocal.
Detecting such a form of multipartite nonlocality is obviously an important issue. As for detecting standard nonlocality, it is possible to write down Bell inequalities, the violation of which guarantee that the correlations are genuinely multipartite (see Section~\ref{detection multi}). 

Operationally, we can define local correlations as those that can be generated by separated classical observers that have access to share randomness but who cannot communicate, $2$-way correlations as those where arbitrary communication is allowed between two parties, and $3$-way as those where arbitrary communication is allowed between all parties. One can also consider more refined definitions based on more general communication patterns (particularly in the multipartite case with a large number of parties). For instance, we can consider the case where Alice is allowed to communicate to Bob and to Charles, while Bob and Charles cannot communicate to anyone. Such generalizations of Svetlichny's approach were considered in \cite{JLM05} and in \cite{BBGP09}. 

While Svetlichny's notion of genuine multipartite nonlocality is often used in the literature, it has certain drawbacks discussed below.

\subsubsection{Beyond Svetlichny's model}\label{timeorder}
In Svetlichny's definition of genuine multipartite nonlocality, parties that are allowed to share non-local resources can display arbitrary correlations. In particular, this includes signaling probability distributions. For instance, considering again the above tripartite example, the bipartite probability distributions, e.g. $p_{\lambda}(ab|xy)$, entering decomposition \eqref{svet} are unconstrained, apart from normalization.
 In particular, this means that we have not imposed the no-signaling constraints:
\ba\label{NS_BtoA} p_{\lambda}(a|xy)  &=& p_{\lambda}(a|xy') \quad \forall a,x,y,y' \\ \label{NS_AtoB}
p_{\lambda}(b|xy)  &=& p_{\lambda}(b|x'y) \quad \forall b,x,x',y \ea
where $p_{\lambda}(a|xy) = \sum_b p_{\lambda}(ab|xy)$ is Alice's marginal probability distribution, and similarly for Bob. These conditions guarantee that, even given the knowledge of $\lambda$, Alice cannot send a message to Bob by choosing her measurement setting, and vice versa. If at least one of the above constraints is not satisfied, then this allows for signaling. Signaling from Alice to Bob occurs when condition \eqref{NS_AtoB} is not satisfied. Similarly, signaling from Bob to Alice occurs when \eqref{NS_BtoA} is not satisfied.

Such signaling terms in Svetlichny's definition \eqref{svet} are inconsistent from a physical perspective (they lead to grandfather-type paradoxes) as well as from an operational point of view \cite{BPBG11,GWAN11b}.
To give a rough idea of why this is so (see \cite{BPBG11,GWAN11b}) for more details), consider, for instance, Svetlichny's definition from the perspective of classical simulations of quantum correlations in terms of shared random data and communication. The decomposition (\ref{svet}) correspond to simulation models where all parties receive their measurement setting at the same time, then there are several rounds of communication between only two of the parties, say Alice and Bob, and finally, all parties produce a measurement outcome. During the communication step, Alice and Bob can establish arbitrary correlations in Svetlichny's model, in particular they can violate the two above no-signaling conditions. But consider now a slightly different simulation model where measurements are given to the parties in a sequence that is arbitrary and not fixed in advance.  Upon receiving a measurement setting, a party must produce
an output immediately, as it happens when measuring a real quantum state. But then if Alice received her measurement choice before Bob, she must determine her output without having received any communication from Bob and thus the condition (\ref{NS_BtoA}), imposing no-signaling from Bob to Alice, cannot be violated. If in an other round, it is Bob that receives his measurement before Alice, then it is the condition (\ref{NS_AtoB}), imposing no-signaling from Alice to Bob, that cannot be violated. 

To address such shortcomings of Sveltichny' definition, there are two alternatives. The most immediate one is to require that all bipartite correlations, e.g. $p_{\lambda}(ab|xy)$, appearing in the decomposition (\ref{svet}), to satisfy the no-signalling conditions \cite{BPBG11,ACSA10}. The set of correlations that can be written that admit such a decomposition is denoted $\mathcal{S}_{2|1}^{ns}$. Correlations that cannot be written in this form can then be considered to be genuinely tripartite nonlocal. 

However there is a more interesting definition of genuine multipartite nonlocality based on time-ordering. Basically, one now requires that in the decomposition (\ref{svet}), all bipartite correlations are time-ordered. Specifically, the set $\mathcal{S}_{2|1}^{to}$ of $2$-way time-ordered correlations contains all distributions that can be written in the form
\ba\nonumber\label{TO} p(abc|xyz) = \int d\lambda q(\lambda) p_{\lambda}^{T_{AB}}(ab|xy) p_\lambda (c|z)  \\ + \int d \mu q(\mu) p_{\mu}^{T_{AC}}(ac|xz) p_\mu (b|y)  \\\nonumber +
\int d \nu q(\nu) p_{\nu}^{T_{BC}}(bc|yz) p_\nu (a|x) \ea
where $p_{\lambda}^{T_{AB}}(ab|xy)$ denotes a probability distribution that is time-order dependent: when Alice measure before Bob, we have that $p_{\lambda}^{T_{AB}}(ab|xy)=p_{\lambda}^{A<B}(ab|xy) $; when Bob measures before Alice, we have that $p_{\lambda}^{T_{AB}}(ab|xy)=p_{\lambda}^{B<A}(ab|xy) $. It is then required that $p_{\lambda}^{A<B}(ab|xy) $ and $p_{\lambda}^{B<A}(ab|xy)$ are both (at most) 1-way signaling; $p_{\lambda}^{A<B}(ab|xy)$ is such that only Alice can signal to Bob, while $p_{\lambda}^{B<A}(ab|xy)$ is such that only Bob can signal to Alice. These requirements avoid the problems discussed above. According to this definition, a probability distribution $p(abc|xyz)$ that cannot be written in the form \eqref{TO} is then said to be genuine multipartite nonlocal. 

All three definitions of genuine multipartite nonlocality introduced in this section are nonequivalent \cite{BPBG11,GWAN11b} and we have the strict relations 
\ba \mathcal{L}  \subset {S}_{2|1}^{ns} \subset \mathcal{S}_{2|1}^{to}\subset \mathcal{S}_{2|1}^{Svet} \ea
Thus while violation of Svetlichny's decomposition (\ref{svet}) always guarantee that the correlations $p(abc|xyz)$ are genuinely tripartite nonlocal, there exist some correlations whose tripartite character only manifests itself when considering the weaker definitions$ {S}_{2|1}^{ns}$ and $\mathcal{S}_{2|1}^{to}$.


\subsection{Detecting genuine multipartite nonlocality}\label{detection multi}
After having defined the concept of genuine multipartite nonlocality, we know briefly discuss how one can detect it through the violation of appropriate Bell inequalities. 

\subsubsection{Svetlichny's inequality}

The first inequality for detecting genuine multipartite nonlocality was introduced in \cite{Svetlichny87}. Focusing on a tripartite system, the author derived a Bell-type inequality which holds for any distribution of the form \eqref{svet}. Thus a violation of such inequality implies the presence of genuine tripartite nonlocality. It should be noted that this in turn implies the presence of genuine tripartite entanglement. 

Let us focus on the case where each party $j$ performs one out of two possible measurements denoted $x_j$ and $x'_j$. All measurements are dichotomic, hence their results are denoted by $a_j=\pm1$ and $a'_j=\pm1$. Svetlichny then proved that the inequality
\ba\label{svet_ineq} S_3 &=& a_1a_2a'_3 + a_1a'_2a_3 + a'_1a_2a_3 - a'_1a'_2a'_3 \\\nonumber  & & a'_1a'_2a_3 + a'_1a_2a'_3 + a_1a'_2a'_3 - a_1a_2a_3 \leq 4 \ea
holds for any probability distribution of the form \eqref{svet}. Note that the above polynomial should be understood as a sum of expectation values; for instance $a_1a_2a'_3$ stands for the expectation value of the product of the measurement outcomes when the measurements are $x_1$, $x_2$, and $x'_3$.

To get more intuition about Svetlichny's inequality, and to prove that its violation implies the presence of genuine multipartite nonlocality, we follow the simple approach of \cite{BBBGL11}. We first rewrite the inequality as follows:
\ba\label{trick} S_3 &=& S\, a'_3 + S' \,a_3 \leq 4\ea
where $S=a_1a_2+ a_1a'_2 + a'_1a_2 - a'_1a'_2$ is the CHSH expression, and $S'=a'_1a'_2+ a'_1a_2 + a_1a'_2 - a_1a_2$ one of its equivalent forms
obtained by permuting primed and non-primed measurements.
Now observe that it is the input setting of Charlie that defines which version of the CHSH game Alice and Bob are playing. When C gets the input $x'_3$, then AB play the standard CHSH game; when C gets the input $x_3$, AB play its symmetry. Hence it follows that $S_3\leq4$ holds for any bipartition model of the form \eqref{svet}. Consider the bipartition A$|$BC. B knows which version of the CHSH game he is supposed to play with A, since he is together with C. However, CHSH being a nonlocal game, AB cannot achieve better than the local bound (i.e. $S=2$ or $S'=2$), as they are separated. Thus it follows that $S_3\leq4$ for the bipartition A$|$BC. Note that the same reasoning holds for the bipartition B$|$AC. Finally, since the polynomial is symmetric under permutation of the parties, it follows that $S_3\leq4$ for all bipartitions.
The inequality \eqref{svet_ineq} detects the genuine multipartite nonlocality of important classes of quantum states, such as GHZ and W (see Section~\ref{multi nl quantum}).

\subsubsection{Generalizations to more parties, measurements and dimensions}

Svetlichny's inequality has been generalized to a scenario featuring an arbitrary number of parties $n$ \cite{CGPR+02,SS02}. By repeating the procedure which allowed us to build Svetlichny's inequality from CHSH (see \eqref{trick}) we get
\ba\label{Sm} S_n = S_{n-1}\,a'_n + S'_{n-1}\,a_n \leq 2^{n-1}  \ea
where $S'_{n-1}$ is obtained from $S_{n-1}$ by applying the mapping $a_1 \rightarrow  a'_1$ and $a'_1  \rightarrow a_1$ \cite{BBBGL11}.
Note also that generalizations to the most general scenario, featuring an arbitrary
number of parties, measurements and systems of arbitrary
dimensions, was derived in \cite{BBBGL11}; see also \cite{AGCA12}.

Finally, note that Bell inequalities detecting notions of genuine multipartite nonlocality more refined than that of Sveltichny (see Section~\ref{timeorder}) were presented in \cite{BPBG11}.

\subsection{Monogamy}\label{sec:monogamy}

The monogamy of nonlocal correlations is nicely illustrated by considering the CHSH inequality in a tripartite scenario. 
Let Alice, Bob, and Charlie have two possible dichotomic measurements, represented by observables $A_x$, $B_y$ and $C_z$ with $x,y,z \in \{0,1\}$. We can now evaluate the CHSH expression for Alice-Bob, and Alice-Charlie. Denote by $\mathcal{B}_{AB}$ and $\mathcal{B}_{AC}$ denote the corresponding Bell operators for the CHSH inequality as defined in Section~\ref{sec:bellOpCHSH}. It is important to note that Alice's measurements are the same for both inequalities. It was shown in \cite{SG01} that, for any 3-qubit state shared by the parties, if $\langle\mathcal{B}_{AB}\rangle>2$ then $\langle\mathcal{B}_{AC}\rangle \leq 2$. That is, if the statistics of Alice and Bob violate the CHSH inequality, then the statistics of Alice and Charlie will not. More generally, \cite{TV06} showed that for an arbitrary quantum state shared
by the three parties, we have that
\begin{align}\label{eq:monogamy}
\langle \mathcal{B}_{AB} \rangle^2 + \langle  \mathcal{B}_{AC} \rangle^2 \leq 8\ .
\end{align}
Note again that if Alice and Bob violate their CHSH inequality, then Alice and Charlie do not. Moreover, if Alice and Bob observe maximal CHSH violation (i.e. a CHSH value of $2\sqrt{2}$), then $\langle \mathcal{B}_{AB} \rangle^2 = 8$ and hence by~\eqref{eq:monogamy} the data of $A$ and $C$ are uncorrelated. 
Monogamy of correlations, however, is not specific to the CHSH inequality but applies to essentially all bipartite 
Bell inequalities. In the language of games (Section~\ref{sec:games}), 
this has been used in~\cite{vidick:immunize1, vidick:immunize2} to ''immunize'' a nonlocal game against the use of entanglement. 

It is interesting to note that even no-signaling correlations are monogamous~\cite{BLMP+05,MAG06,PB09} (see Section \ref{properties of ns}). In particular, it has been shown in~\cite{Toner09} that $|\langle \mathcal{B}_{AB} \rangle| + |\langle \mathcal{B}_{AC} \rangle| \leq 4$, which is tight if we consider no-signaling correlations.

The fact that QKD protocols based on nonlocality can be proven secure (see Section~\ref{sec:QKD}) can also be understood as a consequence of the monogamy of quantum correlations among Alice, Bob and Eve, and was indeed one of the factors motivating its study. 

Underlying the monogamy of correlations in the quantum setting is an inherent monogamy of entanglement~\cite{terhal:monogamy}. Understanding the exact relation between both forms of monogamy is an interesting open problem.


\subsection{Nonlocality of multipartite quantum states}\label{multi nl quantum}

\subsubsection{Multipartite nonlocality vs multipartite entanglement}

In this section we discuss the relation between quantum nonlocality and entanglement in the multipartite setting. 
Similarly to the bipartite case, the two concepts are intimately related, although understanding precisely the link is a challenging problem. 

First, note that all pure entangled $n$-partite states are nonlocal \cite{PR92}. That is, their measurement statistics cannot be decomposed in the form \eqref{locality}. This follows from the fact that it is always possible for $n-2$ parties to project (via
a local projection) the remaining two parties in a pure entangled
state. Since the latter is nonlocal, the result follows. It should be stressed that this result does not rely on any form of post-selection. 

In entanglement theory, a concept of particular importance is that
of genuine multipartite entanglement. A quantum state features
genuine multipartite entanglement when it cannot be decomposed as
a convex combination of biseparable states (states which are
separable on at least one bipartition of the parties). Indeed,
this notion is somehow analogous to that of genuine multipartite
nonlocality, and it is not surprising that both are related. In
particular, genuine multipartite quantum nonlocality can be
obtained only if measurements on a genuine multipartite entangled
state are made. Thus, the presence of genuine multipartite
nonlocality witnesses the presence of genuine multipartite
entanglement. Importantly this is achieved in a device-independent
way, that is, genuine multipartite entanglement is here certified
without placing any assumptions about the devices used in the
experiment, contrary to usual methods such as entanglement
witnesses and quantum tomography. Note that it is possible to
design even better device-independent techniques for witnessing genuine
multipartite, the violation of which does not imply the presence
of genuine multipartite nonlocality \cite{BGLP11} (see also \cite{Uffink02,NKI02}).

It is however not known whether all pure genuine multipartite
entangled state are genuine multipartite nonlocal. It has been
shown \cite{ACSA10} that all connected graph states are fully
genuine nonlocal, in the no-signaling approach discussed in Section~\ref{timeorder}. Moreover, it was also shown that the tangle, a specific
measure of multipartite entanglement, is closely related to the
violation of Svetlichny's inequality \cite{AR10,GSDR+09}. In
particular, from this connection it can be shown that there exist
pure entangled states in the GHZ class which do not violate
Svetlichny's inequality.

Finally, it is worth noting that the connection between genuine multipartite entanglement and nonlocality may depend on which definition of genuine multipartite nonlocality is used. Using the definition based on time-ordering (see \eqref{TO}), numerical evidence suggests that all pure genuine tripartite entangled qubit states are genuine tripartite nonlocal \cite{BPBG11}. More recently, \cite{YO13} proved that all pure genuinely tripartite entangled states are tripartite nonlocal with respect to the definition based on no-signaling (see Section \ref{timeorder}). Tripartite nonlocality of Gaussian states was discussed in \cite{AP13}.

\subsubsection{Greenberger-Horne-Zeilinger states}
\label{multiGHZ}

Greenberger-Horne-Zeilinger (GHZ) states are today arguably the most studied---and possibly the best understood---multipartite quantum states from the point of view of entanglement and nonlocality. GHZ states display one of the most striking forms of nonlocality in the context of the Mermin-GHZ paradox (see Section~\ref{multi}). By performing local measurements on a tripartite GHZ state
\ba \ket{\text{GHZ}} = \frac{1}{\sqrt{2}} (\ket{000}+\ket{111}) \ea
one obtains correlations which are maximally nonlocal, since the predictions of quantum mechanics are here in full contradiction with those of local models. Interestingly, it turns out however that these particular GHZ correlations do not feature genuine multipartite nonlocality \cite{Cereceda02,MPR04}, as they can be reproduced by a bi-separable model of the form \eqref{svet}.

It is nevertheless possible to generate genuine multipartite nonlocal correlations from local measurements on a tripartite GHZ state \cite{Svetlichny87}. In particular, one can get violation of Svetlichny's inequality \eqref{svet_ineq} of $S_3=4\sqrt{2}>4$, which turns out to be the largest possible violation in quantum mechanics \cite{MPR04}. This violation can be intuitively understood by considering again the form \eqref{trick} of Svetlichny's inequality. Since it is Charlie's measurement setting that dictates which version of the CHSH game Alice and Bob are playing, the best strategy for C consists in remotely preparing (by performing a measurement on her qubit) a state for AB that is optimal for the violation of the corresponding CHSH game \cite{BBBGL11}.

The nonlocal correlations of generalized GHZ states, of the form
\ba\label{GHZgen}   \ket{\text{GHZ}_n^d} = \frac{1}{\sqrt{d}} \sum_{j=0}^{d-1}\ket{j  }^{\otimes n} \ea
featuring $n$ parties and systems of local dimension $d$, have also been investigated. First, analogues of the Mermin-GHZ paradox were reported \cite{ZK99,CMP02} for certain combinations of $n$ and $d$. More recently, a general construction for arbitrary $n$ and $d$ was given in \cite{RLZL13}. A Mermin-GHZ type paradox was also presented for the case of continuous variable systems \cite{MP01,LB01}.

The genuine multipartite nonlocality of generalized GHZ states has
also been investigated. First it was shown that all qubit GHZ
states (i.e. $\ket{\text{GHZ}_n^2}$) violate the generalization
\eqref{Sm} of Svetlichny's inequality for an arbitrary number of
parties, and hence display genuine multipartite nonlocal
correlations \cite{CGPR+02,SS02}. Recently, it was shown that the
correlations of any state of the form \eqref{GHZgen} are fully
genuinely multipartite nonlocal, as well as monogamous and locally
random \cite{AGCA11}. The robustness of GHZ nonlocality against
local noise was investigated by \cite{CCAA12,LPBZ10}.

\subsubsection{Graph states}

Graph states \cite{HEB04} form an important family of multipartite quantum states (including GHZ and cluster states) useful for applications in quantum information science. In particular, all codeword states used in the standard quantum error correcting codes correspond to graph states, and one-way quantum computation uses graph states as a resource. Here we will discuss the nonlocality of graph states (for GHZ states see Section~\ref{multiGHZ}).

Graph states are defined as follows. Let $G$ be a graph featuring $n$ vertices and a certain number of edges connecting them. For each vertex $i$, we define $\text{neigh}(i)$ as the neighborhood of $i$, which represents the set of vertices which are connected to $i$ by an edge. Next one associates to each vertex $i$ a stabilizing operator
\ba\label{stab_op} g_i = X_i  \bigotimes_{j \in \text{neigh}(i)} Z_j \ea
where $X_i$, $Y_i$, and $Z_i$ denote the Pauli matrices applied to qubit $i$. The graph state $\ket{G}$ associated to graph $G$ is then the unique common eigenvector to all stabilizing operators $g_i$, i.e. $g_i \ket{G} =  \ket{G}$ for all $i \in\{1,...,n\}$. From a physical point of view, the graph $G$ describes all the perfect correlations of the state, since $\bra{G}g_i\ket{G}=1$ for all $i \in\{1,...,n\}$. By considering the set of operators that can be obtained from products of stabilizer operators \eqref{stab_op}, one obtains a commutative group featuring $2^n$ elements. This is the stabilizer group, defined as
\ba S(G) = \{s_j\}_{j=1,...,2^n} \quad \text{where}  \quad s_j =  \prod_{i \in I_j(G)} s_i \ea
where $I_j(G)$ denotes any of the $2^n$ subsets of the vertices of the graph $G$.

Interestingly, this fundamental structure of graph states underpins a strong form of nonlocality \cite{SASA05,GTHB05}. It turns out that all graph states feature nonlocal correlations \cite{GTHB05}. In order to prove this, the main idea consists to construct Bell inequalities by adding all elements of the stabilizer group $S(G)$. Thus we consider the operator
\ba \mathcal{B}(G) = \sum_{i=1}^{2^n} s_i = \sum_{i=1}^{2^n} \bigotimes_{j=1}^{n} O_j^i\ea
where operators $O_j^i \in \{\openone, X_j,Y_j,Z_j \}$ are from the Pauli basis.

It is then possible to define a Bell inequality based on the above Bell operator, and to compute its local bound
\ba L(G) = \max_{\text{LHV}} | \langle \mathcal{B} \rangle | \ea
While the graph state $\ket{G}$ reaches the value of $2^n$ for such Bell inequality (indeed $s_i\ket{G}=\ket{G}$ for all $i \in \{1,...,2^n\}$), it turns out that $L(G)<2^n$ for any graph $G$. Thus, for all graph states it is possible to construct a Bell inequality, which the state then violates maximally. Indeed this demonstrates that nonlocality is a generic feature of all graph states. Moreover, for certain families of graph states, basically states based on tree graphs (featuring no closed loops), the violation of the Bell inequality grows exponentially with the number of vertices \cite{GTHB05,TGB06}.

While the generality of the above approach is remarkable, it is possible for certain important classes of graph states, in particular for cluster states, to derive stronger proofs of nonlocality \cite{SASA05}. Cluster states form a subclass of graph states based on square lattice graphs. For simplicity and clarity we will discuss here the case of a four qubit cluster state on a one-dimensional lattice\footnote{Note that for two and three qubits, the 1D cluster state is equivalent to a Bell state and to a GHZ state, respectively.}, which is locally equivalent to
\ba \ket{\text{Cl}_4} = \frac{1}{2}(\ket{0000}+\ket{0011}+\ket{1100}-\ket{1111}). \ea
The state $\ket{\text{Cl}_4}$ is defined by the stabilizer relations
\ba
\begin{array}{cccccc}
 X & Z & \openone & \openone &=1 & \quad (E1)  \\
 Z & X  & Z & \openone & =1 & \quad (E2)  \\
 \openone & Z  & X & Z & =1 & \quad (E3) \\
 \openone &  \openone & Z  & X  & =1  & \quad (E4)
 \end{array}
\ea
By multiplying certain of these four relations, we get
\ba\label{cluster_paradox}
\begin{array}{rccccc}
 (E1) \,  \times \, (E3) : \quad & X &  \openone & X & Z &=1   \\
 (E2)  \,  \times \,  (E3) : \quad &Z & Y  & Y & Z & =1   \\
(E1)  \,  \times \, (E3)  \,  \times \,  (E4) : \quad & X & \openone & Y  & Y  & =1 \\
 (E2)  \,  \times \, (E3)  \,  \times \,  (E4) : \quad & Z & Y & X  & Y  & =-1
 \end{array}
\ea
Note that here we have used the Pauli algebra, which explains the emergence of a minus sign in the last relation above. It can be readily checked, that for any deterministic local model, i.e. attributing $\pm1$ values to each measurement ($X,Y,Z$), it is impossible to satisfy simultaneously all four relations above; at least one of them will not hold. Check for instance that by simply multiplying (using standard multiplication) the first three relations in \eqref{cluster_paradox}, one obtains the fourth relation. Therefore, we obtain a perfect contradiction between quantum and classical predictions, in the spirit of the GHZ paradox. 

Similarly to the Mermin-GHZ case (see Section~\ref{multi}), this logical contradiction can be rephrased as a Bell inequality. By considering the four relations \eqref{cluster_paradox}, we get
\ba\label{BI_cluster4} | a_1 a'_3 a_4 + a_1 a_3 a'_4  + a'_1 a_2a_3 a_4  - a'_1 a_2a'_3 a'_4 |  \leq 2 \ea
Notice that by grouping the first two parties one obtains the Mermin inequality \eqref{sec2.mermin}. Performing measurements on the state $\ket{\text{Cl}_4}$, the algebraic maximum of 4 can be obtained for the left hand side of \eqref{BI_cluster4}. Finally note that an interesting feature of the Bell inequality \eqref{BI_cluster4} is that it cannot be violated by the four-qubit GHZ state. Thus the inequality is a strong entanglement witness\footnote{Notice however that, as written, the inequality \eqref{BI_cluster4} can also be maximally violated by a three-partite entangled state, since party 2 has only one setting. This deficiency can be overcome by symmetrizing the inequality over the parties.} for the cluster $\ket{\text{Cl}_4}$.
The above construction can be generalized to cluster states of an arbitrary number of qubits and of arbitrary local dimension, as well as to certain classes of graph states \cite{SASA05}.

The nonlocality of graph states can also be revealed by using sets of local measurements that are not stabilizers \cite{GC08}. Also, an interesting issue is to understand whether there exists a link between the nonlocality of cluster states and the computational power they offer. Although such a connection has not been clearly established yet, progress has been made \cite{HCLB11}. Another important class of graph states is codeword states, the nonlocality of which has been discussed in \cite{DP97}.

Finally, it is important to note that all nonlocality proofs discussed in this subsection concern Bell nonlocality. Unfortunately much less is known concerning the genuine multipartite nonlocality of graph states. It is however known that all states based on connected graphs (graphs in which any two vertices are connected, although not necessarily in a direct manner) display fully genuine multipartite nonlocality \cite{ACSA10}.

\subsubsection{Nonlocality of other multipartite quantum states}

In the multipartite case, entanglement displays a rich structure, with many inequivalent classes of states. Although we know that all multipartite entangled
pure states are nonlocal, very few is known beyond the case of
graph states.

An important class of multipartite entangled states are Dicke states, that is, states
with a fixed number of excitations and symmetric under permutation of the parties, which are
central in the context of the interaction of light and matter \cite{Dicke54}. The
symmetric state of $n$ particles with a single excitation, known
as the W state, reads \ba\label{W}    \ket{W_n} =
\frac{1}{\sqrt{n}} \left( \ket{0 \hdots 0 1} +   \hdots + \ket{1 0 \hdots  0} \right)  \ea Such states are
relevant to the description of various physical systems, such as
quantum memories. One possibility for detecting the nonlocality of W states consists in having $n-2$ parties performing a measurement in the logical basis $\{\ket{0},\ket{1}\}$. When all project onto the $\ket{0}$ eigenstate, which happens with a fairly large probability (increasing with $n$), they prepare for the remaining two parties a (two qubit) Bell state, on which the CHSH inequality can then be tested and violated \cite{SSWK+03}.
Another manifestation of the nonlocality of the Dicke states is based on their robustness to losses. Indeed when $k \ll n $ particles are lost, the state remains basically unchanged. For instance, for W states one has that $\text{tr}_k(\ket{W_n}\bra{W_n}) \approx \ket{W_{n-k}} \bra{W_{n-k}}$, where $\text{tr}_k$ denotes the partial trace on the $k$ particles which have been lost. The W state thus has a high ``persistency'' of nonlocality \cite{BV12}, in the sense that a large number of particles must be lost in order to destroy all nonlocal correlations. This appears to be a generic feature of Dicke states.

Another relevant problem is whether one can distinguish different classes of multipartite entangled states via their nonlocal correlations. This can be done using judiciously designed Bell inequalities \cite{BSV12,SKL+08}. For instance, the resistance to losses of W states can be exploited to distinguish their nonlocal correlations from those of GHZ states.

The nonlocal properties of more general classes of states have been discussed. First, the nonlocality of symmetric qubit states was investigated in \cite{WM12}. Exploiting the Majorana representation, the authors could derive Hardy-type nonlocality proofs (see Section~\ref{withoutBI}) for arbitrary symmetric pure entangled states. Also, the resistance to noise 
has been evaluated numerically for a large class of multipartite quantum states \cite{GLZK+10}. 

Finally, the relation between entanglement distillability and nonlocality was also investigated in the multipartite case. It was shown in \cite{Dur01,AH06} that a multi-qubit bound entangled state can violate the Mermin inequalities. However, the states considered in these works become distillable when several parties can group. In fact, it was shown in \cite{Acin01} that the violation of the Mermin inequalities implies that distillibality between groups of parties. More recently, \cite{VB12} presented an example of a fully bound entangled state (for which no entanglement can be distilled even parties are allowed to group) which violates a Bell inequality. This shows that nonlocality does not imply the presence of distillable entanglement, and refutes the Peres conjecture in the multipartite case (see Section~\ref{peres}).

\section{Experimental aspects}
\label{Experimental aspects}
Violations of Bell inequalities have been observed experimentally in a variety of physical systems, giving strong evidence that nature is nonlocal. Nevertheless, all experiments suffer from various loopholes, opened by technical imperfections, which make it in principle possible for a local model to reproduce the experimental data, even if in a highly contrived way. In recent years, an intense research effort has been devoted to the design and realization of a loophole-free Bell experiment, which should be within experimental reach in the near future. Besides its fundamental interest, closing some of these loopholes (in particular the detection loophole) is important from the perspective of practical applications of nonlocality such as device-independent quantum information processing. Indeed, while the idea that nature is exploiting such loopholes to fake non-local correlations may sound conspiratorial, the perspective is entirely different when we consider the possibility that they are exploited by an adversary to break a cryptography protocol. 

In this section we review the main achievements and challenges in this area. For a more exhaustive discussion on Bell experiments, we refer the reader to recent reviews \cite{Genovese05,PCLW+12}.

\subsection{Bell experiments}

\subsubsection{Photons}

Tremendous experimental progress in quantum optics during the 1960s opened the door to possible tests of quantum nonlocality in the laboratory. First, using atomic cascades, it became possible to create pairs of photons entangled in polarization. Second, the polarization of single photons could be measured using polarizers and photomultipliers. Only three years after the proposal of CHSH \cite{CHSH69}, Freedman and Clauser \cite{FC72} reported the first conclusive test of quantum nonlocality, demonstrating a violation of the CHSH Bell inequality by 6 standard deviations.

During the following years, other experiments \cite{FT76,AGR81,AGR82} were performed, giving further confirmation of the predictions of quantum mechanics. However, the main drawback of all these experiments was that they were performed with static setups in which the polarization analyzers were held fixed, so that all four correlations terms had to be estimated one after the other. In principle, the detector on one side could have been aware of the measurement setting chosen on the other side, thus opening a loophole open\footnote{Moreover, by performing Bell tests with all correlation terms measured successively with the settings held fixed, it is not unusual to observe experimentally, because of slow drifts in the set-up, apparent violations of Bell inequalities above Tsirelson's bound or even violation of the no-signaling conditions (\ref{sec2.nosig}) \cite{Afzelius11}.} (see subsection~\ref{sslocloop}).

Crucial progress came in 1982, when Aspect, Dalibard and Roger \cite{ADR82} performed the first Bell experiment with time-varying polarization analyzers. The settings where changed during the flight of the particle and the change of orientation on one side and the detection event on the other side were separated by a space-like interval, thus closing the locality loophole (see section~\ref{sslocloop}). It should be noted though that the choice of measurement settings was based on acousto-optical switches, and thus governed by a quasi-periodic process rather than a truly random one. Nevertheless the two switches on the two sides were driven by different generators at different frequencies and it is very natural to assume that their functioning was uncorrelated. The experimental data turned out to be in excellent agreement with quantum predictions and led to a violation of the CHSH inequality by 5 standard deviations.

The advent of quantum information in the 1990s triggered renewed interest in experimental tests of quantum nonlocality. In 1998, violation of Bell inequalities with photons separated by more than 10km was reported \cite{TBZG98}. The same year, another experiment demonstrated violation of Bell inequalities with the locality loophole closed and using a quantum random number generator to generate the measurement settings \cite{WJSW+98}.
In turn, both of these experiments were adapted to implement quantum key distribution based on nonlocal quantum correlations \cite{JSWW+00,TBZG00}, following Ekert's idea (see Section \ref{sec:QKD}).

Demonstrations of quantum nonlocality in photonic systems have
been reported using various types of encoding apart from
polarization. First, Bell inequality violations based on phase and
momentum of photons have been achieved \cite{RT90}. In 1989,
Franson proposed a test of quantum nonlocality based on the
energy-time uncertainty principle \cite{Franson89}. This encoding,
used for instance in the experiment of \cite{TBZG98}, led to the
concept of time-bin encoding \cite{BGTZ99} which turned out to be
particularly well suited for the distribution of entanglement on
long distances. Bell inequality violation have also been
demonstrated using photons entangled in orbital angular momentum
\cite{MVWZ01}. An important advantage of both time-bin and orbital
angular momentum encodings, is that they allow for the realization
of higher dimensional quantum systems, whereas polarization is
limited to qubits. Nonlocal correlations of qutrits have been
reported with time-bins \cite{TAZG04}, while Bell violation with
orbital angular momentum have recently been reported using systems
of dimensions up to 11 \cite{DLBP+11}. Another possibility for creating higher dimensional entanglement consists in generating
pairs of photons entangled in several degrees of freedom, so
called hyper-entangled photons \cite{Kwiat97}. Bell experiments
have been performed with such systems \cite{BLPK05,CVDM+09},
combining polarization, spatial mode and energy-time degrees of
freedom. Finally, continuous variable systems have also been investigated. 
In particular, \cite{BAL05} demonstrated the nonlocality of a single photon using homodyne measurements.

Other interesting aspects of quantum nonlocality have been investigated experimentally. Notably, the phenomenon of hidden nonlocality (see section
III.A.3) was observed in \cite{KBSG01}, and Hardy's paradox (see Section \ref{withoutBI}) was realized in \cite{WJEK99}. It is also worth mentioning the experiment of \cite{FUHN+09} which demonstrated violation of the CHSH inequality over a free-space link of 144km.

Multipartite quantum nonlocality has also been demonstrated
experimentally. Bell inequality violations were achieved with
three photons, generating both GHZ \cite{PBDW+00} and W
\cite{EKBK+04} states, and with four-photon GHZ states
\cite{EGBK+03,ZYCZ+03} and cluster states \cite{WARZ05}. Genuine
multipartite nonlocality of three-photon GHZ states was demonstrated in \cite{LKR09}.

Note also that nowadays Bell experiments can even be envisaged for pedagogical purposes \cite{DM02}. In particular, ready-to-use setups are available commercially \cite{qutools}, which are fully operational, even from the perspective of research \cite{PBSR+11}.

Finally, it is important to keep in mind that all the Bell experiments discussed above
are plagued by the detection loophole (see subsection~\ref{det loop}).
This is because the photon detection efficiency in these
experiments is low (typically $10-20\%$) which makes it possible,
in principle, for a local model to reproduce the raw data. It is
only under the assumption that the probability of detecting or non-detecting a
photon is independent of the choice of measurement (the so-called
'fair-sampling' assumption, allowing one to discard inconclusive
events) that the experimental data leads to Bell inequality
violations. 

Recently though experimental violation of Bell inequalities with the detection loophole closed were reported in \cite{GMR+12,CMAC+13}. It should be noted however that the data analysis of \cite{GMR+12} is affected by the time-coincidence loophole (see section \ref{det loop}), and is thus not fully satisfactory. This point was subsequently addressed in \cite{LGKW+13}. Since both of these experiments are table-top, using relatively slow detectors, they are still plagued by the locality loophole.

\begin{figure}[b]
\includegraphics[width = 0.4\textwidth]{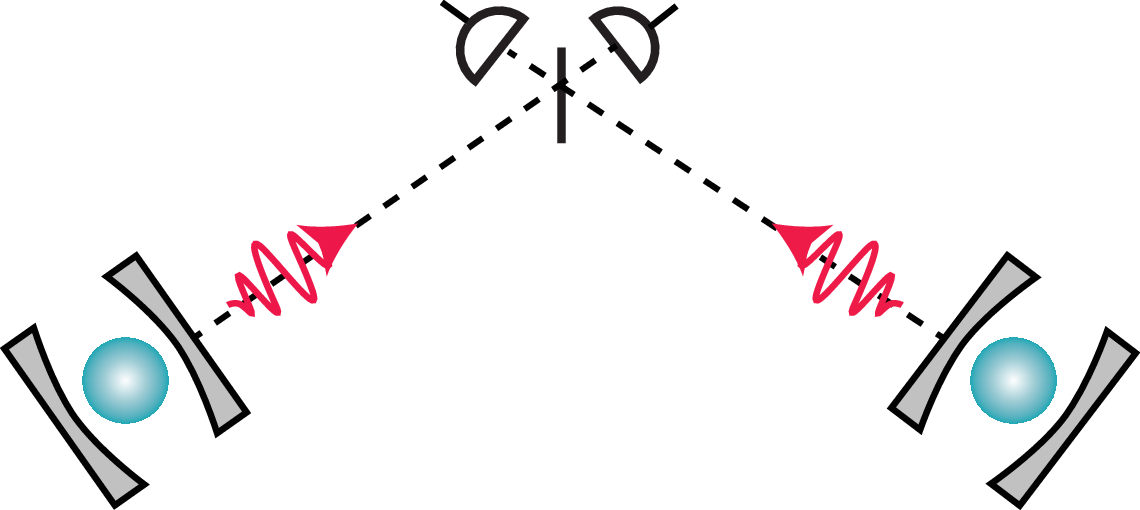}
    \caption{Bell test based on distant entangled atoms. Each atom is entangled with an emitted photon. Upon successful projection of the two photons onto a Bell state, the two atoms become entangled. The scheme is therefore `event-ready', which makes it robust to photon losses in the channel. Moreover, since atomic measurements have an efficiency close to one, this scheme is free of the detection loophole. This setup has been implemented experimentally with a distance of 20m between the atoms \cite{HKO+12}, and used for device-independent randomness expansion \cite{PAMB+10}.}
\label{ExpSetup}
\end{figure}

\subsubsection{Atoms}\label{atoms}

Besides photons, Bell experiments have also been conducted with atomic systems. Such systems offer an important advantage from the point of view of the detection, with efficiencies typically close to unity. Therefore, atomic systems are well-adapted for performing Bell experiments free of the detection loophole. Such an experiment was first realized in \cite{RKMS+01}, using two Be$^+$ ions in a magnetic trap. In this experiment, the two ions were placed in the same trap, separated only by 3 $\mu$m. The locality loophole was thus left wide open, since each ion can feel the light field aimed at measuring the state of the other ion.

More recently, quantum nonlocality was demonstrated between two Yb$^+$ ions sitting in separated traps, one meter apart \cite{MMMO+08}. This was further improved to a distance of 20m using Rubidium atoms \cite{HKO+12}. Although this distance is still insufficient to close the locality loophole---a distance of 300m is required using the fastest procedure to measure the atomic state of the atoms---the crosstalk between the two atoms is now completely suppressed. Here the entanglement between the distant atoms is achieved using an `event-ready' scheme \cite{SI03}, sketched in Fig.~\ref{ExpSetup}, which is based on entanglement swapping \cite{ZZHE93}. First each atom is transferred to an excited state. The ion is de-excited by emitting a photon. The structure of the atomic levels is chosen such that the polarization of the emitted photon is maximally entangled with the state of the atom. The emitted photons are then collected in single mode optical fibers. Finally a partial Bell state measurement is performed on the two photons, using a simple 50:50 beam-splitter followed by single photon detectors. A coincidence detection of two photons at the detectors indicates that the photon were in a given Bell state. In this case entanglement swapping is achieved, that is, the initial atom-photon entanglement has been converted to atom-atom entanglement. Upon successful detection of the photons, local measurements are performed on the atoms. The procedure is repeated until enough data has been taken in order to obtain good statistics. Important advantages of such an `event ready' experiment is its robustness to photon losses and to the coincidence-time loophole \cite{LG04}. Recently, such an experiment was used to conduct a proof-of-principle demonstration of device-independent randomness expansion \cite{PAMB+10} (see Section~\ref{DIrandomness}).

\subsubsection{Hybrid schemes and other systems}
Finally, let us mention that Bell inequality violations have also been reported using atom-photon entanglement \cite{MMBM04}, and entanglement between a photon and a collective atomic excitation \cite{MCBL+05}. Nonlocality was also demonstrated in Josephson phase superconducting qubits. In particular, violation of the CHSH inequality was achieved in \cite{AWBH+09}, whereas the GHZ paradox was demonstrated in \cite{NBLL+10,DRSJ+10}.


\subsection{Loopholes}

\subsubsection{Detection Loophole}\label{det loop}
In a large class of Bell experiments, in particular those carried out with photons, measurements do not always yield conclusive outcomes. This is due either to losses between the source of particles and the detectors or to the fact that the detectors themselves have non-unit efficiency. A measurement apparatus, used e.g. to test the CHSH inequality, has then three outcomes instead of two: it can as usual give the outcomes $-1$ or $+1$, or it can give a `no-click' outcome, denoted $\perp$. The simplest way to deal with such `inconclusive' data is simply to discard them and evaluate the Bell expression on the subset of `valid' $\pm 1$ measurement outcomes. As pointed out by \cite{Pearle70} and \cite{CH74}, this way of analyzing the results is only consistent under the assumption that the set of detected events is a \emph{fair sample}, i.e., that the accepted data is representative of the data that would have been recorded if the detectors had unit efficiency. More generally, one can consider local models where this fair-sampling assumption fails and in which the probability to obtain a `no-click' outcome $\perp$ depends on the choice of measurement \cite{Pearle70,CH74,Santos92}. If the detection efficiency is too low (below a certain threshold), such local models can completely reproduce the observed data, opening the so-called detection loophole. The threshold efficiency required to close this detection loophole is typically high for practical Bell tests. As a consequence, most optical realization of Bell tests performed so far are plagued by the detection loophole.

Another closely related loophole is the time-coincidence loophole \cite{LG04}. This loophole exploits timing issues in Bell tests, which in turn can affect detection efficiency. In \cite{CMAC+13} it is shown how this loophole can affect real experiments.

\paragraph{Faking Bell inequality violations with post-selection.}
Throwing away `no-click' outcomes and keeping only the valid outcomes $\pm 1$ is an example of post-selection. In general, allowing for post-selection in a given theory, allows one to achieve tasks which would be impossible without it. In particular, post-selection makes it possible to fake the violation of a Bell inequality, even in a purely local theory. 

To illustrate this idea, we will see how it is possible for a local model to fake
maximal violation of the CHSH inequality. In particular, we will show how to 
generate Popescu-Rohrlich correlations, $a \oplus b = xy$, where
$x,y,a,b \in \{0,1\}$ (see Section~\ref{NScorrs}), starting from shared randomness, and
allowing the detectors on Alice's side to produce a no-click outcome $\perp$. The model is the following. The shared randomness correspond to two uniform random bits $x_{\text{guess}}$ and
$a$. Given measurement setting $y$, Bob's detector outputs $b=a
\oplus x_{\text{guess}} y$. Alice's detectors outputs $a$ whenever her measurement setting is $x=x_{\text{guess}}$, and outputs $\perp$ when $x \neq x_{\text{guess}}$. Focusing on the conclusive outcomes (e.g. $\pm1$), 
Alice and Bob have achieved maximally nonlocal PR correlations, i.e. achieving a CHSH value of $S=4$. The probability for Alice to obtain a conclusive outcome is 1/2,
which is the probability that $x = x_{\text{guess}}$, while Bob
always obtains a conclusive outcome. With additional shared randomness, it is
possible to symmetrize the above model, such that Alice and Bob's detection probability is 2/3 \cite{MP03}. Therefore, if
the detection efficiency in a CHSH Bell experiment is below 2/3,
no genuine Bell inequality violation can be obtained, since the above
local strategy could have been used by the measurement apparatuses. More generally, the minimum
detection efficiency required for successfully violating a given Bell inequality
depends on the number of parties and measurements involved (see
Section~\ref{noclicks}).

Interestingly, recent experiments demonstrated fake violations of
Bell inequalities using classical optics \cite{GLLS+11}, positive
Wigner function states and quadrature measurements \cite{TWTS09}, a classical amplification scheme \cite{PSSZ+11}, and high-dimensional analyzers \cite{RGTB+13}. These
experiments are performed under the same conditions as standard Bell experiments, but exploit side-channels. This nicely illustrates the importance of closing the detection loophole in Bell tests, in particular for the
perspective of implementing device-independent protocols.

\paragraph{Taking into account no-click events.}\label{noclicks}
The previous discussion shows that in order to close the detection loophole no-click outcomes cannot be discarded without making further assumptions. The most general way to take no-clicks events into account is simply to treat them as an additional outcome and instead of a $\Delta$-outcome Bell inequality (if the number of `conclusive' outcomes is $\Delta$) use a $\Delta+1$-outcome Bell inequality. A possible way to obtain an effective $\Delta+1$-outcome Bell inequality from a $\Delta$-outcome one is simply to merge the no-click outcome with one of the valid outcomes\footnote{Inequalities obtained in this way are \emph{liftings} of the original inequality \cite{Pironio05}. If the original inequality is facet-defining for the $\Delta$-outcome Bell polytope, then the lifted inequality is facet-defining for the $\Delta+1$-outcome polytope. However, the $\Delta+1$-outcome Bell polytope has also in general facets that cannot be viewed as liftings of $\Delta$-outcome inequalities.}, i.e., systematically assign one of the valid outcomes to the no-click events. In particular, the Clauser-Horne inequality \cite{CH74}, which is often used in Bell tests with inefficient detectors, is nothing but the CHSH inequality where the $-1$ outcome and the no-click outcome $\perp$ have been merged into one effective `$-1$' outcome. 

Assigning one of the valid outcomes to the no-click outcome $\perp$ is often the optimal way to treat no-click events, although there is no general proof of this and a counter-example exists for no-signaling correlations \cite{WDAP08}. In the case where $\Delta$ detectors are used to register the $\Delta$ outcomes of a measurement, assigning one of the conclusive outcomes to the no-click events has also the technical advantage that the detector associated to that particular outcome is no longer needed, i.e. only $\Delta-1$ detectors are sufficient since no distinction is being made between obtaining the $\Delta$-th outcome and not detecting anything.

\paragraph{Threshold efficiencies.}\label{the eff}
When treating the no-click outcome as described above, one will generally find that a Bell violation is obtained only if the detector efficiencies are above a certain threshold.
The minimal threshold efficiency $\eta^*$ required to close the detection loophole, depends generally on the number of parties, measurements and outcomes involved in the Bell test. Moreover, $\eta^*$ may also vary depending on the exact set of correlations that is considered. Thus, in quantum Bell tests, $\eta^*$ may also depend on which entangled state and which measurement settings are considered. Below we review the efficiency thresholds for the most important Bell inequalities and for the most common quantum entangled states.

We start by deriving $\eta^*$ for the CHSH Bell inequality using a two-qubit maximally entangled state. Performing judicious local measurement on this state, one obtains a CHSH value of $S=2\sqrt{2}$ (the maximum possible value in quantum mechanics). Now, let us assume that Alice and Bob have imperfect detectors with efficiency $\eta$ and that when a no-click results $\perp$ is obtained, they assign to it the $+1$ outcome. When both detectors click, which happens with probability $\eta^2$, Alice and Bob achieve $S=2\sqrt{2}$. When only one detector clicks, the outcomes are completely uncorrelated leading to $S=0$. Finally, when no detectors click, which happens with probability $(1-\eta)^2$ Alice and Bob both always output $+1$, thus achieving the local bound $S=2$. In order to close the detection loophole, we must ensure that the entire data of the experiment violates the CHSH inequality, i.e. that
\ba \eta^2 2\sqrt{2} +  (1-\eta)^2 2 > 2 \label{condeta}\ea
This leads to the condition that
\ba \eta > \eta^* = \frac{2}{1+\sqrt{2}} \approx 82.8\%. \ea
Therefore, in order to get a detection loophole free CHSH violation with a two-qubit maximally entangled state, it is sufficient to have a detection efficiency larger than $\sim 82.8\%$ \cite{Mermin86}. This efficiency is also necessary: an explicit local model can be built which reproduces the experimental data when $\eta<2/(1+\sqrt{2})$.

Remarkably, it turns out that this threshold efficiency can be lowered by considering partially entangled states, of the form $\ket{\psi_\theta}= \cos{\theta}\ket{00}+\sin{\theta}\ket{11}$, as discovered by \cite{Eberhard93}. In particular, in the limit of a product state (i.e. $\theta \rightarrow 0$) one obtains that $\eta^* \rightarrow 2/3$. This astonishing result was the first demonstration that sometimes less entanglement leads to more nonlocality (see Section~\ref{MoreNL}). 

From an experimental perspective, it is relevant to see how the above results are affected by the presence of background noise. In general this amounts to increase considerably the threshold efficiencies. Even for very low levels of background noise, the threshold efficiency usually increases by a few percents. A detailed analysis can be found in \cite{Eberhard93}. Another point concerns events in which no detection happened on both sides. In certain cases, joint losses are not detrimental for the demonstration of nonlocality \cite{MPRG02}.

Beyond the CHSH case, discussed in detail in \cite{Branciard11}, it is known that lower threshold efficiencies can be tolerated for Bell inequalities featuring more measurement settings. A lower bound for the threshold efficiency is given by
\ba \eta^* \geq \frac{m_A+m_B-2}{m_Am_B-1} \ea
where $m_A$ ($m_B$) denotes the number of settings of Alice (Bob) \cite{MP03}. While it is not known whether this bound can be obtained in general with quantum correlations, improvements over the threshold efficiencies of the CHSH inequalities have been obtained by considering Bell scenarios with more measurement settings. For qubit states only minor improvements were found \cite{MPRG02,BG08,PV09}. However, in \cite{Massar02} it was shown that, when considering systems of higher Hilbert space dimension $d$, the threshold efficiency can become arbitrarily close to zero. Unfortunately, this result is of limited practical interest since improvements over the CHSH case are only obtained for systems with $d\gtrsim 1600$. Also, the number of measurements becomes exponentially large, namely $2^d$. More recently a Bell test involving entangled quqats ($d=4$) and four (binary) measurement settings was shown to tolerate detection efficiencies as low as $\sim 61.8\%$ \cite{VPB10}. Such a scheme could be implemented optically using hyper-entanglement.

Threshold efficiencies have also been derived for certain multipartite Bell tests (with $n$ parties), using qubit GHZ states. Based on a combinatorial study, \cite{BHMR03} showed that an arbitrarily small efficiency can be tolerated as $n$ becomes large. Threshold efficiencies approaching $50\%$ in the limit of a large $n$ where demonstrated for the Mermin inequalities \cite{CRV08} and for a multipartite generalization of the CH inequality \cite{LS01}. More recently, multipartite Bell tests tolerating efficiencies significantly below $50\%$ were reported in \cite{PVB12}.

Finally, detection efficiencies have also been considered in asymmetric Bell experiments. Consider first the case in which Alice and Bob feature different detection efficiencies ($\eta_A$ and $\eta_B$ respectively). In particular results have been obtained for the case where $\eta_A<1$ and $\eta_B=1$, which is relevant for Bell tests based on atom-photon entanglement (Alice holds the photon and Bob the atom). It has been shown that for the CHSH inequality the efficiency of Alice's detector can be lowered to $50\%$ \cite{BGSS07,CL07}. Moreover, this efficiency can be further lowered to $\sim43\%$ by considering a three-setting Bell inequality \cite{BGSS07}. Considering Bell tests with $d$ measurement settings and $d$-dimensional systems, an efficiency as low as $1/d$ can be tolerated, which is optimal \cite{VPB10}. Another asymmetric scenario is the case in which the local measurements have different efficiencies. Let $\eta_{A_0}$ and $\eta_{A_1}$ be the efficiencies of Alice's measurements and $\eta_{B_0}$ and $\eta_{B_1}$ the efficiencies of Bob's measurements. If one of the measurements of each party has unit efficiency (say $\eta_{A_0}=\eta_{B_0}=1$) then the CHSH inequality can be violated for an arbitrarily low efficiency for the other measurement, \ie $\eta_{A_1}=\eta_{B_1}\rightarrow 0$ \cite{Garbarino10}. Such an approach fits Bell tests using hybrid measurements, such as homodyne (high efficiency) and photo-detection (low efficiency); see section \ref{towards} for more details.

\subsubsection{Locality loophole}
\label{sslocloop}

The locality condition (\ref{intro.loc}) is motivated by the absence of communication between the two measurement sites of a Bell experiment. This seems well justified if the two sites are sufficiently separated so that the measurement duration is shorter than the time taken by a signal traveling at the speed of light, to travel from one site to the other. If this condition is not satisfied, one could in principle conceive a purely ``local" mechanism -- \ie, involving slower-than-light speed signals -- underlying the observed correlations \cite{Aspect1975,Aspect76,Bell79}. 

In addition to the requirement that the two measurement sites are space-like separated, it must also be the case that the measurement setting on one side is not determined by an earlier event that could be correlated with the measurement setting on the other side; in particular it should not be correlated with the  hidden variables $\lambda$ characterizing the source of particles. That is, the measurement settings must correspond to ``random" or ``free" choices, which are independent from the other side and from the hidden state of the particle pairs \cite{SHC76,Bell77}. 

Mathematically, the above requirements correspond to the following conditions
\be \label{loc}
p(a|x,y,b,\lambda) = p(a|x,\lambda)\,,\quad  p(b|y,x,a,\lambda) = p(b|y,\lambda)\,,
\ee
and
\be \label{foc}
q(\lambda|xy)=q(\lambda)
\ee
from which the locality decomposition (\ref{intro.loc}) follows. 
Failure to satisfy them is known as the \emph{locality loophole}. The failure to address specifically the independence condition (\ref{foc}) between measurement choices and hidden variables is sometimes called the ``freedom-of-choice" \cite{SUK+10} or ``measurement-independence" loophole \cite{BAG11,Hall11b}.

The experiment of \cite{ADR82} using entangled photons was the first to address convincingly the locality loophole. It involved on each side a switching device for the incoming photons followed by two polarizers aligned along different orientations. A change of direction occurred approximately every $10\,ns$. The distance $L=13\,m$ between the two switches was large enough so that the time of travel of a signal between the switches at light speed $L/c=43\,ns$ was larger than the delay of $10\,ns$ between two switchings. The switching of the polarizers was done through a home built device, based on the acousto-optical interaction of the incoming light with an ultrasonic standing wave in water. Using this mechanism it should be noted that not all photons were submitted to forced switching. In addition, the switches were not truly random, since the acousto-optical were driven by periodic generators. The two generators on the two sides, however, were functioning in a completely 
uncorrelated way, since they were operated by different RF generators at different 
frequencies with uncorrelated frequency drifts.

The experiment of \cite{ADR82} was the only one involving fast changes of the measurement settings until the one of \cite{WJSW+98}, which used high-speed electro-optic modulators to switch between two polarization measurement settings on each side. The two modulators where controlled on a nanosecond timescale by two independent quantum random number generators, excluding any light-speed influence between the two sides, which were separated by a distance of a few hundred meters. Leaving aside the possibility that the outputs of the quantum random number generators are predetermined at some hidden level, this setup is often regarded as having conclusively closed the locality loophole. In \cite{SUK+10}, space-like separation was not only enforced between the outputs of the two random generators, but also between them and the emission of photons from the laser source generating the entangled particles, implying that these three processes are independent from each other, provided that they are not themselves determined by some earlier events. 

At this point, it should be stressed that, contrarily to the detection loophole, the locality loophole can never be ``completely" closed. Strictly speaking, it requires space-like separation between the event determining the choice of measurement setting on one side and the event corresponding to the output of the measurement device on the other side. A first problem is that this requires to know precisely the time at which these two events happen. But if we use some random process that outputs a random value at time, say $t=0$, how do we know that this value was precisely determined at this time and not at some earlier time $t=-\delta$? Similarly, how do we know precisely when a measurement is completed without making some assumptions on the collapse of the wavefunction \cite{Kent05}? This last issue was addressed in \cite{SBBG+08} where the violation of Bell inequalities for events that are space-like separated according to a simple model of gravitational collapse has been reported. 

A second problem is that we may never be sure that the choices of measurements are really ``random" or ``free". For instance, in the experiments \cite{WJSW+98,SUK+10} the measurement choices are decided by processes that are genuinely random according to standard quantum theory. But this need not be the case according to some deeper theory.  Some people have argued that a better experiment for closing the locality loophole would be to arrange  the choice of measurement setting to be determined directly by humans or by photons arriving from distant galaxies from opposite directions, in which case any local explanation would involve a conspiracy on the intergalactic scale \cite{Vaidman01}. 

The point of this discussion is that an experiment ``closing" the locality loophole should be designed in such a way that any theory salvaging locality by exploiting weaknesses of the above type should be sufficiently conspiratorial and contrived that it reasonably not worth considering it.

Finally, it is worth stressing that in device-independent applications of quantum nonlocality, the experimental requirements for satisfying conditions (\ref{loc}) and (\ref{foc}) are sensibly different than in fundamental tests of nonlocality, since one usually assumes the validity of quantum theory, that the quantum systems are confined in well defined measurement devices that can be shielded from the outside world, that the inputs are under the control of the users, and so on. This stance was used for instance in \cite{PAMB+10}, where the atoms were separated from each other, though by no means fulfilling any of the space-like separation prescriptions required for a fundamental, locality loophole-free Bell test. 

\subsubsection{Finite statistics}\label{finitestat}
Since it is expressed in terms of the \emph{probabilities} for the possible measurement outcomes in an experiment, a Bell inequality is formally a constraint on the expected or average behavior of a local model. In an actual experimental test, however, the Bell expression is only estimated from a finite set of data and one must take into account the possibility of statistical deviations from the average behavior. The conclusion that Bell locality is violated is thus necessarily a statistical one. In most experimental papers reporting Bell violations, the statistical relevance of the observed violation is expressed in terms of the number of standard deviations separating the estimated violation from its local bound. Their are several problems with this analysis, however. First it lacks a clear operational significance. Second, it implicitly assumes some underlying Gaussian distribution for the measured systems, which is only justified if the number of trials approaches infinity. It also relies on the assumption that the random process associated to the $k$-th trial is independent from the chosen settings and observed outcomes of the previous $k-1$ trials. In other words, the devices are assumed to have no memory, which is a questionable assumption \cite{AR00}.

A better measure of the strength of the evidence against local models is given by the probability with which the observed data could have been reproduced by a local model. For instance, consider the CHSH test and let $\langle a_xb_y\rangle_\text{obs}$ be the mean of the observed products of $a$ and $b$ when measurement $x$ and $y$ are chosen computed over $N$ trials.  It is shown in \cite{Gill12}, that the probability that two devices behaving according to a local model give rise to a value $S_\text{obs}=\langle a_0b_0\rangle_\text{obs}+\langle a_0b_1\rangle_\text{obs}+\langle a_1b_0\rangle_\text{obs}-\langle a_1b_1\rangle_\text{obs}>2+\epsilon$ is
\be
p\left(S_\text{obs}>2+\epsilon\right)\leq 8e^{-4N(\frac{\epsilon}{16})^2}\,.
\ee
This statement assumes that the behavior of the devices at the $k$-th trial does not depend on the inputs and outputs in previous runs. But this memoryless assumption can be avoided and similar statements taking into account arbitrary memory effects can be obtained \cite{Gill03}. As discussed in section~\ref{repetition}, it is easy to convince oneself that, in the limit of infinitely many runs, devices with memory cannot fake the violation of a Bell inequality \cite{BCHK+02,Gill03}. Indeed, for any given local variable strategy, there is always at least one set of settings for which that strategy fails in the corresponding Bell game. But in any run, the settings are chosen at random, independently of the source and of the past: therefore, even taking the past into account, the local variables cannot avoid the possibility that the wrong settings are chosen. This reasoning can be extended to the finite statistics case through the use of martingales \cite{Gill03}. A better general method to deal with memory effects and finite statistics which is asymptotically optimal in the limit of large trials has been proposed in \cite{ZGK11,ZKG13}.

\subsection{Towards a loophole-free Bell test}\label{towards}

\subsubsection{Photons}

The main drawback of photonic experiments for performing a loophole-free Bell test is the limited detection efficiency of single photon detectors. Nevertheless, considerable progress has been achieved in the last years, in particular with the development of detectors based on superconducting materials, which can achieve detection efficiencies above $95\%$. Such detectors were recently used in Bell-type experiment. First, in \cite{SGAB+1212} a detection loophole-free demonstration of quantum steering (see Section \ref{steering}) was reported. More recently, an experiment demonstrated violation of Clauser-Horne Bell inequality with the detection loophole closed \cite{GMR+12,CMAC+13}. Here total efficiencies of $~75\%$ were achieved for each party. Note however, that these experiments are table-top and did not close the locality loophole. Since the detection process in superconducting detectors is typically slow (of order of $\mu s$), achieving a loophole-free Bell violation will require a separation of the order of 300m.
 
Another possibility was recently suggested in \cite{CS12}, which consists in pre-certifying the presence of a photon before the choice of local measurement is performed. This proposal appears however challenging from an experimental point of view.

\subsubsection{Continuous variable systems.}\label{Continuous variable systems}
An interesting alternative for achieving high detection efficiencies using photon consists in using homodyne measurements, which measure a continuous degree of freedom (often called quadrature) of the optical mode. Such measurements are realized by mixing the optical mode with an intense reference oscillator on a beam splitter, and can reach efficiencies close to unity nowadays. 
First proposal for using homodyne measurements in Bell tests were presented in \cite{GPY88,TWC91} and in \cite{GDR98}. Since then, many alternative proposals were discussed. However, their experimental realization has remained elusive so far.

First of all, it is important to mention that a homodyne measurement has a continuous number of possible outcomes. Since Bell inequalities have typically a discrete number of outcomes (for instance, binary outcomes in the case of CHSH), one has to dichotomize the outcome of the homodyne measurement, a procedure referred to as \emph{binning}. For instance, a natural dichotomization strategy is given by the \emph{sign binning}, where one assigns the values $+1$ if the measurement returns a positive outcome, and $-1$ otherwise. 

Homodyne measurements were shown to be able to detect the nonlocality of certain quantum states. However, all tests proposed so far  present major difficulties for experimental realizations. First, several schemes consider quantum states which cannot be produced using current technology \cite{Munro99,CFRD07,ACFN09,WHGT+03}. Second, the proposals of \cite{GFCW+04,NC04} (see also \cite{PFC05}) use states which could be realized experimentally but lead only to very small Bell inequality violations, hence requiring an extremely low level of noise, currently out of reach from an experimental point of view. Note that an experiment using homodyne measurements demonstrated a violation of the CHSH inequality \cite{BAL05}, but post-selection was involved which thus opened the detection loophole. 

An interesting alternative consists in devising hybrid schemes, which make use of homodyne measurements as well as photo-detection. In \cite{CBSL+11} it was shown that relatively high CHSH violations can be achieved using a state that can be experimentally produced \footnote{Note that the idea of considering hybrid measurements was first discussed in \cite{JKLZ+10}, although the proposed scheme is not a proper Bell test \cite{CS11}.}. Promising developments of hybrid schemes were recently discussed in \cite{QACS+11,AQCS+11,BBCL12,BC12,TAQM+12}.

Finally, several works also considered more complex measurements, such as parity measurements. In particular, in \cite{BW98} (see also \cite{BW99}) it is demonstrated that such measurements can reveal the nonlocality of the famous EPR state, discussed in \cite{EPR35}. The use of measurements based on nonlinear local operations was proposed in \cite{SJR07}. However, realization of such measurements is still out of reach from current technologies.

\subsubsection{Atom-atom and atom-photon entanglement}
A promising avenue towards a loophole-free Bell test is based on the the possibility to generate long-distance entanglement between two trapped atoms \cite{SI03}. The procedure for entangling the two remote atoms consists in the joint detection of two photons, each coming from one of the atoms, in an entangled basis. In this way, the initial atom-photon entanglement is transformed into atom-atom entanglement via entanglement swapping.

This scheme was demonstrated experimentally using two trapped atoms separated by one meter \cite{MMMO+08,PAMB+10} and more recently up to 20 meters \cite{HKO+12}. The detection loophole was closed in these experiments, thanks to the near unit efficiency of atomic measurements. In order to close the locality loophole, a distance of the order of 300 meters would be required using the fastest atomic measurement techniques available today \cite{RWVH09}. This is still currently challenging but the perspectives for a loophole-free test are promising.

Bell tests based on the direct observation of atom-photon entanglement have also been proposed \cite{BGSS07,CL07}. However, the difficulty of efficiently collecting the photons emitted from the atom, and the relatively high detection efficiencies required for the photon detection in order to close the detection loophole make this approach more delicate to implement. To overcome some of these problems, the use of continuous variable degrees of freedom of the light field, combined with efficient homodyne measurements was recently explored \cite{SBGS+11,AQCS+11,TAQM+12}.

More recently, it was proposed to achieve a loophole-free Bell test using spin photon interactions in cavities \cite{SBMG+13,BYHR13}. Here the entanglement of a pair of photons is mapped to two distant spins (e.g. carried by a single atom or a quantum dot). Importantly, this mapping can be achieved in a heralded way. By choosing the measurement settings only after successful heralding, the scheme is immune from the detection loophole, since spin systems can usually be measured with high efficiencies.

\subsection{Bell tests without alignment}

Bell inequality violations in quantum mechanics requires the parties to perform judiciously chosen measurement settings on an entangled state.
Experimentally, this requires a careful calibration of the measurement devices and alignment of a shared reference frame between the distant parties. 
Although such assumptions are typically made implicitly in theoretical works, establishing a common reference frame, as well as aligning and calibrating measurement devices in experimental situations are never trivial issues. For instance, in quantum communications via optical fibres, unavoidable small temperature changes induce strong rotations of the polarization of photons in the fibre, which makes it challenging to maintain a good alignment.
In turn this may considerably degrade the implementation of quantum protocols, such as Bell tests. 

This lead several authors to investigate whether Bell tests, and more generally quantum communication protocols \cite{BRS07}, could be realized without the need of a common reference frame. The first approach proposed was based on decoherence-free subspaces \cite{Cabello03}. The experimental realization of such ideas is challenging as it requires high-dimensional entanglement, although progress was recently reported \cite{DNWA+12}.

A more recent approach investigated Bell tests performed with randomly chosen measurements \cite{LHBR10}. This first theoretical work considered the CHSH Bell scenario, with qubit measurements chosen randomly and uniformly (according to the Haar measure) on the Bloch sphere, on a maximally entangled state of two-qubits. When all four measurements are chosen at random, the probability that the obtained statistics will violate the CHSH inequality is $\sim 28 \% $. When unbiased measurements are used, this probability increases to $\sim 42\%$. More recently it was shown however that if both parties perform three unbiased measurements (i.e. forming an orthogonal triad on the Bloch sphere), the probability of violating a Bell inequality becomes one \cite{SVLB+12,WB12}). This scheme was realized experimentally demonstrating the robustness of the effect to experimental imperfections such as losses and finite statistics \cite{SVLB+12} (see also \cite{PWBP12}). From a more conceptual point of view, these works are interesting as they show that quantum nonlocality is much more generic than previously thought.

\section{Related concepts}
\label{Related concepts}
The last section of this review deals with variations around Bell's theorem, in which different notions of non-locality -- stronger or weaker than Bell's one -- are considered. Note that there are also exist mathematical relations between local models and non-contextual models. We do not review this connection here, see \cite{Mermin93} for a short review of both concepts and their relation.

\subsection{Bi-locality and more general correlation scenarios}

In a tripartite Bell scenario, the standard definition of locality is given by 

\ba \label{loc3}  p(abc|xyz) = \int d \lambda q(\lambda)p_\lambda(a|x)p_\lambda(b|y)p_\lambda(c|z) \ea
where $\lambda$ is a shared local random variable and $ \int d \lambda q(\lambda)=1$. 
This represents the most general model in which the outcome of each each observer is determined by his input and $\lambda$. Since $\lambda$ is shared between all three parties, arbitrary prior correlations can be established between the parties.

In certain quantum experiments involving three separated observers, the distribution $p(abc|xyz)$ is obtained by performing measurements on independent quantum states, originating from different sources. A typical example is the protocol of entanglement swapping \cite{ZZHE93}---also known as teleportation of entanglement---in which two systems that never interacted become entangled. Here, one party (say Bob), shares initially an entangled state with both Alice (denoted system $AB_1$) and Charlie (system $B_2C$). 
That is, Alice and Bob share an entangled pair distributed by a source located between them, while a second source distributes entanglement between Bob and Charlie. Importantly, these two sources are completely independent from each other, hence systems $AB_1$ and $B_2C$ share no prior correlations.
It is then natural to ask whether the observed correlations $p(abc|xyz)$ can be reproduced using a local model with the same feature, that is, in which systems $AB_1$ and $B_2C$ are initially uncorrelated. Formally such models can be written in the following form
\be \label{biloc}  p(abc|xyz)= \int d \lambda d \mu q(\lambda) q(\mu) p_\lambda(a|x)p_{\lambda,\mu}(b|y)p_\mu(c|z)  \ee 
where Alice and Bob share the local random variable $\lambda$, while Bob and Charlie share $\mu$. Since the variables $\lambda$ and $\mu$ are assumed to be independent, their distribution factorizes, i.e. $q(\lambda,\mu)= q(\lambda) q(\mu)$. The above condition is termed \emph{bilocality}, and correlations satisfying it are called bilocal.
It turns out that not all local correlations (i.e. of the form \eqref{loc3}) can be written in the bilocal form. Hence the bilocality condition is a strictly stronger constraint than locality. it is then interesting to ask how to characterize the set of bilocal correlations, as this will lead to more stringent tests for revealing nonlocality in an entanglement swapping experiment.

First exploratory works investigated this question in the context of the detection loophole \cite{GG02,ZGHZ08}. More recently, a systematic approach was taken in \cite{BGP10,BRGP12}. In particular, these works present nonlinear inequalities for testing the bilocality condition, which are much more stringent compared to standard Bell inequalities.
Note that the set of bilocal correlations is not convex in general, hence its characterization requires nonlinear inequalities.

More generally, it is possible to consider an arbitrary correlation scenario, involving an arbitrary number of sources and observers, where certain systems can be initially correlated while others are independent. Similarly to the above discussion, when two systems are assumed to be independent, they are described by a product distribution \cite{BRGP12,Fritz12}. In fact, a typical Bell experiment can be viewed in this picture, featuring three independent sources. These are the source that produces the entangled state, the source generating the measurement settings of Alice, and the source generating the setting of Bob. Indeed, if these sources are not independent, the Bell violation is plagued by the measurement-independence loophole. A related approach for considering locality in general correlation scenarios using the formalism of causal networks was put forward in \cite{WS12}.

\subsection{No-go theorems for nonlocal models}

The study of no-go theorems for \textit{nonlocal models} is reduced to few examples. On the one hand, it is obvious that some of these models will reproduce all observed correlations, so there is no hope of finding a result \`a la Bell which would falsify all of them. On the other hand, one needs to have a good motivation in order to propose a specific example of nonlocal model. In fact, basically two classes of models have been considered so far: we review them briefly here, but refer to the original articles for a detailed justification of the interest of each model.

\subsubsection{Models \`a la Leggett}

Pure entangled states are characterized by the fact that \textit{the properties of the pair are well defined, but those of the individual subsystems are not}. 
Consider for instance a maximally entangled state of two qubits. Although the global state has zero entropy, the state being pure, the reduced state of each qubit is fully mixed thus having maximum entropy. An interesting question is whether one could devise alternative no-signaling models, reproducing quantum correlations, in which the individual properties are well-defined, or at least where the model gives a higher degree of predictability compared to quantum predictions. 
The first work in this direction was presented by \cite{Leggett03}, who discussed a specific nonlocal model and proved its inability to reproduce quantum correlations. Leggett's model was first tested experimentally by \cite{GPKB+07}. A clear discussion of the scope and limitations of this type of models is found in \cite{BBGK+08}. 

In a nutshell, the question is whether the probability distribution predicted by quantum theory $p_Q$ can be seen as a convex combination of more fundamental distributions: $p_Q=\int d\lambda p_\lambda$. Because of Bell's theorem, for some $\lambda$ at least, the distribution $p_\lambda$ must be nonlocal; but let us leave the correlations and their mechanism aside and concentrate on the marginals: we request that the $p_\lambda$ specify well defined individual properties. Focusing on the case of a maximally entangled qubit pair, Leggett proposed a model in which the marginals take the form $p_\lambda (a|\vec{x})=\frac{1}{2}(1+a \vec{u} \cdot\vec{x})$ and $p_\lambda (b|\vec{y})=\frac{1}{2}(1+b \vec{v} \cdot\vec{y})$. Here the the hidden variables consists of two vectors: $\lambda = (\vec{u},\vec{v})$. The intuition behind this model is the following. Locally, say on Alice's side, the system behaves as if it had well defined polarization given by $\vec{u}$. For a measurement direction $\vec{x}$, the marginal distribution is then given by Malus' law. Hence this model makes more definite predictions for individual properties compared to quantum theory. It turns out however that Leggett's model is unable to reproduce quantum correlations. In particular, from the no-signaling condition and the above marginals, one can derive constraints, in the form of inequalities, on the correlations. Quantum predictions violate these inequalities. Note that there is no direct relation between Leggett inequalities and Bell inequalities. In particular, the tests of Leggett inequalities known to date rely on the characterization of the measurement parameters and are therefore not device-independent, contrary to Bell inequalities.

Finally, note that more general models were also discussed and demonstrated to be incompatible with quantum predictions. This shows that quantum correlations cannot be reproduced using no-signaling theories which make more accurate predictions of individual properties compared to quantum theory \cite{CR08,CR12}

\subsubsection{Superluminal signaling models}

A second class of models addresses the possibility of explaining quantum correlation using some explicit \textit{signaling} mechanism. Of course, this is problematic, because the signal should propagate faster than light: these models must thus postulate the existence of a preferred frame in which this signal propagates. Two cases have been considered.

In the first one, the preferred frame is universal. From Bell's theorem, it follows that the speed $v$ of the signal must be infinite. Clearly, one can find a model with $v$ being infinite which reproduces all quantum correlations. On the other hand, the predictions of any signaling model where $v$ is finite will differ from those of quantum mechanics. For instance, in a bipartite Bell test, nonlocal correlations should vanish when the distance between the two observers exceeds a certain bound, since there is then simply not enough time for the signal to propagate. Experimental investigations could place lower bounds on $v$ \cite{SBBG+08}. For a wide class of preferred frames, this bound exceeds the speed of light by several orders of magnitude.

Furthermore, it is in fact possible to rule out theoretically any communication model in which $v$ is finite using certain assumptions. Specifically, consider a model that 1) reproduce the quantum predictions when there is enough time for a signal to propagate at speed $v$ between the parties; 2) the model is no-signalling on average, that is at the level of the observed statistics the no-signaling conditions (\ref{sec2.nosig}) are satisfied (i.e. any explicit signaling only happens at the hidden level). Then by considering a multipartite arrangement, it is shown in \cite{BPALSG11}, building on earlier work in \cite{SG05,CHW11}, that these two conditions are mutually incompatible. In other words,  under the assumption that the observed statistics satisfy the no-signaling principle, quantum correlations cannot be reproduced by a model with finite speed signaling.

In the second type of models, the rest frame of each measurement device is its own preferred frame. In this case, if the measurement devices moves away from one another, a \textit{before-before} configuration can be created, in which each particle perceives that it is the first one to undergo the measurement: then, nonlocal correlations should disappear \cite{SS97}. This prediction has been falsified experimentally \cite{ZBGT01,SZGS02,SZGS03}.

\subsection{Steering}\label{steering}

One of the most common ways of describing the effect of entanglement, noticed already in the seminal paper of \cite{Schrodinger36}, uses the notion of \textit{steering}: by making a measurement on her system, Alice can prepare at a distance Bob's state. More precisely, Alice cannot choose \textit{which} state she prepares in each instance, because this would amount to signaling; however, if she sends to Bob the results of her measurements, Bob can check that indeed his conditional states has been steered.

Though often invoked in the field, this notion had not been the object of detailed studies until the work of \cite{WJD07}. In this and subsequent work, these authors formalized steering as information-theoretic tasks, and pointed out how it differs from nonlocality. Steering can be viewed as the distribution of entanglement from an untrusted party.
Alice wants to convince Bob that she can prepare entanglement. Bob trusts his measurement device, hence he knows what observables he is measuring on his system. However Bob does not trust Alice, whose device is then described by a black box. 
In this sense the task is intermediate between nonlocality (in which both Alice and Bob work with black boxes) and standard entanglement witnessing (in which both parties have perfect control of the observables which are measured). The protocol works as follows. Alice sends a quantum system to Bob, whose state, she claims, is entangled to her system. Upon receiving his system, Bob chooses a measurement setting (from a pre-determined set of measurements) to perform on it. He then informs Alice about his choice of measurement, and asks her to provide a guess for his measurement outcome. After repeating this procedure a sufficiently large number of times, Bob can estimate how strongly his system is correlated to that of Alice. If the correlations are strong enough, Bob concludes that the systems are indeed entangled and that Alice did indeed steer his state.

Interestingly, it turns out that entanglement is necessary but not sufficient for steering, while steering is necessary but not sufficient for nonlocality. Hence, steering represents a novel form of inseparability in quantum mechanics, intermediate between entanglement and nonlocality \cite{WJD07,SJWP10}. The quantitative relation between steering, entanglement, and Bell nonlocality is yet to be fully understood.

Experimentally, steering can be tested using steering inequalities, similar to Bell inequalities. The first steering criterion were derived for continuous variable systems, mostly based on variances of observables \cite{Reid89} and entropic uncertainty relations \cite{WSGT+11}; see \cite{RDBC+09} for a review. More recently, steering inequalities were presented for discrete variables \cite{CJWR09}. All these tests were investigated experimentally. Similarly to Bell tests, experimental steering tests suffer from loopholes. Nevertheless, closing these loopholes appears to be less demanding compared to Bell tests, in particular in terms of detection efficiency. A loophole-free steering experiment was recently reported \cite{WRSL+12}.

\subsection{Semi-quantum games}

In the usual Bell test scenario, distant parties share a quantum state distributed from a source. The local measurements and their outcomes are represented by classical data. In a recent paper, \cite{Buscemi12} proposed a variant of Bell tests, termed semi-quantum games, in which the inputs of each party are given by quantum states. That is, each party holds a black box in which the observer inputs a quantum state. Inside the box, this quantum input is then jointly measured with the quantum system coming from the source, and a classical output is produced. In case the input quantum states are orthogonal (hence perfectly distinguishable), the setup is simply equivalent to a standard Bell test with classical inputs.
However, when the states are not mutually orthogonal, the setup becomes more general. Buscemi showed that, in this case, for any entangled state $\rho$ there exists a semi-quantum game, the statistics of which cannot be reproduced by a local model. Hence, semi-quantum games highlight a nonlocal aspect of quantum states, which is however different from Bell nonlocality. More recently it was shown that semi-quantum games and  entanglement witnesses are intimately related. In particular, for detecting an entangled state $\rho$, a semi-quantum game can be constructed directly from an entanglement witness detecting $\rho$ \cite{BRLG12}. Applications of these ideas for the detection of entanglement \cite{BRLG12}, steering \cite{CHW12}, and the classical simulation of quantum correlations \cite{RBLG13}, were recently discussed.

\section{Conclusion}
\label{Conclusion}
Fifty years after the publication of Bell's theorem \cite{Bell64}, Bell nonlocality is still -- perhaps more than ever --  at the center of an active and intense research activity that spans the foundations of quantum theory, applications in quantum information science, and experimental implementations.

We covered in this review most of the recent developments, some of them happening while this review was being written. To give only three examples of recent progress on longstanding or important questions: Peres conjecture that no bound entangled state can give rise to non-local correlations is now disproved 
in the tripartite case (V´ertesi and Brunner, 2012) (but is still open in the bipartite case); it has been shown that non-local correlations can be exploited to perform arbitrary computations in a device-independent way \cite{RUV12}; on the experimental side, the detection loophole has finally been closed in full-optical implementations \cite{GMR+12,CMAC+13}. We hope that this review paper will motivate further developments on this fascinating topic of Bell non-locality.

\begin{acknowledgments}
We would like to thank all our nonlocal colleagues for enlightening discussions over the recent years. Also, we thank Antonio Ac\'{\i}n, Alain Aspect, Richard Gill, Nicolas Gisin, and Yeong-Cherng Liang, for comments on this manuscript. This work is supported by the National Research Foundation and the Ministry of
Education of Singapore, the UK EPSRC, the Swiss National Science Foundations (grant PP00P2\_138917), the EU CHISTERA projects DIQIP and QCS, the FRS-FNRS project DIQIP, and the Brussels-Capital Region through a BB2B Grant.
\end{acknowledgments}

\begin{appendix}

\section*{Guide to Bell inequalities}
\label{Guide to Bell inequalities}

The goal of this section is to orientate the reader looking for a particular type of Bell inequality. Here we classify Bell inequalities according to the number of parties $n$, the number of measurements for each party $m$, and the number of outcomes $\Delta$ \footnote{Note that for Bell inequalities featuring different number of measurements or outcomes for different parties, $m$ and $\Delta$ represent the maximum values.}. Below, for any Bell test scenario given by the triple $(n,m,\Delta)$, we provide references to articles presenting relevant Bell inequalities. Note that we do not give the inequalities explicitly; some of these can nevertheless be found in parts of this review, in particular in Sections II and V.

\subsection{Bipartite Bell inequalities}

\subsubsection{Binary outputs: $(2,m,2)$}

For the case $m=2$, CHSH is the only tight Bell inequality. Note that if one of the parties has more measurement, i.e. $m_A=2$ and $m_B=m$, CHSH is still known to be the only tight inequality. 

For $m=3$, one additional tight inequality arises: $I_{3322}$ (see Section 2), discovered independently by Froissard \cite{Froissard81} and Collins and Gisin \cite{CG04}. 

For $m=4,5$, the complete list of tight Bell inequalities is unfortunately not known. Incomplete lists can be found in \cite{BG08,PV09,BGP10b}.
Note that for $m_A=4$ and $m_B=3$, the complete list of tight Bell inequalities was given in \cite{CG04}.

For $m\geq 6$, much less is known. Incomplete lists of facets could be derived using sophisticated techniques from convex geometry \cite{aiis04}. 

For arbitrary values of $m$, the family of inequalities $I_{mm22}$ introduced in \cite{CG04} turns out to be tight in general \cite{AI07}.
It is also worth mentioning the family of chained Bell inequalities \cite{Pearle70,BC90}, although these inequalities are not tight for $m\geq3$.

\subsubsection{Arbitrary number of outputs: $(2,m,\Delta)$}

In the case $m=2$ and $\Delta=3$, the inequality of CGLMP \cite{CGL+02} is the only tight inequality additional to CHSH. For $\Delta \geq 4$, the CGLMP inequality is known to be tight, but there exist additional facets in this case \cite{BGP10b}. 

Note that the chained Bell inequalities have been generalized to this scenario \cite{BKP07}. Whether they are tight or not for $\Delta \geq3$ is not known.

\subsection{Multipartite Bell inequalities}

\subsubsection{Binary outputs: $(n,m,2)$}

All tight correlation Bell inequalities are known for the case $m=2$ \cite{WW01,ZB02}. Indeed, this set contains the inequalities of Mermin-Ardehali-Belinskiõ-Klyshko (MABK) \cite{Mermin90,Ardehali92,BK93}. 

For non-correlation Bell inequality, the complete set of tight Bell inequalities in the case $n=3$ and $m=2$ was given in \cite{Sliwa03}. For arbitrary $n$ and $m$, Ref. \cite{LPZB04} provided tight Bell inequalities.

\subsubsection{Arbitrary number of outputs: $(n,m,\Delta)$}

A general family of Bell inequalities based on variance inequalities was derived in \cite{CFRD07}, and further developed in \cite{HCRD09}. More generally these inequalities are particular cases of inequalities discussed in \cite{SCA+10}.
Note that these inequalities are not tight as they are not linear. However, they can be conveniently used for continuous variables (i.e. $\Delta \rightarrow \infty$).

A generalization of the Mermin inequalities for the scenario $(3,2,\Delta)$ was presented in \cite{GLBB+12}; these inequalities are known to be tight for $\Delta \leq 8$. For $n=3$, $m=2$ and $\Delta=3,4,5$, tight inequalities were given in \cite{CWKO08}. 

Note also that there exist functional Bell inequalities which can be defined for an infinite number of settings \cite{SSZ02}.

\subsubsection{Bell inequalities detecting genuine multipartite nonlocality}

Svetlichny derived the first inequality for testing genuine multipartite nonlocality \cite{Svetlichny87} in the case $(3,2,2)$. This inequality was later generalized to an arbitrary number of parties, e.g. to the scenario $(n,2,2)$ \cite{SS02,CGPR+02}. 
\cite{BGP10b} also introduced more inequalities for the simplest case $(3,2,2)$.

Svetlichny's inequality was also generalized to a more general scenario: $(n,2,\Delta)$ in \cite{BBBGL11} and $(n,m,\Delta)$ in \cite{BBBG+12}.

In \cite{BPBG11}, the authors discussed various definitions of the concept of genuine multipartite nonlocality, and introduced inequalities for testing each of these models in the scenario $(3,2,2)$.

\end{appendix}



\begin{thebibliography}{512}
\expandafter\ifx\csname natexlab\endcsname\relax\def\natexlab#1{#1}\fi
\expandafter\ifx\csname bibnamefont\endcsname\relax
  \def\bibnamefont#1{#1}\fi
\expandafter\ifx\csname bibfnamefont\endcsname\relax
  \def\bibfnamefont#1{#1}\fi
\expandafter\ifx\csname citenamefont\endcsname\relax
  \def\citenamefont#1{#1}\fi
\expandafter\ifx\csname url\endcsname\relax
  \def\url#1{\texttt{#1}}\fi
\expandafter\ifx\csname urlprefix\endcsname\relax\def\urlprefix{URL }\fi
\providecommand{\bibinfo}[2]{#2}
\providecommand{\eprint}[2][]{\url{#2}}

\bibitem[{\citenamefont{Accardi and Regoli}(2000)}]{AR00}
\bibinfo{author}{\bibnamefont{Accardi}, \bibfnamefont{L.}}, and
  \bibinfo{author}{\bibfnamefont{M.}~\bibnamefont{Regoli}},
  \bibinfo{year}{2000}, \bibinfo{journal}{Preprint 399, Volterra Institute,
  University of Rome II} .

\bibitem[{\citenamefont{Ac\'in}(2001)}]{Acin01}
\bibinfo{author}{\bibnamefont{Ac\'in}, \bibfnamefont{A.}},
  \bibinfo{year}{2001}, \bibinfo{journal}{Phys. Rev. Lett.}
  \textbf{\bibinfo{volume}{88}}, \bibinfo{pages}{027901}.

\bibitem[{\citenamefont{Ac\'in} \emph{et~al.}(2010)\citenamefont{Ac\'in,
  Augusiak, Cavalcanti, Hadley, Korbicz, Lewenstein, and Piani}}]{AACC+10}
\bibinfo{author}{\bibnamefont{Ac\'in}, \bibfnamefont{A.}},
  \bibinfo{author}{\bibfnamefont{R.}~\bibnamefont{Augusiak}},
  \bibinfo{author}{\bibfnamefont{D.}~\bibnamefont{Cavalcanti}},
  \bibinfo{author}{\bibfnamefont{C.}~\bibnamefont{Hadley}},
  \bibinfo{author}{\bibfnamefont{J.}~\bibnamefont{Korbicz}},
  \bibinfo{author}{\bibfnamefont{M.}~\bibnamefont{Lewenstein}}, and
  \bibinfo{author}{\bibfnamefont{M.}~\bibnamefont{Piani}},
  \bibinfo{year}{2010}, \bibinfo{journal}{Phys. Rev. Lett.}
  \textbf{\bibinfo{volume}{104}}, \bibinfo{pages}{140404}.

\bibitem[{\citenamefont{Ac\'in} \emph{et~al.}(2007)\citenamefont{Ac\'in,
  Brunner, Gisin, Massar, Pironio, and Scarani}}]{ABGM+07}
\bibinfo{author}{\bibnamefont{Ac\'in}, \bibfnamefont{A.}},
  \bibinfo{author}{\bibfnamefont{N.}~\bibnamefont{Brunner}},
  \bibinfo{author}{\bibfnamefont{N.}~\bibnamefont{Gisin}},
  \bibinfo{author}{\bibfnamefont{S.}~\bibnamefont{Massar}},
  \bibinfo{author}{\bibfnamefont{S.}~\bibnamefont{Pironio}}, and
  \bibinfo{author}{\bibfnamefont{V.}~\bibnamefont{Scarani}},
  \bibinfo{year}{2007}, \bibinfo{journal}{Phys. Rev. Lett.}
  \textbf{\bibinfo{volume}{98}}, \bibinfo{pages}{230501}.

\bibitem[{\citenamefont{Ac\'in} \emph{et~al.}(2009)\citenamefont{Ac\'in, Cerf,
  Ferraro, and Niset}}]{ACFN09}
\bibinfo{author}{\bibnamefont{Ac\'in}, \bibfnamefont{A.}},
  \bibinfo{author}{\bibfnamefont{N.~J.} \bibnamefont{Cerf}},
  \bibinfo{author}{\bibfnamefont{A.}~\bibnamefont{Ferraro}}, and
  \bibinfo{author}{\bibfnamefont{J.}~\bibnamefont{Niset}},
  \bibinfo{year}{2009}, \bibinfo{journal}{Phys. Rev.~A}
  \textbf{\bibinfo{volume}{79}}, \bibinfo{pages}{012112}.

\bibitem[{\citenamefont{Ac\'in} \emph{et~al.}(2002)\citenamefont{Ac\'in, Durt,
  Gisin, and Latorre}}]{ADGL02}
\bibinfo{author}{\bibnamefont{Ac\'in}, \bibfnamefont{A.}},
  \bibinfo{author}{\bibfnamefont{T.}~\bibnamefont{Durt}},
  \bibinfo{author}{\bibfnamefont{N.}~\bibnamefont{Gisin}}, and
  \bibinfo{author}{\bibfnamefont{J.}~\bibnamefont{Latorre}},
  \bibinfo{year}{2002}, \bibinfo{journal}{Phys. Rev.~A}
  \textbf{\bibinfo{volume}{65}}, \bibinfo{pages}{052325}.

\bibitem[{\citenamefont{Ac\'in} \emph{et~al.}(2005)\citenamefont{Ac\'in, Gill,
  and Gisin}}]{AGG05}
\bibinfo{author}{\bibnamefont{Ac\'in}, \bibfnamefont{A.}},
  \bibinfo{author}{\bibfnamefont{R.}~\bibnamefont{Gill}}, and
  \bibinfo{author}{\bibfnamefont{N.}~\bibnamefont{Gisin}},
  \bibinfo{year}{2005}, \bibinfo{journal}{Phys. Rev. Lett.}
  \textbf{\bibinfo{volume}{95}}, \bibinfo{pages}{210402}.

\bibitem[{\citenamefont{Ac\'in}
  \emph{et~al.}(2006{\natexlab{a}})\citenamefont{Ac\'in, Gisin, and
  Masanes}}]{AGM06}
\bibinfo{author}{\bibnamefont{Ac\'in}, \bibfnamefont{A.}},
  \bibinfo{author}{\bibfnamefont{N.}~\bibnamefont{Gisin}}, and
  \bibinfo{author}{\bibfnamefont{L.}~\bibnamefont{Masanes}},
  \bibinfo{year}{2006}{\natexlab{a}}, \bibinfo{journal}{Phys. Rev. Lett.}
  \textbf{\bibinfo{volume}{97}}, \bibinfo{pages}{120405}.

\bibitem[{\citenamefont{Ac\'in}
  \emph{et~al.}(2006{\natexlab{b}})\citenamefont{Ac\'in, Gisin, and
  Toner}}]{AGT06}
\bibinfo{author}{\bibnamefont{Ac\'in}, \bibfnamefont{A.}},
  \bibinfo{author}{\bibfnamefont{N.}~\bibnamefont{Gisin}}, and
  \bibinfo{author}{\bibfnamefont{B.}~\bibnamefont{Toner}},
  \bibinfo{year}{2006}{\natexlab{b}}, \bibinfo{journal}{Phys. Rev.~A}
  \textbf{\bibinfo{volume}{73}}, \bibinfo{pages}{062105}.

\bibitem[{\citenamefont{Ac\'in}
  \emph{et~al.}(2006{\natexlab{c}})\citenamefont{Ac\'in, Massar, and
  Pironio}}]{AMP06}
\bibinfo{author}{\bibnamefont{Ac\'in}, \bibfnamefont{A.}},
  \bibinfo{author}{\bibfnamefont{S.}~\bibnamefont{Massar}}, and
  \bibinfo{author}{\bibfnamefont{S.}~\bibnamefont{Pironio}},
  \bibinfo{year}{2006}{\natexlab{c}}, \bibinfo{journal}{New J.~Phys}
  \textbf{\bibinfo{volume}{8}}, \bibinfo{pages}{126}.

\bibitem[{\citenamefont{Ac\'in} \emph{et~al.}(2003)\citenamefont{Ac\'in,
  Scarani, and Wolf}}]{ASW03}
\bibinfo{author}{\bibnamefont{Ac\'in}, \bibfnamefont{A.}},
  \bibinfo{author}{\bibfnamefont{V.}~\bibnamefont{Scarani}}, and
  \bibinfo{author}{\bibfnamefont{M.~M.} \bibnamefont{Wolf}},
  \bibinfo{year}{2003}, \bibinfo{journal}{J.~Phys. A: Math. Theor.}
  \textbf{\bibinfo{volume}{36}}, \bibinfo{pages}{L21}.

\bibitem[{\citenamefont{Ac\'in} \emph{et~al.}(2012)\citenamefont{Ac\'in,
  S.Massar, and Pironio}}]{AMP12}
\bibinfo{author}{\bibnamefont{Ac\'in}, \bibfnamefont{A.}},
  \bibinfo{author}{\bibnamefont{S.Massar}}, and
  \bibinfo{author}{\bibfnamefont{S.}~\bibnamefont{Pironio}},
  \bibinfo{year}{2012}, \bibinfo{journal}{Phys. Rev. Lett.}
  \textbf{\bibinfo{volume}{108}}, \bibinfo{pages}{100402}.

\bibitem[{\citenamefont{Adesso and Piano}(2014)}]{AP13}
\bibinfo{author}{\bibnamefont{Adesso}, \bibfnamefont{G.}}, and
  \bibinfo{author}{\bibfnamefont{S.}~\bibnamefont{Piano}},
  \bibinfo{year}{2014}, \bibinfo{journal}{Phys. Rev. Lett.}
  \textbf{\bibinfo{volume}{112}}, \bibinfo{pages}{010401}.

\bibitem[{\citenamefont{Afzelius}(2011)}]{Afzelius11}
\bibinfo{author}{\bibnamefont{Afzelius}, \bibfnamefont{M.}},
  \bibinfo{year}{2011}, \bibinfo{note}{private communication}.

\bibitem[{\citenamefont{Aharon} \emph{et~al.}(2013)\citenamefont{Aharon,
  Machnes, Reznik, Silman, and Vaidman}}]{AMR+12}
\bibinfo{author}{\bibnamefont{Aharon}, \bibfnamefont{N.}},
  \bibinfo{author}{\bibfnamefont{S.}~\bibnamefont{Machnes}},
  \bibinfo{author}{\bibfnamefont{B.}~\bibnamefont{Reznik}},
  \bibinfo{author}{\bibfnamefont{J.}~\bibnamefont{Silman}}, and
  \bibinfo{author}{\bibfnamefont{L.}~\bibnamefont{Vaidman}},
  \bibinfo{year}{2013}, \bibinfo{journal}{Nat. Comput.}
  \textbf{\bibinfo{volume}{12}}, \bibinfo{pages}{5}.

\bibitem[{\citenamefont{Ajoy and Rungta}(2010)}]{AR10}
\bibinfo{author}{\bibnamefont{Ajoy}, \bibfnamefont{A.}}, and
  \bibinfo{author}{\bibfnamefont{P.}~\bibnamefont{Rungta}},
  \bibinfo{year}{2010}, \bibinfo{journal}{Phys. Rev.~A}
  \textbf{\bibinfo{volume}{81}}, \bibinfo{pages}{052334}.

\bibitem[{\citenamefont{Allcock}
  \emph{et~al.}(2009{\natexlab{a}})\citenamefont{Allcock, Brunner, Linden,
  Popescu, Skrzypczyk, and V\'ertesi}}]{ABLP+09}
\bibinfo{author}{\bibnamefont{Allcock}, \bibfnamefont{J.}},
  \bibinfo{author}{\bibfnamefont{N.}~\bibnamefont{Brunner}},
  \bibinfo{author}{\bibfnamefont{N.}~\bibnamefont{Linden}},
  \bibinfo{author}{\bibfnamefont{S.}~\bibnamefont{Popescu}},
  \bibinfo{author}{\bibfnamefont{P.}~\bibnamefont{Skrzypczyk}}, and
  \bibinfo{author}{\bibfnamefont{T.}~\bibnamefont{V\'ertesi}},
  \bibinfo{year}{2009}{\natexlab{a}}, \bibinfo{journal}{Phys. Rev.~A}
  \textbf{\bibinfo{volume}{80}}, \bibinfo{pages}{062107}.

\bibitem[{\citenamefont{Allcock}
  \emph{et~al.}(2009{\natexlab{b}})\citenamefont{Allcock, Brunner, Pawlowski,
  and Scarani}}]{ABPS09}
\bibinfo{author}{\bibnamefont{Allcock}, \bibfnamefont{J.}},
  \bibinfo{author}{\bibfnamefont{N.}~\bibnamefont{Brunner}},
  \bibinfo{author}{\bibfnamefont{M.}~\bibnamefont{Pawlowski}}, and
  \bibinfo{author}{\bibfnamefont{V.}~\bibnamefont{Scarani}},
  \bibinfo{year}{2009}{\natexlab{b}}, \bibinfo{journal}{Phys. Rev.~A}
  \textbf{\bibinfo{volume}{80}}, \bibinfo{pages}{040103}.

\bibitem[{\citenamefont{Allcock}
  \emph{et~al.}(2009{\natexlab{c}})\citenamefont{Allcock, Buhrman, and
  Linden}}]{ABL09}
\bibinfo{author}{\bibnamefont{Allcock}, \bibfnamefont{J.}},
  \bibinfo{author}{\bibfnamefont{H.}~\bibnamefont{Buhrman}}, and
  \bibinfo{author}{\bibfnamefont{N.}~\bibnamefont{Linden}},
  \bibinfo{year}{2009}{\natexlab{c}}, \bibinfo{journal}{Phys. Rev.~A}
  \textbf{\bibinfo{volume}{80}}, \bibinfo{pages}{032105}.

\bibitem[{\citenamefont{Almeida}
  \emph{et~al.}(2010{\natexlab{a}})\citenamefont{Almeida, Bancal, Brunner,
  Ac\'in, Gisin, and Pironio}}]{ABBA+10}
\bibinfo{author}{\bibnamefont{Almeida}, \bibfnamefont{M.}},
  \bibinfo{author}{\bibfnamefont{J.-D.} \bibnamefont{Bancal}},
  \bibinfo{author}{\bibfnamefont{N.}~\bibnamefont{Brunner}},
  \bibinfo{author}{\bibfnamefont{A.}~\bibnamefont{Ac\'in}},
  \bibinfo{author}{\bibfnamefont{N.}~\bibnamefont{Gisin}}, and
  \bibinfo{author}{\bibfnamefont{S.}~\bibnamefont{Pironio}},
  \bibinfo{year}{2010}{\natexlab{a}}, \bibinfo{journal}{Phys. Rev. Lett.}
  \textbf{\bibinfo{volume}{104}}, \bibinfo{pages}{230404}.

\bibitem[{\citenamefont{Almeida}
  \emph{et~al.}(2010{\natexlab{b}})\citenamefont{Almeida, Cavalcanti, Scarani,
  and Ac\'in}}]{ACSA10}
\bibinfo{author}{\bibnamefont{Almeida}, \bibfnamefont{M.}},
  \bibinfo{author}{\bibfnamefont{D.}~\bibnamefont{Cavalcanti}},
  \bibinfo{author}{\bibfnamefont{V.}~\bibnamefont{Scarani}}, and
  \bibinfo{author}{\bibfnamefont{A.}~\bibnamefont{Ac\'in}},
  \bibinfo{year}{2010}{\natexlab{b}}, \bibinfo{journal}{Phys. Rev.~A}
  \textbf{\bibinfo{volume}{81}}, \bibinfo{pages}{052111}.

\bibitem[{\citenamefont{Almeida} \emph{et~al.}(2007)\citenamefont{Almeida,
  Pironio, Barrett, T\'oth, and Ac\'in}}]{APBT+07}
\bibinfo{author}{\bibnamefont{Almeida}, \bibfnamefont{M.~L.}},
  \bibinfo{author}{\bibfnamefont{S.}~\bibnamefont{Pironio}},
  \bibinfo{author}{\bibfnamefont{J.}~\bibnamefont{Barrett}},
  \bibinfo{author}{\bibfnamefont{G.}~\bibnamefont{T\'oth}}, and
  \bibinfo{author}{\bibfnamefont{A.}~\bibnamefont{Ac\'in}},
  \bibinfo{year}{2007}, \bibinfo{journal}{Phys. Rev. Lett.}
  \textbf{\bibinfo{volume}{99}}, \bibinfo{pages}{040403}.

\bibitem[{\citenamefont{Ansmann} \emph{et~al.}(2009)\citenamefont{Ansmann,
  Wang, Bialczak, Hofheinz, Lucero, Neeley, O'Connell, Sank, Weides, Wenner,
  Cleland, and Martinis}}]{AWBH+09}
\bibinfo{author}{\bibnamefont{Ansmann}, \bibfnamefont{M.}},
  \bibinfo{author}{\bibfnamefont{H.}~\bibnamefont{Wang}},
  \bibinfo{author}{\bibfnamefont{R.}~\bibnamefont{Bialczak}},
  \bibinfo{author}{\bibfnamefont{M.}~\bibnamefont{Hofheinz}},
  \bibinfo{author}{\bibfnamefont{E.}~\bibnamefont{Lucero}},
  \bibinfo{author}{\bibfnamefont{M.}~\bibnamefont{Neeley}},
  \bibinfo{author}{\bibfnamefont{A.}~\bibnamefont{O'Connell}},
  \bibinfo{author}{\bibfnamefont{D.}~\bibnamefont{Sank}},
  \bibinfo{author}{\bibfnamefont{M.}~\bibnamefont{Weides}},
  \bibinfo{author}{\bibfnamefont{J.}~\bibnamefont{Wenner}},
  \bibinfo{author}{\bibfnamefont{A.}~\bibnamefont{Cleland}}, and
  \bibinfo{author}{\bibfnamefont{J.}~\bibnamefont{Martinis}},
  \bibinfo{year}{2009}, \bibinfo{journal}{Nature}
  \textbf{\bibinfo{volume}{461}}, \bibinfo{pages}{504}.

\bibitem[{\citenamefont{Aolita}
  \emph{et~al.}(2012{\natexlab{a}})\citenamefont{Aolita, Gallego, Ac\'in,
  Chiuri, Vallone, Mataloni, and Cabello}}]{AGA+12}
\bibinfo{author}{\bibnamefont{Aolita}, \bibfnamefont{L.}},
  \bibinfo{author}{\bibfnamefont{R.}~\bibnamefont{Gallego}},
  \bibinfo{author}{\bibfnamefont{A.}~\bibnamefont{Ac\'in}},
  \bibinfo{author}{\bibfnamefont{A.}~\bibnamefont{Chiuri}},
  \bibinfo{author}{\bibfnamefont{G.}~\bibnamefont{Vallone}},
  \bibinfo{author}{\bibfnamefont{P.}~\bibnamefont{Mataloni}}, and
  \bibinfo{author}{\bibfnamefont{A.}~\bibnamefont{Cabello}},
  \bibinfo{year}{2012}{\natexlab{a}}, \bibinfo{journal}{Phys. Rev.~A}
  \textbf{\bibinfo{volume}{85}}, \bibinfo{pages}{032107}.

\bibitem[{\citenamefont{Aolita}
  \emph{et~al.}(2012{\natexlab{b}})\citenamefont{Aolita, Gallego, Cabello, and
  Ac\'in}}]{AGCA12}
\bibinfo{author}{\bibnamefont{Aolita}, \bibfnamefont{L.}},
  \bibinfo{author}{\bibfnamefont{R.}~\bibnamefont{Gallego}},
  \bibinfo{author}{\bibfnamefont{A.}~\bibnamefont{Cabello}}, and
  \bibinfo{author}{\bibfnamefont{A.}~\bibnamefont{Ac\'in}},
  \bibinfo{year}{2012}{\natexlab{b}}, \bibinfo{journal}{Phys. Rev. Lett.}
  \textbf{\bibinfo{volume}{108}}, \bibinfo{pages}{100401}.

\bibitem[{\citenamefont{Aolita}
  \emph{et~al.}(2012{\natexlab{c}})\citenamefont{Aolita, Gallego, Cabello, and
  Ac\'in}}]{AGCA11}
\bibinfo{author}{\bibnamefont{Aolita}, \bibfnamefont{L.}},
  \bibinfo{author}{\bibfnamefont{R.}~\bibnamefont{Gallego}},
  \bibinfo{author}{\bibfnamefont{A.}~\bibnamefont{Cabello}}, and
  \bibinfo{author}{\bibfnamefont{A.}~\bibnamefont{Ac\'in}},
  \bibinfo{year}{2012}{\natexlab{c}}, \bibinfo{journal}{Phys. Rev. Lett.}
  \textbf{\bibinfo{volume}{108}}, \bibinfo{pages}{100401}.

\bibitem[{\citenamefont{Ara\'ujo} \emph{et~al.}(2012)\citenamefont{Ara\'ujo,
  Quintino, Cavalcanti, Santos, Cabello, and Cunha}}]{AQCS+11}
\bibinfo{author}{\bibnamefont{Ara\'ujo}, \bibfnamefont{M.}},
  \bibinfo{author}{\bibfnamefont{M.~T.} \bibnamefont{Quintino}},
  \bibinfo{author}{\bibfnamefont{D.}~\bibnamefont{Cavalcanti}},
  \bibinfo{author}{\bibfnamefont{M.~F.} \bibnamefont{Santos}},
  \bibinfo{author}{\bibfnamefont{A.}~\bibnamefont{Cabello}}, and
  \bibinfo{author}{\bibfnamefont{M.~T.} \bibnamefont{Cunha}},
  \bibinfo{year}{2012}, \bibinfo{journal}{Phys. Rev.~A}
  \textbf{\bibinfo{volume}{86}}, \bibinfo{pages}{030101(R)}.

\bibitem[{\citenamefont{Aravind}(2002)}]{Aravind02}
\bibinfo{author}{\bibnamefont{Aravind}, \bibfnamefont{P.~K.}},
  \bibinfo{year}{2002}, \bibinfo{journal}{Found. Phys. Lett.}
  \textbf{\bibinfo{volume}{15}}, \bibinfo{pages}{397}.

\bibitem[{\citenamefont{Ardehali}(1992)}]{Ardehali92}
\bibinfo{author}{\bibnamefont{Ardehali}, \bibfnamefont{M.}},
  \bibinfo{year}{1992}, \bibinfo{journal}{Phys. Rev.~A}
  \textbf{\bibinfo{volume}{46}}, \bibinfo{pages}{5375}.

\bibitem[{\citenamefont{Arnon-Friedman}
  \emph{et~al.}(2012)\citenamefont{Arnon-Friedman, H{\"a}nggi, and
  Ta-Shma}}]{AHT12}
\bibinfo{author}{\bibnamefont{Arnon-Friedman}, \bibfnamefont{R.}},
  \bibinfo{author}{\bibfnamefont{E.}~\bibnamefont{H{\"a}nggi}}, and
  \bibinfo{author}{\bibfnamefont{A.}~\bibnamefont{Ta-Shma}},
  \bibinfo{year}{2012}, \bibinfo{journal}{arXiv:1205.3736} .

\bibitem[{\citenamefont{Arnon-Friedman and Ta-Shma}(2012)}]{RT12}
\bibinfo{author}{\bibnamefont{Arnon-Friedman}, \bibfnamefont{R.}}, and
  \bibinfo{author}{\bibfnamefont{A.}~\bibnamefont{Ta-Shma}},
  \bibinfo{year}{2012}, \bibinfo{journal}{Phys. Rev.~A}
  \textbf{\bibinfo{volume}{86}}, \bibinfo{pages}{062333}.

\bibitem[{\citenamefont{Aspect}(1975)}]{Aspect1975}
\bibinfo{author}{\bibnamefont{Aspect}, \bibfnamefont{A.}},
  \bibinfo{year}{1975}, \bibinfo{journal}{Physics Letters A}
  \textbf{\bibinfo{volume}{54}}(\bibinfo{number}{2}), \bibinfo{pages}{117}.

\bibitem[{\citenamefont{Aspect}(1976)}]{Aspect76}
\bibinfo{author}{\bibnamefont{Aspect}, \bibfnamefont{A.}},
  \bibinfo{year}{1976}, \bibinfo{journal}{Physical Review D}
  \textbf{\bibinfo{volume}{14}}, \bibinfo{pages}{1944}.

\bibitem[{\citenamefont{Aspect}
  \emph{et~al.}(1982{\natexlab{a}})\citenamefont{Aspect, Dalibard, and
  Roger}}]{ADR82}
\bibinfo{author}{\bibnamefont{Aspect}, \bibfnamefont{A.}},
  \bibinfo{author}{\bibfnamefont{J.}~\bibnamefont{Dalibard}}, and
  \bibinfo{author}{\bibfnamefont{G.}~\bibnamefont{Roger}},
  \bibinfo{year}{1982}{\natexlab{a}}, \bibinfo{journal}{Phys. Rev. Lett.}
  \textbf{\bibinfo{volume}{49}}, \bibinfo{pages}{1804}.

\bibitem[{\citenamefont{Aspect} \emph{et~al.}(1981)\citenamefont{Aspect,
  Grangier, and Roger}}]{AGR81}
\bibinfo{author}{\bibnamefont{Aspect}, \bibfnamefont{A.}},
  \bibinfo{author}{\bibfnamefont{P.}~\bibnamefont{Grangier}}, and
  \bibinfo{author}{\bibfnamefont{G.}~\bibnamefont{Roger}},
  \bibinfo{year}{1981}, \bibinfo{journal}{Phys. Rev. Lett.}
  \textbf{\bibinfo{volume}{47}}, \bibinfo{pages}{460}.

\bibitem[{\citenamefont{Aspect}
  \emph{et~al.}(1982{\natexlab{b}})\citenamefont{Aspect, Grangier, and
  Roger}}]{AGR82}
\bibinfo{author}{\bibnamefont{Aspect}, \bibfnamefont{A.}},
  \bibinfo{author}{\bibfnamefont{P.}~\bibnamefont{Grangier}}, and
  \bibinfo{author}{\bibfnamefont{G.}~\bibnamefont{Roger}},
  \bibinfo{year}{1982}{\natexlab{b}}, \bibinfo{journal}{Phys. Rev. Lett.}
  \textbf{\bibinfo{volume}{49}}, \bibinfo{pages}{91}.

\bibitem[{\citenamefont{Augusiak} \emph{et~al.}(2011)\citenamefont{Augusiak,
  Augusiak, Stasinska, Hadley, Korbicz, Lewenstein, and Ac\'in}}]{ASHK+11}
\bibinfo{author}{\bibnamefont{Augusiak}, \bibfnamefont{R.}},
  \bibinfo{author}{\bibfnamefont{R.}~\bibnamefont{Augusiak}},
  \bibinfo{author}{\bibfnamefont{J.}~\bibnamefont{Stasinska}},
  \bibinfo{author}{\bibfnamefont{C.}~\bibnamefont{Hadley}},
  \bibinfo{author}{\bibfnamefont{J.}~\bibnamefont{Korbicz}},
  \bibinfo{author}{\bibfnamefont{M.}~\bibnamefont{Lewenstein}}, and
  \bibinfo{author}{\bibfnamefont{A.}~\bibnamefont{Ac\'in}},
  \bibinfo{year}{2011}, \bibinfo{journal}{Phys. Rev. Lett.}
  \textbf{\bibinfo{volume}{107}}, \bibinfo{pages}{070401}.

\bibitem[{\citenamefont{Augusiak and Horodecki}(2006)}]{AH06}
\bibinfo{author}{\bibnamefont{Augusiak}, \bibfnamefont{R.}}, and
  \bibinfo{author}{\bibfnamefont{P.}~\bibnamefont{Horodecki}},
  \bibinfo{year}{2006}, \bibinfo{journal}{Phys. Rev.~A}
  \textbf{\bibinfo{volume}{74}}, \bibinfo{pages}{010305}.

\bibitem[{\citenamefont{Avis} \emph{et~al.}(2004)\citenamefont{Avis, Imai, Ito,
  and Sasaki}}]{aiis04}
\bibinfo{author}{\bibnamefont{Avis}, \bibfnamefont{D.}},
  \bibinfo{author}{\bibfnamefont{H.}~\bibnamefont{Imai}},
  \bibinfo{author}{\bibfnamefont{T.}~\bibnamefont{Ito}}, and
  \bibinfo{author}{\bibfnamefont{Y.}~\bibnamefont{Sasaki}},
  \bibinfo{year}{2004}, \bibinfo{journal}{quant-ph/0404014} .

\bibitem[{\citenamefont{Avis} \emph{et~al.}(2005)\citenamefont{Avis, Imai, Ito,
  and Sasaki}}]{AIIS05}
\bibinfo{author}{\bibnamefont{Avis}, \bibfnamefont{D.}},
  \bibinfo{author}{\bibfnamefont{H.}~\bibnamefont{Imai}},
  \bibinfo{author}{\bibfnamefont{T.}~\bibnamefont{Ito}}, and
  \bibinfo{author}{\bibfnamefont{Y.}~\bibnamefont{Sasaki}},
  \bibinfo{year}{2005}, \bibinfo{journal}{J. Phys. A}
  \textbf{\bibinfo{volume}{38}}, \bibinfo{pages}{10971}.

\bibitem[{\citenamefont{Avis and Ito}(2007)}]{AI07}
\bibinfo{author}{\bibnamefont{Avis}, \bibfnamefont{D.}}, and
  \bibinfo{author}{\bibfnamefont{T.}~\bibnamefont{Ito}}, \bibinfo{year}{2007},
  \bibinfo{journal}{Discrete Applied Mathematics}
  \textbf{\bibinfo{volume}{155}}, \bibinfo{pages}{1689}.

\bibitem[{\citenamefont{Avis} \emph{et~al.}(2009)\citenamefont{Avis, Moriyama,
  and Owari}}]{AMO09}
\bibinfo{author}{\bibnamefont{Avis}, \bibfnamefont{D.}},
  \bibinfo{author}{\bibfnamefont{S.}~\bibnamefont{Moriyama}}, and
  \bibinfo{author}{\bibfnamefont{M.}~\bibnamefont{Owari}},
  \bibinfo{year}{2009}, \bibinfo{journal}{IEICE Trans. on Fundamentals}
  \textbf{\bibinfo{volume}{E92-A}}, \bibinfo{pages}{1254}.

\bibitem[{\citenamefont{Babai} \emph{et~al.}(1991)\citenamefont{Babai, Fortnow,
  and Lund}}]{fortnow}
\bibinfo{author}{\bibnamefont{Babai}, \bibfnamefont{L.}},
  \bibinfo{author}{\bibfnamefont{L.}~\bibnamefont{Fortnow}}, and
  \bibinfo{author}{\bibfnamefont{C.}~\bibnamefont{Lund}}, \bibinfo{year}{1991},
  \bibinfo{journal}{Computational Complexity}
  \textbf{\bibinfo{volume}{1}}(\bibinfo{number}{1}), \bibinfo{pages}{3}.

\bibitem[{\citenamefont{Babichev} \emph{et~al.}(2005)\citenamefont{Babichev,
  Appel, and Lvovsky}}]{BAL05}
\bibinfo{author}{\bibnamefont{Babichev}, \bibfnamefont{S.~A.}},
  \bibinfo{author}{\bibfnamefont{J.}~\bibnamefont{Appel}}, and
  \bibinfo{author}{\bibfnamefont{A.~I.} \bibnamefont{Lvovsky}},
  \bibinfo{year}{2005}, \bibinfo{journal}{Phys. Rev. Lett.}
  \textbf{\bibinfo{volume}{92}}, \bibinfo{pages}{193601}.

\bibitem[{\citenamefont{Bacon and Toner}(2003)}]{BT03}
\bibinfo{author}{\bibnamefont{Bacon}, \bibfnamefont{D.}}, and
  \bibinfo{author}{\bibfnamefont{B.}~\bibnamefont{Toner}},
  \bibinfo{year}{2003}, \bibinfo{journal}{Phys. Rev. Lett.}
  \textbf{\bibinfo{volume}{90}}, \bibinfo{pages}{157904}.

\bibitem[{\citenamefont{Banaszek and W\'odkiewicz}(1998)}]{BW98}
\bibinfo{author}{\bibnamefont{Banaszek}, \bibfnamefont{K.}}, and
  \bibinfo{author}{\bibfnamefont{K.}~\bibnamefont{W\'odkiewicz}},
  \bibinfo{year}{1998}, \bibinfo{journal}{Phys. Rev.~A}
  \textbf{\bibinfo{volume}{58}}, \bibinfo{pages}{4345}.

\bibitem[{\citenamefont{Banaszek and W\'odkiewicz}(1999)}]{BW99}
\bibinfo{author}{\bibnamefont{Banaszek}, \bibfnamefont{K.}}, and
  \bibinfo{author}{\bibfnamefont{K.}~\bibnamefont{W\'odkiewicz}},
  \bibinfo{year}{1999}, \bibinfo{journal}{Phys. Rev. Lett.}
  \textbf{\bibinfo{volume}{82}}, \bibinfo{pages}{2009}.

\bibitem[{\citenamefont{Bancal}
  \emph{et~al.}(2012{\natexlab{a}})\citenamefont{Bancal, Branciard, Brunner,
  Gisin, and Liang}}]{BBBG+12}
\bibinfo{author}{\bibnamefont{Bancal}, \bibfnamefont{J.-D.}},
  \bibinfo{author}{\bibfnamefont{C.}~\bibnamefont{Branciard}},
  \bibinfo{author}{\bibfnamefont{N.}~\bibnamefont{Brunner}},
  \bibinfo{author}{\bibfnamefont{N.}~\bibnamefont{Gisin}}, and
  \bibinfo{author}{\bibfnamefont{Y.-C.} \bibnamefont{Liang}},
  \bibinfo{year}{2012}{\natexlab{a}}, \bibinfo{journal}{J.~Phys. A: Math.
  Theor.} \textbf{\bibinfo{volume}{45}}, \bibinfo{pages}{125301}.

\bibitem[{\citenamefont{Bancal} \emph{et~al.}(2008)\citenamefont{Bancal,
  Branciard, Brunner, Gisin, Popescu, and Simon}}]{BBBG+08}
\bibinfo{author}{\bibnamefont{Bancal}, \bibfnamefont{J.-D.}},
  \bibinfo{author}{\bibfnamefont{C.}~\bibnamefont{Branciard}},
  \bibinfo{author}{\bibfnamefont{N.}~\bibnamefont{Brunner}},
  \bibinfo{author}{\bibfnamefont{N.}~\bibnamefont{Gisin}},
  \bibinfo{author}{\bibfnamefont{S.}~\bibnamefont{Popescu}}, and
  \bibinfo{author}{\bibfnamefont{C.}~\bibnamefont{Simon}},
  \bibinfo{year}{2008}, \bibinfo{journal}{Phys. Rev.~A}
  \textbf{\bibinfo{volume}{78}}, \bibinfo{pages}{062110}.

\bibitem[{\citenamefont{Bancal} \emph{et~al.}(2009)\citenamefont{Bancal,
  Branciard, Gisin, and Pironio}}]{BBGP09}
\bibinfo{author}{\bibnamefont{Bancal}, \bibfnamefont{J.-D.}},
  \bibinfo{author}{\bibfnamefont{C.}~\bibnamefont{Branciard}},
  \bibinfo{author}{\bibfnamefont{N.}~\bibnamefont{Gisin}}, and
  \bibinfo{author}{\bibfnamefont{S.}~\bibnamefont{Pironio}},
  \bibinfo{year}{2009}, \bibinfo{journal}{Phys. Rev. Lett.}
  \textbf{\bibinfo{volume}{103}}, \bibinfo{pages}{090503}.

\bibitem[{\citenamefont{Bancal}
  \emph{et~al.}(2011{\natexlab{a}})\citenamefont{Bancal, Brunner, Gisin, and
  Liang}}]{BBBGL11}
\bibinfo{author}{\bibnamefont{Bancal}, \bibfnamefont{J.-D.}},
  \bibinfo{author}{\bibfnamefont{N.}~\bibnamefont{Brunner}},
  \bibinfo{author}{\bibfnamefont{N.}~\bibnamefont{Gisin}}, and
  \bibinfo{author}{\bibfnamefont{Y.-C.} \bibnamefont{Liang}},
  \bibinfo{year}{2011}{\natexlab{a}}, \bibinfo{journal}{Phys. Rev. Lett.}
  \textbf{\bibinfo{volume}{106}}, \bibinfo{pages}{020405}.

\bibitem[{\citenamefont{Bancal}
  \emph{et~al.}(2011{\natexlab{b}})\citenamefont{Bancal, Gisin, Liang, and
  Pironio}}]{BGLP11}
\bibinfo{author}{\bibnamefont{Bancal}, \bibfnamefont{J.-D.}},
  \bibinfo{author}{\bibfnamefont{N.}~\bibnamefont{Gisin}},
  \bibinfo{author}{\bibfnamefont{Y.-C.} \bibnamefont{Liang}}, and
  \bibinfo{author}{\bibfnamefont{S.}~\bibnamefont{Pironio}},
  \bibinfo{year}{2011}{\natexlab{b}}, \bibinfo{journal}{Phys. Rev. Lett.}
  \textbf{\bibinfo{volume}{106}}, \bibinfo{pages}{250404}.

\bibitem[{\citenamefont{Bancal} \emph{et~al.}(2010)\citenamefont{Bancal, Gisin,
  and Pironio}}]{BGP10b}
\bibinfo{author}{\bibnamefont{Bancal}, \bibfnamefont{J.-D.}},
  \bibinfo{author}{\bibfnamefont{N.}~\bibnamefont{Gisin}}, and
  \bibinfo{author}{\bibfnamefont{S.}~\bibnamefont{Pironio}},
  \bibinfo{year}{2010}, \bibinfo{journal}{J.~Phys. A: Math. Theor.}
  \textbf{\bibinfo{volume}{43}}, \bibinfo{pages}{385303}.

\bibitem[{\citenamefont{Bancal}
  \emph{et~al.}(2012{\natexlab{b}})\citenamefont{Bancal, Pironio, Ac\'{\i}n,
  Liang, Scarani, and Gisin}}]{BPALSG11}
\bibinfo{author}{\bibnamefont{Bancal}, \bibfnamefont{J.-D.}},
  \bibinfo{author}{\bibfnamefont{S.}~\bibnamefont{Pironio}},
  \bibinfo{author}{\bibfnamefont{A.}~\bibnamefont{Ac\'{\i}n}},
  \bibinfo{author}{\bibfnamefont{Y.-C.} \bibnamefont{Liang}},
  \bibinfo{author}{\bibfnamefont{V.}~\bibnamefont{Scarani}}, and
  \bibinfo{author}{\bibfnamefont{N.}~\bibnamefont{Gisin}},
  \bibinfo{year}{2012}{\natexlab{b}}, \bibinfo{journal}{Nat. Phys.}
  \textbf{\bibinfo{volume}{8}}, \bibinfo{pages}{867}.

\bibitem[{\citenamefont{Bardyn} \emph{et~al.}(2009)\citenamefont{Bardyn, Liew,
  Massar, McKague, and Scarani}}]{BLM+09}
\bibinfo{author}{\bibnamefont{Bardyn}, \bibfnamefont{C.~E.}},
  \bibinfo{author}{\bibfnamefont{T.~C.~H.} \bibnamefont{Liew}},
  \bibinfo{author}{\bibfnamefont{S.}~\bibnamefont{Massar}},
  \bibinfo{author}{\bibfnamefont{M.}~\bibnamefont{McKague}}, and
  \bibinfo{author}{\bibfnamefont{V.}~\bibnamefont{Scarani}},
  \bibinfo{year}{2009}, \bibinfo{journal}{Phys. Rev.~A}
  \textbf{\bibinfo{volume}{80}}, \bibinfo{pages}{062337}.

\bibitem[{\citenamefont{Barnum} \emph{et~al.}(2007)\citenamefont{Barnum,
  Barrett, Leifer, and Wilce}}]{BBLW07}
\bibinfo{author}{\bibnamefont{Barnum}, \bibfnamefont{H.}},
  \bibinfo{author}{\bibfnamefont{J.}~\bibnamefont{Barrett}},
  \bibinfo{author}{\bibfnamefont{M.}~\bibnamefont{Leifer}}, and
  \bibinfo{author}{\bibfnamefont{A.}~\bibnamefont{Wilce}},
  \bibinfo{year}{2007}, \bibinfo{journal}{Phys. Rev. Lett.}
  \textbf{\bibinfo{volume}{99}}, \bibinfo{pages}{240501}.

\bibitem[{\citenamefont{Barnum} \emph{et~al.}(2010)\citenamefont{Barnum, Beigi,
  Boixo, Elliot, and Wehner}}]{bbb+09}
\bibinfo{author}{\bibnamefont{Barnum}, \bibfnamefont{H.}},
  \bibinfo{author}{\bibfnamefont{S.}~\bibnamefont{Beigi}},
  \bibinfo{author}{\bibfnamefont{S.}~\bibnamefont{Boixo}},
  \bibinfo{author}{\bibfnamefont{M.}~\bibnamefont{Elliot}}, and
  \bibinfo{author}{\bibfnamefont{S.}~\bibnamefont{Wehner}},
  \bibinfo{year}{2010}, \bibinfo{journal}{Phys. Rev. Lett.}
  \textbf{\bibinfo{volume}{104}}, \bibinfo{pages}{140401}.

\bibitem[{\citenamefont{Barnum and Wilce}(2012)}]{BW12}
\bibinfo{author}{\bibnamefont{Barnum}, \bibfnamefont{H.}}, and
  \bibinfo{author}{\bibfnamefont{A.}~\bibnamefont{Wilce}},
  \bibinfo{year}{2012}, \bibinfo{journal}{arXiv:1205.3833} .

\bibitem[{\citenamefont{Barreiro} \emph{et~al.}(2005)\citenamefont{Barreiro,
  Langford, Peters, and Kwiat}}]{BLPK05}
\bibinfo{author}{\bibnamefont{Barreiro}, \bibfnamefont{J.}},
  \bibinfo{author}{\bibfnamefont{N.}~\bibnamefont{Langford}},
  \bibinfo{author}{\bibfnamefont{N.}~\bibnamefont{Peters}}, and
  \bibinfo{author}{\bibfnamefont{P.}~\bibnamefont{Kwiat}},
  \bibinfo{year}{2005}, \bibinfo{journal}{J. Mod. Opt.}
  \textbf{\bibinfo{volume}{95}}, \bibinfo{pages}{260501}.

\bibitem[{\citenamefont{Barrett}(2002)}]{Barrett02}
\bibinfo{author}{\bibnamefont{Barrett}, \bibfnamefont{J.}},
  \bibinfo{year}{2002}, \bibinfo{journal}{Phys. Rev.~A}
  \textbf{\bibinfo{volume}{65}}, \bibinfo{pages}{042302}.

\bibitem[{\citenamefont{Barrett}(2007)}]{Barrett07}
\bibinfo{author}{\bibnamefont{Barrett}, \bibfnamefont{J.}},
  \bibinfo{year}{2007}, \bibinfo{journal}{Phys. Rev.~A}
  \textbf{\bibinfo{volume}{75}}, \bibinfo{pages}{032304}.

\bibitem[{\citenamefont{Barrett} \emph{et~al.}(2012)\citenamefont{Barrett,
  Colbeck, and Kent}}]{BCK12}
\bibinfo{author}{\bibnamefont{Barrett}, \bibfnamefont{J.}},
  \bibinfo{author}{\bibfnamefont{R.}~\bibnamefont{Colbeck}}, and
  \bibinfo{author}{\bibfnamefont{A.}~\bibnamefont{Kent}}, \bibinfo{year}{2012},
  \bibinfo{journal}{Phys. Rev.~A} \textbf{\bibinfo{volume}{86}},
  \bibinfo{pages}{062326}.

\bibitem[{\citenamefont{Barrett}
  \emph{et~al.}(2013{\natexlab{a}})\citenamefont{Barrett, Colbeck, and
  Kent}}]{BCK12p}
\bibinfo{author}{\bibnamefont{Barrett}, \bibfnamefont{J.}},
  \bibinfo{author}{\bibfnamefont{R.}~\bibnamefont{Colbeck}}, and
  \bibinfo{author}{\bibfnamefont{A.}~\bibnamefont{Kent}},
  \bibinfo{year}{2013}{\natexlab{a}}, \bibinfo{journal}{Phys. Rev. Lett.}
  \textbf{\bibinfo{volume}{110}}, \bibinfo{pages}{010503}.

\bibitem[{\citenamefont{Barrett} \emph{et~al.}(2002)\citenamefont{Barrett,
  Collins, Hardy, Kent, and Popescu}}]{BCHK+02}
\bibinfo{author}{\bibnamefont{Barrett}, \bibfnamefont{J.}},
  \bibinfo{author}{\bibfnamefont{D.}~\bibnamefont{Collins}},
  \bibinfo{author}{\bibfnamefont{L.}~\bibnamefont{Hardy}},
  \bibinfo{author}{\bibfnamefont{A.}~\bibnamefont{Kent}}, and
  \bibinfo{author}{\bibfnamefont{S.}~\bibnamefont{Popescu}},
  \bibinfo{year}{2002}, \bibinfo{journal}{Phys. Rev.~A}
  \textbf{\bibinfo{volume}{66}}, \bibinfo{pages}{042111}.

\bibitem[{\citenamefont{Barrett and Gisin}(2011)}]{BAG11}
\bibinfo{author}{\bibnamefont{Barrett}, \bibfnamefont{J.}}, and
  \bibinfo{author}{\bibfnamefont{N.}~\bibnamefont{Gisin}},
  \bibinfo{year}{2011}, \bibinfo{journal}{Phys. Rev. Lett.}
  \textbf{\bibinfo{volume}{106}}, \bibinfo{pages}{4}.

\bibitem[{\citenamefont{Barrett}
  \emph{et~al.}(2005{\natexlab{a}})\citenamefont{Barrett, Hardy, and
  Kent}}]{BHK05}
\bibinfo{author}{\bibnamefont{Barrett}, \bibfnamefont{J.}},
  \bibinfo{author}{\bibfnamefont{L.}~\bibnamefont{Hardy}}, and
  \bibinfo{author}{\bibfnamefont{A.}~\bibnamefont{Kent}},
  \bibinfo{year}{2005}{\natexlab{a}}, \bibinfo{journal}{Phys. Rev. Lett.}
  \textbf{\bibinfo{volume}{95}}, \bibinfo{pages}{010503}.

\bibitem[{\citenamefont{Barrett} \emph{et~al.}(2007)\citenamefont{Barrett,
  Kent, and Pironio}}]{BKP07}
\bibinfo{author}{\bibnamefont{Barrett}, \bibfnamefont{J.}},
  \bibinfo{author}{\bibfnamefont{A.}~\bibnamefont{Kent}}, and
  \bibinfo{author}{\bibfnamefont{S.}~\bibnamefont{Pironio}},
  \bibinfo{year}{2007}, \bibinfo{journal}{Phys. Rev. Lett.}
  \textbf{\bibinfo{volume}{97}}, \bibinfo{pages}{170409}.

\bibitem[{\citenamefont{Barrett}
  \emph{et~al.}(2005{\natexlab{b}})\citenamefont{Barrett, Linden, Massar,
  Pironio, Popescu, and Roberts}}]{BLMP+05}
\bibinfo{author}{\bibnamefont{Barrett}, \bibfnamefont{J.}},
  \bibinfo{author}{\bibfnamefont{N.}~\bibnamefont{Linden}},
  \bibinfo{author}{\bibfnamefont{S.}~\bibnamefont{Massar}},
  \bibinfo{author}{\bibfnamefont{S.}~\bibnamefont{Pironio}},
  \bibinfo{author}{\bibfnamefont{S.}~\bibnamefont{Popescu}}, and
  \bibinfo{author}{\bibfnamefont{D.}~\bibnamefont{Roberts}},
  \bibinfo{year}{2005}{\natexlab{b}}, \bibinfo{journal}{Phys. Rev.~A}
  \textbf{\bibinfo{volume}{71}}, \bibinfo{pages}{022101}.

\bibitem[{\citenamefont{Barrett and Pironio}(2005)}]{BP05}
\bibinfo{author}{\bibnamefont{Barrett}, \bibfnamefont{J.}}, and
  \bibinfo{author}{\bibfnamefont{S.}~\bibnamefont{Pironio}},
  \bibinfo{year}{2005}, \bibinfo{journal}{Phys. Rev. Lett.}
  \textbf{\bibinfo{volume}{95}}, \bibinfo{pages}{140401}.

\bibitem[{\citenamefont{Barrett}
  \emph{et~al.}(2013{\natexlab{b}})\citenamefont{Barrett, Pironio, Bancal, and
  Gisin}}]{BPBG11}
\bibinfo{author}{\bibnamefont{Barrett}, \bibfnamefont{J.}},
  \bibinfo{author}{\bibfnamefont{S.}~\bibnamefont{Pironio}},
  \bibinfo{author}{\bibfnamefont{J.-D.} \bibnamefont{Bancal}}, and
  \bibinfo{author}{\bibfnamefont{N.}~\bibnamefont{Gisin}},
  \bibinfo{year}{2013}{\natexlab{b}}, \bibinfo{journal}{Phys. Rev.~A}
  \textbf{\bibinfo{volume}{88}}, \bibinfo{pages}{014102}.

\bibitem[{\citenamefont{Bartlett} \emph{et~al.}(2007)\citenamefont{Bartlett,
  Rudolph, and Spekkens}}]{BRS07}
\bibinfo{author}{\bibnamefont{Bartlett}, \bibfnamefont{S.}},
  \bibinfo{author}{\bibfnamefont{T.}~\bibnamefont{Rudolph}}, and
  \bibinfo{author}{\bibfnamefont{R.}~\bibnamefont{Spekkens}},
  \bibinfo{year}{2007}, \bibinfo{journal}{Rev. Mod. Phys.}
  \textbf{\bibinfo{volume}{79}}, \bibinfo{pages}{555}.

\bibitem[{\citenamefont{Belinskii and Klyshko}(1993)}]{BK93}
\bibinfo{author}{\bibnamefont{Belinskii}, \bibfnamefont{A.}}, and
  \bibinfo{author}{\bibfnamefont{D.}~\bibnamefont{Klyshko}},
  \bibinfo{year}{1993}, \bibinfo{journal}{Phys. Usp.}
  \textbf{\bibinfo{volume}{36}}, \bibinfo{pages}{653}.

\bibitem[{\citenamefont{Bell}(1964)}]{Bell64}
\bibinfo{author}{\bibnamefont{Bell}, \bibfnamefont{J.}}, \bibinfo{year}{1964},
  \bibinfo{journal}{Physics} \textbf{\bibinfo{volume}{1}},
  \bibinfo{pages}{195}.

\bibitem[{\citenamefont{Bell}(1975)}]{Bell75}
\bibinfo{author}{\bibnamefont{Bell}, \bibfnamefont{J.~S.}},
  \bibinfo{year}{1975}, \bibinfo{note}{\emph{The theory of local beables},
  reprinted in \cite{Bell04}}.

\bibitem[{\citenamefont{Bell}(1977{\natexlab{a}})}]{Bell79}
\bibinfo{author}{\bibnamefont{Bell}, \bibfnamefont{J.~S.}},
  \bibinfo{year}{1977}{\natexlab{a}}, \bibinfo{note}{\emph{Atomic-Cascade
  Photons and Quantum-Mechanical Nonlocality}, reprinted in \cite{Bell04}}.

\bibitem[{\citenamefont{Bell}(1977{\natexlab{b}})}]{Bell77}
\bibinfo{author}{\bibnamefont{Bell}, \bibfnamefont{J.~S.}},
  \bibinfo{year}{1977}{\natexlab{b}}, \bibinfo{note}{\emph{Free Variables and
  Local Causality}, reprinted in \cite{Bell04}}.

\bibitem[{\citenamefont{Bell}(1990)}]{Bell90}
\bibinfo{author}{\bibnamefont{Bell}, \bibfnamefont{J.~S.}},
  \bibinfo{year}{1990}, \bibinfo{note}{\emph{La nouvelle cuisine}, reprinted in
  \cite{Bell04}}.

\bibitem[{\citenamefont{Bell}(2004)}]{Bell04}
\bibinfo{author}{\bibnamefont{Bell}, \bibfnamefont{J.~S.}},
  \bibinfo{year}{2004}, \emph{\bibinfo{title}{Speakable and Unspeakable in
  Quantum Mechanics}} (\bibinfo{publisher}{Cambridge University Press}).

\bibitem[{\citenamefont{Bennett} \emph{et~al.}(1993)\citenamefont{Bennett,
  Brassard, Cr\'epeau, Jozsa, Peres, and Wootters}}]{BBC+93}
\bibinfo{author}{\bibnamefont{Bennett}, \bibfnamefont{C.~H.}},
  \bibinfo{author}{\bibfnamefont{G.}~\bibnamefont{Brassard}},
  \bibinfo{author}{\bibfnamefont{C.}~\bibnamefont{Cr\'epeau}},
  \bibinfo{author}{\bibfnamefont{R.}~\bibnamefont{Jozsa}},
  \bibinfo{author}{\bibfnamefont{A.}~\bibnamefont{Peres}}, and
  \bibinfo{author}{\bibfnamefont{W.~K.} \bibnamefont{Wootters}},
  \bibinfo{year}{1993}, \bibinfo{journal}{Phys. Rev. Lett.}
  \textbf{\bibinfo{volume}{70}}, \bibinfo{pages}{1895}.

\bibitem[{\citenamefont{Bennett} \emph{et~al.}(1992)\citenamefont{Bennett,
  Brassard, and Mermin}}]{BBM92}
\bibinfo{author}{\bibnamefont{Bennett}, \bibfnamefont{C.~H.}},
  \bibinfo{author}{\bibfnamefont{G.}~\bibnamefont{Brassard}}, and
  \bibinfo{author}{\bibfnamefont{N.~D.} \bibnamefont{Mermin}},
  \bibinfo{year}{1992}, \bibinfo{journal}{Phys. Rev. Lett.}
  \textbf{\bibinfo{volume}{68}}, \bibinfo{pages}{557}.

\bibitem[{\citenamefont{Bennett} \emph{et~al.}(2002)\citenamefont{Bennett,
  Shor, Smolin, and Thapliyal}}]{noIncreaseEntangled}
\bibinfo{author}{\bibnamefont{Bennett}, \bibfnamefont{C.~H.}},
  \bibinfo{author}{\bibfnamefont{P.~W.} \bibnamefont{Shor}},
  \bibinfo{author}{\bibfnamefont{J.~A.} \bibnamefont{Smolin}}, and
  \bibinfo{author}{\bibfnamefont{A.~V.} \bibnamefont{Thapliyal}},
  \bibinfo{year}{2002}, \bibinfo{journal}{IEEE Trans. Inf. Theory}
  \textbf{\bibinfo{volume}{48}}, \bibinfo{pages}{2637}.

\bibitem[{\citenamefont{Berta} \emph{et~al.}(2012)\citenamefont{Berta, Fawzi,
  and Wehner}}]{bfw:random}
\bibinfo{author}{\bibnamefont{Berta}, \bibfnamefont{M.}},
  \bibinfo{author}{\bibfnamefont{O.}~\bibnamefont{Fawzi}}, and
  \bibinfo{author}{\bibfnamefont{S.}~\bibnamefont{Wehner}},
  \bibinfo{year}{2012}, \textbf{\bibinfo{volume}{7417}}, \bibinfo{pages}{776}.

\bibitem[{\citenamefont{Boyd and Vandenberghe}(2004)}]{boyd:book}
\bibinfo{author}{\bibnamefont{Boyd}, \bibfnamefont{S.}}, and
  \bibinfo{author}{\bibfnamefont{L.}~\bibnamefont{Vandenberghe}},
  \bibinfo{year}{2004}, \emph{\bibinfo{title}{Convex Optimization}}
  (\bibinfo{publisher}{Cambridge University Press}).

\bibitem[{\citenamefont{Branciard}(2011)}]{Branciard11}
\bibinfo{author}{\bibnamefont{Branciard}, \bibfnamefont{C.}},
  \bibinfo{year}{2011}, \bibinfo{journal}{Phys. Rev.~A}
  \textbf{\bibinfo{volume}{83}}, \bibinfo{pages}{032123}.

\bibitem[{\citenamefont{Branciard}
  \emph{et~al.}(2012{\natexlab{a}})\citenamefont{Branciard, Brunner, Buhrman,
  Cleve, Gisin, Portmann, Rosset, and Szegedy}}]{BBBC+12}
\bibinfo{author}{\bibnamefont{Branciard}, \bibfnamefont{C.}},
  \bibinfo{author}{\bibfnamefont{N.}~\bibnamefont{Brunner}},
  \bibinfo{author}{\bibfnamefont{H.}~\bibnamefont{Buhrman}},
  \bibinfo{author}{\bibfnamefont{R.}~\bibnamefont{Cleve}},
  \bibinfo{author}{\bibfnamefont{N.}~\bibnamefont{Gisin}},
  \bibinfo{author}{\bibfnamefont{S.}~\bibnamefont{Portmann}},
  \bibinfo{author}{\bibfnamefont{D.}~\bibnamefont{Rosset}}, and
  \bibinfo{author}{\bibfnamefont{M.}~\bibnamefont{Szegedy}},
  \bibinfo{year}{2012}{\natexlab{a}}, \bibinfo{journal}{Phys. Rev. Lett.}
  \textbf{\bibinfo{volume}{109}}, \bibinfo{pages}{100401}.

\bibitem[{\citenamefont{Branciard} \emph{et~al.}(2008)\citenamefont{Branciard,
  Brunner, Gisin, Kurtsiefer, Linares, Ling, and Scarani}}]{BBGK+08}
\bibinfo{author}{\bibnamefont{Branciard}, \bibfnamefont{C.}},
  \bibinfo{author}{\bibfnamefont{N.}~\bibnamefont{Brunner}},
  \bibinfo{author}{\bibfnamefont{N.}~\bibnamefont{Gisin}},
  \bibinfo{author}{\bibfnamefont{C.}~\bibnamefont{Kurtsiefer}},
  \bibinfo{author}{\bibfnamefont{A.}~\bibnamefont{Linares}},
  \bibinfo{author}{\bibfnamefont{A.}~\bibnamefont{Ling}}, and
  \bibinfo{author}{\bibfnamefont{V.}~\bibnamefont{Scarani}},
  \bibinfo{year}{2008}, \bibinfo{journal}{Nature Physics}
  \textbf{\bibinfo{volume}{4}}, \bibinfo{pages}{681}.

\bibitem[{\citenamefont{Branciard}
  \emph{et~al.}(2012{\natexlab{b}})\citenamefont{Branciard, Cavalcanti,
  Walborn, Scarani, and Wiseman}}]{BCW+12}
\bibinfo{author}{\bibnamefont{Branciard}, \bibfnamefont{C.}},
  \bibinfo{author}{\bibfnamefont{E.}~\bibnamefont{Cavalcanti}},
  \bibinfo{author}{\bibfnamefont{S.}~\bibnamefont{Walborn}},
  \bibinfo{author}{\bibfnamefont{V.}~\bibnamefont{Scarani}}, and
  \bibinfo{author}{\bibfnamefont{H.}~\bibnamefont{Wiseman}},
  \bibinfo{year}{2012}{\natexlab{b}}, \bibinfo{journal}{Phys. Rev.~A}
  \textbf{\bibinfo{volume}{85}}, \bibinfo{pages}{010301}.

\bibitem[{\citenamefont{Branciard and Gisin}(2011)}]{BG11}
\bibinfo{author}{\bibnamefont{Branciard}, \bibfnamefont{C.}}, and
  \bibinfo{author}{\bibfnamefont{N.}~\bibnamefont{Gisin}},
  \bibinfo{year}{2011}, \bibinfo{journal}{Phys. Rev. Lett.}
  \textbf{\bibinfo{volume}{107}}, \bibinfo{pages}{020401}.

\bibitem[{\citenamefont{Branciard} \emph{et~al.}(2010)\citenamefont{Branciard,
  Gisin, and Pironio}}]{BGP10}
\bibinfo{author}{\bibnamefont{Branciard}, \bibfnamefont{C.}},
  \bibinfo{author}{\bibfnamefont{N.}~\bibnamefont{Gisin}}, and
  \bibinfo{author}{\bibfnamefont{S.}~\bibnamefont{Pironio}},
  \bibinfo{year}{2010}, \bibinfo{journal}{Phys. Rev. Lett.}
  \textbf{\bibinfo{volume}{104}}, \bibinfo{pages}{170401}.

\bibitem[{\citenamefont{Branciard}
  \emph{et~al.}(2012{\natexlab{c}})\citenamefont{Branciard, Rosset, Gisin, and
  Pironio}}]{BRGP12}
\bibinfo{author}{\bibnamefont{Branciard}, \bibfnamefont{C.}},
  \bibinfo{author}{\bibfnamefont{D.}~\bibnamefont{Rosset}},
  \bibinfo{author}{\bibfnamefont{N.}~\bibnamefont{Gisin}}, and
  \bibinfo{author}{\bibfnamefont{S.}~\bibnamefont{Pironio}},
  \bibinfo{year}{2012}{\natexlab{c}}, \bibinfo{journal}{Phys. Rev.~A}
  \textbf{\bibinfo{volume}{85}}, \bibinfo{pages}{032119}.

\bibitem[{\citenamefont{Branciard} \emph{et~al.}(2013)\citenamefont{Branciard,
  Rosset, Liang, and Gisin}}]{BRLG12}
\bibinfo{author}{\bibnamefont{Branciard}, \bibfnamefont{C.}},
  \bibinfo{author}{\bibfnamefont{D.}~\bibnamefont{Rosset}},
  \bibinfo{author}{\bibfnamefont{Y.-C.} \bibnamefont{Liang}}, and
  \bibinfo{author}{\bibfnamefont{N.}~\bibnamefont{Gisin}},
  \bibinfo{year}{2013}, \bibinfo{journal}{Phys. Rev. Lett.}
  \textbf{\bibinfo{volume}{110}}, \bibinfo{pages}{060405}.

\bibitem[{\citenamefont{Brask} \emph{et~al.}(2012)\citenamefont{Brask, Brunner,
  Cavalcanti, and Leverrier}}]{BBCL12}
\bibinfo{author}{\bibnamefont{Brask}, \bibfnamefont{J.~B.}},
  \bibinfo{author}{\bibfnamefont{N.}~\bibnamefont{Brunner}},
  \bibinfo{author}{\bibfnamefont{D.}~\bibnamefont{Cavalcanti}}, and
  \bibinfo{author}{\bibfnamefont{A.}~\bibnamefont{Leverrier}},
  \bibinfo{year}{2012}, \bibinfo{journal}{Phys. Rev.~A}
  \textbf{\bibinfo{volume}{85}}, \bibinfo{pages}{042116}.

\bibitem[{\citenamefont{Brask and Chaves}(2012)}]{BC12}
\bibinfo{author}{\bibnamefont{Brask}, \bibfnamefont{J.~B.}}, and
  \bibinfo{author}{\bibfnamefont{R.}~\bibnamefont{Chaves}},
  \bibinfo{year}{2012}, \bibinfo{journal}{Phys. Rev.~A}
  \textbf{\bibinfo{volume}{86}}, \bibinfo{pages}{010103}.

\bibitem[{\citenamefont{Brassard} \emph{et~al.}(2005)\citenamefont{Brassard,
  Broadbent, and Tapp}}]{broadbent:survey}
\bibinfo{author}{\bibnamefont{Brassard}, \bibfnamefont{G.}},
  \bibinfo{author}{\bibfnamefont{A.}~\bibnamefont{Broadbent}}, and
  \bibinfo{author}{\bibfnamefont{A.}~\bibnamefont{Tapp}}, \bibinfo{year}{2005},
  \bibinfo{journal}{Found. Phys.}
  \textbf{\bibinfo{volume}{35}}(\bibinfo{number}{11}), \bibinfo{pages}{1877}.

\bibitem[{\citenamefont{Brassard} \emph{et~al.}(2006)\citenamefont{Brassard,
  Buhrman, Linden, M\'ethot, Tapp, and Unger}}]{BBLM+06}
\bibinfo{author}{\bibnamefont{Brassard}, \bibfnamefont{G.}},
  \bibinfo{author}{\bibfnamefont{H.}~\bibnamefont{Buhrman}},
  \bibinfo{author}{\bibfnamefont{N.}~\bibnamefont{Linden}},
  \bibinfo{author}{\bibfnamefont{A.}~\bibnamefont{M\'ethot}},
  \bibinfo{author}{\bibfnamefont{A.}~\bibnamefont{Tapp}}, and
  \bibinfo{author}{\bibfnamefont{F.}~\bibnamefont{Unger}},
  \bibinfo{year}{2006}, \bibinfo{journal}{Phys. Rev. Lett.}
  \textbf{\bibinfo{volume}{96}}, \bibinfo{pages}{250401}.

\bibitem[{\citenamefont{Brassard} \emph{et~al.}(1999)\citenamefont{Brassard,
  Cleve, and Tapp}}]{BCT99}
\bibinfo{author}{\bibnamefont{Brassard}, \bibfnamefont{G.}},
  \bibinfo{author}{\bibfnamefont{R.}~\bibnamefont{Cleve}}, and
  \bibinfo{author}{\bibfnamefont{A.}~\bibnamefont{Tapp}}, \bibinfo{year}{1999},
  \bibinfo{journal}{Phys. Rev. Lett.} \textbf{\bibinfo{volume}{83}},
  \bibinfo{pages}{1874}.

\bibitem[{\citenamefont{Braunstein and Caves}(1988)}]{BC88}
\bibinfo{author}{\bibnamefont{Braunstein}, \bibfnamefont{S.}}, and
  \bibinfo{author}{\bibfnamefont{C.}~\bibnamefont{Caves}},
  \bibinfo{year}{1988}, \bibinfo{journal}{Phys. Rev. Lett.}
  \textbf{\bibinfo{volume}{61}}, \bibinfo{pages}{662}.

\bibitem[{\citenamefont{Braunstein and Pirandola}(2012)}]{BP12}
\bibinfo{author}{\bibnamefont{Braunstein}, \bibfnamefont{S.}}, and
  \bibinfo{author}{\bibfnamefont{S.}~\bibnamefont{Pirandola}},
  \bibinfo{year}{2012}, \bibinfo{journal}{Phys. Rev. Lett.}
  \textbf{\bibinfo{volume}{108}}, \bibinfo{pages}{130502}.

\bibitem[{\citenamefont{Braunstein and Caves}(1990)}]{BC90}
\bibinfo{author}{\bibnamefont{Braunstein}, \bibfnamefont{S.~L.}}, and
  \bibinfo{author}{\bibfnamefont{C.~M.} \bibnamefont{Caves}},
  \bibinfo{year}{1990}, \bibinfo{journal}{Ann. Phys.}
  \textbf{\bibinfo{volume}{202}}, \bibinfo{pages}{22}.

\bibitem[{\citenamefont{Braunstein}
  \emph{et~al.}(1992)\citenamefont{Braunstein, Mann, and Revzen}}]{BMR92}
\bibinfo{author}{\bibnamefont{Braunstein}, \bibfnamefont{S.~L.}},
  \bibinfo{author}{\bibfnamefont{A.}~\bibnamefont{Mann}}, and
  \bibinfo{author}{\bibfnamefont{M.}~\bibnamefont{Revzen}},
  \bibinfo{year}{1992}, \bibinfo{journal}{Phys. Rev. Lett.}
  \textbf{\bibinfo{volume}{68}}, \bibinfo{pages}{3259}.

\bibitem[{\citenamefont{Bri\"et} \emph{et~al.}(2011)\citenamefont{Bri\"et,
  Buhrman, and Toner}}]{BBT11}
\bibinfo{author}{\bibnamefont{Bri\"et}, \bibfnamefont{J.}},
  \bibinfo{author}{\bibfnamefont{H.}~\bibnamefont{Buhrman}}, and
  \bibinfo{author}{\bibfnamefont{B.}~\bibnamefont{Toner}},
  \bibinfo{year}{2011}, \bibinfo{journal}{Comm. Math. Phys.}
  \textbf{\bibinfo{volume}{305}}, \bibinfo{pages}{827}.

\bibitem[{\citenamefont{Briet and Vidick}(2013)}]{thomas:xorGame}
\bibinfo{author}{\bibnamefont{Briet}, \bibfnamefont{J.}}, and
  \bibinfo{author}{\bibfnamefont{T.}~\bibnamefont{Vidick}},
  \bibinfo{year}{2013}, \bibinfo{journal}{Comm. Math. Phys.}
  \textbf{\bibinfo{volume}{321}}, \bibinfo{pages}{181}.

\bibitem[{\citenamefont{Brunner} \emph{et~al.}(2011)\citenamefont{Brunner,
  Cavalcanti, Salles, and Skrzypczyk}}]{BCSS11}
\bibinfo{author}{\bibnamefont{Brunner}, \bibfnamefont{N.}},
  \bibinfo{author}{\bibfnamefont{D.}~\bibnamefont{Cavalcanti}},
  \bibinfo{author}{\bibfnamefont{A.}~\bibnamefont{Salles}}, and
  \bibinfo{author}{\bibfnamefont{P.}~\bibnamefont{Skrzypczyk}},
  \bibinfo{year}{2011}, \bibinfo{journal}{Phys. Rev. Lett.}
  \textbf{\bibinfo{volume}{106}}, \bibinfo{pages}{020402}.

\bibitem[{\citenamefont{Brunner and Gisin}(2008)}]{BG08}
\bibinfo{author}{\bibnamefont{Brunner}, \bibfnamefont{N.}}, and
  \bibinfo{author}{\bibfnamefont{N.}~\bibnamefont{Gisin}},
  \bibinfo{year}{2008}, \bibinfo{journal}{Phys. Lett.~A}
  \textbf{\bibinfo{volume}{372}}, \bibinfo{pages}{3162}.

\bibitem[{\citenamefont{Brunner}
  \emph{et~al.}(2008{\natexlab{a}})\citenamefont{Brunner, Gisin, Popescu, and
  Scarani}}]{BGPS08}
\bibinfo{author}{\bibnamefont{Brunner}, \bibfnamefont{N.}},
  \bibinfo{author}{\bibfnamefont{N.}~\bibnamefont{Gisin}},
  \bibinfo{author}{\bibfnamefont{S.}~\bibnamefont{Popescu}}, and
  \bibinfo{author}{\bibfnamefont{V.}~\bibnamefont{Scarani}},
  \bibinfo{year}{2008}{\natexlab{a}}, \bibinfo{journal}{Phys. Rev.~A}
  \textbf{\bibinfo{volume}{78}}, \bibinfo{pages}{052111}.

\bibitem[{\citenamefont{Brunner} \emph{et~al.}(2005)\citenamefont{Brunner,
  Gisin, and Scarani}}]{BGS05}
\bibinfo{author}{\bibnamefont{Brunner}, \bibfnamefont{N.}},
  \bibinfo{author}{\bibfnamefont{N.}~\bibnamefont{Gisin}}, and
  \bibinfo{author}{\bibfnamefont{V.}~\bibnamefont{Scarani}},
  \bibinfo{year}{2005}, \bibinfo{journal}{New J.~Phys}
  \textbf{\bibinfo{volume}{7}}, \bibinfo{pages}{88}.

\bibitem[{\citenamefont{Brunner} \emph{et~al.}(2007)\citenamefont{Brunner,
  Gisin, Scarani, and Simon}}]{BGSS07}
\bibinfo{author}{\bibnamefont{Brunner}, \bibfnamefont{N.}},
  \bibinfo{author}{\bibfnamefont{N.}~\bibnamefont{Gisin}},
  \bibinfo{author}{\bibfnamefont{V.}~\bibnamefont{Scarani}}, and
  \bibinfo{author}{\bibfnamefont{C.}~\bibnamefont{Simon}},
  \bibinfo{year}{2007}, \bibinfo{journal}{Phys. Rev. Lett.}
  \textbf{\bibinfo{volume}{98}}, \bibinfo{pages}{220403}.

\bibitem[{\citenamefont{Brunner}
  \emph{et~al.}(2008{\natexlab{b}})\citenamefont{Brunner, Pironio, Ac\'in,
  Gisin, M\'ethot, and Scarani}}]{BPAG+08}
\bibinfo{author}{\bibnamefont{Brunner}, \bibfnamefont{N.}},
  \bibinfo{author}{\bibfnamefont{S.}~\bibnamefont{Pironio}},
  \bibinfo{author}{\bibfnamefont{A.}~\bibnamefont{Ac\'in}},
  \bibinfo{author}{\bibfnamefont{N.}~\bibnamefont{Gisin}},
  \bibinfo{author}{\bibfnamefont{A.}~\bibnamefont{M\'ethot}}, and
  \bibinfo{author}{\bibfnamefont{V.}~\bibnamefont{Scarani}},
  \bibinfo{year}{2008}{\natexlab{b}}, \bibinfo{journal}{Phys. Rev. Lett.}
  \textbf{\bibinfo{volume}{100}}, \bibinfo{pages}{210503}.

\bibitem[{\citenamefont{Brunner} \emph{et~al.}(2012)\citenamefont{Brunner,
  Sharam, and Vertesi}}]{BSV12}
\bibinfo{author}{\bibnamefont{Brunner}, \bibfnamefont{N.}},
  \bibinfo{author}{\bibfnamefont{J.}~\bibnamefont{Sharam}}, and
  \bibinfo{author}{\bibfnamefont{T.}~\bibnamefont{Vertesi}},
  \bibinfo{year}{2012}, \bibinfo{journal}{Phys. Rev. Lett.}
  \textbf{\bibinfo{volume}{108}}, \bibinfo{pages}{110501}.

\bibitem[{\citenamefont{Brunner and Skrzypczyk}(2009)}]{BS09}
\bibinfo{author}{\bibnamefont{Brunner}, \bibfnamefont{N.}}, and
  \bibinfo{author}{\bibfnamefont{P.}~\bibnamefont{Skrzypczyk}},
  \bibinfo{year}{2009}, \bibinfo{journal}{Phys. Rev. Lett.}
  \textbf{\bibinfo{volume}{102}}, \bibinfo{pages}{160403}.

\bibitem[{\citenamefont{Brunner and Vertesi}(2012)}]{BV12}
\bibinfo{author}{\bibnamefont{Brunner}, \bibfnamefont{N.}}, and
  \bibinfo{author}{\bibfnamefont{T.}~\bibnamefont{Vertesi}},
  \bibinfo{year}{2012}, \bibinfo{journal}{Phys. Rev.~A}
  \textbf{\bibinfo{volume}{86}}, \bibinfo{pages}{042113}.

\bibitem[{\citenamefont{Brunner} \emph{et~al.}(2013)\citenamefont{Brunner,
  Young, Hu, and Rarity}}]{BYHR13}
\bibinfo{author}{\bibnamefont{Brunner}, \bibfnamefont{N.}},
  \bibinfo{author}{\bibfnamefont{A.}~\bibnamefont{Young}},
  \bibinfo{author}{\bibfnamefont{C.}~\bibnamefont{Hu}}, and
  \bibinfo{author}{\bibfnamefont{J.}~\bibnamefont{Rarity}},
  \bibinfo{year}{2013}, \bibinfo{journal}{New J.~Phys} 
  \textbf{\bibinfo{volume}{15}}, \bibinfo{pages}{105006}.

\bibitem[{\citenamefont{Buhrman} \emph{et~al.}(2010)\citenamefont{Buhrman,
  Cleve, Massar, and de~Wolf}}]{BCMW10}
\bibinfo{author}{\bibnamefont{Buhrman}, \bibfnamefont{H.}},
  \bibinfo{author}{\bibfnamefont{R.}~\bibnamefont{Cleve}},
  \bibinfo{author}{\bibfnamefont{S.}~\bibnamefont{Massar}}, and
  \bibinfo{author}{\bibfnamefont{R.}~\bibnamefont{de~Wolf}},
  \bibinfo{year}{2010}, \bibinfo{journal}{Rev. Mod. Phys.}
  \textbf{\bibinfo{volume}{82}}, \bibinfo{pages}{665}.

\bibitem[{\citenamefont{Buhrman} \emph{et~al.}(2003)\citenamefont{Buhrman,
  Hoyer, Massar, and R\"ohrig}}]{BHMR03}
\bibinfo{author}{\bibnamefont{Buhrman}, \bibfnamefont{H.}},
  \bibinfo{author}{\bibfnamefont{P.}~\bibnamefont{Hoyer}},
  \bibinfo{author}{\bibfnamefont{S.}~\bibnamefont{Massar}}, and
  \bibinfo{author}{\bibfnamefont{H.}~\bibnamefont{R\"ohrig}},
  \bibinfo{year}{2003}, \bibinfo{journal}{Phys. Rev. Lett.}
  \textbf{\bibinfo{volume}{91}}, \bibinfo{pages}{047903}.

\bibitem[{\citenamefont{Buhrman} \emph{et~al.}(2011)\citenamefont{Buhrman,
  Regev, Scarpa, and de~Wolf}}]{BRSW11}
\bibinfo{author}{\bibnamefont{Buhrman}, \bibfnamefont{H.}},
  \bibinfo{author}{\bibfnamefont{O.}~\bibnamefont{Regev}},
  \bibinfo{author}{\bibfnamefont{G.}~\bibnamefont{Scarpa}}, and
  \bibinfo{author}{\bibfnamefont{R.}~\bibnamefont{de~Wolf}},
  \bibinfo{year}{2011}, \bibinfo{journal}{Proc. of 26th IEEE Annual Conference
  on Computational Complexity (CCC)} , \bibinfo{pages}{157}.

\bibitem[{\citenamefont{Buscemi}(2012)}]{Buscemi12}
\bibinfo{author}{\bibnamefont{Buscemi}, \bibfnamefont{F.}},
  \bibinfo{year}{2012}, \bibinfo{journal}{Phys. Rev. Lett.}
  \textbf{\bibinfo{volume}{108}}, \bibinfo{pages}{200401}.

\bibitem[{\citenamefont{Cabello}(2001)}]{Cabello01}
\bibinfo{author}{\bibnamefont{Cabello}, \bibfnamefont{A.}},
  \bibinfo{year}{2001}, \bibinfo{journal}{Phys. Rev. Lett.}
  \textbf{\bibinfo{volume}{86}}, \bibinfo{pages}{1911}.

\bibitem[{\citenamefont{Cabello}(2003)}]{Cabello03}
\bibinfo{author}{\bibnamefont{Cabello}, \bibfnamefont{A.}},
  \bibinfo{year}{2003}, \bibinfo{journal}{Phys. Rev. Lett.}
  \textbf{\bibinfo{volume}{91}}, \bibinfo{pages}{230403}.

\bibitem[{\citenamefont{Cabello and Larsson}(2007)}]{CL07}
\bibinfo{author}{\bibnamefont{Cabello}, \bibfnamefont{A.}}, and
  \bibinfo{author}{\bibfnamefont{J.-A.} \bibnamefont{Larsson}},
  \bibinfo{year}{2007}, \bibinfo{journal}{Phys. Rev. Lett.}
  \textbf{\bibinfo{volume}{98}}, \bibinfo{pages}{220402}.

\bibitem[{\citenamefont{Cabello} \emph{et~al.}(2008)\citenamefont{Cabello,
  Rodriguez, and Villanueva}}]{CRV08}
\bibinfo{author}{\bibnamefont{Cabello}, \bibfnamefont{A.}},
  \bibinfo{author}{\bibfnamefont{D.}~\bibnamefont{Rodriguez}}, and
  \bibinfo{author}{\bibfnamefont{I.}~\bibnamefont{Villanueva}},
  \bibinfo{year}{2008}, \bibinfo{journal}{Phys. Rev. Lett.}
  \textbf{\bibinfo{volume}{101}}, \bibinfo{pages}{120402}.

\bibitem[{\citenamefont{Cabello and Sciarrino}(2012)}]{CS12}
\bibinfo{author}{\bibnamefont{Cabello}, \bibfnamefont{A.}}, and
  \bibinfo{author}{\bibfnamefont{F.}~\bibnamefont{Sciarrino}},
  \bibinfo{year}{2012}, \bibinfo{journal}{Phys. Rev. X}
  \textbf{\bibinfo{volume}{2}}, \bibinfo{pages}{021010}.

\bibitem[{\citenamefont{Capasso} \emph{et~al.}(1973)\citenamefont{Capasso,
  Fortunato, and Selleri}}]{CFS73}
\bibinfo{author}{\bibnamefont{Capasso}, \bibfnamefont{V.}},
  \bibinfo{author}{\bibfnamefont{D.}~\bibnamefont{Fortunato}}, and
  \bibinfo{author}{\bibfnamefont{F.}~\bibnamefont{Selleri}},
  \bibinfo{year}{1973}, \bibinfo{journal}{Int. J. Mod. Phys.}
  \textbf{\bibinfo{volume}{7}}, \bibinfo{pages}{319}.

\bibitem[{\citenamefont{Cavalcanti}
  \emph{et~al.}(2013{\natexlab{a}})\citenamefont{Cavalcanti, Ac\'in, Brunner,
  and Vert\'esi}}]{CABV12}
\bibinfo{author}{\bibnamefont{Cavalcanti}, \bibfnamefont{D.}},
  \bibinfo{author}{\bibfnamefont{A.}~\bibnamefont{Ac\'in}},
  \bibinfo{author}{\bibfnamefont{N.}~\bibnamefont{Brunner}}, and
  \bibinfo{author}{\bibfnamefont{T.}~\bibnamefont{Vert\'esi}},
  \bibinfo{year}{2013}{\natexlab{a}}, \bibinfo{journal}{Phys. Rev.~A}
  \textbf{\bibinfo{volume}{87}}, \bibinfo{pages}{042104}.

\bibitem[{\citenamefont{Cavalcanti}
  \emph{et~al.}(2011{\natexlab{a}})\citenamefont{Cavalcanti, Almeida, Scarani,
  and Ac\'in}}]{CASA11}
\bibinfo{author}{\bibnamefont{Cavalcanti}, \bibfnamefont{D.}},
  \bibinfo{author}{\bibfnamefont{M.~L.} \bibnamefont{Almeida}},
  \bibinfo{author}{\bibfnamefont{V.}~\bibnamefont{Scarani}}, and
  \bibinfo{author}{\bibfnamefont{A.}~\bibnamefont{Ac\'in}},
  \bibinfo{year}{2011}{\natexlab{a}}, \bibinfo{journal}{Nat. Commun.}
  \textbf{\bibinfo{volume}{2}}, \bibinfo{pages}{184}.

\bibitem[{\citenamefont{Cavalcanti}
  \emph{et~al.}(2011{\natexlab{b}})\citenamefont{Cavalcanti, Brunner,
  Skrzypczyk, Salles, and Scarani}}]{CBSL+11}
\bibinfo{author}{\bibnamefont{Cavalcanti}, \bibfnamefont{D.}},
  \bibinfo{author}{\bibfnamefont{N.}~\bibnamefont{Brunner}},
  \bibinfo{author}{\bibfnamefont{P.}~\bibnamefont{Skrzypczyk}},
  \bibinfo{author}{\bibfnamefont{A.}~\bibnamefont{Salles}}, and
  \bibinfo{author}{\bibfnamefont{V.}~\bibnamefont{Scarani}},
  \bibinfo{year}{2011}{\natexlab{b}}, \bibinfo{journal}{Phys. Rev.~A}
  \textbf{\bibinfo{volume}{84}}, \bibinfo{pages}{022105}.

\bibitem[{\citenamefont{Cavalcanti}
  \emph{et~al.}(2012)\citenamefont{Cavalcanti, Rabelo, and Scarani}}]{CRS12}
\bibinfo{author}{\bibnamefont{Cavalcanti}, \bibfnamefont{D.}},
  \bibinfo{author}{\bibfnamefont{R.}~\bibnamefont{Rabelo}}, and
  \bibinfo{author}{\bibfnamefont{V.}~\bibnamefont{Scarani}},
  \bibinfo{year}{2012}, \bibinfo{journal}{Phys. Rev. Lett.}
  \textbf{\bibinfo{volume}{108}}, \bibinfo{pages}{040402}.

\bibitem[{\citenamefont{Cavalcanti}
  \emph{et~al.}(2010)\citenamefont{Cavalcanti, Salles, and Scarani}}]{CSS10}
\bibinfo{author}{\bibnamefont{Cavalcanti}, \bibfnamefont{D.}},
  \bibinfo{author}{\bibfnamefont{A.}~\bibnamefont{Salles}}, and
  \bibinfo{author}{\bibfnamefont{V.}~\bibnamefont{Scarani}},
  \bibinfo{year}{2010}, \bibinfo{journal}{Nat. Commun.}
  \textbf{\bibinfo{volume}{1}}, \bibinfo{pages}{136}.

\bibitem[{\citenamefont{Cavalcanti and Scarani}(2011)}]{CS11}
\bibinfo{author}{\bibnamefont{Cavalcanti}, \bibfnamefont{D.}}, and
  \bibinfo{author}{\bibfnamefont{V.}~\bibnamefont{Scarani}},
  \bibinfo{year}{2011}, \bibinfo{journal}{Phys. Rev. Lett.}
  \textbf{\bibinfo{volume}{106}}, \bibinfo{pages}{208901}.

\bibitem[{\citenamefont{Cavalcanti}
  \emph{et~al.}(2007)\citenamefont{Cavalcanti, Foster, Reid, and
  Drummond}}]{CFRD07}
\bibinfo{author}{\bibnamefont{Cavalcanti}, \bibfnamefont{E.}},
  \bibinfo{author}{\bibfnamefont{C.}~\bibnamefont{Foster}},
  \bibinfo{author}{\bibfnamefont{M.}~\bibnamefont{Reid}}, and
  \bibinfo{author}{\bibfnamefont{P.}~\bibnamefont{Drummond}},
  \bibinfo{year}{2007}, \bibinfo{journal}{Phys. Rev. Lett.}
  \textbf{\bibinfo{volume}{99}}, \bibinfo{pages}{210405}.

\bibitem[{\citenamefont{Cavalcanti}
  \emph{et~al.}(2013{\natexlab{b}})\citenamefont{Cavalcanti, Hall, and
  Wiseman}}]{CHW12}
\bibinfo{author}{\bibnamefont{Cavalcanti}, \bibfnamefont{E.}},
  \bibinfo{author}{\bibfnamefont{M.}~\bibnamefont{Hall}}, and
  \bibinfo{author}{\bibfnamefont{H.}~\bibnamefont{Wiseman}},
  \bibinfo{year}{2013}{\natexlab{b}}, \bibinfo{journal}{Phys. Rev.~A}
  \textbf{\bibinfo{volume}{87}}, \bibinfo{pages}{032306}.

\bibitem[{\citenamefont{Cavalcanti}
  \emph{et~al.}(2009)\citenamefont{Cavalcanti, Jones, Wiseman, and
  Reid}}]{CJWR09}
\bibinfo{author}{\bibnamefont{Cavalcanti}, \bibfnamefont{E.}},
  \bibinfo{author}{\bibfnamefont{S.}~\bibnamefont{Jones}},
  \bibinfo{author}{\bibfnamefont{H.}~\bibnamefont{Wiseman}}, and
  \bibinfo{author}{\bibfnamefont{M.}~\bibnamefont{Reid}}, \bibinfo{year}{2009},
  \bibinfo{journal}{Phys. Rev.~A} \textbf{\bibinfo{volume}{80}},
  \bibinfo{pages}{032112}.

\bibitem[{\citenamefont{Ceccarelli}
  \emph{et~al.}(2009)\citenamefont{Ceccarelli, Vallone, Martini, Mataloni, and
  Cabello}}]{CVDM+09}
\bibinfo{author}{\bibnamefont{Ceccarelli}, \bibfnamefont{R.}},
  \bibinfo{author}{\bibfnamefont{G.}~\bibnamefont{Vallone}},
  \bibinfo{author}{\bibfnamefont{F.~D.} \bibnamefont{Martini}},
  \bibinfo{author}{\bibfnamefont{P.}~\bibnamefont{Mataloni}}, and
  \bibinfo{author}{\bibfnamefont{A.}~\bibnamefont{Cabello}},
  \bibinfo{year}{2009}, \bibinfo{journal}{Phys. Rev. Lett.}
  \textbf{\bibinfo{volume}{103}}, \bibinfo{pages}{160401}.

\bibitem[{\citenamefont{Cereceda}(2002)}]{Cereceda02}
\bibinfo{author}{\bibnamefont{Cereceda}, \bibfnamefont{J.}},
  \bibinfo{year}{2002}, \bibinfo{journal}{Phys. Rev.~A}
  \textbf{\bibinfo{volume}{66}}, \bibinfo{pages}{024102}.

\bibitem[{\citenamefont{Cerf and Adami}(1997)}]{CA97}
\bibinfo{author}{\bibnamefont{Cerf}, \bibfnamefont{N.}}, and
  \bibinfo{author}{\bibfnamefont{C.}~\bibnamefont{Adami}},
  \bibinfo{year}{1997}, \bibinfo{journal}{Phys. Rev.~A}
  \textbf{\bibinfo{volume}{55}}, \bibinfo{pages}{3371}.

\bibitem[{\citenamefont{Cerf} \emph{et~al.}(2005)\citenamefont{Cerf, Gisin,
  Massar, and Popescu}}]{CGMP05}
\bibinfo{author}{\bibnamefont{Cerf}, \bibfnamefont{N.}},
  \bibinfo{author}{\bibfnamefont{N.}~\bibnamefont{Gisin}},
  \bibinfo{author}{\bibfnamefont{S.}~\bibnamefont{Massar}}, and
  \bibinfo{author}{\bibfnamefont{S.}~\bibnamefont{Popescu}},
  \bibinfo{year}{2005}, \bibinfo{journal}{Phys. Rev. Lett.}
  \textbf{\bibinfo{volume}{94}}, \bibinfo{pages}{220403}.

\bibitem[{\citenamefont{Cerf} \emph{et~al.}(2002)\citenamefont{Cerf, Massar,
  and Pironio}}]{CMP02}
\bibinfo{author}{\bibnamefont{Cerf}, \bibfnamefont{N.}},
  \bibinfo{author}{\bibfnamefont{S.}~\bibnamefont{Massar}}, and
  \bibinfo{author}{\bibfnamefont{S.}~\bibnamefont{Pironio}},
  \bibinfo{year}{2002}, \bibinfo{journal}{Phys. Rev. Lett.}
  \textbf{\bibinfo{volume}{89}}, \bibinfo{pages}{080402}.

\bibitem[{\citenamefont{Chaves} \emph{et~al.}(2012)\citenamefont{Chaves,
  Cavalcanti, Aolita, and Ac\'in}}]{CCAA12}
\bibinfo{author}{\bibnamefont{Chaves}, \bibfnamefont{R.}},
  \bibinfo{author}{\bibfnamefont{D.}~\bibnamefont{Cavalcanti}},
  \bibinfo{author}{\bibfnamefont{L.}~\bibnamefont{Aolita}}, and
  \bibinfo{author}{\bibfnamefont{A.}~\bibnamefont{Ac\'in}},
  \bibinfo{year}{2012}, \bibinfo{journal}{Phys. Rev.~A}
  \textbf{\bibinfo{volume}{86}}, \bibinfo{pages}{012108}.

\bibitem[{\citenamefont{Chaves and Fritz}(2012)}]{CF12}
\bibinfo{author}{\bibnamefont{Chaves}, \bibfnamefont{R.}}, and
  \bibinfo{author}{\bibfnamefont{T.}~\bibnamefont{Fritz}},
  \bibinfo{year}{2012}, \bibinfo{journal}{Phys. Rev.~A}
  \textbf{\bibinfo{volume}{85}}, \bibinfo{pages}{032113}.

\bibitem[{\citenamefont{Chen} \emph{et~al.}(2008)\citenamefont{Chen, Wu, Kwek,
  and Oh}}]{CWKO08}
\bibinfo{author}{\bibnamefont{Chen}, \bibfnamefont{J.-L.}},
  \bibinfo{author}{\bibfnamefont{C.}~\bibnamefont{Wu}},
  \bibinfo{author}{\bibfnamefont{L.}~\bibnamefont{Kwek}}, and
  \bibinfo{author}{\bibfnamefont{C.}~\bibnamefont{Oh}}, \bibinfo{year}{2008},
  \bibinfo{journal}{Phys. Rev.~A} \textbf{\bibinfo{volume}{78}},
  \bibinfo{pages}{032107}.

\bibitem[{\citenamefont{Christensen}
  \emph{et~al.}(2013)\citenamefont{Christensen, McCusker, Altepeter, Calkins,
  Gerrits, Lita, Miller, Shalm, Zhang, Nam, Brunner, Lim}
  \emph{et~al.}}]{CMAC+13}
\bibinfo{author}{\bibnamefont{Christensen}, \bibfnamefont{B.}},
  \bibinfo{author}{\bibfnamefont{K.}~\bibnamefont{McCusker}},
  \bibinfo{author}{\bibfnamefont{J.}~\bibnamefont{Altepeter}},
  \bibinfo{author}{\bibfnamefont{B.}~\bibnamefont{Calkins}},
  \bibinfo{author}{\bibfnamefont{T.}~\bibnamefont{Gerrits}},
  \bibinfo{author}{\bibfnamefont{A.}~\bibnamefont{Lita}},
  \bibinfo{author}{\bibfnamefont{A.}~\bibnamefont{Miller}},
  \bibinfo{author}{\bibfnamefont{L.}~\bibnamefont{Shalm}},
  \bibinfo{author}{\bibfnamefont{Y.}~\bibnamefont{Zhang}},
  \bibinfo{author}{\bibfnamefont{S.~W.} \bibnamefont{Nam}},
  \bibinfo{author}{\bibfnamefont{N.}~\bibnamefont{Brunner}},
  \bibinfo{author}{\bibfnamefont{C.}~\bibnamefont{Lim}}, \emph{et~al.},
  \bibinfo{year}{2013}, \bibinfo{journal}{Phys. Rev. Lett.}
  \textbf{\bibinfo{volume}{111}}, \bibinfo{pages}{130406}.

\bibitem[{\citenamefont{Christof and Lobel}(1997)}]{porta}
\bibinfo{author}{\bibnamefont{Christof}, \bibfnamefont{T.}}, and
  \bibinfo{author}{\bibfnamefont{A.}~\bibnamefont{Lobel}},
  \bibinfo{year}{1997}, \bibinfo{title}{porta},
  \urlprefix\url{http://typo.zib.de/opt-long_projects/Software/Porta/}.

\bibitem[{\citenamefont{Cirel'son}(1980)}]{Cirelson80}
\bibinfo{author}{\bibnamefont{Cirel'son}, \bibfnamefont{B.~S.}},
  \bibinfo{year}{1980}, \bibinfo{journal}{Letters in Mathematical Physics}
  \textbf{\bibinfo{volume}{4}}, \bibinfo{pages}{93}.

\bibitem[{\citenamefont{Clauser and Horne}(1974)}]{CH74}
\bibinfo{author}{\bibnamefont{Clauser}, \bibfnamefont{J.~F.}}, and
  \bibinfo{author}{\bibfnamefont{M.~A.} \bibnamefont{Horne}},
  \bibinfo{year}{1974}, \bibinfo{journal}{Phys. Rev. D}
  \textbf{\bibinfo{volume}{10}}, \bibinfo{pages}{526}.

\bibitem[{\citenamefont{Clauser} \emph{et~al.}(1969)\citenamefont{Clauser,
  Horne, Shimony, and Holt}}]{CHSH69}
\bibinfo{author}{\bibnamefont{Clauser}, \bibfnamefont{J.~F.}},
  \bibinfo{author}{\bibfnamefont{M.~A.} \bibnamefont{Horne}},
  \bibinfo{author}{\bibfnamefont{A.}~\bibnamefont{Shimony}}, and
  \bibinfo{author}{\bibfnamefont{R.~A.} \bibnamefont{Holt}},
  \bibinfo{year}{1969}, \bibinfo{journal}{Phys. Rev. Lett.}
  \textbf{\bibinfo{volume}{23}}, \bibinfo{pages}{880}.

\bibitem[{\citenamefont{Clauser and Shimony}(1978)}]{CS78}
\bibinfo{author}{\bibnamefont{Clauser}, \bibfnamefont{J.~F.}}, and
  \bibinfo{author}{\bibfnamefont{A.}~\bibnamefont{Shimony}},
  \bibinfo{year}{1978}, \bibinfo{journal}{Rep. Prog. Phys}
  \textbf{\bibinfo{volume}{41}}, \bibinfo{pages}{1881}.

\bibitem[{\citenamefont{Cleve and Buhrman}(1997)}]{CB97}
\bibinfo{author}{\bibnamefont{Cleve}, \bibfnamefont{R.}}, and
  \bibinfo{author}{\bibfnamefont{H.}~\bibnamefont{Buhrman}},
  \bibinfo{year}{1997}, \bibinfo{journal}{Phys. Rev.~A}
  \textbf{\bibinfo{volume}{56}}, \bibinfo{pages}{1201}.

\bibitem[{\citenamefont{Cleve} \emph{et~al.}(1998)\citenamefont{Cleve, van Dam,
  Nielsen, and Tapp}}]{CDNT98}
\bibinfo{author}{\bibnamefont{Cleve}, \bibfnamefont{R.}},
  \bibinfo{author}{\bibfnamefont{W.}~\bibnamefont{van Dam}},
  \bibinfo{author}{\bibfnamefont{M.}~\bibnamefont{Nielsen}}, and
  \bibinfo{author}{\bibfnamefont{A.}~\bibnamefont{Tapp}}, \bibinfo{year}{1998},
  \bibinfo{journal}{Lect. Notes Comput. Sci.} \textbf{\bibinfo{volume}{1509}},
  \bibinfo{pages}{61}.

\bibitem[{\citenamefont{Cleve} \emph{et~al.}(2004)\citenamefont{Cleve, Hoyer,
  Toner, and Watrous}}]{CHT+04}
\bibinfo{author}{\bibnamefont{Cleve}, \bibfnamefont{R.}},
  \bibinfo{author}{\bibfnamefont{P.}~\bibnamefont{Hoyer}},
  \bibinfo{author}{\bibfnamefont{B.}~\bibnamefont{Toner}}, and
  \bibinfo{author}{\bibfnamefont{J.}~\bibnamefont{Watrous}},
  \bibinfo{year}{2004}, in \emph{\bibinfo{booktitle}{19th IEEE Conference on
  Computational Complexity}}, p. \bibinfo{pages}{236}.

\bibitem[{\citenamefont{Cleve} \emph{et~al.}(2007)\citenamefont{Cleve,
  Slofstra, Unger, and Upadhyay}}]{cleve:xorParallel}
\bibinfo{author}{\bibnamefont{Cleve}, \bibfnamefont{R.}},
  \bibinfo{author}{\bibfnamefont{W.}~\bibnamefont{Slofstra}},
  \bibinfo{author}{\bibfnamefont{F.}~\bibnamefont{Unger}}, and
  \bibinfo{author}{\bibfnamefont{S.}~\bibnamefont{Upadhyay}},
  \bibinfo{year}{2007}, in \emph{\bibinfo{booktitle}{Proceedings of the 22nd
  Annual Conference on Computational Complexity}} (\bibinfo{publisher}{IEEE}),
  pp. \bibinfo{pages}{109--114}.

\bibitem[{\citenamefont{Colbeck}(2007)}]{Colbeck07}
\bibinfo{author}{\bibnamefont{Colbeck}, \bibfnamefont{R.}},
  \bibinfo{year}{2007}, \emph{\bibinfo{title}{Quantum and relativistic
  protocols for secure multi-party computation}}, Ph.D. thesis,
  \bibinfo{school}{University of Cambridge}.

\bibitem[{\citenamefont{Colbeck and Kent}(2011)}]{CK11}
\bibinfo{author}{\bibnamefont{Colbeck}, \bibfnamefont{R.}}, and
  \bibinfo{author}{\bibfnamefont{A.}~\bibnamefont{Kent}}, \bibinfo{year}{2011},
  \bibinfo{journal}{J.~Phys. A: Math. Theor.} \textbf{\bibinfo{volume}{44}},
  \bibinfo{pages}{095305}.

\bibitem[{\citenamefont{Colbeck and Renner}(2008)}]{CR08}
\bibinfo{author}{\bibnamefont{Colbeck}, \bibfnamefont{R.}}, and
  \bibinfo{author}{\bibfnamefont{R.}~\bibnamefont{Renner}},
  \bibinfo{year}{2008}, \bibinfo{journal}{Phys. Rev. Lett.}
  \textbf{\bibinfo{volume}{101}}, \bibinfo{pages}{050403}.

\bibitem[{\citenamefont{Colbeck and Renner}(2011)}]{CR11}
\bibinfo{author}{\bibnamefont{Colbeck}, \bibfnamefont{R.}}, and
  \bibinfo{author}{\bibfnamefont{R.}~\bibnamefont{Renner}},
  \bibinfo{year}{2011}, \bibinfo{journal}{Nat. Commun.}
  \textbf{\bibinfo{volume}{2}}, \bibinfo{pages}{411}.

\bibitem[{\citenamefont{Colbeck and Renner}(2012)}]{CR12}
\bibinfo{author}{\bibnamefont{Colbeck}, \bibfnamefont{R.}}, and
  \bibinfo{author}{\bibfnamefont{R.}~\bibnamefont{Renner}},
  \bibinfo{year}{2012}, \bibinfo{journal}{Nat. Phys.}
  \textbf{\bibinfo{volume}{8}}, \bibinfo{pages}{450}.

\bibitem[{\citenamefont{Collins and Gisin}(2004)}]{CG04}
\bibinfo{author}{\bibnamefont{Collins}, \bibfnamefont{D.}}, and
  \bibinfo{author}{\bibfnamefont{N.}~\bibnamefont{Gisin}},
  \bibinfo{year}{2004}, \bibinfo{journal}{J.~Phys. A: Math. Theor.}
  \textbf{\bibinfo{volume}{37}}, \bibinfo{pages}{1775}.

\bibitem[{\citenamefont{Collins}
  \emph{et~al.}(2002{\natexlab{a}})\citenamefont{Collins, Gisin, Linden,
  Massar, and Popescu}}]{CGL+02}
\bibinfo{author}{\bibnamefont{Collins}, \bibfnamefont{D.}},
  \bibinfo{author}{\bibfnamefont{N.}~\bibnamefont{Gisin}},
  \bibinfo{author}{\bibfnamefont{N.}~\bibnamefont{Linden}},
  \bibinfo{author}{\bibfnamefont{S.}~\bibnamefont{Massar}}, and
  \bibinfo{author}{\bibfnamefont{S.}~\bibnamefont{Popescu}},
  \bibinfo{year}{2002}{\natexlab{a}}, \bibinfo{journal}{Phys. Rev. Lett.}
  \textbf{\bibinfo{volume}{88}}, \bibinfo{pages}{040404}.

\bibitem[{\citenamefont{Collins}
  \emph{et~al.}(2002{\natexlab{b}})\citenamefont{Collins, Gisin, Popescu,
  Roberts, and Scarani}}]{CGPR+02}
\bibinfo{author}{\bibnamefont{Collins}, \bibfnamefont{D.}},
  \bibinfo{author}{\bibfnamefont{N.}~\bibnamefont{Gisin}},
  \bibinfo{author}{\bibfnamefont{S.}~\bibnamefont{Popescu}},
  \bibinfo{author}{\bibfnamefont{D.}~\bibnamefont{Roberts}}, and
  \bibinfo{author}{\bibfnamefont{V.}~\bibnamefont{Scarani}},
  \bibinfo{year}{2002}{\natexlab{b}}, \bibinfo{journal}{Phys. Rev. Lett.}
  \textbf{\bibinfo{volume}{88}}, \bibinfo{pages}{170405}.

\bibitem[{\citenamefont{Condon}(1989)}]{condon:oldSurvey}
\bibinfo{author}{\bibnamefont{Condon}, \bibfnamefont{A.}},
  \bibinfo{year}{1989}, in \emph{\bibinfo{booktitle}{Proceedings of 30th Annual
  Symposium on Foundations of Computer Science}} (\bibinfo{publisher}{IEEE}),
  pp. \bibinfo{pages}{462--467}.

\bibitem[{\citenamefont{Coretti} \emph{et~al.}(2011)\citenamefont{Coretti,
  H\"anggi, and Wolf}}]{CHW11}
\bibinfo{author}{\bibnamefont{Coretti}, \bibfnamefont{S.}},
  \bibinfo{author}{\bibfnamefont{E.}~\bibnamefont{H\"anggi}}, and
  \bibinfo{author}{\bibfnamefont{S.}~\bibnamefont{Wolf}}, \bibinfo{year}{2011},
  \bibinfo{journal}{Phys. Rev. Lett.} \textbf{\bibinfo{volume}{107}},
  \bibinfo{pages}{100402}.

\bibitem[{\citenamefont{Cubitt} \emph{et~al.}(2010)\citenamefont{Cubitt, Leung,
  Matthews, and Winter}}]{zeroError1}
\bibinfo{author}{\bibnamefont{Cubitt}, \bibfnamefont{T.~S.}},
  \bibinfo{author}{\bibfnamefont{D.}~\bibnamefont{Leung}},
  \bibinfo{author}{\bibfnamefont{W.}~\bibnamefont{Matthews}}, and
  \bibinfo{author}{\bibfnamefont{A.}~\bibnamefont{Winter}},
  \bibinfo{year}{2010}, \bibinfo{journal}{Phys. Rev. Lett.}
  \textbf{\bibinfo{volume}{104}}, \bibinfo{pages}{230503}.

\bibitem[{\citenamefont{Cubitt} \emph{et~al.}(2011)\citenamefont{Cubitt, Leung,
  Matthews, and Winter}}]{noIncreaseNSig}
\bibinfo{author}{\bibnamefont{Cubitt}, \bibfnamefont{T.~S.}},
  \bibinfo{author}{\bibfnamefont{D.}~\bibnamefont{Leung}},
  \bibinfo{author}{\bibfnamefont{W.}~\bibnamefont{Matthews}}, and
  \bibinfo{author}{\bibfnamefont{A.}~\bibnamefont{Winter}},
  \bibinfo{year}{2011}, \bibinfo{journal}{IEEE Trans. Inf. Theory}
  \textbf{\bibinfo{volume}{57}}(\bibinfo{number}{8}), \bibinfo{pages}{5509}.

\bibitem[{\citenamefont{Curty and Moroder}(2011)}]{CM11}
\bibinfo{author}{\bibnamefont{Curty}, \bibfnamefont{M.}}, and
  \bibinfo{author}{\bibfnamefont{T.}~\bibnamefont{Moroder}},
  \bibinfo{year}{2011}, \bibinfo{journal}{Phys. Rev. A}
  \textbf{\bibinfo{volume}{84}}, \bibinfo{pages}{010304}.

\bibitem[{\citenamefont{Dada} \emph{et~al.}(2011)\citenamefont{Dada, Leach,
  Buller, Padgett, and Andersson}}]{DLBP+11}
\bibinfo{author}{\bibnamefont{Dada}, \bibfnamefont{A.}},
  \bibinfo{author}{\bibfnamefont{J.}~\bibnamefont{Leach}},
  \bibinfo{author}{\bibfnamefont{G.}~\bibnamefont{Buller}},
  \bibinfo{author}{\bibfnamefont{M.}~\bibnamefont{Padgett}}, and
  \bibinfo{author}{\bibfnamefont{E.}~\bibnamefont{Andersson}},
  \bibinfo{year}{2011}, \bibinfo{journal}{Nat. Phys.}
  \textbf{\bibinfo{volume}{7}}, \bibinfo{pages}{677}.

\bibitem[{\citenamefont{van Dam}(2005)}]{vanDam05}
\bibinfo{author}{\bibnamefont{van Dam}, \bibfnamefont{W.}},
  \bibinfo{year}{2005}, \bibinfo{journal}{quant-ph/0501159v1} .

\bibitem[{\citenamefont{van Dam} \emph{et~al.}(2005)\citenamefont{van Dam,
  Grunwald, and Gill}}]{DGG05}
\bibinfo{author}{\bibnamefont{van Dam}, \bibfnamefont{W.}},
  \bibinfo{author}{\bibfnamefont{P.}~\bibnamefont{Grunwald}}, and
  \bibinfo{author}{\bibfnamefont{R.}~\bibnamefont{Gill}}, \bibinfo{year}{2005},
  \bibinfo{journal}{IEEE-Trans. on Information Theory}
  \textbf{\bibinfo{volume}{51}}, \bibinfo{pages}{2812}.

\bibitem[{\citenamefont{van Dam and Hayden}(2003)}]{patrick:embezzle}
\bibinfo{author}{\bibnamefont{van Dam}, \bibfnamefont{W.}}, and
  \bibinfo{author}{\bibfnamefont{P.}~\bibnamefont{Hayden}},
  \bibinfo{year}{2003}, \bibinfo{journal}{Phys. Rev. A}
  \textbf{\bibinfo{volume}{67}}(\bibinfo{number}{6}),
  \bibinfo{pages}{060302(R)}.

\bibitem[{\citenamefont{D'Ambrosio}
  \emph{et~al.}(2012)\citenamefont{D'Ambrosio, Nagali, Walborn, Aolita,
  Slussarenko, Marrucci, and Sciarrino}}]{DNWA+12}
\bibinfo{author}{\bibnamefont{D'Ambrosio}, \bibfnamefont{V.}},
  \bibinfo{author}{\bibfnamefont{E.}~\bibnamefont{Nagali}},
  \bibinfo{author}{\bibfnamefont{S.}~\bibnamefont{Walborn}},
  \bibinfo{author}{\bibfnamefont{L.}~\bibnamefont{Aolita}},
  \bibinfo{author}{\bibfnamefont{S.}~\bibnamefont{Slussarenko}},
  \bibinfo{author}{\bibfnamefont{L.}~\bibnamefont{Marrucci}}, and
  \bibinfo{author}{\bibfnamefont{F.}~\bibnamefont{Sciarrino}},
  \bibinfo{year}{2012}, \bibinfo{journal}{Nat. Commun.}
  \textbf{\bibinfo{volume}{3}}, \bibinfo{pages}{961}.

\bibitem[{\citenamefont{De} \emph{et~al.}(2009)\citenamefont{De, Portmann,
  Vidick, and Renner}}]{DPV+09}
\bibinfo{author}{\bibnamefont{De}, \bibfnamefont{A.}},
  \bibinfo{author}{\bibfnamefont{C.}~\bibnamefont{Portmann}},
  \bibinfo{author}{\bibfnamefont{T.}~\bibnamefont{Vidick}}, and
  \bibinfo{author}{\bibfnamefont{R.}~\bibnamefont{Renner}},
  \bibinfo{year}{2009}, \bibinfo{journal}{arXiv:0912.5514} .

\bibitem[{\citenamefont{De} \emph{et~al.}(2005)\citenamefont{De, Sen, Brukner,
  Buzek, and Zukowski}}]{SSB+05}
\bibinfo{author}{\bibnamefont{De}, \bibfnamefont{A.~S.}},
  \bibinfo{author}{\bibfnamefont{U.}~\bibnamefont{Sen}},
  \bibinfo{author}{\bibfnamefont{C.}~\bibnamefont{Brukner}},
  \bibinfo{author}{\bibfnamefont{V.}~\bibnamefont{Buzek}}, and
  \bibinfo{author}{\bibfnamefont{M.}~\bibnamefont{Zukowski}},
  \bibinfo{year}{2005}, \bibinfo{journal}{pra} \textbf{\bibinfo{volume}{72}},
  \bibinfo{pages}{042310}.

\bibitem[{\citenamefont{Degorre} \emph{et~al.}(2011)\citenamefont{Degorre,
  Kaplan, Laplante, and Roland}}]{DKLR11}
\bibinfo{author}{\bibnamefont{Degorre}, \bibfnamefont{J.}},
  \bibinfo{author}{\bibfnamefont{M.}~\bibnamefont{Kaplan}},
  \bibinfo{author}{\bibfnamefont{S.}~\bibnamefont{Laplante}}, and
  \bibinfo{author}{\bibfnamefont{J.}~\bibnamefont{Roland}},
  \bibinfo{year}{2011}, \bibinfo{journal}{Quant. Inf. and Comp.}
  \textbf{\bibinfo{volume}{11}}, \bibinfo{pages}{649}.

\bibitem[{\citenamefont{Degorre} \emph{et~al.}(2005)\citenamefont{Degorre,
  Laplante, and Roland}}]{DLR05}
\bibinfo{author}{\bibnamefont{Degorre}, \bibfnamefont{J.}},
  \bibinfo{author}{\bibfnamefont{S.}~\bibnamefont{Laplante}}, and
  \bibinfo{author}{\bibfnamefont{J.}~\bibnamefont{Roland}},
  \bibinfo{year}{2005}, \bibinfo{journal}{Phys. Rev.~A}
  \textbf{\bibinfo{volume}{72}}, \bibinfo{pages}{062314}.

\bibitem[{\citenamefont{Dehlinger and Mitchell}(2002)}]{DM02}
\bibinfo{author}{\bibnamefont{Dehlinger}, \bibfnamefont{D.}}, and
  \bibinfo{author}{\bibfnamefont{M.}~\bibnamefont{Mitchell}},
  \bibinfo{year}{2002}, \bibinfo{journal}{American Journal of Physics}
  \textbf{\bibinfo{volume}{70}}, \bibinfo{pages}{903}.

\bibitem[{\citenamefont{Dhara} \emph{et~al.}(2013)\citenamefont{Dhara,
  Prettico, and Ac\'in}}]{DPA12}
\bibinfo{author}{\bibnamefont{Dhara}, \bibfnamefont{C.}},
  \bibinfo{author}{\bibfnamefont{G.}~\bibnamefont{Prettico}}, and
  \bibinfo{author}{\bibfnamefont{A.}~\bibnamefont{Ac\'in}},
  \bibinfo{year}{2013}, \bibinfo{journal}{Phys. Rev. A}  
  \textbf{\bibinfo{volume}{88}}, \bibinfo{pages}{052116}.
  

\bibitem[{\citenamefont{DiCarlo} \emph{et~al.}(2010)\citenamefont{DiCarlo,
  Reed, Sun, Johnson, Chow, Gambetta, Frunzio, Girvin, Devoret, and
  Schoelkopf}}]{DRSJ+10}
\bibinfo{author}{\bibnamefont{DiCarlo}, \bibfnamefont{L.}},
  \bibinfo{author}{\bibfnamefont{M.}~\bibnamefont{Reed}},
  \bibinfo{author}{\bibfnamefont{L.}~\bibnamefont{Sun}},
  \bibinfo{author}{\bibfnamefont{B.}~\bibnamefont{Johnson}},
  \bibinfo{author}{\bibfnamefont{J.}~\bibnamefont{Chow}},
  \bibinfo{author}{\bibfnamefont{J.}~\bibnamefont{Gambetta}},
  \bibinfo{author}{\bibfnamefont{L.}~\bibnamefont{Frunzio}},
  \bibinfo{author}{\bibfnamefont{S.}~\bibnamefont{Girvin}},
  \bibinfo{author}{\bibfnamefont{M.}~\bibnamefont{Devoret}}, and
  \bibinfo{author}{\bibfnamefont{R.}~\bibnamefont{Schoelkopf}},
  \bibinfo{year}{2010}, \bibinfo{journal}{Nature}
  \textbf{\bibinfo{volume}{467}}, \bibinfo{pages}{574}.

\bibitem[{\citenamefont{Dicke}(1954)}]{Dicke54}
\bibinfo{author}{\bibnamefont{Dicke}, \bibfnamefont{R.}}, \bibinfo{year}{1954},
  \bibinfo{journal}{Phys. Rev.} \textbf{\bibinfo{volume}{93}},
  \bibinfo{pages}{99}.

\bibitem[{\citenamefont{Dinur and Reingold}(2006)}]{dinur:parallel}
\bibinfo{author}{\bibnamefont{Dinur}, \bibfnamefont{I.}}, and
  \bibinfo{author}{\bibfnamefont{O.}~\bibnamefont{Reingold}},
  \bibinfo{year}{2006}, \bibinfo{journal}{SIAM Journal on Computing}
  \textbf{\bibinfo{volume}{36}}(\bibinfo{number}{4}), \bibinfo{pages}{975}.

\bibitem[{\citenamefont{DiVincenzo and Peres}(1997)}]{DP97}
\bibinfo{author}{\bibnamefont{DiVincenzo}, \bibfnamefont{D.}}, and
  \bibinfo{author}{\bibfnamefont{A.}~\bibnamefont{Peres}},
  \bibinfo{year}{1997}, \bibinfo{journal}{Phys. Rev.~A}
  \textbf{\bibinfo{volume}{55}}, \bibinfo{pages}{4089}.

\bibitem[{\citenamefont{Doherty and Wehner}(2011)}]{DW:njp}
\bibinfo{author}{\bibnamefont{Doherty}, \bibfnamefont{A.}}, and
  \bibinfo{author}{\bibfnamefont{S.}~\bibnamefont{Wehner}},
  \bibinfo{year}{2011}, \bibinfo{journal}{New Journal of Physics}
  \textbf{\bibinfo{volume}{13}}, \bibinfo{pages}{073033}.

\bibitem[{\citenamefont{Doherty} \emph{et~al.}(2008)\citenamefont{Doherty,
  Liang, Toner, and Wehner}}]{DLTW08}
\bibinfo{author}{\bibnamefont{Doherty}, \bibfnamefont{A.~C.}},
  \bibinfo{author}{\bibfnamefont{Y.~C.} \bibnamefont{Liang}},
  \bibinfo{author}{\bibfnamefont{B.}~\bibnamefont{Toner}}, and
  \bibinfo{author}{\bibfnamefont{S.}~\bibnamefont{Wehner}},
  \bibinfo{year}{2008}, in \emph{\bibinfo{booktitle}{IEEE Conference on
  Computational Complexity}}, p. \bibinfo{pages}{199}.

\bibitem[{\citenamefont{Doherty} \emph{et~al.}(2002)\citenamefont{Doherty,
  Parrilo, and Spedalieri}}]{doherty1}
\bibinfo{author}{\bibnamefont{Doherty}, \bibfnamefont{A.~C.}},
  \bibinfo{author}{\bibfnamefont{P.~A.} \bibnamefont{Parrilo}}, and
  \bibinfo{author}{\bibfnamefont{F.~M.} \bibnamefont{Spedalieri}},
  \bibinfo{year}{2002}, \bibinfo{journal}{Phys. Rev. Lett.}
  \textbf{\bibinfo{volume}{88}}(\bibinfo{number}{18}), \bibinfo{pages}{187904}.

\bibitem[{\citenamefont{Doherty} \emph{et~al.}(2004)\citenamefont{Doherty,
  Parrilo, and Spedalieri}}]{DPS04}
\bibinfo{author}{\bibnamefont{Doherty}, \bibfnamefont{A.~C.}},
  \bibinfo{author}{\bibfnamefont{P.~A.} \bibnamefont{Parrilo}}, and
  \bibinfo{author}{\bibfnamefont{F.~M.} \bibnamefont{Spedalieri}},
  \bibinfo{year}{2004}, \bibinfo{journal}{Phys. Rev.~A}
  \textbf{\bibinfo{volume}{69}}, \bibinfo{pages}{022308}.

\bibitem[{\citenamefont{Dukaric and Wolf}(2008)}]{DW08}
\bibinfo{author}{\bibnamefont{Dukaric}, \bibfnamefont{D.}}, and
  \bibinfo{author}{\bibfnamefont{S.}~\bibnamefont{Wolf}}, \bibinfo{year}{2008},
  \bibinfo{journal}{arXiv:0808.3317} .

\bibitem[{\citenamefont{D\"ur}(2001)}]{Dur01}
\bibinfo{author}{\bibnamefont{D\"ur}, \bibfnamefont{W.}}, \bibinfo{year}{2001},
  \bibinfo{journal}{Phys. Rev. Lett.} \textbf{\bibinfo{volume}{87}},
  \bibinfo{pages}{230402}.

\bibitem[{\citenamefont{Eberhard}(1993)}]{Eberhard93}
\bibinfo{author}{\bibnamefont{Eberhard}, \bibfnamefont{P.}},
  \bibinfo{year}{1993}, \bibinfo{journal}{Phys. Rev.~A}
  \textbf{\bibinfo{volume}{47}}, \bibinfo{pages}{747}.

\bibitem[{\citenamefont{Eibl} \emph{et~al.}(2003)\citenamefont{Eibl, Gaertner,
  Bourennane, Kurtsiefer, Zukowski, and Weinfurter}}]{EGBK+03}
\bibinfo{author}{\bibnamefont{Eibl}, \bibfnamefont{M.}},
  \bibinfo{author}{\bibfnamefont{S.}~\bibnamefont{Gaertner}},
  \bibinfo{author}{\bibfnamefont{M.}~\bibnamefont{Bourennane}},
  \bibinfo{author}{\bibfnamefont{C.}~\bibnamefont{Kurtsiefer}},
  \bibinfo{author}{\bibfnamefont{M.}~\bibnamefont{Zukowski}}, and
  \bibinfo{author}{\bibfnamefont{H.}~\bibnamefont{Weinfurter}},
  \bibinfo{year}{2003}, \bibinfo{journal}{Phys. Rev. Lett.}
  \textbf{\bibinfo{volume}{90}}, \bibinfo{pages}{200403}.

\bibitem[{\citenamefont{Eibl} \emph{et~al.}(2004)\citenamefont{Eibl, Kiesel,
  Bourennane, Kurtsiefer, and Weinfurter}}]{EKBK+04}
\bibinfo{author}{\bibnamefont{Eibl}, \bibfnamefont{M.}},
  \bibinfo{author}{\bibfnamefont{N.}~\bibnamefont{Kiesel}},
  \bibinfo{author}{\bibfnamefont{M.}~\bibnamefont{Bourennane}},
  \bibinfo{author}{\bibfnamefont{C.}~\bibnamefont{Kurtsiefer}}, and
  \bibinfo{author}{\bibfnamefont{H.}~\bibnamefont{Weinfurter}},
  \bibinfo{year}{2004}, \bibinfo{journal}{Phys. Rev. Lett.}
  \textbf{\bibinfo{volume}{92}}, \bibinfo{pages}{077901}.

\bibitem[{\citenamefont{Einstein} \emph{et~al.}(1935)\citenamefont{Einstein,
  Podolsky, and Rosen}}]{EPR35}
\bibinfo{author}{\bibnamefont{Einstein}, \bibfnamefont{A.}},
  \bibinfo{author}{\bibfnamefont{B.}~\bibnamefont{Podolsky}}, and
  \bibinfo{author}{\bibfnamefont{N.}~\bibnamefont{Rosen}},
  \bibinfo{year}{1935}, \bibinfo{journal}{Phys. Rev.}
  \textbf{\bibinfo{volume}{47}}, \bibinfo{pages}{777}.

\bibitem[{\citenamefont{Ekert}(1991)}]{Ekert91}
\bibinfo{author}{\bibnamefont{Ekert}, \bibfnamefont{A.}}, \bibinfo{year}{1991},
  \bibinfo{journal}{Phys. Rev. Lett.} \textbf{\bibinfo{volume}{67}},
  \bibinfo{pages}{661}.

\bibitem[{\citenamefont{Elitzur} \emph{et~al.}(1992)\citenamefont{Elitzur,
  Popescu, and Rohrlich}}]{EPR92}
\bibinfo{author}{\bibnamefont{Elitzur}, \bibfnamefont{A.}},
  \bibinfo{author}{\bibfnamefont{S.}~\bibnamefont{Popescu}}, and
  \bibinfo{author}{\bibfnamefont{D.}~\bibnamefont{Rohrlich}},
  \bibinfo{year}{1992}, \bibinfo{journal}{Phys. Lett.~A}
  \textbf{\bibinfo{volume}{162}}, \bibinfo{pages}{25}.

\bibitem[{\citenamefont{Fedrizzi} \emph{et~al.}(2009)\citenamefont{Fedrizzi,
  Ursin, Herbst, Nespoli, Prevedel, Scheidl, Tiefenbacher, Jennewein, and
  Zeilinger}}]{FUHN+09}
\bibinfo{author}{\bibnamefont{Fedrizzi}, \bibfnamefont{A.}},
  \bibinfo{author}{\bibfnamefont{R.}~\bibnamefont{Ursin}},
  \bibinfo{author}{\bibfnamefont{T.}~\bibnamefont{Herbst}},
  \bibinfo{author}{\bibfnamefont{M.}~\bibnamefont{Nespoli}},
  \bibinfo{author}{\bibfnamefont{R.}~\bibnamefont{Prevedel}},
  \bibinfo{author}{\bibfnamefont{T.}~\bibnamefont{Scheidl}},
  \bibinfo{author}{\bibfnamefont{F.}~\bibnamefont{Tiefenbacher}},
  \bibinfo{author}{\bibfnamefont{T.}~\bibnamefont{Jennewein}}, and
  \bibinfo{author}{\bibfnamefont{A.}~\bibnamefont{Zeilinger}},
  \bibinfo{year}{2009}, \bibinfo{journal}{Nat. Phys.}
  \textbf{\bibinfo{volume}{5}}, \bibinfo{pages}{389}.

\bibitem[{\citenamefont{Fehr} \emph{et~al.}(2013)\citenamefont{Fehr, Gelles,
  and Schaffner}}]{FGS11}
\bibinfo{author}{\bibnamefont{Fehr}, \bibfnamefont{S.}},
  \bibinfo{author}{\bibfnamefont{R.}~\bibnamefont{Gelles}}, and
  \bibinfo{author}{\bibfnamefont{C.}~\bibnamefont{Schaffner}},
  \bibinfo{year}{2013}, \bibinfo{journal}{Phys. Rev.~A}
  \textbf{\bibinfo{volume}{87}}, \bibinfo{pages}{012335}.

\bibitem[{\citenamefont{Feige and Kilian}(2000)}]{feige:parallel}
\bibinfo{author}{\bibnamefont{Feige}, \bibfnamefont{U.}}, and
  \bibinfo{author}{\bibfnamefont{J.}~\bibnamefont{Kilian}},
  \bibinfo{year}{2000}, \bibinfo{journal}{SIAM Journal on Computing}
  \textbf{\bibinfo{volume}{30}}(\bibinfo{number}{1}), \bibinfo{pages}{324}.

\bibitem[{\citenamefont{Fine}(1982)}]{Fine82}
\bibinfo{author}{\bibnamefont{Fine}, \bibfnamefont{A.}}, \bibinfo{year}{1982},
  \bibinfo{journal}{Phys. Rev. Lett.} \textbf{\bibinfo{volume}{48}},
  \bibinfo{pages}{291}.

\bibitem[{\citenamefont{Forster}(2011)}]{Forster11}
\bibinfo{author}{\bibnamefont{Forster}, \bibfnamefont{M.}},
  \bibinfo{year}{2011}, \bibinfo{journal}{Phys. Rev.~A}
  \textbf{\bibinfo{volume}{83}}, \bibinfo{pages}{062114}.

\bibitem[{\citenamefont{Forster} \emph{et~al.}(2009)\citenamefont{Forster,
  Winkler, and Wolf}}]{FWW09}
\bibinfo{author}{\bibnamefont{Forster}, \bibfnamefont{M.}},
  \bibinfo{author}{\bibfnamefont{S.}~\bibnamefont{Winkler}}, and
  \bibinfo{author}{\bibfnamefont{S.}~\bibnamefont{Wolf}}, \bibinfo{year}{2009},
  \bibinfo{journal}{Phys. Rev. Lett.} \textbf{\bibinfo{volume}{102}},
  \bibinfo{pages}{120401}.

\bibitem[{\citenamefont{Forster and Wolf}(2011)}]{FW11}
\bibinfo{author}{\bibnamefont{Forster}, \bibfnamefont{M.}}, and
  \bibinfo{author}{\bibfnamefont{S.}~\bibnamefont{Wolf}}, \bibinfo{year}{2011},
  \bibinfo{journal}{Phys. Rev.~A} \textbf{\bibinfo{volume}{84}},
  \bibinfo{pages}{042112}.

\bibitem[{\citenamefont{Franson}(1989)}]{Franson89}
\bibinfo{author}{\bibnamefont{Franson}, \bibfnamefont{J.}},
  \bibinfo{year}{1989}, \bibinfo{journal}{Phys. Rev. Lett.}
  \textbf{\bibinfo{volume}{62}}, \bibinfo{pages}{2205}.

\bibitem[{\citenamefont{Franz} \emph{et~al.}(2011)\citenamefont{Franz, Furrer,
  and Werner}}]{FFW11}
\bibinfo{author}{\bibnamefont{Franz}, \bibfnamefont{T.}},
  \bibinfo{author}{\bibfnamefont{F.}~\bibnamefont{Furrer}}, and
  \bibinfo{author}{\bibfnamefont{R.}~\bibnamefont{Werner}},
  \bibinfo{year}{2011}, \bibinfo{journal}{Phys. Rev. Lett.}
  \textbf{\bibinfo{volume}{106}}, \bibinfo{pages}{250502}.

\bibitem[{\citenamefont{Freedman and Clauser}(1972)}]{FC72}
\bibinfo{author}{\bibnamefont{Freedman}, \bibfnamefont{S.}}, and
  \bibinfo{author}{\bibfnamefont{J.}~\bibnamefont{Clauser}},
  \bibinfo{year}{1972}, \bibinfo{journal}{Phys. Rev. Lett.}
  \textbf{\bibinfo{volume}{28}}, \bibinfo{pages}{938}.

\bibitem[{\citenamefont{Fritz}(2011)}]{Fritz11}
\bibinfo{author}{\bibnamefont{Fritz}, \bibfnamefont{T.}}, \bibinfo{year}{2011},
  \bibinfo{journal}{Found. Phys.} \textbf{\bibinfo{volume}{41}},
  \bibinfo{pages}{1493}.

\bibitem[{\citenamefont{Fritz}(2012{\natexlab{a}})}]{Fritz10}
\bibinfo{author}{\bibnamefont{Fritz}, \bibfnamefont{T.}},
  \bibinfo{year}{2012}{\natexlab{a}}, \bibinfo{journal}{Rev. Math. Phys.}
  \textbf{\bibinfo{volume}{24}}, \bibinfo{pages}{1250012}.

\bibitem[{\citenamefont{Fritz}(2012{\natexlab{b}})}]{Fritz12}
\bibinfo{author}{\bibnamefont{Fritz}, \bibfnamefont{T.}},
  \bibinfo{year}{2012}{\natexlab{b}}, \bibinfo{journal}{J. Math. Phys.}
  \textbf{\bibinfo{volume}{53}}, \bibinfo{pages}{072202}.

\bibitem[{\citenamefont{Fritz} \emph{et~al.}(2013)\citenamefont{Fritz, Sainz,
  Augusiak, Brask, Chaves, Leverrier, and Ac\'i'n}}]{FSAB+12}
\bibinfo{author}{\bibnamefont{Fritz}, \bibfnamefont{T.}},
  \bibinfo{author}{\bibfnamefont{A.~B.} \bibnamefont{Sainz}},
  \bibinfo{author}{\bibfnamefont{R.}~\bibnamefont{Augusiak}},
  \bibinfo{author}{\bibfnamefont{J.~B.} \bibnamefont{Brask}},
  \bibinfo{author}{\bibfnamefont{R.}~\bibnamefont{Chaves}},
  \bibinfo{author}{\bibfnamefont{A.}~\bibnamefont{Leverrier}}, and
  \bibinfo{author}{\bibfnamefont{A.}~\bibnamefont{Ac\'i'n}},
  \bibinfo{year}{2013}, \bibinfo{journal}{Nat. Commun.}
  \textbf{\bibinfo{volume}{4}}, \bibinfo{pages}{2263}.

\bibitem[{\citenamefont{Froissard}(1981)}]{Froissard81}
\bibinfo{author}{\bibnamefont{Froissard}, \bibfnamefont{M.}},
  \bibinfo{year}{1981}, \bibinfo{journal}{Nuovo Cimento B}
  \textbf{\bibinfo{volume}{64}}, \bibinfo{pages}{241}.

\bibitem[{\citenamefont{Fry and Thompson}(1976)}]{FT76}
\bibinfo{author}{\bibnamefont{Fry}, \bibfnamefont{E.}}, and
  \bibinfo{author}{\bibfnamefont{R.}~\bibnamefont{Thompson}},
  \bibinfo{year}{1976}, \bibinfo{journal}{Phys. Rev. Lett.}
  \textbf{\bibinfo{volume}{37}}, \bibinfo{pages}{465}.

\bibitem[{\citenamefont{Fukuda}(2003)}]{cdd}
\bibinfo{author}{\bibnamefont{Fukuda}, \bibfnamefont{K.}},
  \bibinfo{year}{2003}, \bibinfo{title}{cdd},
  \urlprefix\url{http://www.ifor.math.ethz.ch/~fukuda/cdd_home/cdd.html}.

\bibitem[{\citenamefont{Gallego} \emph{et~al.}(2010)\citenamefont{Gallego,
  Brunner, Hadley, and Ac\'in}}]{GBHA10}
\bibinfo{author}{\bibnamefont{Gallego}, \bibfnamefont{R.}},
  \bibinfo{author}{\bibfnamefont{N.}~\bibnamefont{Brunner}},
  \bibinfo{author}{\bibfnamefont{C.}~\bibnamefont{Hadley}}, and
  \bibinfo{author}{\bibfnamefont{A.}~\bibnamefont{Ac\'in}},
  \bibinfo{year}{2010}, \bibinfo{journal}{Phys. Rev. Lett.}
  \textbf{\bibinfo{volume}{105}}, \bibinfo{pages}{230501}.

\bibitem[{\citenamefont{Gallego}
  \emph{et~al.}(2013{\natexlab{a}})\citenamefont{Gallego, Masanes, de~la Torre,
  Dhara, Aolita, and Ac\'in}}]{GMTD+12}
\bibinfo{author}{\bibnamefont{Gallego}, \bibfnamefont{R.}},
  \bibinfo{author}{\bibfnamefont{L.}~\bibnamefont{Masanes}},
  \bibinfo{author}{\bibfnamefont{G.}~\bibnamefont{de~la Torre}},
  \bibinfo{author}{\bibfnamefont{C.}~\bibnamefont{Dhara}},
  \bibinfo{author}{\bibfnamefont{L.}~\bibnamefont{Aolita}}, and
  \bibinfo{author}{\bibfnamefont{A.}~\bibnamefont{Ac\'in}},
  \bibinfo{year}{2013}{\natexlab{a}}, \bibinfo{journal}{Nat. Commun.}
  \textbf{\bibinfo{volume}{4}}, \bibinfo{pages}{2654}.

\bibitem[{\citenamefont{Gallego} \emph{et~al.}(2011)\citenamefont{Gallego,
  W\"ulfliger, Ac\'in, and Navascues}}]{GWAN11}
\bibinfo{author}{\bibnamefont{Gallego}, \bibfnamefont{R.}},
  \bibinfo{author}{\bibfnamefont{L.}~\bibnamefont{W\"ulfliger}},
  \bibinfo{author}{\bibfnamefont{A.}~\bibnamefont{Ac\'in}}, and
  \bibinfo{author}{\bibfnamefont{M.}~\bibnamefont{Navascues}},
  \bibinfo{year}{2011}, \bibinfo{journal}{Phys. Rev. Lett.}
  \textbf{\bibinfo{volume}{107}}, \bibinfo{pages}{210403}.

\bibitem[{\citenamefont{Gallego}
  \emph{et~al.}(2012{\natexlab{b}})\citenamefont{Gallego, W\"ulfliger, Ac\'in,
  and Navascues}}]{GWAN11b}
\bibinfo{author}{\bibnamefont{Gallego}, \bibfnamefont{R.}},
  \bibinfo{author}{\bibfnamefont{L.}~\bibnamefont{W\"ulfliger}},
  \bibinfo{author}{\bibfnamefont{A.}~\bibnamefont{Ac\'in}}, and
  \bibinfo{author}{\bibfnamefont{M.}~\bibnamefont{Navascues}},
  \bibinfo{year}{2012}{\natexlab{b}}, \bibinfo{journal}{Phys. Rev. Lett.}
  \textbf{\bibinfo{volume}{109}}, \bibinfo{pages}{070401}.

\bibitem[{\citenamefont{Gallego} \emph{et~al.}(2014)\citenamefont{Gallego,
  Wurflinger, Chaves, Acin, and Navascues}}]{GWCA+13}
\bibinfo{author}{\bibnamefont{Gallego}, \bibfnamefont{R.}},
  \bibinfo{author}{\bibfnamefont{L.}~\bibnamefont{Wurflinger}},
  \bibinfo{author}{\bibfnamefont{R.}~\bibnamefont{Chaves}},
  \bibinfo{author}{\bibfnamefont{A.}~\bibnamefont{Acin}}, and
  \bibinfo{author}{\bibfnamefont{M.}~\bibnamefont{Navascues}},
  \bibinfo{year}{2014}, \bibinfo{journal}{New J. Phys.} 
  \textbf{\bibinfo{volume}{16}}, \bibinfo{pages}{033037}.

\bibitem[{\citenamefont{Garbarino}(2010)}]{Garbarino10}
\bibinfo{author}{\bibnamefont{Garbarino}, \bibfnamefont{G.}},
  \bibinfo{year}{2010}, \bibinfo{journal}{Phys. Rev.~A}
  \textbf{\bibinfo{volume}{81}}, \bibinfo{pages}{032106}.

\bibitem[{\citenamefont{Garcia-Patron}
  \emph{et~al.}(2005)\citenamefont{Garcia-Patron, Fiurasek, and Cerf}}]{PFC05}
\bibinfo{author}{\bibnamefont{Garcia-Patron}, \bibfnamefont{R.}},
  \bibinfo{author}{\bibfnamefont{J.}~\bibnamefont{Fiurasek}}, and
  \bibinfo{author}{\bibfnamefont{N.}~\bibnamefont{Cerf}}, \bibinfo{year}{2005},
  \bibinfo{journal}{Phys. Rev.~A} \textbf{\bibinfo{volume}{71}},
  \bibinfo{pages}{022105}.

\bibitem[{\citenamefont{Garc\'{\i}a-Patron}
  \emph{et~al.}(2004)\citenamefont{Garc\'{\i}a-Patron, Fiurasek, Cerf, Wenger,
  Tualle-Brouri, and Grangier}}]{GFCW+04}
\bibinfo{author}{\bibnamefont{Garc\'{\i}a-Patron}, \bibfnamefont{R.}},
  \bibinfo{author}{\bibfnamefont{J.}~\bibnamefont{Fiurasek}},
  \bibinfo{author}{\bibfnamefont{N.}~\bibnamefont{Cerf}},
  \bibinfo{author}{\bibfnamefont{J.}~\bibnamefont{Wenger}},
  \bibinfo{author}{\bibfnamefont{R.}~\bibnamefont{Tualle-Brouri}}, and
  \bibinfo{author}{\bibfnamefont{P.}~\bibnamefont{Grangier}},
  \bibinfo{year}{2004}, \bibinfo{journal}{Phys. Rev. Lett.}
  \textbf{\bibinfo{volume}{93}}, \bibinfo{pages}{130409}.

\bibitem[{\citenamefont{Garg and Mermin}(1984)}]{GM84}
\bibinfo{author}{\bibnamefont{Garg}, \bibfnamefont{A.}}, and
  \bibinfo{author}{\bibfnamefont{N.~D.} \bibnamefont{Mermin}},
  \bibinfo{year}{1984}, \bibinfo{journal}{Foundations of Physics}
  \textbf{\bibinfo{volume}{14}}, \bibinfo{pages}{1}.

\bibitem[{\citenamefont{Genovese}(2005)}]{Genovese05}
\bibinfo{author}{\bibnamefont{Genovese}, \bibfnamefont{M.}},
  \bibinfo{year}{2005}, \bibinfo{journal}{Physics Reports}
  \textbf{\bibinfo{volume}{413}}, \bibinfo{pages}{419}.

\bibitem[{\citenamefont{Gerhardt} \emph{et~al.}(2011)\citenamefont{Gerhardt,
  Liu, Lamas-Linares, Skaar, Scarani, Makarov, and Kurtsiefer}}]{GLLS+11}
\bibinfo{author}{\bibnamefont{Gerhardt}, \bibfnamefont{I.}},
  \bibinfo{author}{\bibfnamefont{Q.}~\bibnamefont{Liu}},
  \bibinfo{author}{\bibfnamefont{A.}~\bibnamefont{Lamas-Linares}},
  \bibinfo{author}{\bibfnamefont{J.}~\bibnamefont{Skaar}},
  \bibinfo{author}{\bibfnamefont{V.}~\bibnamefont{Scarani}},
  \bibinfo{author}{\bibfnamefont{V.}~\bibnamefont{Makarov}}, and
  \bibinfo{author}{\bibfnamefont{C.}~\bibnamefont{Kurtsiefer}},
  \bibinfo{year}{2011}, \bibinfo{journal}{Phys. Rev. Lett.}
  \textbf{\bibinfo{volume}{107}}, \bibinfo{pages}{170404}.

\bibitem[{\citenamefont{Ghose} \emph{et~al.}(2009)\citenamefont{Ghose,
  Sinclair, Debnath, Rungta, and Stock}}]{GSDR+09}
\bibinfo{author}{\bibnamefont{Ghose}, \bibfnamefont{S.}},
  \bibinfo{author}{\bibfnamefont{N.}~\bibnamefont{Sinclair}},
  \bibinfo{author}{\bibfnamefont{S.}~\bibnamefont{Debnath}},
  \bibinfo{author}{\bibfnamefont{P.}~\bibnamefont{Rungta}}, and
  \bibinfo{author}{\bibfnamefont{R.}~\bibnamefont{Stock}},
  \bibinfo{year}{2009}, \bibinfo{journal}{Phys. Rev. Lett.}
  \textbf{\bibinfo{volume}{102}}, \bibinfo{pages}{250404}.

\bibitem[{\citenamefont{Gilchrist} \emph{et~al.}(1998)\citenamefont{Gilchrist,
  Deuar, and Reid}}]{GDR98}
\bibinfo{author}{\bibnamefont{Gilchrist}, \bibfnamefont{A.}},
  \bibinfo{author}{\bibfnamefont{P.}~\bibnamefont{Deuar}}, and
  \bibinfo{author}{\bibfnamefont{M.~D.} \bibnamefont{Reid}},
  \bibinfo{year}{1998}, \bibinfo{journal}{Phys. Rev. Lett.}
  \textbf{\bibinfo{volume}{80}}, \bibinfo{pages}{3169}.

\bibitem[{\citenamefont{Gill}(2003)}]{Gill03}
\bibinfo{author}{\bibnamefont{Gill}, \bibfnamefont{R.}}, \bibinfo{year}{2003},
  \bibinfo{journal}{Mathematical Statistics and Applications: Festschrift for
  Constance van Eeden. Eds: M. Moore, S. Froda and C. L\'eger. IMS Lecture
  Notes -- Monograph Series} \textbf{\bibinfo{volume}{42}},
  \bibinfo{pages}{133}.

\bibitem[{\citenamefont{Gill}(2012)}]{Gill12}
\bibinfo{author}{\bibnamefont{Gill}, \bibfnamefont{R.}}, \bibinfo{year}{2012},
  \bibinfo{journal}{arXiv:1207.5103} .

\bibitem[{\citenamefont{Gisin}(1991)}]{Gisin91}
\bibinfo{author}{\bibnamefont{Gisin}, \bibfnamefont{N.}}, \bibinfo{year}{1991},
  \bibinfo{journal}{Phys. Lett.~A} \textbf{\bibinfo{volume}{154}},
  \bibinfo{pages}{201}.

\bibitem[{\citenamefont{Gisin}(1996)}]{Gisin96}
\bibinfo{author}{\bibnamefont{Gisin}, \bibfnamefont{N.}}, \bibinfo{year}{1996},
  \bibinfo{journal}{Phys. Lett.~A} \textbf{\bibinfo{volume}{210}},
  \bibinfo{pages}{151}.

\bibitem[{\citenamefont{Gisin and Gisin}(1999)}]{GG99}
\bibinfo{author}{\bibnamefont{Gisin}, \bibfnamefont{N.}}, and
  \bibinfo{author}{\bibfnamefont{B.}~\bibnamefont{Gisin}},
  \bibinfo{year}{1999}, \bibinfo{journal}{Phys. Lett.~A}
  \textbf{\bibinfo{volume}{260}}, \bibinfo{pages}{323}.

\bibitem[{\citenamefont{Gisin and Gisin}(2002)}]{GG02}
\bibinfo{author}{\bibnamefont{Gisin}, \bibfnamefont{N.}}, and
  \bibinfo{author}{\bibfnamefont{B.}~\bibnamefont{Gisin}},
  \bibinfo{year}{2002}, \bibinfo{journal}{Phys. Lett.~A}
  \textbf{\bibinfo{volume}{297}}, \bibinfo{pages}{279}.

\bibitem[{\citenamefont{Gisin} \emph{et~al.}(2010)\citenamefont{Gisin, Pironio,
  and Sangouard}}]{GPS10}
\bibinfo{author}{\bibnamefont{Gisin}, \bibfnamefont{N.}},
  \bibinfo{author}{\bibfnamefont{S.}~\bibnamefont{Pironio}}, and
  \bibinfo{author}{\bibfnamefont{N.}~\bibnamefont{Sangouard}},
  \bibinfo{year}{2010}, \bibinfo{journal}{prl} \textbf{\bibinfo{volume}{105}},
  \bibinfo{pages}{070501}.

\bibitem[{\citenamefont{Giustina} \emph{et~al.}(2013)\citenamefont{Giustina,
  Mech, Ramelow, Wittmann, Kofler, Beyer, Lita, Calkins, Gerrits, Nam, Ursin,
  and Zeilinger}}]{GMR+12}
\bibinfo{author}{\bibnamefont{Giustina}, \bibfnamefont{M.}},
  \bibinfo{author}{\bibfnamefont{A.}~\bibnamefont{Mech}},
  \bibinfo{author}{\bibfnamefont{S.}~\bibnamefont{Ramelow}},
  \bibinfo{author}{\bibfnamefont{B.}~\bibnamefont{Wittmann}},
  \bibinfo{author}{\bibfnamefont{J.}~\bibnamefont{Kofler}},
  \bibinfo{author}{\bibfnamefont{J.}~\bibnamefont{Beyer}},
  \bibinfo{author}{\bibfnamefont{A.}~\bibnamefont{Lita}},
  \bibinfo{author}{\bibfnamefont{B.}~\bibnamefont{Calkins}},
  \bibinfo{author}{\bibfnamefont{T.}~\bibnamefont{Gerrits}},
  \bibinfo{author}{\bibfnamefont{S.~W.} \bibnamefont{Nam}},
  \bibinfo{author}{\bibfnamefont{R.}~\bibnamefont{Ursin}}, and
  \bibinfo{author}{\bibfnamefont{A.}~\bibnamefont{Zeilinger}},
  \bibinfo{year}{2013}, \bibinfo{journal}{Nature}
  \textbf{\bibinfo{volume}{497}}, \bibinfo{pages}{227}.

\bibitem[{\citenamefont{Goldstein} \emph{et~al.}(2011)\citenamefont{Goldstein,
  Norsen, Tausk, and Zanghi}}]{Goldstein2011}
\bibinfo{author}{\bibnamefont{Goldstein}, \bibfnamefont{S.}},
  \bibinfo{author}{\bibfnamefont{T.}~\bibnamefont{Norsen}},
  \bibinfo{author}{\bibfnamefont{D.~V.} \bibnamefont{Tausk}}, and
  \bibinfo{author}{\bibfnamefont{N.}~\bibnamefont{Zanghi}},
  \bibinfo{year}{2011}, \bibinfo{journal}{Scholarpedia}
  \textbf{\bibinfo{volume}{6}}(\bibinfo{number}{3}), \bibinfo{pages}{8378}.

\bibitem[{\citenamefont{Grandjean} \emph{et~al.}(2012)\citenamefont{Grandjean,
  Liang, Bancal, Brunner, and Gisin}}]{GLBB+12}
\bibinfo{author}{\bibnamefont{Grandjean}, \bibfnamefont{B.}},
  \bibinfo{author}{\bibfnamefont{Y.-C.} \bibnamefont{Liang}},
  \bibinfo{author}{\bibfnamefont{J.-D.} \bibnamefont{Bancal}},
  \bibinfo{author}{\bibfnamefont{N.}~\bibnamefont{Brunner}}, and
  \bibinfo{author}{\bibfnamefont{N.}~\bibnamefont{Gisin}},
  \bibinfo{year}{2012}, \bibinfo{journal}{Phys. Rev.~A}
  \textbf{\bibinfo{volume}{85}}, \bibinfo{pages}{052113}.

\bibitem[{\citenamefont{Grangier} \emph{et~al.}(1988)\citenamefont{Grangier,
  Potasek, and Yurke}}]{GPY88}
\bibinfo{author}{\bibnamefont{Grangier}, \bibfnamefont{P.}},
  \bibinfo{author}{\bibfnamefont{M.}~\bibnamefont{Potasek}}, and
  \bibinfo{author}{\bibfnamefont{B.}~\bibnamefont{Yurke}},
  \bibinfo{year}{1988}, \bibinfo{journal}{Phys. Rev.~A}
  \textbf{\bibinfo{volume}{38}}, \bibinfo{pages}{3132}.

\bibitem[{\citenamefont{Greenberger}
  \emph{et~al.}(1990)\citenamefont{Greenberger, Horne, Shimony, and
  Zeilinger}}]{GHSZ90}
\bibinfo{author}{\bibnamefont{Greenberger}, \bibfnamefont{D.}},
  \bibinfo{author}{\bibfnamefont{M.}~\bibnamefont{Horne}},
  \bibinfo{author}{\bibfnamefont{A.}~\bibnamefont{Shimony}}, and
  \bibinfo{author}{\bibfnamefont{A.}~\bibnamefont{Zeilinger}},
  \bibinfo{year}{1990}, \bibinfo{journal}{Am. J.~Phys.}
  \textbf{\bibinfo{volume}{58}}, \bibinfo{pages}{1131}.

\bibitem[{\citenamefont{Greenberger}
  \emph{et~al.}(1989)\citenamefont{Greenberger, Horne, and Zeilinger}}]{GHZ89}
\bibinfo{author}{\bibnamefont{Greenberger}, \bibfnamefont{D.~M.}},
  \bibinfo{author}{\bibfnamefont{M.}~\bibnamefont{Horne}}, and
  \bibinfo{author}{\bibfnamefont{A.}~\bibnamefont{Zeilinger}},
  \bibinfo{year}{1989}, \bibinfo{journal}{Bell's Theorem, Quantum Theory, and
  Conceptions of the Universe (Kluwer Academic, Dordrecht, 1989)} ,
  \bibinfo{pages}{73}.

\bibitem[{\citenamefont{Gr\"oblacher}
  \emph{et~al.}(2007)\citenamefont{Gr\"oblacher, Paterek, Kaltenbaek, Brukner,
  Zukowski, Aspelmeyer, and Zeilinger}}]{GPKB+07}
\bibinfo{author}{\bibnamefont{Gr\"oblacher}, \bibfnamefont{S.}},
  \bibinfo{author}{\bibfnamefont{T.}~\bibnamefont{Paterek}},
  \bibinfo{author}{\bibfnamefont{R.}~\bibnamefont{Kaltenbaek}},
  \bibinfo{author}{\bibfnamefont{C.}~\bibnamefont{Brukner}},
  \bibinfo{author}{\bibfnamefont{M.}~\bibnamefont{Zukowski}},
  \bibinfo{author}{\bibfnamefont{M.}~\bibnamefont{Aspelmeyer}}, and
  \bibinfo{author}{\bibfnamefont{A.}~\bibnamefont{Zeilinger}},
  \bibinfo{year}{2007}, \bibinfo{journal}{Nature}
  \textbf{\bibinfo{volume}{446}}, \bibinfo{pages}{871}.

\bibitem[{\citenamefont{Gruca} \emph{et~al.}(2010)\citenamefont{Gruca,
  Laskowski, Zukowski, Kiesel, Wieczorek, Schmid, and Weinfurter}}]{GLZK+10}
\bibinfo{author}{\bibnamefont{Gruca}, \bibfnamefont{J.}},
  \bibinfo{author}{\bibfnamefont{W.}~\bibnamefont{Laskowski}},
  \bibinfo{author}{\bibfnamefont{M.}~\bibnamefont{Zukowski}},
  \bibinfo{author}{\bibfnamefont{N.}~\bibnamefont{Kiesel}},
  \bibinfo{author}{\bibfnamefont{W.}~\bibnamefont{Wieczorek}},
  \bibinfo{author}{\bibfnamefont{C.}~\bibnamefont{Schmid}}, and
  \bibinfo{author}{\bibfnamefont{H.}~\bibnamefont{Weinfurter}},
  \bibinfo{year}{2010}, \bibinfo{journal}{Phys. Rev.~A}
  \textbf{\bibinfo{volume}{82}}, \bibinfo{pages}{012118}.

\bibitem[{\citenamefont{G\"uhne and Cabello}(2008)}]{GC08}
\bibinfo{author}{\bibnamefont{G\"uhne}, \bibfnamefont{O.}}, and
  \bibinfo{author}{\bibfnamefont{A.}~\bibnamefont{Cabello}},
  \bibinfo{year}{2008}, \bibinfo{journal}{Phys. Rev.~A}
  \textbf{\bibinfo{volume}{77}}, \bibinfo{pages}{032108}.

\bibitem[{\citenamefont{G\"uhne} \emph{et~al.}(2005)\citenamefont{G\"uhne,
  Toth, Hyllus, and Briegel}}]{GTHB05}
\bibinfo{author}{\bibnamefont{G\"uhne}, \bibfnamefont{O.}},
  \bibinfo{author}{\bibfnamefont{G.}~\bibnamefont{Toth}},
  \bibinfo{author}{\bibfnamefont{P.}~\bibnamefont{Hyllus}}, and
  \bibinfo{author}{\bibfnamefont{H.}~\bibnamefont{Briegel}},
  \bibinfo{year}{2005}, \bibinfo{journal}{Phys. Rev. Lett.}
  \textbf{\bibinfo{volume}{95}}, \bibinfo{pages}{120405}.

\bibitem[{\citenamefont{Hall}(2010)}]{Hall11b}
\bibinfo{author}{\bibnamefont{Hall}, \bibfnamefont{M.}}, \bibinfo{year}{2010},
  \bibinfo{journal}{Phys. Rev. Lett.} \textbf{\bibinfo{volume}{105}},
  \bibinfo{pages}{250404}.

\bibitem[{\citenamefont{Hall}(2011)}]{Hall11}
\bibinfo{author}{\bibnamefont{Hall}, \bibfnamefont{M.}}, \bibinfo{year}{2011},
  \bibinfo{journal}{Phys. Rev.~A} \textbf{\bibinfo{volume}{84}},
  \bibinfo{pages}{022102}.

\bibitem[{\citenamefont{H{\"a}nggi and Renner}(2010)}]{HR10}
\bibinfo{author}{\bibnamefont{H{\"a}nggi}, \bibfnamefont{E.}}, and
  \bibinfo{author}{\bibfnamefont{R.}~\bibnamefont{Renner}},
  \bibinfo{year}{2010}, \bibinfo{journal}{arXiv:1009.1833} .

\bibitem[{\citenamefont{H{\"a}nggi}
  \emph{et~al.}(2009)\citenamefont{H{\"a}nggi, Renner, and Wolf}}]{HRW09}
\bibinfo{author}{\bibnamefont{H{\"a}nggi}, \bibfnamefont{E.}},
  \bibinfo{author}{\bibfnamefont{R.}~\bibnamefont{Renner}}, and
  \bibinfo{author}{\bibfnamefont{S.}~\bibnamefont{Wolf}}, \bibinfo{year}{2009},
  \bibinfo{journal}{arXiv:0906.4760} .

\bibitem[{\citenamefont{H{\"a}nggi}
  \emph{et~al.}(2010)\citenamefont{H{\"a}nggi, Renner, and Wolf}}]{HRW10}
\bibinfo{author}{\bibnamefont{H{\"a}nggi}, \bibfnamefont{E.}},
  \bibinfo{author}{\bibfnamefont{R.}~\bibnamefont{Renner}}, and
  \bibinfo{author}{\bibfnamefont{S.}~\bibnamefont{Wolf}}, \bibinfo{year}{2010},
  \bibinfo{journal}{Proceedings of Advances in Cryptology - EUROCRYPT 2010} ,
  \bibinfo{pages}{216}.

\bibitem[{\citenamefont{Hardy}(1993)}]{Hardy93}
\bibinfo{author}{\bibnamefont{Hardy}, \bibfnamefont{L.}}, \bibinfo{year}{1993},
  \bibinfo{journal}{Phys. Rev. Lett.} \textbf{\bibinfo{volume}{71}},
  \bibinfo{pages}{1993}.

\bibitem[{\citenamefont{Hardy}(2001)}]{Hardy01}
\bibinfo{author}{\bibnamefont{Hardy}, \bibfnamefont{L.}}, \bibinfo{year}{2001},
  \bibinfo{journal}{arXiv:quant-ph/0101012} .

\bibitem[{\citenamefont{He} \emph{et~al.}(2009)\citenamefont{He, Cavalcanti,
  Reid, and Drummond}}]{HCRD09}
\bibinfo{author}{\bibnamefont{He}, \bibfnamefont{Q.}},
  \bibinfo{author}{\bibfnamefont{E.}~\bibnamefont{Cavalcanti}},
  \bibinfo{author}{\bibfnamefont{M.}~\bibnamefont{Reid}}, and
  \bibinfo{author}{\bibfnamefont{P.}~\bibnamefont{Drummond}},
  \bibinfo{year}{2009}, \bibinfo{journal}{Phys. Rev. Lett.}
  \textbf{\bibinfo{volume}{103}}, \bibinfo{pages}{180402}.

\bibitem[{\citenamefont{Hein} \emph{et~al.}(2004)\citenamefont{Hein, Eisert,
  and Briegel}}]{HEB04}
\bibinfo{author}{\bibnamefont{Hein}, \bibfnamefont{M.}},
  \bibinfo{author}{\bibfnamefont{J.}~\bibnamefont{Eisert}}, and
  \bibinfo{author}{\bibfnamefont{H.}~\bibnamefont{Briegel}},
  \bibinfo{year}{2004}, \bibinfo{journal}{Phys. Rev.~A}
  \textbf{\bibinfo{volume}{69}}, \bibinfo{pages}{062311}.

\bibitem[{\citenamefont{Herbert}(1975)}]{Herbert75}
\bibinfo{author}{\bibnamefont{Herbert}, \bibfnamefont{N.}},
  \bibinfo{year}{1975}, \bibinfo{journal}{Am. J. Phys.}
  \textbf{\bibinfo{volume}{43}}, \bibinfo{pages}{315}.

\bibitem[{\citenamefont{Hirsch} \emph{et~al.}(2013)\citenamefont{Hirsch,
  Quintino, Bowles, and Brunner}}]{HQBB13}
\bibinfo{author}{\bibnamefont{Hirsch}, \bibfnamefont{F.}},
  \bibinfo{author}{\bibfnamefont{M.}~\bibnamefont{Quintino}},
  \bibinfo{author}{\bibfnamefont{J.}~\bibnamefont{Bowles}}, and
  \bibinfo{author}{\bibfnamefont{N.}~\bibnamefont{Brunner}},
  \bibinfo{year}{2013}, \bibinfo{journal}{Phys. Rev. Lett.} 
   \textbf{\bibinfo{volume}{111}}, \bibinfo{pages}{160402}.

\bibitem[{\citenamefont{Ho} \emph{et~al.}(2013)\citenamefont{Ho, Bancal, and
  Scarani}}]{ho}
\bibinfo{author}{\bibnamefont{Ho}, \bibfnamefont{M.}},
  \bibinfo{author}{\bibfnamefont{J.-D.} \bibnamefont{Bancal}}, and
  \bibinfo{author}{\bibfnamefont{V.}~\bibnamefont{Scarani}},
  \bibinfo{year}{2013}, \bibinfo{journal}{Phys. Rev. A} 
   \textbf{\bibinfo{volume}{88}}, \bibinfo{pages}{052318}.
  

\bibitem[{\citenamefont{Hoban} \emph{et~al.}(2011)\citenamefont{Hoban,
  Campbell, Loukopoulos, and Browne}}]{HCLB11}
\bibinfo{author}{\bibnamefont{Hoban}, \bibfnamefont{M.}},
  \bibinfo{author}{\bibfnamefont{E.}~\bibnamefont{Campbell}},
  \bibinfo{author}{\bibfnamefont{K.}~\bibnamefont{Loukopoulos}}, and
  \bibinfo{author}{\bibfnamefont{D.}~\bibnamefont{Browne}},
  \bibinfo{year}{2011}, \bibinfo{journal}{New J.~Phys}
  \textbf{\bibinfo{volume}{13}}, \bibinfo{pages}{023014}.

\bibitem[{\citenamefont{Hofmann} \emph{et~al.}(2012)\citenamefont{Hofmann,
  Krug, Ortegel, GŽrard, Weber, Rosenfeld, and Weinfurter}}]{HKO+12}
\bibinfo{author}{\bibnamefont{Hofmann}, \bibfnamefont{J.}},
  \bibinfo{author}{\bibfnamefont{M.}~\bibnamefont{Krug}},
  \bibinfo{author}{\bibfnamefont{N.}~\bibnamefont{Ortegel}},
  \bibinfo{author}{\bibfnamefont{L.}~\bibnamefont{GŽrard}},
  \bibinfo{author}{\bibfnamefont{M.}~\bibnamefont{Weber}},
  \bibinfo{author}{\bibfnamefont{W.}~\bibnamefont{Rosenfeld}}, and
  \bibinfo{author}{\bibfnamefont{H.}~\bibnamefont{Weinfurter}},
  \bibinfo{year}{2012}, \bibinfo{journal}{Science}
  \textbf{\bibinfo{volume}{337}}, \bibinfo{pages}{72}.

\bibitem[{\citenamefont{Holenstein}(2007)}]{holenstein}
\bibinfo{author}{\bibnamefont{Holenstein}, \bibfnamefont{T.}},
  \bibinfo{year}{2007}, in \emph{\bibinfo{booktitle}{Proceedings of 39th
  Symposium on the Theory of Computing}} (\bibinfo{publisher}{ACM}).

\bibitem[{\citenamefont{Home and Selleri}(1991)}]{HS91}
\bibinfo{author}{\bibnamefont{Home}, \bibfnamefont{D.}}, and
  \bibinfo{author}{\bibfnamefont{F.}~\bibnamefont{Selleri}},
  \bibinfo{year}{1991}, \bibinfo{journal}{Rivista del Nuovo Cimento}
  \textbf{\bibinfo{volume}{14}}(\bibinfo{number}{9}), \bibinfo{pages}{1}.

\bibitem[{\citenamefont{Horodecki} \emph{et~al.}(1998)\citenamefont{Horodecki,
  Horodecki, and Horodecki}}]{HHH98}
\bibinfo{author}{\bibnamefont{Horodecki}, \bibfnamefont{M.}},
  \bibinfo{author}{\bibfnamefont{P.}~\bibnamefont{Horodecki}}, and
  \bibinfo{author}{\bibfnamefont{R.}~\bibnamefont{Horodecki}},
  \bibinfo{year}{1998}, \bibinfo{journal}{Phys. Rev. Lett.}
  \textbf{\bibinfo{volume}{80}}, \bibinfo{pages}{5239}.

\bibitem[{\citenamefont{Horodecki} \emph{et~al.}(1999)\citenamefont{Horodecki,
  Horodecki, and Horodecki}}]{HHH99}
\bibinfo{author}{\bibnamefont{Horodecki}, \bibfnamefont{M.}},
  \bibinfo{author}{\bibfnamefont{P.}~\bibnamefont{Horodecki}}, and
  \bibinfo{author}{\bibfnamefont{R.}~\bibnamefont{Horodecki}},
  \bibinfo{year}{1999}, \bibinfo{journal}{Phys. Rev.~A}
  \textbf{\bibinfo{volume}{60}}, \bibinfo{pages}{1888}.

\bibitem[{\citenamefont{Horodecki} \emph{et~al.}(1996)\citenamefont{Horodecki,
  Horodecki, and Horodecki}}]{HHH96}
\bibinfo{author}{\bibnamefont{Horodecki}, \bibfnamefont{R.}},
  \bibinfo{author}{\bibfnamefont{M.}~\bibnamefont{Horodecki}}, and
  \bibinfo{author}{\bibfnamefont{P.}~\bibnamefont{Horodecki}},
  \bibinfo{year}{1996}, \bibinfo{journal}{Phys. Lett.~A}
  \textbf{\bibinfo{volume}{222}}, \bibinfo{pages}{21}.

\bibitem[{\citenamefont{Horodecki} \emph{et~al.}(1995)\citenamefont{Horodecki,
  Horodecki, and Horodecki}}]{HHH95}
\bibinfo{author}{\bibnamefont{Horodecki}, \bibfnamefont{R.}},
  \bibinfo{author}{\bibfnamefont{P.}~\bibnamefont{Horodecki}}, and
  \bibinfo{author}{\bibfnamefont{M.}~\bibnamefont{Horodecki}},
  \bibinfo{year}{1995}, \bibinfo{journal}{Phys. Lett.~A}
  \textbf{\bibinfo{volume}{200}}, \bibinfo{pages}{340}.

\bibitem[{\citenamefont{Impagliazzo}
  \emph{et~al.}(1989)\citenamefont{Impagliazzo, Levin, and Luby}}]{impagliazzo}
\bibinfo{author}{\bibnamefont{Impagliazzo}, \bibfnamefont{R.}},
  \bibinfo{author}{\bibfnamefont{L.}~\bibnamefont{Levin}}, and
  \bibinfo{author}{\bibfnamefont{M.}~\bibnamefont{Luby}}, \bibinfo{year}{1989},
  in \emph{\bibinfo{booktitle}{Proceedings of the 21st Symposium on the Theory
  of Computing}} (\bibinfo{publisher}{ACM}), p.~\bibinfo{pages}{12}.

\bibitem[{\citenamefont{Ito and Vidick}(2012)}]{vidick:immunize2}
\bibinfo{author}{\bibnamefont{Ito}, \bibfnamefont{T.}}, and
  \bibinfo{author}{\bibfnamefont{T.}~\bibnamefont{Vidick}},
  \bibinfo{year}{2012}, in \emph{\bibinfo{booktitle}{Proceedings of the 53rd
  Annual Symposium on the Foundations of Computer Science}}
  (\bibinfo{publisher}{IEEE}), pp. \bibinfo{pages}{243--252}.

\bibitem[{\citenamefont{Jain} \emph{et~al.}(2010)\citenamefont{Jain, Ji,
  Upadhyay, and Watrous}}]{qipPspace}
\bibinfo{author}{\bibnamefont{Jain}, \bibfnamefont{R.}},
  \bibinfo{author}{\bibfnamefont{Z.}~\bibnamefont{Ji}},
  \bibinfo{author}{\bibfnamefont{S.}~\bibnamefont{Upadhyay}}, and
  \bibinfo{author}{\bibfnamefont{J.}~\bibnamefont{Watrous}},
  \bibinfo{year}{2010}, in \emph{\bibinfo{booktitle}{Proceedings of the 42nd
  symposium on Theory of computing}} (\bibinfo{publisher}{ACM}).

\bibitem[{\citenamefont{Janotta} \emph{et~al.}(2011)\citenamefont{Janotta,
  Gogolin, Barrett, and Brunner}}]{JGBB11}
\bibinfo{author}{\bibnamefont{Janotta}, \bibfnamefont{P.}},
  \bibinfo{author}{\bibfnamefont{C.}~\bibnamefont{Gogolin}},
  \bibinfo{author}{\bibfnamefont{J.}~\bibnamefont{Barrett}}, and
  \bibinfo{author}{\bibfnamefont{N.}~\bibnamefont{Brunner}},
  \bibinfo{year}{2011}, \bibinfo{journal}{New J.~Phys}
  \textbf{\bibinfo{volume}{13}}, \bibinfo{pages}{063024}.

\bibitem[{\citenamefont{Jennewein} \emph{et~al.}(2000)\citenamefont{Jennewein,
  Simon, Weihs, Weinfurter, and Zeilinger}}]{JSWW+00}
\bibinfo{author}{\bibnamefont{Jennewein}, \bibfnamefont{T.}},
  \bibinfo{author}{\bibfnamefont{C.}~\bibnamefont{Simon}},
  \bibinfo{author}{\bibfnamefont{G.}~\bibnamefont{Weihs}},
  \bibinfo{author}{\bibfnamefont{H.}~\bibnamefont{Weinfurter}}, and
  \bibinfo{author}{\bibfnamefont{A.}~\bibnamefont{Zeilinger}},
  \bibinfo{year}{2000}, \bibinfo{journal}{Phys. Rev. Lett.}
  \textbf{\bibinfo{volume}{84}}, \bibinfo{pages}{4729}.

\bibitem[{\citenamefont{Ji} \emph{et~al.}(2010)\citenamefont{Ji, Kim, Lee, and
  Nha}}]{JKLZ+10}
\bibinfo{author}{\bibnamefont{Ji}, \bibfnamefont{S.-W.}},
  \bibinfo{author}{\bibfnamefont{J.}~\bibnamefont{Kim}},
  \bibinfo{author}{\bibfnamefont{H.-W.} \bibnamefont{Lee}}, and
  \bibinfo{author}{\bibfnamefont{M.~Z.~H.} \bibnamefont{Nha}},
  \bibinfo{year}{2010}, \bibinfo{journal}{Phys. Rev. Lett.}
  \textbf{\bibinfo{volume}{105}}, \bibinfo{pages}{170404}.

\bibitem[{\citenamefont{Jones and Masanes}(2005)}]{JM05}
\bibinfo{author}{\bibnamefont{Jones}, \bibfnamefont{N.}}, and
  \bibinfo{author}{\bibfnamefont{L.}~\bibnamefont{Masanes}},
  \bibinfo{year}{2005}, \bibinfo{journal}{Phys. Rev.~A}
  \textbf{\bibinfo{volume}{72}}, \bibinfo{pages}{052312}.

\bibitem[{\citenamefont{Jones} \emph{et~al.}(2005)\citenamefont{Jones, Linden,
  and Massar}}]{JLM05}
\bibinfo{author}{\bibnamefont{Jones}, \bibfnamefont{N.~S.}},
  \bibinfo{author}{\bibfnamefont{N.}~\bibnamefont{Linden}}, and
  \bibinfo{author}{\bibfnamefont{S.}~\bibnamefont{Massar}},
  \bibinfo{year}{2005}, \bibinfo{journal}{Phys. Rev.~A}
  \textbf{\bibinfo{volume}{71}}, \bibinfo{pages}{042329}.

\bibitem[{\citenamefont{Joshi} \emph{et~al.}(2013)\citenamefont{Joshi, Grudka,
  Horodecki, Horodecki, Horodecki, and Horodecki}}]{JGHH+13}
\bibinfo{author}{\bibnamefont{Joshi}, \bibfnamefont{P.}},
  \bibinfo{author}{\bibfnamefont{A.}~\bibnamefont{Grudka}},
  \bibinfo{author}{\bibfnamefont{K.}~\bibnamefont{Horodecki}},
  \bibinfo{author}{\bibfnamefont{M.}~\bibnamefont{Horodecki}},
  \bibinfo{author}{\bibfnamefont{P.}~\bibnamefont{Horodecki}}, and
  \bibinfo{author}{\bibfnamefont{R.}~\bibnamefont{Horodecki}},
  \bibinfo{year}{2013}, \bibinfo{journal}{Quant. Inf. and Comp.}
  \textbf{\bibinfo{volume}{7/8}}, \bibinfo{pages}{567}.

\bibitem[{\citenamefont{Junge}
  \emph{et~al.}(2010{\natexlab{a}})\citenamefont{Junge, Navascues, Palazuelos,
  Perez-Garcia, Scholz, and Werner}}]{JNP+10}
\bibinfo{author}{\bibnamefont{Junge}, \bibfnamefont{M.}},
  \bibinfo{author}{\bibfnamefont{M.}~\bibnamefont{Navascues}},
  \bibinfo{author}{\bibfnamefont{C.}~\bibnamefont{Palazuelos}},
  \bibinfo{author}{\bibfnamefont{D.}~\bibnamefont{Perez-Garcia}},
  \bibinfo{author}{\bibfnamefont{V.}~\bibnamefont{Scholz}}, and
  \bibinfo{author}{\bibfnamefont{R.}~\bibnamefont{Werner}},
  \bibinfo{year}{2010}{\natexlab{a}}, \bibinfo{journal}{J. Math. Phys.}
  \textbf{\bibinfo{volume}{52}}, \bibinfo{pages}{012102}.

\bibitem[{\citenamefont{Junge and Palazuelos}(2011)}]{JP11}
\bibinfo{author}{\bibnamefont{Junge}, \bibfnamefont{M.}}, and
  \bibinfo{author}{\bibfnamefont{C.}~\bibnamefont{Palazuelos}},
  \bibinfo{year}{2011}, \bibinfo{journal}{Comm. Math. Phys.}
  \textbf{\bibinfo{volume}{306}}, \bibinfo{pages}{695}.

\bibitem[{\citenamefont{Junge}
  \emph{et~al.}(2010{\natexlab{b}})\citenamefont{Junge, Palazuelos,
  P\'erez-Garcia, Villanueva, and Wolf}}]{JPPV+10}
\bibinfo{author}{\bibnamefont{Junge}, \bibfnamefont{M.}},
  \bibinfo{author}{\bibfnamefont{C.}~\bibnamefont{Palazuelos}},
  \bibinfo{author}{\bibfnamefont{D.}~\bibnamefont{P\'erez-Garcia}},
  \bibinfo{author}{\bibfnamefont{I.}~\bibnamefont{Villanueva}}, and
  \bibinfo{author}{\bibfnamefont{M.}~\bibnamefont{Wolf}},
  \bibinfo{year}{2010}{\natexlab{b}}, \bibinfo{journal}{Phys. Rev. Lett.}
  \textbf{\bibinfo{volume}{104}}, \bibinfo{pages}{170405}.

\bibitem[{\citenamefont{Kaszlikowski}
  \emph{et~al.}(2000)\citenamefont{Kaszlikowski, GnAc\'inski, Zukowski,
  Miklaszewski, and Zeilinger}}]{KGZMZ01}
\bibinfo{author}{\bibnamefont{Kaszlikowski}, \bibfnamefont{D.}},
  \bibinfo{author}{\bibfnamefont{P.}~\bibnamefont{GnAc\'inski}},
  \bibinfo{author}{\bibfnamefont{M.}~\bibnamefont{Zukowski}},
  \bibinfo{author}{\bibfnamefont{W.}~\bibnamefont{Miklaszewski}}, and
  \bibinfo{author}{\bibfnamefont{A.}~\bibnamefont{Zeilinger}},
  \bibinfo{year}{2000}, \bibinfo{journal}{Phys. Rev. Lett.}
  \textbf{\bibinfo{volume}{85}}, \bibinfo{pages}{4418}.

\bibitem[{\citenamefont{Kaszlikowski and Zukowski}(1999)}]{KZ99}
\bibinfo{author}{\bibnamefont{Kaszlikowski}, \bibfnamefont{D.}}, and
  \bibinfo{author}{\bibfnamefont{M.}~\bibnamefont{Zukowski}},
  \bibinfo{year}{1999}, \bibinfo{journal}{Phys. Rev.~A}
  \textbf{\bibinfo{volume}{61}}, \bibinfo{pages}{022114}.

\bibitem[{\citenamefont{Kempe} \emph{et~al.}(2008)\citenamefont{Kempe,
  Kobayashi, Matsumoto, Toner, and Vidick}}]{vidick:immunize1}
\bibinfo{author}{\bibnamefont{Kempe}, \bibfnamefont{J.}},
  \bibinfo{author}{\bibfnamefont{H.}~\bibnamefont{Kobayashi}},
  \bibinfo{author}{\bibfnamefont{K.}~\bibnamefont{Matsumoto}},
  \bibinfo{author}{\bibfnamefont{B.}~\bibnamefont{Toner}}, and
  \bibinfo{author}{\bibfnamefont{T.}~\bibnamefont{Vidick}},
  \bibinfo{year}{2008}, in \emph{\bibinfo{booktitle}{Proceedings of 49th Annual
  Symposium on the Foundations of Computer Science}}
  (\bibinfo{publisher}{IEEE}), pp. \bibinfo{pages}{447--456}.

\bibitem[{\citenamefont{Kempe} \emph{et~al.}(2011)\citenamefont{Kempe,
  Kobayashi, Matsumoto, Toner, and Vidick}}]{KKV07}
\bibinfo{author}{\bibnamefont{Kempe}, \bibfnamefont{J.}},
  \bibinfo{author}{\bibfnamefont{H.}~\bibnamefont{Kobayashi}},
  \bibinfo{author}{\bibfnamefont{K.}~\bibnamefont{Matsumoto}},
  \bibinfo{author}{\bibfnamefont{B.}~\bibnamefont{Toner}}, and
  \bibinfo{author}{\bibfnamefont{T.}~\bibnamefont{Vidick}},
  \bibinfo{year}{2011}, \bibinfo{journal}{SIAM Journal on Computing}
  \textbf{\bibinfo{volume}{40}}(\bibinfo{number}{3}), \bibinfo{pages}{848}.

\bibitem[{\citenamefont{Kempe} \emph{et~al.}(2010)\citenamefont{Kempe, Regev,
  and Toner}}]{KRT07}
\bibinfo{author}{\bibnamefont{Kempe}, \bibfnamefont{J.}},
  \bibinfo{author}{\bibfnamefont{O.}~\bibnamefont{Regev}}, and
  \bibinfo{author}{\bibfnamefont{B.}~\bibnamefont{Toner}},
  \bibinfo{year}{2010}, \bibinfo{journal}{SIAM Journal on Computing}
  \textbf{\bibinfo{volume}{39}}(\bibinfo{number}{7}), \bibinfo{pages}{3207}.

\bibitem[{\citenamefont{Kempe and
  Vidick}(2011{\natexlab{a}})}]{kempeVidick:parallel}
\bibinfo{author}{\bibnamefont{Kempe}, \bibfnamefont{J.}}, and
  \bibinfo{author}{\bibfnamefont{T.}~\bibnamefont{Vidick}},
  \bibinfo{year}{2011}{\natexlab{a}}, in \emph{\bibinfo{booktitle}{Proceedings
  of 43rd Annual Symposium on the Theory of Computing}}
  (\bibinfo{publisher}{ACM}), pp. \bibinfo{pages}{353--362}.

\bibitem[{\citenamefont{Kempe and
  Vidick}(2011{\natexlab{b}})}]{vidick:parallel}
\bibinfo{author}{\bibnamefont{Kempe}, \bibfnamefont{J.}}, and
  \bibinfo{author}{\bibfnamefont{T.}~\bibnamefont{Vidick}},
  \bibinfo{year}{2011}{\natexlab{b}}, in \emph{\bibinfo{booktitle}{Proceedings
  of 43rd Symposium on Theory of Computing}} (\bibinfo{publisher}{ACM}).

\bibitem[{\citenamefont{Kent}(2005)}]{Kent05}
\bibinfo{author}{\bibnamefont{Kent}, \bibfnamefont{A.}}, \bibinfo{year}{2005},
  \bibinfo{journal}{Phys. Rev.~A} \textbf{\bibinfo{volume}{72}},
  \bibinfo{pages}{012107}.

\bibitem[{\citenamefont{Khalfin and Tsirelson}(1992)}]{KT92}
\bibinfo{author}{\bibnamefont{Khalfin}, \bibfnamefont{L.}}, and
  \bibinfo{author}{\bibfnamefont{B.}~\bibnamefont{Tsirelson}},
  \bibinfo{year}{1992}, \bibinfo{journal}{Found. Phys.}
  \textbf{\bibinfo{volume}{22}}(\bibinfo{number}{7}), \bibinfo{pages}{879}.

\bibitem[{\citenamefont{Khalfin and Tsirelson}(1985)}]{KT85}
\bibinfo{author}{\bibnamefont{Khalfin}, \bibfnamefont{L.~A.}}, and
  \bibinfo{author}{\bibfnamefont{B.~S.} \bibnamefont{Tsirelson}},
  \bibinfo{year}{1985}, in \emph{\bibinfo{booktitle}{Symposium on the
  Foundations of Modern Physics}}, edited by
  \bibinfo{editor}{\bibfnamefont{P.}~\bibnamefont{Lahti}} and
  \bibinfo{editor}{\bibfnamefont{P.}~\bibnamefont{Mittelstaedt}}.

\bibitem[{\citenamefont{Klobus} \emph{et~al.}(2012)\citenamefont{Klobus,
  Laskowski, Markiewicz, and Grudka}}]{KLMG12}
\bibinfo{author}{\bibnamefont{Klobus}, \bibfnamefont{W.}},
  \bibinfo{author}{\bibfnamefont{W.}~\bibnamefont{Laskowski}},
  \bibinfo{author}{\bibfnamefont{M.}~\bibnamefont{Markiewicz}}, and
  \bibinfo{author}{\bibfnamefont{A.}~\bibnamefont{Grudka}},
  \bibinfo{year}{2012}, \bibinfo{journal}{pra} \textbf{\bibinfo{volume}{86}},
  \bibinfo{pages}{020302(R)}.

\bibitem[{\citenamefont{Koh} \emph{et~al.}(2012)\citenamefont{Koh, Hall,
  Setiawan, Pope, Marletto, Kay, Scarani, and Ekert}}]{KHS+12}
\bibinfo{author}{\bibnamefont{Koh}, \bibfnamefont{D.~E.}},
  \bibinfo{author}{\bibfnamefont{M.~W.~H.} \bibnamefont{Hall}},
  \bibinfo{author}{\bibnamefont{Setiawan}},
  \bibinfo{author}{\bibfnamefont{J.~E.} \bibnamefont{Pope}},
  \bibinfo{author}{\bibfnamefont{C.}~\bibnamefont{Marletto}},
  \bibinfo{author}{\bibfnamefont{A.}~\bibnamefont{Kay}},
  \bibinfo{author}{\bibfnamefont{V.}~\bibnamefont{Scarani}}, and
  \bibinfo{author}{\bibfnamefont{A.}~\bibnamefont{Ekert}},
  \bibinfo{year}{2012}, \bibinfo{journal}{Phys. Rev. Lett.}
  \textbf{\bibinfo{volume}{109}}, \bibinfo{pages}{160404}.

\bibitem[{\citenamefont{K\"onig} \emph{et~al.}(2009)\citenamefont{K\"onig,
  Renner, and Schaffner}}]{krs:operational}
\bibinfo{author}{\bibnamefont{K\"onig}, \bibfnamefont{R.}},
  \bibinfo{author}{\bibfnamefont{R.}~\bibnamefont{Renner}}, and
  \bibinfo{author}{\bibfnamefont{C.}~\bibnamefont{Schaffner}},
  \bibinfo{year}{2009}, \bibinfo{journal}{IEEE Trans. Inf. Theory}
  \textbf{\bibinfo{volume}{55}}, \bibinfo{pages}{4674}.

\bibitem[{\citenamefont{Kwiat}(1997)}]{Kwiat97}
\bibinfo{author}{\bibnamefont{Kwiat}, \bibfnamefont{P.}}, \bibinfo{year}{1997},
  \bibinfo{journal}{J. Mod. Opt.} \textbf{\bibinfo{volume}{44}},
  \bibinfo{pages}{2173}.

\bibitem[{\citenamefont{Kwiat} \emph{et~al.}(2001)\citenamefont{Kwiat,
  Barraza-Lopez, Stefanov, and Gisin}}]{KBSG01}
\bibinfo{author}{\bibnamefont{Kwiat}, \bibfnamefont{P.~G.}},
  \bibinfo{author}{\bibfnamefont{S.}~\bibnamefont{Barraza-Lopez}},
  \bibinfo{author}{\bibfnamefont{A.}~\bibnamefont{Stefanov}}, and
  \bibinfo{author}{\bibfnamefont{N.}~\bibnamefont{Gisin}},
  \bibinfo{year}{2001}, \bibinfo{journal}{Nature}
  \textbf{\bibinfo{volume}{409}}, \bibinfo{pages}{1014}.

\bibitem[{\citenamefont{Landau}(1988)}]{Landau88}
\bibinfo{author}{\bibnamefont{Landau}, \bibfnamefont{L.}},
  \bibinfo{year}{1988}, \bibinfo{journal}{Foundations of Physics}
  \textbf{\bibinfo{volume}{18}}, \bibinfo{pages}{449}, ISSN
  \bibinfo{issn}{0015-9018}.

\bibitem[{\citenamefont{Landau}(1987)}]{Landau87}
\bibinfo{author}{\bibnamefont{Landau}, \bibfnamefont{L.~J.}},
  \bibinfo{year}{1987}, \bibinfo{journal}{Physics Letters A}
  \textbf{\bibinfo{volume}{123}}(\bibinfo{number}{3}), \bibinfo{pages}{115 },
  ISSN \bibinfo{issn}{0375-9601}.

\bibitem[{\citenamefont{Larsson and Gill}(2004)}]{LG04}
\bibinfo{author}{\bibnamefont{Larsson}, \bibfnamefont{J.-A.}}, and
  \bibinfo{author}{\bibfnamefont{R.}~\bibnamefont{Gill}}, \bibinfo{year}{2004},
  \bibinfo{journal}{Europhys. Lett.} \textbf{\bibinfo{volume}{67}},
  \bibinfo{pages}{707}.

\bibitem[{\citenamefont{Larsson} \emph{et~al.}(2013)\citenamefont{Larsson,
  Giustina, Kofler, Wittmann, Ursin, and Ramelow}}]{LGKW+13}
\bibinfo{author}{\bibnamefont{Larsson}, \bibfnamefont{J.-A.}},
  \bibinfo{author}{\bibfnamefont{M.}~\bibnamefont{Giustina}},
  \bibinfo{author}{\bibfnamefont{J.}~\bibnamefont{Kofler}},
  \bibinfo{author}{\bibfnamefont{B.}~\bibnamefont{Wittmann}},
  \bibinfo{author}{\bibfnamefont{R.}~\bibnamefont{Ursin}}, and
  \bibinfo{author}{\bibfnamefont{S.}~\bibnamefont{Ramelow}},
  \bibinfo{year}{2013}, \bibinfo{journal}{arXiv:1309.0712} .

\bibitem[{\citenamefont{Larsson and Semitecolos}(2001)}]{LS01}
\bibinfo{author}{\bibnamefont{Larsson}, \bibfnamefont{J.-A.}}, and
  \bibinfo{author}{\bibfnamefont{J.}~\bibnamefont{Semitecolos}},
  \bibinfo{year}{2001}, \bibinfo{journal}{Phys. Rev.~A}
  \textbf{\bibinfo{volume}{63}}, \bibinfo{pages}{022117}.

\bibitem[{\citenamefont{Laskowski} \emph{et~al.}(2010)\citenamefont{Laskowski,
  Paterek, Brukner, and Zukowski}}]{LPBZ10}
\bibinfo{author}{\bibnamefont{Laskowski}, \bibfnamefont{W.}},
  \bibinfo{author}{\bibfnamefont{T.}~\bibnamefont{Paterek}},
  \bibinfo{author}{\bibfnamefont{C.}~\bibnamefont{Brukner}}, and
  \bibinfo{author}{\bibfnamefont{M.}~\bibnamefont{Zukowski}},
  \bibinfo{year}{2010}, \bibinfo{journal}{Phys. Rev.~A}
  \textbf{\bibinfo{volume}{81}}, \bibinfo{pages}{042101}.

\bibitem[{\citenamefont{Laskowski} \emph{et~al.}(2004)\citenamefont{Laskowski,
  Paterek, Zukowski, and Brukner}}]{LPZB04}
\bibinfo{author}{\bibnamefont{Laskowski}, \bibfnamefont{W.}},
  \bibinfo{author}{\bibfnamefont{T.}~\bibnamefont{Paterek}},
  \bibinfo{author}{\bibfnamefont{M.}~\bibnamefont{Zukowski}}, and
  \bibinfo{author}{\bibfnamefont{C.}~\bibnamefont{Brukner}},
  \bibinfo{year}{2004}, \bibinfo{journal}{Phys. Rev. Lett.}
  \textbf{\bibinfo{volume}{93}}, \bibinfo{pages}{200401}.

\bibitem[{\citenamefont{Lasserre}(2001)}]{lasserre}
\bibinfo{author}{\bibnamefont{Lasserre}, \bibfnamefont{J.~B.}},
  \bibinfo{year}{2001}, \bibinfo{journal}{{SIAM} Journal of Optimization}
  \textbf{\bibinfo{volume}{11}}(\bibinfo{number}{3}), \bibinfo{pages}{796¿}.

\bibitem[{\citenamefont{Lavoie} \emph{et~al.}(2009)\citenamefont{Lavoie,
  Kaltenbaek, and Resch}}]{LKR09}
\bibinfo{author}{\bibnamefont{Lavoie}, \bibfnamefont{J.}},
  \bibinfo{author}{\bibfnamefont{R.}~\bibnamefont{Kaltenbaek}}, and
  \bibinfo{author}{\bibfnamefont{K.}~\bibnamefont{Resch}},
  \bibinfo{year}{2009}, \bibinfo{journal}{New J.~Phys}
  \textbf{\bibinfo{volume}{11}}, \bibinfo{pages}{073051}.

\bibitem[{\citenamefont{Leggett}(2003)}]{Leggett03}
\bibinfo{author}{\bibnamefont{Leggett}, \bibfnamefont{A.}},
  \bibinfo{year}{2003}, \bibinfo{journal}{Found. Phys.}
  \textbf{\bibinfo{volume}{33}}, \bibinfo{pages}{1469}.

\bibitem[{\citenamefont{Liang and Doherty}(2006)}]{LD06}
\bibinfo{author}{\bibnamefont{Liang}, \bibfnamefont{Y.-C.}}, and
  \bibinfo{author}{\bibfnamefont{A.}~\bibnamefont{Doherty}},
  \bibinfo{year}{2006}, \bibinfo{journal}{Phys. Rev.~A}
  \textbf{\bibinfo{volume}{73}}, \bibinfo{pages}{052116}.

\bibitem[{\citenamefont{Liang and Doherty}(2007)}]{LD07}
\bibinfo{author}{\bibnamefont{Liang}, \bibfnamefont{Y.-C.}}, and
  \bibinfo{author}{\bibfnamefont{A.~C.} \bibnamefont{Doherty}},
  \bibinfo{year}{2007}, \bibinfo{journal}{Phys. Rev.~A}
  \textbf{\bibinfo{volume}{75}}, \bibinfo{pages}{042103}.

\bibitem[{\citenamefont{Liang} \emph{et~al.}(2010)\citenamefont{Liang,
  Harrigan, Bartlett, and Rudolph}}]{LHBR10}
\bibinfo{author}{\bibnamefont{Liang}, \bibfnamefont{Y.-C.}},
  \bibinfo{author}{\bibfnamefont{N.}~\bibnamefont{Harrigan}},
  \bibinfo{author}{\bibfnamefont{S.}~\bibnamefont{Bartlett}}, and
  \bibinfo{author}{\bibfnamefont{T.~G.} \bibnamefont{Rudolph}},
  \bibinfo{year}{2010}, \bibinfo{journal}{Phys. Rev. Lett.}
  \textbf{\bibinfo{volume}{104}}, \bibinfo{pages}{050401}.

\bibitem[{\citenamefont{Liang} \emph{et~al.}(2012)\citenamefont{Liang, Masanes,
  and Rosset}}]{LMR12}
\bibinfo{author}{\bibnamefont{Liang}, \bibfnamefont{Y.-C.}},
  \bibinfo{author}{\bibfnamefont{L.}~\bibnamefont{Masanes}}, and
  \bibinfo{author}{\bibfnamefont{D.}~\bibnamefont{Rosset}},
  \bibinfo{year}{2012}, \bibinfo{journal}{Phys. Rev.~A}
  \textbf{\bibinfo{volume}{86}}, \bibinfo{pages}{052115}.

\bibitem[{\citenamefont{Liang} \emph{et~al.}(2011)\citenamefont{Liang,
  V\'ertesi, and Brunner}}]{LVB11}
\bibinfo{author}{\bibnamefont{Liang}, \bibfnamefont{Y.-C.}},
  \bibinfo{author}{\bibfnamefont{T.}~\bibnamefont{V\'ertesi}}, and
  \bibinfo{author}{\bibfnamefont{N.}~\bibnamefont{Brunner}},
  \bibinfo{year}{2011}, \bibinfo{journal}{Phys. Rev.~A}
  \textbf{\bibinfo{volume}{83}}, \bibinfo{pages}{052325}.

\bibitem[{\citenamefont{Linden} \emph{et~al.}(2007)\citenamefont{Linden,
  Popescu, Short, and Winter}}]{LPSW07}
\bibinfo{author}{\bibnamefont{Linden}, \bibfnamefont{N.}},
  \bibinfo{author}{\bibfnamefont{S.}~\bibnamefont{Popescu}},
  \bibinfo{author}{\bibfnamefont{A.~J.} \bibnamefont{Short}}, and
  \bibinfo{author}{\bibfnamefont{A.}~\bibnamefont{Winter}},
  \bibinfo{year}{2007}, \bibinfo{journal}{Phys. Rev. Lett.}
  \textbf{\bibinfo{volume}{99}}, \bibinfo{pages}{180502}.

\bibitem[{\citenamefont{Lo} \emph{et~al.}(2012)\citenamefont{Lo, Curty, and
  Qi}}]{LCQ12}
\bibinfo{author}{\bibnamefont{Lo}, \bibfnamefont{H.-K.}},
  \bibinfo{author}{\bibfnamefont{M.}~\bibnamefont{Curty}}, and
  \bibinfo{author}{\bibfnamefont{B.}~\bibnamefont{Qi}}, \bibinfo{year}{2012},
  \bibinfo{journal}{Phys. Rev. Lett.} \textbf{\bibinfo{volume}{108}},
  \bibinfo{pages}{130503}.

\bibitem[{\citenamefont{van Loock and Braunstein}(2001)}]{LB01}
\bibinfo{author}{\bibnamefont{van Loock}, \bibfnamefont{P.}}, and
  \bibinfo{author}{\bibfnamefont{S.~L.} \bibnamefont{Braunstein}},
  \bibinfo{year}{2001}, \bibinfo{journal}{Phys. Rev.~A}
  \textbf{\bibinfo{volume}{63}}, \bibinfo{pages}{022106}.

\bibitem[{\citenamefont{Lydersen} \emph{et~al.}(2010)\citenamefont{Lydersen,
  Wiechers, Wittmann, Elser, Skaar, and Makarov}}]{LWW+11}
\bibinfo{author}{\bibnamefont{Lydersen}, \bibfnamefont{L.}},
  \bibinfo{author}{\bibfnamefont{C.}~\bibnamefont{Wiechers}},
  \bibinfo{author}{\bibfnamefont{C.}~\bibnamefont{Wittmann}},
  \bibinfo{author}{\bibfnamefont{D.}~\bibnamefont{Elser}},
  \bibinfo{author}{\bibfnamefont{J.}~\bibnamefont{Skaar}}, and
  \bibinfo{author}{\bibfnamefont{V.}~\bibnamefont{Makarov}},
  \bibinfo{year}{2010}, \bibinfo{journal}{Nat. Photonics}
  \textbf{\bibinfo{volume}{4}}, \bibinfo{pages}{686}.

\bibitem[{\citenamefont{Ma and L\"utkenhaus}(2012)}]{ML12}
\bibinfo{author}{\bibnamefont{Ma}, \bibfnamefont{X.}}, and
  \bibinfo{author}{\bibfnamefont{N.}~\bibnamefont{L\"utkenhaus}},
  \bibinfo{year}{2012}, \bibinfo{journal}{Quant. Inf. and Comp.}
  \textbf{\bibinfo{volume}{12}}, \bibinfo{pages}{0203}.

\bibitem[{\citenamefont{Magniez} \emph{et~al.}(2006)\citenamefont{Magniez,
  Mayers, Mosca, , and Ollivier}}]{MMMO06}
\bibinfo{author}{\bibnamefont{Magniez}, \bibfnamefont{F.}},
  \bibinfo{author}{\bibfnamefont{D.}~\bibnamefont{Mayers}},
  \bibinfo{author}{\bibfnamefont{M.}~\bibnamefont{Mosca}}, , and
  \bibinfo{author}{\bibfnamefont{H.}~\bibnamefont{Ollivier}},
  \bibinfo{year}{2006}, \bibinfo{journal}{Proceedings of 33rd International
  Colloquium on Automata, Languages and Programming, volume 4051 of Lecture
  Notes in Computer Science (Springer Verlag)} \textbf{\bibinfo{volume}{4}},
  \bibinfo{pages}{72}.

\bibitem[{\citenamefont{Mair} \emph{et~al.}(2001)\citenamefont{Mair, Vaziri,
  Weihs, and Zeilinger}}]{MVWZ01}
\bibinfo{author}{\bibnamefont{Mair}, \bibfnamefont{A.}},
  \bibinfo{author}{\bibfnamefont{A.}~\bibnamefont{Vaziri}},
  \bibinfo{author}{\bibfnamefont{G.}~\bibnamefont{Weihs}}, and
  \bibinfo{author}{\bibfnamefont{A.}~\bibnamefont{Zeilinger}},
  \bibinfo{year}{2001}, \bibinfo{journal}{Nature}
  \textbf{\bibinfo{volume}{412}}, \bibinfo{pages}{313}.

\bibitem[{\citenamefont{Masanes}(2003)}]{Masanes03}
\bibinfo{author}{\bibnamefont{Masanes}, \bibfnamefont{L.}},
  \bibinfo{year}{2003}, \bibinfo{journal}{Quantum Information and Computation}
  \textbf{\bibinfo{volume}{3}}, \bibinfo{pages}{345}.

\bibitem[{\citenamefont{Masanes}(2006)}]{Masanes06}
\bibinfo{author}{\bibnamefont{Masanes}, \bibfnamefont{L.}},
  \bibinfo{year}{2006}, \bibinfo{journal}{Phys. Rev. Lett.}
  \textbf{\bibinfo{volume}{97}}, \bibinfo{pages}{050503}.

\bibitem[{\citenamefont{Masanes}(2009)}]{Masanes09}
\bibinfo{author}{\bibnamefont{Masanes}, \bibfnamefont{L.}},
  \bibinfo{year}{2009}, \bibinfo{journal}{Phys. Rev. Lett.}
  \textbf{\bibinfo{volume}{102}}, \bibinfo{pages}{140501}.

\bibitem[{\citenamefont{Masanes} \emph{et~al.}(2006)\citenamefont{Masanes,
  Ac\'in, and Gisin}}]{MAG06}
\bibinfo{author}{\bibnamefont{Masanes}, \bibfnamefont{L.}},
  \bibinfo{author}{\bibfnamefont{A.}~\bibnamefont{Ac\'in}}, and
  \bibinfo{author}{\bibfnamefont{N.}~\bibnamefont{Gisin}},
  \bibinfo{year}{2006}, \bibinfo{journal}{Phys. Rev.~A}
  \textbf{\bibinfo{volume}{73}}, \bibinfo{pages}{012112}.

\bibitem[{\citenamefont{Masanes} \emph{et~al.}(2008)\citenamefont{Masanes,
  Liang, and Doherty}}]{MLD08}
\bibinfo{author}{\bibnamefont{Masanes}, \bibfnamefont{L.}},
  \bibinfo{author}{\bibfnamefont{Y.-C.} \bibnamefont{Liang}}, and
  \bibinfo{author}{\bibfnamefont{A.}~\bibnamefont{Doherty}},
  \bibinfo{year}{2008}, \bibinfo{journal}{Phys. Rev. Lett.}
  \textbf{\bibinfo{volume}{100}}, \bibinfo{pages}{090403}.

\bibitem[{\citenamefont{Masanes} \emph{et~al.}(2011)\citenamefont{Masanes,
  Pironio, and Ac\'in}}]{MPA11}
\bibinfo{author}{\bibnamefont{Masanes}, \bibfnamefont{L.}},
  \bibinfo{author}{\bibfnamefont{S.}~\bibnamefont{Pironio}}, and
  \bibinfo{author}{\bibfnamefont{A.}~\bibnamefont{Ac\'in}},
  \bibinfo{year}{2011}, \bibinfo{journal}{Nat. Commun.}
  \textbf{\bibinfo{volume}{2}}, \bibinfo{pages}{238}.

\bibitem[{\citenamefont{Masanes} \emph{et~al.}(2009)\citenamefont{Masanes,
  Renner, Christandl, Winter, and Barrett}}]{MRC+09}
\bibinfo{author}{\bibnamefont{Masanes}, \bibfnamefont{L.}},
  \bibinfo{author}{\bibfnamefont{R.}~\bibnamefont{Renner}},
  \bibinfo{author}{\bibfnamefont{M.}~\bibnamefont{Christandl}},
  \bibinfo{author}{\bibfnamefont{A.}~\bibnamefont{Winter}}, and
  \bibinfo{author}{\bibfnamefont{J.}~\bibnamefont{Barrett}},
  \bibinfo{year}{2009}, \bibinfo{journal}{arXiv:quant-ph/0606049} .

\bibitem[{\citenamefont{Massar}(2002)}]{Massar02}
\bibinfo{author}{\bibnamefont{Massar}, \bibfnamefont{S.}},
  \bibinfo{year}{2002}, \bibinfo{journal}{Phys. Rev.~A}
  \textbf{\bibinfo{volume}{65}}, \bibinfo{pages}{032121}.

\bibitem[{\citenamefont{Massar} \emph{et~al.}(2001)\citenamefont{Massar, Bacon,
  Cerf, and Cleve}}]{MBCC01}
\bibinfo{author}{\bibnamefont{Massar}, \bibfnamefont{S.}},
  \bibinfo{author}{\bibfnamefont{D.}~\bibnamefont{Bacon}},
  \bibinfo{author}{\bibfnamefont{N.}~\bibnamefont{Cerf}}, and
  \bibinfo{author}{\bibfnamefont{R.}~\bibnamefont{Cleve}},
  \bibinfo{year}{2001}, \bibinfo{journal}{Phys. Rev.~A}
  \textbf{\bibinfo{volume}{63}}, \bibinfo{pages}{052305}.

\bibitem[{\citenamefont{Massar and Pironio}(2001)}]{MP01}
\bibinfo{author}{\bibnamefont{Massar}, \bibfnamefont{S.}}, and
  \bibinfo{author}{\bibfnamefont{S.}~\bibnamefont{Pironio}},
  \bibinfo{year}{2001}, \bibinfo{journal}{Phys. Rev.~A}
  \textbf{\bibinfo{volume}{64}}, \bibinfo{pages}{062108}.

\bibitem[{\citenamefont{Massar and Pironio}(2003)}]{MP03}
\bibinfo{author}{\bibnamefont{Massar}, \bibfnamefont{S.}}, and
  \bibinfo{author}{\bibfnamefont{S.}~\bibnamefont{Pironio}},
  \bibinfo{year}{2003}, \bibinfo{journal}{Phys. Rev.~A}
  \textbf{\bibinfo{volume}{68}}, \bibinfo{pages}{062109}.

\bibitem[{\citenamefont{Massar} \emph{et~al.}(2002)\citenamefont{Massar,
  Pironio, Roland, and Gisin}}]{MPRG02}
\bibinfo{author}{\bibnamefont{Massar}, \bibfnamefont{S.}},
  \bibinfo{author}{\bibfnamefont{S.}~\bibnamefont{Pironio}},
  \bibinfo{author}{\bibfnamefont{J.}~\bibnamefont{Roland}}, and
  \bibinfo{author}{\bibfnamefont{B.}~\bibnamefont{Gisin}},
  \bibinfo{year}{2002}, \bibinfo{journal}{Phys. Rev.~A}
  \textbf{\bibinfo{volume}{66}}, \bibinfo{pages}{052112}.

\bibitem[{\citenamefont{Matsukevich}
  \emph{et~al.}(2008)\citenamefont{Matsukevich, Maunz, Moehring, Olmschenk, and
  Monroe}}]{MMMO+08}
\bibinfo{author}{\bibnamefont{Matsukevich}, \bibfnamefont{D.}},
  \bibinfo{author}{\bibfnamefont{P.}~\bibnamefont{Maunz}},
  \bibinfo{author}{\bibfnamefont{D.}~\bibnamefont{Moehring}},
  \bibinfo{author}{\bibfnamefont{S.}~\bibnamefont{Olmschenk}}, and
  \bibinfo{author}{\bibfnamefont{C.}~\bibnamefont{Monroe}},
  \bibinfo{year}{2008}, \bibinfo{journal}{Phys. Rev. Lett.}
  \textbf{\bibinfo{volume}{100}}, \bibinfo{pages}{150404}.

\bibitem[{\citenamefont{Matsukevich}
  \emph{et~al.}(2005)\citenamefont{Matsukevich, Chaneli\`ere, Bhattacharya,
  Lan, Jenkins, Kennedy, and Kuzmich}}]{MCBL+05}
\bibinfo{author}{\bibnamefont{Matsukevich}, \bibfnamefont{D.~N.}},
  \bibinfo{author}{\bibfnamefont{T.}~\bibnamefont{Chaneli\`ere}},
  \bibinfo{author}{\bibfnamefont{M.}~\bibnamefont{Bhattacharya}},
  \bibinfo{author}{\bibfnamefont{S.-Y.} \bibnamefont{Lan}},
  \bibinfo{author}{\bibfnamefont{S.}~\bibnamefont{Jenkins}},
  \bibinfo{author}{\bibfnamefont{T.}~\bibnamefont{Kennedy}}, and
  \bibinfo{author}{\bibfnamefont{A.}~\bibnamefont{Kuzmich}},
  \bibinfo{year}{2005}, \bibinfo{journal}{Phys. Rev. Lett.}
  \textbf{\bibinfo{volume}{95}}, \bibinfo{pages}{040405}.

\bibitem[{\citenamefont{Mattar} \emph{et~al.}(2013)\citenamefont{Mattar, Brask,
  and Acin}}]{MBA13}
\bibinfo{author}{\bibnamefont{Mattar}, \bibfnamefont{A.}},
  \bibinfo{author}{\bibfnamefont{J.}~\bibnamefont{Brask}}, and
  \bibinfo{author}{\bibfnamefont{A.}~\bibnamefont{Acin}}, \bibinfo{year}{2013},
  \bibinfo{journal}{arXiv:1306.2571} .

\bibitem[{\citenamefont{Maudlin}(1992)}]{Maudlin92}
\bibinfo{author}{\bibnamefont{Maudlin}, \bibfnamefont{T.}},
  \bibinfo{year}{1992}, \bibinfo{journal}{Proceedings of the 1992 Meeting of
  the Philosophy of Science Association} \textbf{\bibinfo{volume}{1}},
  \bibinfo{pages}{404}.

\bibitem[{\citenamefont{Maudlin}(2002)}]{Maudlin02}
\bibinfo{author}{\bibnamefont{Maudlin}, \bibfnamefont{T.}},
  \bibinfo{year}{2002}, \emph{\bibinfo{title}{Quantum Non-Locality and
  Relativity: Metaphysical Intimations of Modern Physics}}
  (\bibinfo{publisher}{Blackwell}).

\bibitem[{\citenamefont{Mayers and Yao}(1998)}]{MY98}
\bibinfo{author}{\bibnamefont{Mayers}, \bibfnamefont{D.}}, and
  \bibinfo{author}{\bibfnamefont{A.}~\bibnamefont{Yao}}, \bibinfo{year}{1998},
  \bibinfo{journal}{FOCS} , \bibinfo{pages}{503}.

\bibitem[{\citenamefont{Mayers and Yao}(2004)}]{MY04}
\bibinfo{author}{\bibnamefont{Mayers}, \bibfnamefont{D.}}, and
  \bibinfo{author}{\bibfnamefont{A.}~\bibnamefont{Yao}}, \bibinfo{year}{2004},
  \bibinfo{journal}{Quant. Inf. and Comp.} \textbf{\bibinfo{volume}{4}},
  \bibinfo{pages}{273}.

\bibitem[{\citenamefont{McKague}(2010{\natexlab{a}})}]{McKague10b}
\bibinfo{author}{\bibnamefont{McKague}, \bibfnamefont{M.}},
  \bibinfo{year}{2010}{\natexlab{a}}, \bibinfo{journal}{arXiv:1010.1989} .

\bibitem[{\citenamefont{McKague}(2010{\natexlab{b}})}]{McKague10}
\bibinfo{author}{\bibnamefont{McKague}, \bibfnamefont{M.}},
  \bibinfo{year}{2010}{\natexlab{b}}, \emph{\bibinfo{title}{Quantum Information
  Processing with Adversarial Devices}}, Ph.D. thesis,
  \bibinfo{school}{Universtiy of Waterloo}, \bibinfo{note}{arXiv:1006.2352}.

\bibitem[{\citenamefont{McKague and Mosca}(2011)}]{MM11}
\bibinfo{author}{\bibnamefont{McKague}, \bibfnamefont{M.}}, and
  \bibinfo{author}{\bibfnamefont{M.}~\bibnamefont{Mosca}},
  \bibinfo{year}{2011}, in \emph{\bibinfo{booktitle}{Proceedings of 5th Conf.
  on Theory of Quantum Computation, Communication \& Cryptography (TQC2010),
  Lecture Notes in Computer Science}}, p. \bibinfo{pages}{113}.

\bibitem[{\citenamefont{McKague and Sheridan}(2012)}]{MS12}
\bibinfo{author}{\bibnamefont{McKague}, \bibfnamefont{M.}}, and
  \bibinfo{author}{\bibfnamefont{L.}~\bibnamefont{Sheridan}},
  \bibinfo{year}{2012}, \bibinfo{journal}{arXiv:1209.4696} .

\bibitem[{\citenamefont{McKague} \emph{et~al.}(2012)\citenamefont{McKague,
  Yang, and Scarani}}]{MYS12}
\bibinfo{author}{\bibnamefont{McKague}, \bibfnamefont{M.}},
  \bibinfo{author}{\bibfnamefont{T.}~\bibnamefont{Yang}}, and
  \bibinfo{author}{\bibfnamefont{V.}~\bibnamefont{Scarani}},
  \bibinfo{year}{2012}, \bibinfo{journal}{J.~Phys. A: Math. Theor.}
  \textbf{\bibinfo{volume}{45}}, \bibinfo{pages}{455304}.

\bibitem[{\citenamefont{Mermin}(1990{\natexlab{a}})}]{Mermin90}
\bibinfo{author}{\bibnamefont{Mermin}, \bibfnamefont{D.}},
  \bibinfo{year}{1990}{\natexlab{a}}, \bibinfo{journal}{Phys. Rev. Lett.}
  \textbf{\bibinfo{volume}{65}}, \bibinfo{pages}{1838}.

\bibitem[{\citenamefont{Mermin}(1986)}]{Mermin86}
\bibinfo{author}{\bibnamefont{Mermin}, \bibfnamefont{N.}},
  \bibinfo{year}{1986}, \bibinfo{journal}{in `Techniques and and Ideas in
  Quanturn Measurement Theory', edited by D.M. Greenberger (New York Academy of
  Science, New York)} , \bibinfo{pages}{422}.

\bibitem[{\citenamefont{Mermin}(1990{\natexlab{b}})}]{Mermin90pt}
\bibinfo{author}{\bibnamefont{Mermin}, \bibfnamefont{N.~D.}},
  \bibinfo{year}{1990}{\natexlab{b}}, \bibinfo{journal}{Phys. Today}
  \textbf{\bibinfo{volume}{43}}, \bibinfo{pages}{9}.

\bibitem[{\citenamefont{Mermin}(1993)}]{Mermin93}
\bibinfo{author}{\bibnamefont{Mermin}, \bibfnamefont{N.~D.}},
  \bibinfo{year}{1993}, \bibinfo{journal}{Rev. Mod. Phys.}
  \textbf{\bibinfo{volume}{65}}, \bibinfo{pages}{803}.

\bibitem[{\citenamefont{M\'ethot and Scarani}(2007)}]{MS07}
\bibinfo{author}{\bibnamefont{M\'ethot}, \bibfnamefont{A.}}, and
  \bibinfo{author}{\bibfnamefont{V.}~\bibnamefont{Scarani}},
  \bibinfo{year}{2007}, \bibinfo{journal}{Quant. Inf. and Comp.}
  \textbf{\bibinfo{volume}{7}}, \bibinfo{pages}{157}.

\bibitem[{\citenamefont{Miller and Shi}(2012)}]{MS12b}
\bibinfo{author}{\bibnamefont{Miller}, \bibfnamefont{C.}}, and
  \bibinfo{author}{\bibfnamefont{Y.}~\bibnamefont{Shi}}, \bibinfo{year}{2012},
  \bibinfo{journal}{arXiv:1207.1819} .

\bibitem[{\citenamefont{Mitchell} \emph{et~al.}(2004)\citenamefont{Mitchell,
  Popescu, and Roberts}}]{MPR04}
\bibinfo{author}{\bibnamefont{Mitchell}, \bibfnamefont{P.}},
  \bibinfo{author}{\bibfnamefont{S.}~\bibnamefont{Popescu}}, and
  \bibinfo{author}{\bibfnamefont{D.}~\bibnamefont{Roberts}},
  \bibinfo{year}{2004}, \bibinfo{journal}{Phys. Rev.~A}
  \textbf{\bibinfo{volume}{70}}, \bibinfo{pages}{060101}.

\bibitem[{\citenamefont{Moehring} \emph{et~al.}(2004)\citenamefont{Moehring,
  Madsen, Blinov, and Monroe}}]{MMBM04}
\bibinfo{author}{\bibnamefont{Moehring}, \bibfnamefont{D.}},
  \bibinfo{author}{\bibfnamefont{M.}~\bibnamefont{Madsen}},
  \bibinfo{author}{\bibfnamefont{B.}~\bibnamefont{Blinov}}, and
  \bibinfo{author}{\bibfnamefont{C.}~\bibnamefont{Monroe}},
  \bibinfo{year}{2004}, \bibinfo{journal}{Phys. Rev. Lett.}
  \textbf{\bibinfo{volume}{93}}, \bibinfo{pages}{090410}.

\bibitem[{\citenamefont{Moroder} \emph{et~al.}(2013)\citenamefont{Moroder,
  Bancal, Liang, Hofmann, and G\"uhne}}]{MBLH+13}
\bibinfo{author}{\bibnamefont{Moroder}, \bibfnamefont{T.}},
  \bibinfo{author}{\bibfnamefont{J.-D.} \bibnamefont{Bancal}},
  \bibinfo{author}{\bibfnamefont{Y.-C.} \bibnamefont{Liang}},
  \bibinfo{author}{\bibfnamefont{M.}~\bibnamefont{Hofmann}}, and
  \bibinfo{author}{\bibfnamefont{O.}~\bibnamefont{G\"uhne}},
  \bibinfo{year}{2013}, \bibinfo{journal}{Phys. Rev. Lett.}
  \textbf{\bibinfo{volume}{111}}, \bibinfo{pages}{030501}.

\bibitem[{\citenamefont{Munro}(1999)}]{Munro99}
\bibinfo{author}{\bibnamefont{Munro}, \bibfnamefont{W.~J.}},
  \bibinfo{year}{1999}, \bibinfo{journal}{Phys. Rev.~A}
  \textbf{\bibinfo{volume}{59}}, \bibinfo{pages}{4197}.

\bibitem[{\citenamefont{Nagata} \emph{et~al.}(2002)\citenamefont{Nagata,
  Koashi, and Imoto}}]{NKI02}
\bibinfo{author}{\bibnamefont{Nagata}, \bibfnamefont{K.}},
  \bibinfo{author}{\bibfnamefont{M.}~\bibnamefont{Koashi}}, and
  \bibinfo{author}{\bibfnamefont{N.}~\bibnamefont{Imoto}},
  \bibinfo{year}{2002}, \bibinfo{journal}{Phys. Rev. Lett.}
  \textbf{\bibinfo{volume}{89}}, \bibinfo{pages}{260401}.

\bibitem[{\citenamefont{Navascues} \emph{et~al.}(2011)\citenamefont{Navascues,
  Cooney, Perez-Garcia, and Villanueva}}]{NCP+11}
\bibinfo{author}{\bibnamefont{Navascues}, \bibfnamefont{M.}},
  \bibinfo{author}{\bibfnamefont{T.}~\bibnamefont{Cooney}},
  \bibinfo{author}{\bibfnamefont{D.}~\bibnamefont{Perez-Garcia}}, and
  \bibinfo{author}{\bibfnamefont{I.}~\bibnamefont{Villanueva}},
  \bibinfo{year}{2011}, \bibinfo{journal}{arXiv:1105.3373} .

\bibitem[{\citenamefont{Navascues} \emph{et~al.}(2007)\citenamefont{Navascues,
  Pironio, and Ac\'in}}]{NPA07}
\bibinfo{author}{\bibnamefont{Navascues}, \bibfnamefont{M.}},
  \bibinfo{author}{\bibfnamefont{S.}~\bibnamefont{Pironio}}, and
  \bibinfo{author}{\bibfnamefont{A.}~\bibnamefont{Ac\'in}},
  \bibinfo{year}{2007}, \bibinfo{journal}{Phys. Rev. Lett.}
  \textbf{\bibinfo{volume}{98}}, \bibinfo{pages}{010401}.

\bibitem[{\citenamefont{Navascues} \emph{et~al.}(2008)\citenamefont{Navascues,
  Pironio, and Ac\'in}}]{NPA08}
\bibinfo{author}{\bibnamefont{Navascues}, \bibfnamefont{M.}},
  \bibinfo{author}{\bibfnamefont{S.}~\bibnamefont{Pironio}}, and
  \bibinfo{author}{\bibfnamefont{A.}~\bibnamefont{Ac\'in}},
  \bibinfo{year}{2008}, \bibinfo{journal}{New J.~Phys}
  \textbf{\bibinfo{volume}{10}}, \bibinfo{pages}{073013}.

\bibitem[{\citenamefont{Navascu\'es and Vert\'esi}(2011)}]{NV11}
\bibinfo{author}{\bibnamefont{Navascu\'es}, \bibfnamefont{M.}}, and
  \bibinfo{author}{\bibfnamefont{T.}~\bibnamefont{Vert\'esi}},
  \bibinfo{year}{2011}, \bibinfo{journal}{Phys. Rev. Lett.}
  \textbf{\bibinfo{volume}{106}}, \bibinfo{pages}{060403}.

\bibitem[{\citenamefont{Navascu\'es and Wunderlich}(2009)}]{NW09}
\bibinfo{author}{\bibnamefont{Navascu\'es}, \bibfnamefont{M.}}, and
  \bibinfo{author}{\bibfnamefont{H.}~\bibnamefont{Wunderlich}},
  \bibinfo{year}{2009}, \bibinfo{journal}{Proc. Roy. Soc. Lond. A}
  \textbf{\bibinfo{volume}{881}}, \bibinfo{pages}{466}.

\bibitem[{\citenamefont{Nayak}(1999)}]{nayak99}
\bibinfo{author}{\bibnamefont{Nayak}, \bibfnamefont{A.}}, \bibinfo{year}{1999},
  in \emph{\bibinfo{booktitle}{Proceedings of the 40th Annual Symposium on
  Foundations of Computer Science}} (\bibinfo{publisher}{IEEE}), pp.
  \bibinfo{pages}{369¿--376}.

\bibitem[{\citenamefont{Neeley} \emph{et~al.}(2010)\citenamefont{Neeley,
  Bialczak, Lenander, Lucero, Mariantoni, OÕConnell, Sank, Wang, Weides,
  Wenner, Yin, Yamamoto} \emph{et~al.}}]{NBLL+10}
\bibinfo{author}{\bibnamefont{Neeley}, \bibfnamefont{M.}},
  \bibinfo{author}{\bibfnamefont{R.}~\bibnamefont{Bialczak}},
  \bibinfo{author}{\bibfnamefont{M.}~\bibnamefont{Lenander}},
  \bibinfo{author}{\bibfnamefont{E.}~\bibnamefont{Lucero}},
  \bibinfo{author}{\bibfnamefont{M.}~\bibnamefont{Mariantoni}},
  \bibinfo{author}{\bibfnamefont{A.}~\bibnamefont{OÕConnell}},
  \bibinfo{author}{\bibfnamefont{D.}~\bibnamefont{Sank}},
  \bibinfo{author}{\bibfnamefont{H.}~\bibnamefont{Wang}},
  \bibinfo{author}{\bibfnamefont{M.}~\bibnamefont{Weides}},
  \bibinfo{author}{\bibfnamefont{J.}~\bibnamefont{Wenner}},
  \bibinfo{author}{\bibfnamefont{Y.}~\bibnamefont{Yin}},
  \bibinfo{author}{\bibfnamefont{T.}~\bibnamefont{Yamamoto}}, \emph{et~al.},
  \bibinfo{year}{2010}, \bibinfo{journal}{Nature}
  \textbf{\bibinfo{volume}{467}}, \bibinfo{pages}{570}.

\bibitem[{\citenamefont{Nha and Carmichael}(2004)}]{NC04}
\bibinfo{author}{\bibnamefont{Nha}, \bibfnamefont{H.}}, and
  \bibinfo{author}{\bibfnamefont{H.}~\bibnamefont{Carmichael}},
  \bibinfo{year}{2004}, \bibinfo{journal}{Phys. Rev. Lett.}
  \textbf{\bibinfo{volume}{93}}, \bibinfo{pages}{020401}.

\bibitem[{\citenamefont{Norsen}(2007)}]{Norsen07}
\bibinfo{author}{\bibnamefont{Norsen}, \bibfnamefont{T.}},
  \bibinfo{year}{2007}, \bibinfo{journal}{arXiv:0707.0401} .

\bibitem[{\citenamefont{Norsen}(2009)}]{Norsen09}
\bibinfo{author}{\bibnamefont{Norsen}, \bibfnamefont{T.}},
  \bibinfo{year}{2009}, \bibinfo{journal}{Found. Phys.}
  \textbf{\bibinfo{volume}{39}}, \bibinfo{pages}{273}.

\bibitem[{\citenamefont{de~Oliveira~Oliveira}(2010)}]{embezzle:games}
\bibinfo{author}{\bibnamefont{de~Oliveira~Oliveira}, \bibfnamefont{M.}},
  \bibinfo{year}{2010}, \bibinfo{note}{ar{X}iv:1009.0771}.

\bibitem[{\citenamefont{Oppenheim and Wehner}(2010)}]{OW10}
\bibinfo{author}{\bibnamefont{Oppenheim}, \bibfnamefont{J.}}, and
  \bibinfo{author}{\bibfnamefont{S.}~\bibnamefont{Wehner}},
  \bibinfo{year}{2010}, \bibinfo{journal}{Science}
  \textbf{\bibinfo{volume}{330}}, \bibinfo{pages}{1072}.

\bibitem[{\citenamefont{Pal} \emph{et~al.}(2012)\citenamefont{Pal, Vertesi, and
  Brunner}}]{PVB12}
\bibinfo{author}{\bibnamefont{Pal}, \bibfnamefont{K.}},
  \bibinfo{author}{\bibfnamefont{T.}~\bibnamefont{Vertesi}}, and
  \bibinfo{author}{\bibfnamefont{N.}~\bibnamefont{Brunner}},
  \bibinfo{year}{2012}, \bibinfo{journal}{Phys. Rev.~A}
  \textbf{\bibinfo{volume}{86}}, \bibinfo{pages}{062111}.

\bibitem[{\citenamefont{Pal and Vertesi}(2009)}]{PV09}
\bibinfo{author}{\bibnamefont{Pal}, \bibfnamefont{K.~F.}}, and
  \bibinfo{author}{\bibfnamefont{T.}~\bibnamefont{Vertesi}},
  \bibinfo{year}{2009}, \bibinfo{journal}{Phys. Rev.~A}
  \textbf{\bibinfo{volume}{79}}, \bibinfo{pages}{022120}.

\bibitem[{\citenamefont{P\'{a}l and V\'{e}rtesi}(2010)}]{VP10}
\bibinfo{author}{\bibnamefont{P\'{a}l}, \bibfnamefont{K.~F.}}, and
  \bibinfo{author}{\bibfnamefont{T.}~\bibnamefont{V\'{e}rtesi}},
  \bibinfo{year}{2010}, \bibinfo{journal}{Phys. Rev.~A}
  \textbf{\bibinfo{volume}{82}}, \bibinfo{pages}{02216}.

\bibitem[{\citenamefont{Palazuelos}(2012{\natexlab{a}})}]{Palazuelos12}
\bibinfo{author}{\bibnamefont{Palazuelos}, \bibfnamefont{C.}},
  \bibinfo{year}{2012}{\natexlab{a}}, \bibinfo{journal}{Phys. Rev. Lett.}
  \textbf{\bibinfo{volume}{109}}, \bibinfo{pages}{190401}.

\bibitem[{\citenamefont{Palazuelos}(2012{\natexlab{b}})}]{Palazuelos12b}
\bibinfo{author}{\bibnamefont{Palazuelos}, \bibfnamefont{C.}},
  \bibinfo{year}{2012}{\natexlab{b}}, \bibinfo{journal}{arXiv:1206.3695} .

\bibitem[{\citenamefont{Palsson} \emph{et~al.}(2012)\citenamefont{Palsson,
  Wallman, Bennet, and Pryde}}]{PWBP12}
\bibinfo{author}{\bibnamefont{Palsson}, \bibfnamefont{M.}},
  \bibinfo{author}{\bibfnamefont{J.}~\bibnamefont{Wallman}},
  \bibinfo{author}{\bibfnamefont{A.}~\bibnamefont{Bennet}}, and
  \bibinfo{author}{\bibfnamefont{G.~J.} \bibnamefont{Pryde}},
  \bibinfo{year}{2012}, \bibinfo{journal}{Phys. Rev.~A}
  \textbf{\bibinfo{volume}{86}}, \bibinfo{pages}{032322}.

\bibitem[{\citenamefont{Pan} \emph{et~al.}(2000)\citenamefont{Pan, Bouwmeester,
  Daniell, Weinfurter, and Zeilinger}}]{PBDW+00}
\bibinfo{author}{\bibnamefont{Pan}, \bibfnamefont{J.-W.}},
  \bibinfo{author}{\bibfnamefont{D.}~\bibnamefont{Bouwmeester}},
  \bibinfo{author}{\bibfnamefont{M.}~\bibnamefont{Daniell}},
  \bibinfo{author}{\bibfnamefont{H.}~\bibnamefont{Weinfurter}}, and
  \bibinfo{author}{\bibfnamefont{A.}~\bibnamefont{Zeilinger}},
  \bibinfo{year}{2000}, \bibinfo{journal}{Nature}
  \textbf{\bibinfo{volume}{403}}, \bibinfo{pages}{515}.

\bibitem[{\citenamefont{Pan} \emph{et~al.}(2012)\citenamefont{Pan, Chen, Lu,
  Weinfurter, Zeilinger, and Zukowski}}]{PCLW+12}
\bibinfo{author}{\bibnamefont{Pan}, \bibfnamefont{J.-W.}},
  \bibinfo{author}{\bibfnamefont{Z.-B.} \bibnamefont{Chen}},
  \bibinfo{author}{\bibfnamefont{C.-Y.} \bibnamefont{Lu}},
  \bibinfo{author}{\bibfnamefont{H.}~\bibnamefont{Weinfurter}},
  \bibinfo{author}{\bibfnamefont{A.}~\bibnamefont{Zeilinger}}, and
  \bibinfo{author}{\bibfnamefont{M.}~\bibnamefont{Zukowski}},
  \bibinfo{year}{2012}, \bibinfo{journal}{Rev. Mod. Phys.}
  \textbf{\bibinfo{volume}{84}}, \bibinfo{pages}{777}.

\bibitem[{\citenamefont{Parrilo}(2003)}]{parillo:sdp}
\bibinfo{author}{\bibnamefont{Parrilo}, \bibfnamefont{P.}},
  \bibinfo{year}{2003}, \bibinfo{journal}{Mathematical Programming Ser. B}
  \textbf{\bibinfo{volume}{96}}(\bibinfo{number}{2}), \bibinfo{pages}{293}.

\bibitem[{\citenamefont{Pawlowski and Brukner}(2009)}]{PB09}
\bibinfo{author}{\bibnamefont{Pawlowski}, \bibfnamefont{M.}}, and
  \bibinfo{author}{\bibfnamefont{C.}~\bibnamefont{Brukner}},
  \bibinfo{year}{2009}, \bibinfo{journal}{Phys. Rev. Lett.}
  \textbf{\bibinfo{volume}{102}}, \bibinfo{pages}{030403}.

\bibitem[{\citenamefont{Pawlowski} \emph{et~al.}(2009)\citenamefont{Pawlowski,
  Paterek, Kaszlikowski, Scarani, Winter, and Zukowski}}]{PPKS+09}
\bibinfo{author}{\bibnamefont{Pawlowski}, \bibfnamefont{M.}},
  \bibinfo{author}{\bibfnamefont{T.}~\bibnamefont{Paterek}},
  \bibinfo{author}{\bibfnamefont{D.}~\bibnamefont{Kaszlikowski}},
  \bibinfo{author}{\bibfnamefont{V.}~\bibnamefont{Scarani}},
  \bibinfo{author}{\bibfnamefont{A.}~\bibnamefont{Winter}}, and
  \bibinfo{author}{\bibfnamefont{M.}~\bibnamefont{Zukowski}},
  \bibinfo{year}{2009}, \bibinfo{journal}{Nature}
  \textbf{\bibinfo{volume}{461}}, \bibinfo{pages}{1101}.

\bibitem[{\citenamefont{Pearle}(1970)}]{Pearle70}
\bibinfo{author}{\bibnamefont{Pearle}, \bibfnamefont{P.}},
  \bibinfo{year}{1970}, \bibinfo{journal}{Phys. Rev.~D}
  \textbf{\bibinfo{volume}{2}}, \bibinfo{pages}{1418}.

\bibitem[{\citenamefont{Peres}(1996)}]{Peres96}
\bibinfo{author}{\bibnamefont{Peres}, \bibfnamefont{A.}}, \bibinfo{year}{1996},
  \bibinfo{journal}{Phys. Rev.~A} \textbf{\bibinfo{volume}{54}},
  \bibinfo{pages}{2685}.

\bibitem[{\citenamefont{Peres}(1999)}]{Peres99}
\bibinfo{author}{\bibnamefont{Peres}, \bibfnamefont{A.}}, \bibinfo{year}{1999},
  \bibinfo{journal}{Foundations of Physics} \textbf{\bibinfo{volume}{29}},
  \bibinfo{pages}{589}.

\bibitem[{\citenamefont{P\'erez-Garcia}
  \emph{et~al.}(2008)\citenamefont{P\'erez-Garcia, Wolf, Palazuelos,
  Villanueva, and Junge}}]{PWPV+08}
\bibinfo{author}{\bibnamefont{P\'erez-Garcia}, \bibfnamefont{D.}},
  \bibinfo{author}{\bibfnamefont{M.}~\bibnamefont{Wolf}},
  \bibinfo{author}{\bibfnamefont{C.}~\bibnamefont{Palazuelos}},
  \bibinfo{author}{\bibfnamefont{I.}~\bibnamefont{Villanueva}}, and
  \bibinfo{author}{\bibfnamefont{M.}~\bibnamefont{Junge}},
  \bibinfo{year}{2008}, \bibinfo{journal}{Comm. Math. Phys.}
  \textbf{\bibinfo{volume}{279}}, \bibinfo{pages}{455}.

\bibitem[{\citenamefont{Pironio}(2003)}]{Pironio03}
\bibinfo{author}{\bibnamefont{Pironio}, \bibfnamefont{S.}},
  \bibinfo{year}{2003}, \bibinfo{journal}{Phys. Rev.~A}
  \textbf{\bibinfo{volume}{68}}, \bibinfo{pages}{062102}.

\bibitem[{\citenamefont{Pironio}(2004)}]{Pironio04}
\bibinfo{author}{\bibnamefont{Pironio}, \bibfnamefont{S.}},
  \bibinfo{year}{2004}, \emph{\bibinfo{title}{Aspects of Quantum
  Non-locality}}, Ph.D. thesis, \bibinfo{school}{Universit\'e Libre de
  Bruxelles}.

\bibitem[{\citenamefont{Pironio}(2005)}]{Pironio05}
\bibinfo{author}{\bibnamefont{Pironio}, \bibfnamefont{S.}},
  \bibinfo{year}{2005}, \bibinfo{journal}{J. Math. Phys.}
  \textbf{\bibinfo{volume}{46}}, \bibinfo{pages}{062112}.

\bibitem[{\citenamefont{Pironio} \emph{et~al.}(2009)\citenamefont{Pironio,
  Ac\'in, Brunner, Gisin, Massar, and Scarani}}]{PABG+09}
\bibinfo{author}{\bibnamefont{Pironio}, \bibfnamefont{S.}},
  \bibinfo{author}{\bibfnamefont{A.}~\bibnamefont{Ac\'in}},
  \bibinfo{author}{\bibfnamefont{N.}~\bibnamefont{Brunner}},
  \bibinfo{author}{\bibfnamefont{N.}~\bibnamefont{Gisin}},
  \bibinfo{author}{\bibfnamefont{S.}~\bibnamefont{Massar}}, and
  \bibinfo{author}{\bibfnamefont{V.}~\bibnamefont{Scarani}},
  \bibinfo{year}{2009}, \bibinfo{journal}{New J.~Phys}
  \textbf{\bibinfo{volume}{11}}, \bibinfo{pages}{045021}.

\bibitem[{\citenamefont{Pironio}
  \emph{et~al.}(2010{\natexlab{a}})\citenamefont{Pironio, Ac\'in, Massar, de~la
  Giroday, Matsukevich, Maunz, Olmschenk, Hayes, Luo, Manning, and
  Monroe}}]{PAMB+10}
\bibinfo{author}{\bibnamefont{Pironio}, \bibfnamefont{S.}},
  \bibinfo{author}{\bibfnamefont{A.}~\bibnamefont{Ac\'in}},
  \bibinfo{author}{\bibfnamefont{S.}~\bibnamefont{Massar}},
  \bibinfo{author}{\bibfnamefont{A.~B.} \bibnamefont{de~la Giroday}},
  \bibinfo{author}{\bibfnamefont{D.}~\bibnamefont{Matsukevich}},
  \bibinfo{author}{\bibfnamefont{P.}~\bibnamefont{Maunz}},
  \bibinfo{author}{\bibfnamefont{S.}~\bibnamefont{Olmschenk}},
  \bibinfo{author}{\bibfnamefont{D.}~\bibnamefont{Hayes}},
  \bibinfo{author}{\bibfnamefont{L.}~\bibnamefont{Luo}},
  \bibinfo{author}{\bibfnamefont{T.}~\bibnamefont{Manning}}, and
  \bibinfo{author}{\bibfnamefont{C.}~\bibnamefont{Monroe}},
  \bibinfo{year}{2010}{\natexlab{a}}, \bibinfo{journal}{Nature}
  \textbf{\bibinfo{volume}{464}}, \bibinfo{pages}{1021}.

\bibitem[{\citenamefont{Pironio} \emph{et~al.}(2011)\citenamefont{Pironio,
  Bancal, and Scarani}}]{PBS11}
\bibinfo{author}{\bibnamefont{Pironio}, \bibfnamefont{S.}},
  \bibinfo{author}{\bibfnamefont{J.-D.} \bibnamefont{Bancal}}, and
  \bibinfo{author}{\bibfnamefont{V.}~\bibnamefont{Scarani}},
  \bibinfo{year}{2011}, \bibinfo{journal}{J. Phys. A: Math. Theor.}
  \textbf{\bibinfo{volume}{44}}, \bibinfo{pages}{065303}.

\bibitem[{\citenamefont{Pironio} \emph{et~al.}(2013)\citenamefont{Pironio,
  Masanes, Leverrier, and Ac\'in}}]{PMA+12}
\bibinfo{author}{\bibnamefont{Pironio}, \bibfnamefont{S.}},
  \bibinfo{author}{\bibfnamefont{L.}~\bibnamefont{Masanes}},
  \bibinfo{author}{\bibfnamefont{A.}~\bibnamefont{Leverrier}}, and
  \bibinfo{author}{\bibfnamefont{A.}~\bibnamefont{Ac\'in}},
  \bibinfo{year}{2013}, \bibinfo{journal}{Phys. Rev. X}
  \textbf{\bibinfo{volume}{3}}, \bibinfo{pages}{031007}.

\bibitem[{\citenamefont{Pironio and Massar}(2013)}]{PM11}
\bibinfo{author}{\bibnamefont{Pironio}, \bibfnamefont{S.}}, and
  \bibinfo{author}{\bibfnamefont{S.}~\bibnamefont{Massar}},
  \bibinfo{year}{2013}, \bibinfo{journal}{Phys. Rev.~A}
  \textbf{\bibinfo{volume}{87}}, \bibinfo{pages}{012336}.

\bibitem[{\citenamefont{Pironio}
  \emph{et~al.}(2010{\natexlab{b}})\citenamefont{Pironio, Navascues, and
  Ac\'in}}]{PNA10}
\bibinfo{author}{\bibnamefont{Pironio}, \bibfnamefont{S.}},
  \bibinfo{author}{\bibfnamefont{M.}~\bibnamefont{Navascues}}, and
  \bibinfo{author}{\bibfnamefont{A.}~\bibnamefont{Ac\'in}},
  \bibinfo{year}{2010}{\natexlab{b}}, \bibinfo{journal}{SIAM J. Optim.}
  \textbf{\bibinfo{volume}{20}}, \bibinfo{pages}{2157}.

\bibitem[{\citenamefont{Pitalua-Garc'a}(2013)}]{Pitula13}
\bibinfo{author}{\bibnamefont{Pitalua-Garc'a}, \bibfnamefont{D.}},
  \bibinfo{year}{2013}, \bibinfo{journal}{Phys. Rev. Lett.}
  \textbf{\bibinfo{volume}{110}}, \bibinfo{pages}{210402}.

\bibitem[{\citenamefont{Pitkanen} \emph{et~al.}(2011)\citenamefont{Pitkanen,
  Ma, Wickert, van Loock, and L\"utkenhaus}}]{PMW+11}
\bibinfo{author}{\bibnamefont{Pitkanen}, \bibfnamefont{D.}},
  \bibinfo{author}{\bibfnamefont{X.}~\bibnamefont{Ma}},
  \bibinfo{author}{\bibfnamefont{R.}~\bibnamefont{Wickert}},
  \bibinfo{author}{\bibfnamefont{P.}~\bibnamefont{van Loock}}, and
  \bibinfo{author}{\bibfnamefont{N.}~\bibnamefont{L\"utkenhaus}},
  \bibinfo{year}{2011}, \bibinfo{journal}{Phys. Rev. A}
  \textbf{\bibinfo{volume}{84}}, \bibinfo{pages}{022325}.

\bibitem[{\citenamefont{Pitowsky}(1986)}]{Pitowsky86}
\bibinfo{author}{\bibnamefont{Pitowsky}, \bibfnamefont{I.}},
  \bibinfo{year}{1986}, \bibinfo{journal}{Journal of Mathematical Physics}
  \textbf{\bibinfo{volume}{27}}, \bibinfo{pages}{1556}.

\bibitem[{\citenamefont{Pitowsky}(1989)}]{Pitowsky89}
\bibinfo{author}{\bibnamefont{Pitowsky}, \bibfnamefont{I.}},
  \bibinfo{year}{1989}, \emph{\bibinfo{title}{Quantum {P}robability, {Q}uantum
  {L}ogic}}, volume \bibinfo{volume}{321} of \emph{\bibinfo{series}{Lecture
  Notes in Physics}} (\bibinfo{publisher}{Springer},
  \bibinfo{address}{Heidelberg}).

\bibitem[{\citenamefont{Pomarico}
  \emph{et~al.}(2011{\natexlab{a}})\citenamefont{Pomarico, Bancal, Sanguinetti,
  Rochdi, and Gisin}}]{PBSR+11}
\bibinfo{author}{\bibnamefont{Pomarico}, \bibfnamefont{E.}},
  \bibinfo{author}{\bibfnamefont{J.-D.} \bibnamefont{Bancal}},
  \bibinfo{author}{\bibfnamefont{B.}~\bibnamefont{Sanguinetti}},
  \bibinfo{author}{\bibfnamefont{A.}~\bibnamefont{Rochdi}}, and
  \bibinfo{author}{\bibfnamefont{N.}~\bibnamefont{Gisin}},
  \bibinfo{year}{2011}{\natexlab{a}}, \bibinfo{journal}{Phys. Rev.~A}
  \textbf{\bibinfo{volume}{83}}, \bibinfo{pages}{052104}.

\bibitem[{\citenamefont{Pomarico}
  \emph{et~al.}(2011{\natexlab{b}})\citenamefont{Pomarico, Sanguinetti,
  Sekatski, Zbinden, and Gisin}}]{PSSZ+11}
\bibinfo{author}{\bibnamefont{Pomarico}, \bibfnamefont{E.}},
  \bibinfo{author}{\bibfnamefont{B.}~\bibnamefont{Sanguinetti}},
  \bibinfo{author}{\bibfnamefont{P.}~\bibnamefont{Sekatski}},
  \bibinfo{author}{\bibfnamefont{H.}~\bibnamefont{Zbinden}}, and
  \bibinfo{author}{\bibfnamefont{N.}~\bibnamefont{Gisin}},
  \bibinfo{year}{2011}{\natexlab{b}}, \bibinfo{journal}{New J.~Phys}
  \textbf{\bibinfo{volume}{13}}, \bibinfo{pages}{063031}.

\bibitem[{\citenamefont{Popescu}(1994)}]{Popescu94}
\bibinfo{author}{\bibnamefont{Popescu}, \bibfnamefont{S.}},
  \bibinfo{year}{1994}, \bibinfo{journal}{Phys. Rev. Lett.}
  \textbf{\bibinfo{volume}{72}}, \bibinfo{pages}{797}.

\bibitem[{\citenamefont{Popescu}(1995)}]{popescu95}
\bibinfo{author}{\bibnamefont{Popescu}, \bibfnamefont{S.}},
  \bibinfo{year}{1995}, \bibinfo{journal}{Phys. Rev. Lett.}
  \textbf{\bibinfo{volume}{74}}, \bibinfo{pages}{2619}.

\bibitem[{\citenamefont{Popescu and Rohrlich}(1992)}]{PR92}
\bibinfo{author}{\bibnamefont{Popescu}, \bibfnamefont{S.}}, and
  \bibinfo{author}{\bibfnamefont{D.}~\bibnamefont{Rohrlich}},
  \bibinfo{year}{1992}, \bibinfo{journal}{Phys. Lett.~A}
  \textbf{\bibinfo{volume}{166}}, \bibinfo{pages}{293}.

\bibitem[{\citenamefont{Popescu and Rohrlich}(1994)}]{PR94}
\bibinfo{author}{\bibnamefont{Popescu}, \bibfnamefont{S.}}, and
  \bibinfo{author}{\bibfnamefont{D.}~\bibnamefont{Rohrlich}},
  \bibinfo{year}{1994}, \bibinfo{journal}{Found. Phys}
  \textbf{\bibinfo{volume}{24}}, \bibinfo{pages}{379}.

\bibitem[{\citenamefont{Portmann} \emph{et~al.}(2012)\citenamefont{Portmann,
  Branciard, and Gisin}}]{PBG12}
\bibinfo{author}{\bibnamefont{Portmann}, \bibfnamefont{S.}},
  \bibinfo{author}{\bibfnamefont{C.}~\bibnamefont{Branciard}}, and
  \bibinfo{author}{\bibfnamefont{N.}~\bibnamefont{Gisin}},
  \bibinfo{year}{2012}, \bibinfo{journal}{Phys. Rev.~A}
  \textbf{\bibinfo{volume}{86}}, \bibinfo{pages}{012104}.

\bibitem[{\citenamefont{Quintino} \emph{et~al.}(2012)\citenamefont{Quintino,
  Ara\'ujo, Cavalcanti, Santos, and Cunha}}]{QACS+11}
\bibinfo{author}{\bibnamefont{Quintino}, \bibfnamefont{M.~T.}},
  \bibinfo{author}{\bibfnamefont{M.}~\bibnamefont{Ara\'ujo}},
  \bibinfo{author}{\bibfnamefont{D.}~\bibnamefont{Cavalcanti}},
  \bibinfo{author}{\bibfnamefont{M.~F.} \bibnamefont{Santos}}, and
  \bibinfo{author}{\bibfnamefont{M.~T.} \bibnamefont{Cunha}},
  \bibinfo{year}{2012}, \bibinfo{journal}{J. Phys. A: Mathematical and
  Theoretical} \textbf{\bibinfo{volume}{45}}, \bibinfo{pages}{215308}.

\bibitem[{\citenamefont{Qutools}(2005)}]{qutools}
\bibinfo{author}{\bibnamefont{Qutools}}, \bibinfo{year}{2005},
  \bibinfo{note}{see http://www.qutools.com.}

\bibitem[{\citenamefont{Rabelo} \emph{et~al.}(2011)\citenamefont{Rabelo, Ho,
  Cavalcanti, Brunner, and Scarani}}]{RHCB+11}
\bibinfo{author}{\bibnamefont{Rabelo}, \bibfnamefont{R.}},
  \bibinfo{author}{\bibfnamefont{M.}~\bibnamefont{Ho}},
  \bibinfo{author}{\bibfnamefont{D.}~\bibnamefont{Cavalcanti}},
  \bibinfo{author}{\bibfnamefont{N.}~\bibnamefont{Brunner}}, and
  \bibinfo{author}{\bibfnamefont{V.}~\bibnamefont{Scarani}},
  \bibinfo{year}{2011}, \bibinfo{journal}{Phys. Rev. Lett.}
  \textbf{\bibinfo{volume}{107}}, \bibinfo{pages}{050502}.

\bibitem[{\citenamefont{Rabelo} \emph{et~al.}(2012)\citenamefont{Rabelo, Zhi,
  and Scarani}}]{RZS12}
\bibinfo{author}{\bibnamefont{Rabelo}, \bibfnamefont{R.}},
  \bibinfo{author}{\bibfnamefont{L.}~\bibnamefont{Zhi}}, and
  \bibinfo{author}{\bibfnamefont{V.}~\bibnamefont{Scarani}},
  \bibinfo{year}{2012}, \bibinfo{journal}{Phys. Rev. Lett.}
  \textbf{\bibinfo{volume}{109}}, \bibinfo{pages}{180401}.

\bibitem[{\citenamefont{Rao}(2008)}]{rao:parallel}
\bibinfo{author}{\bibnamefont{Rao}, \bibfnamefont{A.}}, \bibinfo{year}{2008},
  in \emph{\bibinfo{booktitle}{Proceedings of STOC}}
  (\bibinfo{publisher}{ACM}), pp. \bibinfo{pages}{1--10}.

\bibitem[{\citenamefont{Rarity and Tapster}(1990)}]{RT90}
\bibinfo{author}{\bibnamefont{Rarity}, \bibfnamefont{J.}}, and
  \bibinfo{author}{\bibfnamefont{P.}~\bibnamefont{Tapster}},
  \bibinfo{year}{1990}, \bibinfo{journal}{Phys. Rev. Lett.}
  \textbf{\bibinfo{volume}{64}}, \bibinfo{pages}{2495}.

\bibitem[{\citenamefont{Rastall}(1985)}]{Rastall85}
\bibinfo{author}{\bibnamefont{Rastall}, \bibfnamefont{P.}},
  \bibinfo{year}{1985}, \bibinfo{journal}{Found. Phys.}
  \textbf{\bibinfo{volume}{15}}, \bibinfo{pages}{963}.

\bibitem[{\citenamefont{Raz}(1998)}]{raz:parallel}
\bibinfo{author}{\bibnamefont{Raz}, \bibfnamefont{R.}}, \bibinfo{year}{1998},
  \bibinfo{journal}{SIAM Journal on Computing} \textbf{\bibinfo{volume}{27}},
  \bibinfo{pages}{763}.

\bibitem[{\citenamefont{Raz}(2008)}]{raz:counterexample}
\bibinfo{author}{\bibnamefont{Raz}, \bibfnamefont{R.}}, \bibinfo{year}{2008}.

\bibitem[{\citenamefont{Regev}(2012)}]{Regev12}
\bibinfo{author}{\bibnamefont{Regev}, \bibfnamefont{O.}}, \bibinfo{year}{2012},
  \bibinfo{journal}{Quant. Inf. and Comp.} \textbf{\bibinfo{volume}{12}},
  \bibinfo{pages}{9}.

\bibitem[{\citenamefont{Regev and Toner}(2007)}]{RT07}
\bibinfo{author}{\bibnamefont{Regev}, \bibfnamefont{O.}}, and
  \bibinfo{author}{\bibfnamefont{B.}~\bibnamefont{Toner}},
  \bibinfo{year}{2007}, \bibinfo{journal}{IEEE FOCS} .

\bibitem[{\citenamefont{Reichardt} \emph{et~al.}(2012)\citenamefont{Reichardt,
  Unger, and Vazirani}}]{RUV12l}
\bibinfo{author}{\bibnamefont{Reichardt}, \bibfnamefont{B.~W.}},
  \bibinfo{author}{\bibfnamefont{F.}~\bibnamefont{Unger}}, and
  \bibinfo{author}{\bibfnamefont{U.}~\bibnamefont{Vazirani}},
  \bibinfo{year}{2012}, \bibinfo{journal}{arXiv:1209.0448} .

\bibitem[{\citenamefont{Reichardt} \emph{et~al.}(2013)\citenamefont{Reichardt,
  Unger, and Vazirani}}]{RUV12}
\bibinfo{author}{\bibnamefont{Reichardt}, \bibfnamefont{B.~W.}},
  \bibinfo{author}{\bibfnamefont{F.}~\bibnamefont{Unger}}, and
  \bibinfo{author}{\bibfnamefont{U.}~\bibnamefont{Vazirani}},
  \bibinfo{year}{2013}, \bibinfo{journal}{Nature}
  \textbf{\bibinfo{volume}{496}}, \bibinfo{pages}{456}.

\bibitem[{\citenamefont{Reid}(1989)}]{Reid89}
\bibinfo{author}{\bibnamefont{Reid}, \bibfnamefont{M.}}, \bibinfo{year}{1989},
  \bibinfo{journal}{Phys. Rev.~A} \textbf{\bibinfo{volume}{40}},
  \bibinfo{pages}{913}.

\bibitem[{\citenamefont{Reid} \emph{et~al.}(2009)\citenamefont{Reid, Drummond,
  Bowen, Cavalcanti, Lam, Bachor, Andersen, and Leuchs}}]{RDBC+09}
\bibinfo{author}{\bibnamefont{Reid}, \bibfnamefont{M.~D.}},
  \bibinfo{author}{\bibfnamefont{P.~D.} \bibnamefont{Drummond}},
  \bibinfo{author}{\bibfnamefont{W.~P.} \bibnamefont{Bowen}},
  \bibinfo{author}{\bibfnamefont{E.~G.} \bibnamefont{Cavalcanti}},
  \bibinfo{author}{\bibfnamefont{P.~K.} \bibnamefont{Lam}},
  \bibinfo{author}{\bibfnamefont{H.~A.} \bibnamefont{Bachor}},
  \bibinfo{author}{\bibfnamefont{U.~L.} \bibnamefont{Andersen}}, and
  \bibinfo{author}{\bibfnamefont{G.}~\bibnamefont{Leuchs}},
  \bibinfo{year}{2009}, \bibinfo{journal}{Rev. Mod. Phys.}
  \textbf{\bibinfo{volume}{81}}, \bibinfo{pages}{1727}.

\bibitem[{\citenamefont{Renner}(2008)}]{renner:diss}
\bibinfo{author}{\bibnamefont{Renner}, \bibfnamefont{R.}},
  \bibinfo{year}{2008}, \bibinfo{journal}{International Journal of Quantum
  Information} \textbf{\bibinfo{volume}{6}}, \bibinfo{pages}{1}.

\bibitem[{\citenamefont{Renner and Wolf}(2004)}]{renatoStefan:KSset}
\bibinfo{author}{\bibnamefont{Renner}, \bibfnamefont{R.}}, and
  \bibinfo{author}{\bibfnamefont{S.}~\bibnamefont{Wolf}}, \bibinfo{year}{2004},
  in \emph{\bibinfo{booktitle}{Proceedings of International Symposium on
  Information Theory (ISIT)}} (\bibinfo{publisher}{IEEE}).

\bibitem[{\citenamefont{Romero} \emph{et~al.}(2013)\citenamefont{Romero,
  Giovannini, Tasca, Barnett, and Padgett}}]{RGTB+13}
\bibinfo{author}{\bibnamefont{Romero}, \bibfnamefont{J.}},
  \bibinfo{author}{\bibnamefont{Giovannini}},
  \bibinfo{author}{\bibfnamefont{D.}~\bibnamefont{Tasca}},
  \bibinfo{author}{\bibfnamefont{S.}~\bibnamefont{Barnett}}, and
  \bibinfo{author}{\bibnamefont{Padgett}}, \bibinfo{year}{2013},
  \bibinfo{journal}{New J.~Phys} \textbf{\bibinfo{volume}{15}},
  \bibinfo{pages}{083047}.

\bibitem[{\citenamefont{Rosenfeld} \emph{et~al.}(2009)\citenamefont{Rosenfeld,
  Weber, Volz, Henkel, Krug, Cabello, Zukowski, and Weinfurter}}]{RWVH09}
\bibinfo{author}{\bibnamefont{Rosenfeld}, \bibfnamefont{W.}},
  \bibinfo{author}{\bibfnamefont{M.}~\bibnamefont{Weber}},
  \bibinfo{author}{\bibfnamefont{J.}~\bibnamefont{Volz}},
  \bibinfo{author}{\bibfnamefont{F.}~\bibnamefont{Henkel}},
  \bibinfo{author}{\bibfnamefont{M.}~\bibnamefont{Krug}},
  \bibinfo{author}{\bibfnamefont{A.}~\bibnamefont{Cabello}},
  \bibinfo{author}{\bibfnamefont{M.}~\bibnamefont{Zukowski}}, and
  \bibinfo{author}{\bibfnamefont{H.}~\bibnamefont{Weinfurter}},
  \bibinfo{year}{2009}, \bibinfo{journal}{Adv. Sci. Lett.}
  \textbf{\bibinfo{volume}{2}}, \bibinfo{pages}{469}.

\bibitem[{\citenamefont{Rosset} \emph{et~al.}(2013)\citenamefont{Rosset,
  Branciard, Liang, and Gisin}}]{RBLG13}
\bibinfo{author}{\bibnamefont{Rosset}, \bibfnamefont{D.}},
  \bibinfo{author}{\bibfnamefont{C.}~\bibnamefont{Branciard}},
  \bibinfo{author}{\bibfnamefont{Y.-C.} \bibnamefont{Liang}}, and
  \bibinfo{author}{\bibfnamefont{N.}~\bibnamefont{Gisin}},
  \bibinfo{year}{2013}, \bibinfo{journal}{New J.~Phys}
  \textbf{\bibinfo{volume}{15}}, \bibinfo{pages}{053025}.

\bibitem[{\citenamefont{Rowe} \emph{et~al.}(2001)\citenamefont{Rowe,
  Kielpinski, Meyer, Sackett, Itano, Monroe, and Wineland}}]{RKMS+01}
\bibinfo{author}{\bibnamefont{Rowe}, \bibfnamefont{M.}},
  \bibinfo{author}{\bibfnamefont{D.}~\bibnamefont{Kielpinski}},
  \bibinfo{author}{\bibfnamefont{V.}~\bibnamefont{Meyer}},
  \bibinfo{author}{\bibfnamefont{C.}~\bibnamefont{Sackett}},
  \bibinfo{author}{\bibfnamefont{W.}~\bibnamefont{Itano}},
  \bibinfo{author}{\bibfnamefont{C.}~\bibnamefont{Monroe}}, and
  \bibinfo{author}{\bibfnamefont{D.}~\bibnamefont{Wineland}},
  \bibinfo{year}{2001}, \bibinfo{journal}{Nature}
  \textbf{\bibinfo{volume}{409}}, \bibinfo{pages}{791}.

\bibitem[{\citenamefont{Ryu} \emph{et~al.}(2013)\citenamefont{Ryu, Lee,
  Zukowski, and Lee}}]{RLZL13}
\bibinfo{author}{\bibnamefont{Ryu}, \bibfnamefont{J.}},
  \bibinfo{author}{\bibfnamefont{C.}~\bibnamefont{Lee}},
  \bibinfo{author}{\bibfnamefont{M.}~\bibnamefont{Zukowski}}, and
  \bibinfo{author}{\bibfnamefont{J.}~\bibnamefont{Lee}}, \bibinfo{year}{2013},
  \bibinfo{journal}{Phys. Rev. A} 
   \textbf{\bibinfo{volume}{88}}, \bibinfo{pages}{042101}.
  

\bibitem[{\citenamefont{Salart} \emph{et~al.}(2008)\citenamefont{Salart, Baas,
  Branciard, Gisin, and Zbinden}}]{SBBG+08}
\bibinfo{author}{\bibnamefont{Salart}, \bibfnamefont{D.}},
  \bibinfo{author}{\bibfnamefont{A.}~\bibnamefont{Baas}},
  \bibinfo{author}{\bibfnamefont{C.}~\bibnamefont{Branciard}},
  \bibinfo{author}{\bibfnamefont{N.}~\bibnamefont{Gisin}}, and
  \bibinfo{author}{\bibfnamefont{H.}~\bibnamefont{Zbinden}},
  \bibinfo{year}{2008}, \bibinfo{journal}{Nature}
  \textbf{\bibinfo{volume}{454}}, \bibinfo{pages}{861}.

\bibitem[{\citenamefont{Salles} \emph{et~al.}(2008)\citenamefont{Salles,
  Cavalcanti, and Ac\'in}}]{SCA08}
\bibinfo{author}{\bibnamefont{Salles}, \bibfnamefont{A.}},
  \bibinfo{author}{\bibfnamefont{D.}~\bibnamefont{Cavalcanti}}, and
  \bibinfo{author}{\bibfnamefont{A.}~\bibnamefont{Ac\'in}},
  \bibinfo{year}{2008}, \bibinfo{journal}{Phys. Rev. Lett.}
  \textbf{\bibinfo{volume}{101}}, \bibinfo{pages}{040404}.

\bibitem[{\citenamefont{Salles} \emph{et~al.}(2010)\citenamefont{Salles,
  Cavalcanti, Ac\'in, Perez-Garc\'ia, and Wolf}}]{SCA+10}
\bibinfo{author}{\bibnamefont{Salles}, \bibfnamefont{A.}},
  \bibinfo{author}{\bibfnamefont{D.}~\bibnamefont{Cavalcanti}},
  \bibinfo{author}{\bibfnamefont{A.}~\bibnamefont{Ac\'in}},
  \bibinfo{author}{\bibfnamefont{D.}~\bibnamefont{Perez-Garc\'ia}}, and
  \bibinfo{author}{\bibfnamefont{M.}~\bibnamefont{Wolf}}, \bibinfo{year}{2010},
  \bibinfo{journal}{Quant. Inf. and Comp.} \textbf{\bibinfo{volume}{10}},
  \bibinfo{pages}{0703}.

\bibitem[{\citenamefont{Sangouard} \emph{et~al.}(2011)\citenamefont{Sangouard,
  Bancal, Gisin, Rosenfeld, Sekatski, Weber, and Weinfurter}}]{SBGS+11}
\bibinfo{author}{\bibnamefont{Sangouard}, \bibfnamefont{N.}},
  \bibinfo{author}{\bibfnamefont{J.-D.} \bibnamefont{Bancal}},
  \bibinfo{author}{\bibfnamefont{N.}~\bibnamefont{Gisin}},
  \bibinfo{author}{\bibfnamefont{W.}~\bibnamefont{Rosenfeld}},
  \bibinfo{author}{\bibfnamefont{P.}~\bibnamefont{Sekatski}},
  \bibinfo{author}{\bibfnamefont{M.}~\bibnamefont{Weber}}, and
  \bibinfo{author}{\bibfnamefont{H.}~\bibnamefont{Weinfurter}},
  \bibinfo{year}{2011}, \bibinfo{journal}{Phys. Rev.~A}
  \textbf{\bibinfo{volume}{84}}, \bibinfo{pages}{052122}.

\bibitem[{\citenamefont{Sangouard} \emph{et~al.}(2013)\citenamefont{Sangouard,
  Bancal, MŸller, Ghosh, and Eschner}}]{SBMG+13}
\bibinfo{author}{\bibnamefont{Sangouard}, \bibfnamefont{N.}},
  \bibinfo{author}{\bibfnamefont{J.-D.} \bibnamefont{Bancal}},
  \bibinfo{author}{\bibfnamefont{P.}~\bibnamefont{MŸller}},
  \bibinfo{author}{\bibfnamefont{J.}~\bibnamefont{Ghosh}}, and
  \bibinfo{author}{\bibfnamefont{J.}~\bibnamefont{Eschner}},
  \bibinfo{year}{2013}, \bibinfo{journal}{New J.~Phys}
  \textbf{\bibinfo{volume}{15}}, \bibinfo{pages}{085004}.

\bibitem[{\citenamefont{Santos}(1992)}]{Santos92}
\bibinfo{author}{\bibnamefont{Santos}, \bibfnamefont{E.}},
  \bibinfo{year}{1992}, \bibinfo{journal}{Phys. Rev.~A}
  \textbf{\bibinfo{volume}{46}}, \bibinfo{pages}{3646}.

\bibitem[{\citenamefont{Saunders} \emph{et~al.}(2010)\citenamefont{Saunders,
  Jones, Wiseman, and Pryde}}]{SJWP10}
\bibinfo{author}{\bibnamefont{Saunders}, \bibfnamefont{D.~J.}},
  \bibinfo{author}{\bibfnamefont{S.~J.} \bibnamefont{Jones}},
  \bibinfo{author}{\bibfnamefont{H.~M.} \bibnamefont{Wiseman}}, and
  \bibinfo{author}{\bibfnamefont{G.~J.} \bibnamefont{Pryde}},
  \bibinfo{year}{2010}, \bibinfo{journal}{Nat. Phys.}
  \textbf{\bibinfo{volume}{6}}, \bibinfo{pages}{845}.

\bibitem[{\citenamefont{Scarani}(2008)}]{Scarani08}
\bibinfo{author}{\bibnamefont{Scarani}, \bibfnamefont{V.}},
  \bibinfo{year}{2008}, \bibinfo{journal}{Phys. Rev.~A}
  \textbf{\bibinfo{volume}{77}}, \bibinfo{pages}{042112}.

\bibitem[{\citenamefont{Scarani} \emph{et~al.}(2005)\citenamefont{Scarani,
  Ac\'in, Schenk, and Aspelmeyer}}]{SASA05}
\bibinfo{author}{\bibnamefont{Scarani}, \bibfnamefont{V.}},
  \bibinfo{author}{\bibfnamefont{A.}~\bibnamefont{Ac\'in}},
  \bibinfo{author}{\bibfnamefont{E.}~\bibnamefont{Schenk}}, and
  \bibinfo{author}{\bibfnamefont{M.}~\bibnamefont{Aspelmeyer}},
  \bibinfo{year}{2005}, \bibinfo{journal}{Phys. Rev.~A}
  \textbf{\bibinfo{volume}{71}}, \bibinfo{pages}{042325}.

\bibitem[{\citenamefont{Scarani and Gisin}(2001)}]{SG01}
\bibinfo{author}{\bibnamefont{Scarani}, \bibfnamefont{V.}}, and
  \bibinfo{author}{\bibfnamefont{N.}~\bibnamefont{Gisin}},
  \bibinfo{year}{2001}, \bibinfo{journal}{Phys. Rev. Lett.}
  \textbf{\bibinfo{volume}{87}}, \bibinfo{pages}{117901}.

\bibitem[{\citenamefont{Scarani and Gisin}(2005)}]{SG05}
\bibinfo{author}{\bibnamefont{Scarani}, \bibfnamefont{V.}}, and
  \bibinfo{author}{\bibfnamefont{N.}~\bibnamefont{Gisin}},
  \bibinfo{year}{2005}, \bibinfo{journal}{Braz. J. Phys.}
  \textbf{\bibinfo{volume}{35}}, \bibinfo{pages}{2A}.

\bibitem[{\citenamefont{Scarani} \emph{et~al.}(2006)\citenamefont{Scarani,
  Gisin, Brunner, Masanes, Pino, and Ac\'in}}]{SGB+06}
\bibinfo{author}{\bibnamefont{Scarani}, \bibfnamefont{V.}},
  \bibinfo{author}{\bibfnamefont{N.}~\bibnamefont{Gisin}},
  \bibinfo{author}{\bibfnamefont{N.}~\bibnamefont{Brunner}},
  \bibinfo{author}{\bibfnamefont{L.}~\bibnamefont{Masanes}},
  \bibinfo{author}{\bibfnamefont{S.}~\bibnamefont{Pino}}, and
  \bibinfo{author}{\bibfnamefont{A.}~\bibnamefont{Ac\'in}},
  \bibinfo{year}{2006}, \bibinfo{journal}{Phys. Rev.~A}
  \textbf{\bibinfo{volume}{74}}, \bibinfo{pages}{042339}.

\bibitem[{\citenamefont{Scheidl} \emph{et~al.}(2010)\citenamefont{Scheidl,
  Ursin, Kofler, Ramelow, Ma, Herbst, Ratschbacher, Fedrizzi, Langford,
  Jennewein, and Zeilinger}}]{SUK+10}
\bibinfo{author}{\bibnamefont{Scheidl}, \bibfnamefont{T.}},
  \bibinfo{author}{\bibfnamefont{R.}~\bibnamefont{Ursin}},
  \bibinfo{author}{\bibfnamefont{J.}~\bibnamefont{Kofler}},
  \bibinfo{author}{\bibfnamefont{S.}~\bibnamefont{Ramelow}},
  \bibinfo{author}{\bibfnamefont{X.-S.} \bibnamefont{Ma}},
  \bibinfo{author}{\bibfnamefont{T.}~\bibnamefont{Herbst}},
  \bibinfo{author}{\bibfnamefont{L.}~\bibnamefont{Ratschbacher}},
  \bibinfo{author}{\bibfnamefont{A.}~\bibnamefont{Fedrizzi}},
  \bibinfo{author}{\bibfnamefont{N.~K.} \bibnamefont{Langford}},
  \bibinfo{author}{\bibfnamefont{T.}~\bibnamefont{Jennewein}}, and
  \bibinfo{author}{\bibfnamefont{A.}~\bibnamefont{Zeilinger}},
  \bibinfo{year}{2010}, \bibinfo{journal}{Proc. Natl. Acad. Sci. USA}
  \textbf{\bibinfo{volume}{107}}, \bibinfo{pages}{19708}.

\bibitem[{\citenamefont{Schmidt} \emph{et~al.}(2008)\citenamefont{Schmidt,
  Kiesel, Laskowski, Wieczorek, Zukowski, and Weinfurter}}]{SKL+08}
\bibinfo{author}{\bibnamefont{Schmidt}, \bibfnamefont{C.}},
  \bibinfo{author}{\bibfnamefont{N.}~\bibnamefont{Kiesel}},
  \bibinfo{author}{\bibfnamefont{W.}~\bibnamefont{Laskowski}},
  \bibinfo{author}{\bibfnamefont{W.}~\bibnamefont{Wieczorek}},
  \bibinfo{author}{\bibfnamefont{M.}~\bibnamefont{Zukowski}}, and
  \bibinfo{author}{\bibfnamefont{H.}~\bibnamefont{Weinfurter}},
  \bibinfo{year}{2008}, \bibinfo{journal}{Phys. Rev. Lett.}
  \textbf{\bibinfo{volume}{100}}, \bibinfo{pages}{200407}.

\bibitem[{\citenamefont{Scholz and Werner}(2008)}]{SW08}
\bibinfo{author}{\bibnamefont{Scholz}, \bibfnamefont{V.~B.}}, and
  \bibinfo{author}{\bibfnamefont{R.~F.} \bibnamefont{Werner}},
  \bibinfo{year}{2008}, \bibinfo{journal}{arXiv:0812.4305} .

\bibitem[{\citenamefont{Schrijver}(1989)}]{Schrijver89}
\bibinfo{author}{\bibnamefont{Schrijver}, \bibfnamefont{A.}},
  \bibinfo{year}{1989}, \emph{\bibinfo{title}{Theory of linear and integer
  programming}}, Wiley-Interscience series in discrete mathematics
  (\bibinfo{publisher}{John Wiley \& Sons}).

\bibitem[{\citenamefont{Schr\"odinger}(1936)}]{Schrodinger36}
\bibinfo{author}{\bibnamefont{Schr\"odinger}, \bibfnamefont{E.}},
  \bibinfo{year}{1936}, \bibinfo{journal}{Proc. Camb. Phil. Soc.}
  \textbf{\bibinfo{volume}{32}}, \bibinfo{pages}{446}.

\bibitem[{\citenamefont{Seevinck and Svetlichny}(2002)}]{SS02}
\bibinfo{author}{\bibnamefont{Seevinck}, \bibfnamefont{M.}}, and
  \bibinfo{author}{\bibfnamefont{G.}~\bibnamefont{Svetlichny}},
  \bibinfo{year}{2002}, \bibinfo{journal}{Phys. Rev. Lett.}
  \textbf{\bibinfo{volume}{89}}, \bibinfo{pages}{060401}.

\bibitem[{\citenamefont{Sen(De)} \emph{et~al.}(2003)\citenamefont{Sen(De), Sen,
  Wiesniak, Kaszlikowski, and Zukowski}}]{SSWK+03}
\bibinfo{author}{\bibnamefont{Sen(De)}, \bibfnamefont{A.}},
  \bibinfo{author}{\bibfnamefont{U.}~\bibnamefont{Sen}},
  \bibinfo{author}{\bibfnamefont{M.}~\bibnamefont{Wiesniak}},
  \bibinfo{author}{\bibfnamefont{D.}~\bibnamefont{Kaszlikowski}}, and
  \bibinfo{author}{\bibfnamefont{M.}~\bibnamefont{Zukowski}},
  \bibinfo{year}{2003}, \bibinfo{journal}{Phys. Rev.~A}
  \textbf{\bibinfo{volume}{68}}, \bibinfo{pages}{062306}.

\bibitem[{\citenamefont{Sen(De)} \emph{et~al.}(2002)\citenamefont{Sen(De), Sen,
  and Zukowski}}]{SSZ02}
\bibinfo{author}{\bibnamefont{Sen(De)}, \bibfnamefont{A.}},
  \bibinfo{author}{\bibfnamefont{U.}~\bibnamefont{Sen}}, and
  \bibinfo{author}{\bibfnamefont{M.}~\bibnamefont{Zukowski}},
  \bibinfo{year}{2002}, \bibinfo{journal}{Phys. Rev.~A}
  \textbf{\bibinfo{volume}{66}}, \bibinfo{pages}{062318}.

\bibitem[{\citenamefont{Shadbolt} \emph{et~al.}(2012)\citenamefont{Shadbolt,
  V\'ertesi, Liang, Branciard, Brunner, and O'Brien}}]{SVLB+12}
\bibinfo{author}{\bibnamefont{Shadbolt}, \bibfnamefont{P.}},
  \bibinfo{author}{\bibfnamefont{T.}~\bibnamefont{V\'ertesi}},
  \bibinfo{author}{\bibfnamefont{Y.-C.} \bibnamefont{Liang}},
  \bibinfo{author}{\bibfnamefont{C.}~\bibnamefont{Branciard}},
  \bibinfo{author}{\bibfnamefont{N.}~\bibnamefont{Brunner}}, and
  \bibinfo{author}{\bibfnamefont{J.}~\bibnamefont{O'Brien}},
  \bibinfo{year}{2012}, \bibinfo{journal}{Scientific Reports}
  \textbf{\bibinfo{volume}{2}}, \bibinfo{pages}{470}.

\bibitem[{\citenamefont{Shimony} \emph{et~al.}(1976)\citenamefont{Shimony,
  Horne, and Clauser}}]{SHC76}
\bibinfo{author}{\bibnamefont{Shimony}, \bibfnamefont{A.}},
  \bibinfo{author}{\bibfnamefont{M.}~\bibnamefont{Horne}}, and
  \bibinfo{author}{\bibfnamefont{J.}~\bibnamefont{Clauser}},
  \bibinfo{year}{1976}, \bibinfo{journal}{Epistemological Letters}
  \textbf{\bibinfo{volume}{13}}, \bibinfo{pages}{17.1}.

\bibitem[{\citenamefont{Short}(2009)}]{Short09}
\bibinfo{author}{\bibnamefont{Short}, \bibfnamefont{A.}}, \bibinfo{year}{2009},
  \bibinfo{journal}{Phys. Rev. Lett.} \textbf{\bibinfo{volume}{102}},
  \bibinfo{pages}{180502}.

\bibitem[{\citenamefont{Silman} \emph{et~al.}(2011)\citenamefont{Silman,
  Chailloux, Aharon, Kerenidis, Pironio, and Massar}}]{SCA+11}
\bibinfo{author}{\bibnamefont{Silman}, \bibfnamefont{J.}},
  \bibinfo{author}{\bibfnamefont{A.}~\bibnamefont{Chailloux}},
  \bibinfo{author}{\bibfnamefont{N.}~\bibnamefont{Aharon}},
  \bibinfo{author}{\bibfnamefont{I.}~\bibnamefont{Kerenidis}},
  \bibinfo{author}{\bibfnamefont{S.}~\bibnamefont{Pironio}}, and
  \bibinfo{author}{\bibfnamefont{S.}~\bibnamefont{Massar}},
  \bibinfo{year}{2011}, \bibinfo{journal}{Phys. Rev. Lett.}
  \textbf{\bibinfo{volume}{106}}, \bibinfo{pages}{220501}.

\bibitem[{\citenamefont{Silman} \emph{et~al.}(2013)\citenamefont{Silman,
  Pironio, and Massar}}]{SPM12}
\bibinfo{author}{\bibnamefont{Silman}, \bibfnamefont{J.}},
  \bibinfo{author}{\bibfnamefont{S.}~\bibnamefont{Pironio}}, and
  \bibinfo{author}{\bibfnamefont{S.}~\bibnamefont{Massar}},
  \bibinfo{year}{2013}, \bibinfo{journal}{Phys. Rev. Lett.}
  \textbf{\bibinfo{volume}{110}}, \bibinfo{pages}{100504}.

\bibitem[{\citenamefont{Simon and Irvine}(2003)}]{SI03}
\bibinfo{author}{\bibnamefont{Simon}, \bibfnamefont{C.}}, and
  \bibinfo{author}{\bibfnamefont{W.}~\bibnamefont{Irvine}},
  \bibinfo{year}{2003}, \bibinfo{journal}{Phys. Rev. Lett.}
  \textbf{\bibinfo{volume}{91}}, \bibinfo{pages}{110405}.

\bibitem[{\citenamefont{Sliwa}(2003)}]{Sliwa03}
\bibinfo{author}{\bibnamefont{Sliwa}, \bibfnamefont{C.}}, \bibinfo{year}{2003},
  \bibinfo{journal}{Phys. Lett.~A} \textbf{\bibinfo{volume}{317}},
  \bibinfo{pages}{165}.

\bibitem[{\citenamefont{Smith} \emph{et~al.}(2012)\citenamefont{Smith,
  G.~Gillett, Branciard, Fedrizzi, Weinhold, Lita, Calkins, Gerrits, Wiseman,
  Nam, and White}}]{SGAB+1212}
\bibinfo{author}{\bibnamefont{Smith}, \bibfnamefont{D.}},
  \bibinfo{author}{\bibfnamefont{M.~d.~A.} \bibnamefont{G.~Gillett}},
  \bibinfo{author}{\bibfnamefont{C.}~\bibnamefont{Branciard}},
  \bibinfo{author}{\bibfnamefont{A.}~\bibnamefont{Fedrizzi}},
  \bibinfo{author}{\bibfnamefont{T.}~\bibnamefont{Weinhold}},
  \bibinfo{author}{\bibfnamefont{A.}~\bibnamefont{Lita}},
  \bibinfo{author}{\bibfnamefont{B.}~\bibnamefont{Calkins}},
  \bibinfo{author}{\bibfnamefont{T.}~\bibnamefont{Gerrits}},
  \bibinfo{author}{\bibfnamefont{H.}~\bibnamefont{Wiseman}},
  \bibinfo{author}{\bibfnamefont{S.}~\bibnamefont{Nam}}, and
  \bibinfo{author}{\bibfnamefont{A.}~\bibnamefont{White}},
  \bibinfo{year}{2012}, \bibinfo{journal}{Nat. Commun.}
  \textbf{\bibinfo{volume}{3}}, \bibinfo{pages}{625}.

\bibitem[{\citenamefont{Steeg and Wehner}(2009)}]{SW09}
\bibinfo{author}{\bibnamefont{Steeg}, \bibfnamefont{G.~V.}}, and
  \bibinfo{author}{\bibfnamefont{S.}~\bibnamefont{Wehner}},
  \bibinfo{year}{2009}, \bibinfo{journal}{Quant. Inf. and Comp.}
  \textbf{\bibinfo{volume}{9}}, \bibinfo{pages}{801}.

\bibitem[{\citenamefont{Stefanov} \emph{et~al.}(2002)\citenamefont{Stefanov,
  Zbinden, Gisin, and Suarez}}]{SZGS02}
\bibinfo{author}{\bibnamefont{Stefanov}, \bibfnamefont{A.}},
  \bibinfo{author}{\bibfnamefont{H.}~\bibnamefont{Zbinden}},
  \bibinfo{author}{\bibfnamefont{N.}~\bibnamefont{Gisin}}, and
  \bibinfo{author}{\bibfnamefont{A.}~\bibnamefont{Suarez}},
  \bibinfo{year}{2002}, \bibinfo{journal}{Phys. Rev. Lett.}
  \textbf{\bibinfo{volume}{88}}, \bibinfo{pages}{120404}.

\bibitem[{\citenamefont{Stefanov} \emph{et~al.}(2003)\citenamefont{Stefanov,
  Zbinden, Gisin, and Suarez}}]{SZGS03}
\bibinfo{author}{\bibnamefont{Stefanov}, \bibfnamefont{A.}},
  \bibinfo{author}{\bibfnamefont{H.}~\bibnamefont{Zbinden}},
  \bibinfo{author}{\bibfnamefont{N.}~\bibnamefont{Gisin}}, and
  \bibinfo{author}{\bibfnamefont{A.}~\bibnamefont{Suarez}},
  \bibinfo{year}{2003}, \bibinfo{journal}{Phys. Rev.~A}
  \textbf{\bibinfo{volume}{67}}, \bibinfo{pages}{042115}.

\bibitem[{\citenamefont{Steiner}(2000)}]{Steiner00}
\bibinfo{author}{\bibnamefont{Steiner}, \bibfnamefont{M.}},
  \bibinfo{year}{2000}, \bibinfo{journal}{Phys. Lett.~A}
  \textbf{\bibinfo{volume}{270}}, \bibinfo{pages}{239}.

\bibitem[{\citenamefont{Stobi\'nska}
  \emph{et~al.}(2007)\citenamefont{Stobi\'nska, Jeong, , and Ralph}}]{SJR07}
\bibinfo{author}{\bibnamefont{Stobi\'nska}, \bibfnamefont{M.}},
  \bibinfo{author}{\bibfnamefont{H.}~\bibnamefont{Jeong}}, , and
  \bibinfo{author}{\bibfnamefont{T.}~\bibnamefont{Ralph}},
  \bibinfo{year}{2007}, \bibinfo{journal}{Phys. Rev.~A}
  \textbf{\bibinfo{volume}{75}}, \bibinfo{pages}{052105}.

\bibitem[{\citenamefont{Suarez and Scarani}(1997)}]{SS97}
\bibinfo{author}{\bibnamefont{Suarez}, \bibfnamefont{A.}}, and
  \bibinfo{author}{\bibfnamefont{V.}~\bibnamefont{Scarani}},
  \bibinfo{year}{1997}, \bibinfo{journal}{Phys. Lett.~A}
  \textbf{\bibinfo{volume}{232}}, \bibinfo{pages}{9}.

\bibitem[{\citenamefont{Svetlichny}(1987)}]{Svetlichny87}
\bibinfo{author}{\bibnamefont{Svetlichny}, \bibfnamefont{G.}},
  \bibinfo{year}{1987}, \bibinfo{journal}{Phys. Rev.~D}
  \textbf{\bibinfo{volume}{35}}, \bibinfo{pages}{3066}.

\bibitem[{\citenamefont{Ta-Shma}(2009)}]{Tashma09}
\bibinfo{author}{\bibnamefont{Ta-Shma}, \bibfnamefont{A.}},
  \bibinfo{year}{2009}, \bibinfo{journal}{In Proceedings of 41st ACM STOC} ,
  \bibinfo{pages}{401}.

\bibitem[{\citenamefont{Tan} \emph{et~al.}(1991)\citenamefont{Tan, Walls, and
  Collett}}]{TWC91}
\bibinfo{author}{\bibnamefont{Tan}, \bibfnamefont{S.}},
  \bibinfo{author}{\bibfnamefont{D.}~\bibnamefont{Walls}}, and
  \bibinfo{author}{\bibfnamefont{M.}~\bibnamefont{Collett}},
  \bibinfo{year}{1991}, \bibinfo{journal}{Phys. Rev. Lett.}
  \textbf{\bibinfo{volume}{66}}, \bibinfo{pages}{252}.

\bibitem[{\citenamefont{Tasca} \emph{et~al.}(2009)\citenamefont{Tasca, Walborn,
  Toscano, and Ribeiro}}]{TWTS09}
\bibinfo{author}{\bibnamefont{Tasca}, \bibfnamefont{D.}},
  \bibinfo{author}{\bibfnamefont{S.}~\bibnamefont{Walborn}},
  \bibinfo{author}{\bibfnamefont{F.}~\bibnamefont{Toscano}}, and
  \bibinfo{author}{\bibfnamefont{P.~S.} \bibnamefont{Ribeiro}},
  \bibinfo{year}{2009}, \bibinfo{journal}{Phys. Rev.~A}
  \textbf{\bibinfo{volume}{80}}, \bibinfo{pages}{030101(R)}.

\bibitem[{\citenamefont{Teo} \emph{et~al.}(2013)\citenamefont{Teo, Ara\'ujo,
  Quintino, Min\'ar, Cavalcanti, Scarani, Cunha, and Santos}}]{TAQM+12}
\bibinfo{author}{\bibnamefont{Teo}, \bibfnamefont{C.}},
  \bibinfo{author}{\bibfnamefont{M.}~\bibnamefont{Ara\'ujo}},
  \bibinfo{author}{\bibfnamefont{M.~T.} \bibnamefont{Quintino}},
  \bibinfo{author}{\bibfnamefont{J.}~\bibnamefont{Min\'ar}},
  \bibinfo{author}{\bibfnamefont{D.}~\bibnamefont{Cavalcanti}},
  \bibinfo{author}{\bibfnamefont{V.}~\bibnamefont{Scarani}},
  \bibinfo{author}{\bibfnamefont{M.~T.} \bibnamefont{Cunha}}, and
  \bibinfo{author}{\bibfnamefont{M.~F.} \bibnamefont{Santos}},
  \bibinfo{year}{2013}, \bibinfo{journal}{Nat. Commun.}
  \textbf{\bibinfo{volume}{4}}, \bibinfo{pages}{2104}.

\bibitem[{\citenamefont{Terhal}(2004)}]{terhal:monogamy}
\bibinfo{author}{\bibnamefont{Terhal}, \bibfnamefont{B.}},
  \bibinfo{year}{2004}, \bibinfo{journal}{{IBM} Journal of Research and
  Development} \textbf{\bibinfo{volume}{48}}(\bibinfo{number}{1}),
  \bibinfo{pages}{71}.

\bibitem[{\citenamefont{Terhal} \emph{et~al.}(2003)\citenamefont{Terhal,
  Doherty, and Schwab}}]{TDS03}
\bibinfo{author}{\bibnamefont{Terhal}, \bibfnamefont{B.}},
  \bibinfo{author}{\bibfnamefont{A.}~\bibnamefont{Doherty}}, and
  \bibinfo{author}{\bibfnamefont{D.}~\bibnamefont{Schwab}},
  \bibinfo{year}{2003}, \bibinfo{journal}{Phys. Rev. Lett.}
  \textbf{\bibinfo{volume}{90}}, \bibinfo{pages}{157903}.

\bibitem[{\citenamefont{Teufel} \emph{et~al.}(1997)\citenamefont{Teufel,
  Berndl, D\"{u}rr, Goldstein, and Zangh\`{\i}}}]{TBD+97}
\bibinfo{author}{\bibnamefont{Teufel}, \bibfnamefont{S.}},
  \bibinfo{author}{\bibfnamefont{K.}~\bibnamefont{Berndl}},
  \bibinfo{author}{\bibfnamefont{D.}~\bibnamefont{D\"{u}rr}},
  \bibinfo{author}{\bibfnamefont{S.}~\bibnamefont{Goldstein}}, and
  \bibinfo{author}{\bibfnamefont{N.}~\bibnamefont{Zangh\`{\i}}},
  \bibinfo{year}{1997}, \bibinfo{journal}{Phys. Rev.~A}
  \textbf{\bibinfo{volume}{56}}, \bibinfo{pages}{1217}.

\bibitem[{\citenamefont{Thew} \emph{et~al.}(2004)\citenamefont{Thew, Ac\'in,
  Zbinden, and Gisin}}]{TAZG04}
\bibinfo{author}{\bibnamefont{Thew}, \bibfnamefont{R.}},
  \bibinfo{author}{\bibfnamefont{A.}~\bibnamefont{Ac\'in}},
  \bibinfo{author}{\bibfnamefont{H.}~\bibnamefont{Zbinden}}, and
  \bibinfo{author}{\bibfnamefont{N.}~\bibnamefont{Gisin}},
  \bibinfo{year}{2004}, \bibinfo{journal}{Phys. Rev. Lett.}
  \textbf{\bibinfo{volume}{93}}, \bibinfo{pages}{010503}.

\bibitem[{\citenamefont{Tittel} \emph{et~al.}(1998)\citenamefont{Tittel,
  Brendel, Zbinden, and Gisin}}]{TBZG98}
\bibinfo{author}{\bibnamefont{Tittel}, \bibfnamefont{W.}},
  \bibinfo{author}{\bibfnamefont{J.}~\bibnamefont{Brendel}},
  \bibinfo{author}{\bibfnamefont{H.}~\bibnamefont{Zbinden}}, and
  \bibinfo{author}{\bibfnamefont{N.}~\bibnamefont{Gisin}},
  \bibinfo{year}{1998}, \bibinfo{journal}{Phys. Rev. Lett.}
  \textbf{\bibinfo{volume}{81}}, \bibinfo{pages}{3563}.

\bibitem[{\citenamefont{Tittel} \emph{et~al.}(1999)\citenamefont{Tittel,
  Brendel, Zbinden, and Gisin}}]{BGTZ99}
\bibinfo{author}{\bibnamefont{Tittel}, \bibfnamefont{W.}},
  \bibinfo{author}{\bibfnamefont{J.}~\bibnamefont{Brendel}},
  \bibinfo{author}{\bibfnamefont{H.}~\bibnamefont{Zbinden}}, and
  \bibinfo{author}{\bibfnamefont{N.}~\bibnamefont{Gisin}},
  \bibinfo{year}{1999}, \bibinfo{journal}{Phys. Rev. Lett.}
  \textbf{\bibinfo{volume}{82}}, \bibinfo{pages}{2594}.

\bibitem[{\citenamefont{Tittel} \emph{et~al.}(2000)\citenamefont{Tittel,
  Brendel, Zbinden, and Gisin}}]{TBZG00}
\bibinfo{author}{\bibnamefont{Tittel}, \bibfnamefont{W.}},
  \bibinfo{author}{\bibfnamefont{J.}~\bibnamefont{Brendel}},
  \bibinfo{author}{\bibfnamefont{H.}~\bibnamefont{Zbinden}}, and
  \bibinfo{author}{\bibfnamefont{N.}~\bibnamefont{Gisin}},
  \bibinfo{year}{2000}, \bibinfo{journal}{Phys. Rev. Lett.}
  \textbf{\bibinfo{volume}{84}}, \bibinfo{pages}{4737}.

\bibitem[{\citenamefont{Tomamichel}
  \emph{et~al.}(2012)\citenamefont{Tomamichel, Fehr, Kaniewski, and
  Wehner}}]{TFK+12}
\bibinfo{author}{\bibnamefont{Tomamichel}, \bibfnamefont{M.}},
  \bibinfo{author}{\bibfnamefont{S.}~\bibnamefont{Fehr}},
  \bibinfo{author}{\bibfnamefont{J.}~\bibnamefont{Kaniewski}}, and
  \bibinfo{author}{\bibfnamefont{S.}~\bibnamefont{Wehner}},
  \bibinfo{year}{2012}, \bibinfo{journal}{arXiv:1210.4359} .

\bibitem[{\citenamefont{Toner}(2009)}]{Toner09}
\bibinfo{author}{\bibnamefont{Toner}, \bibfnamefont{B.}}, \bibinfo{year}{2009},
  \bibinfo{journal}{Proc. R. Soc. A} \textbf{\bibinfo{volume}{465}},
  \bibinfo{pages}{59}.

\bibitem[{\citenamefont{Toner and Bacon}(2003)}]{TB03}
\bibinfo{author}{\bibnamefont{Toner}, \bibfnamefont{B.}}, and
  \bibinfo{author}{\bibfnamefont{D.}~\bibnamefont{Bacon}},
  \bibinfo{year}{2003}, \bibinfo{journal}{Phys. Rev. Lett.}
  \textbf{\bibinfo{volume}{91}}, \bibinfo{pages}{187904}.

\bibitem[{\citenamefont{Toner and Verstraete}(2006)}]{TV06}
\bibinfo{author}{\bibnamefont{Toner}, \bibfnamefont{B.}}, and
  \bibinfo{author}{\bibfnamefont{F.}~\bibnamefont{Verstraete}},
  \bibinfo{year}{2006}, \bibinfo{journal}{arXiv:quant-ph/0611001} .

\bibitem[{\citenamefont{Toth} \emph{et~al.}(2006)\citenamefont{Toth, G\"uhne,
  and Briegel}}]{TGB06}
\bibinfo{author}{\bibnamefont{Toth}, \bibfnamefont{G.}},
  \bibinfo{author}{\bibfnamefont{O.}~\bibnamefont{G\"uhne}}, and
  \bibinfo{author}{\bibfnamefont{H.}~\bibnamefont{Briegel}},
  \bibinfo{year}{2006}, \bibinfo{journal}{Phys. Rev.~A}
  \textbf{\bibinfo{volume}{73}}, \bibinfo{pages}{022303}.

\bibitem[{\citenamefont{Tsirelson}(1987)}]{Tsirelson87}
\bibinfo{author}{\bibnamefont{Tsirelson}, \bibfnamefont{B.}},
  \bibinfo{year}{1987}, \bibinfo{journal}{J. Soviet Math.}
  \textbf{\bibinfo{volume}{36}}, \bibinfo{pages}{557}.

\bibitem[{\citenamefont{Tsirelson}(1993)}]{Tsirelson93}
\bibinfo{author}{\bibnamefont{Tsirelson}, \bibfnamefont{B.~S.}},
  \bibinfo{year}{1993}, \bibinfo{journal}{Hadronic Journal Supplement}
  \textbf{\bibinfo{volume}{8}}, \bibinfo{pages}{329}.

\bibitem[{\citenamefont{Uffink}(2002)}]{Uffink02}
\bibinfo{author}{\bibnamefont{Uffink}, \bibfnamefont{J.}},
  \bibinfo{year}{2002}, \bibinfo{journal}{Phys. Rev. Lett.}
  \textbf{\bibinfo{volume}{88}}, \bibinfo{pages}{230406}.

\bibitem[{\citenamefont{Vadhan}(2012)}]{vadhan:survey}
\bibinfo{author}{\bibnamefont{Vadhan}, \bibfnamefont{S.}},
  \bibinfo{year}{2012}, \bibinfo{note}{\url{http://people.seas.harvard.edu/\~
  salil/pseudorandomness/pseudorandomness-Aug12.pdf}}.

\bibitem[{\citenamefont{Vaidman}(2001)}]{Vaidman01}
\bibinfo{author}{\bibnamefont{Vaidman}, \bibfnamefont{L.}},
  \bibinfo{year}{2001}, \bibinfo{journal}{Phys. Lett.~A}
  \textbf{\bibinfo{volume}{286}}, \bibinfo{pages}{241}.

\bibitem[{\citenamefont{Vazirani and Vidick}(2012{\natexlab{a}})}]{VV12}
\bibinfo{author}{\bibnamefont{Vazirani}, \bibfnamefont{U.}}, and
  \bibinfo{author}{\bibfnamefont{T.}~\bibnamefont{Vidick}},
  \bibinfo{year}{2012}{\natexlab{a}}, \bibinfo{journal}{STOC '12 Proceedings of
  the 44th symposium on Theory of Computing} , \bibinfo{pages}{61}.

\bibitem[{\citenamefont{Vazirani and Vidick}(2012{\natexlab{b}})}]{VV12b}
\bibinfo{author}{\bibnamefont{Vazirani}, \bibfnamefont{U.}}, and
  \bibinfo{author}{\bibfnamefont{T.}~\bibnamefont{Vidick}},
  \bibinfo{year}{2012}{\natexlab{b}}, \bibinfo{journal}{Phil. Trans. R. Soc. A}
  \textbf{\bibinfo{volume}{370}}, \bibinfo{pages}{3432}.

\bibitem[{\citenamefont{Verstraete and Wolf}(2002)}]{VW02}
\bibinfo{author}{\bibnamefont{Verstraete}, \bibfnamefont{F.}}, and
  \bibinfo{author}{\bibfnamefont{M.}~\bibnamefont{Wolf}}, \bibinfo{year}{2002},
  \bibinfo{journal}{Phys. Rev. Lett.} \textbf{\bibinfo{volume}{89}},
  \bibinfo{pages}{170401}.

\bibitem[{\citenamefont{Vert\'esi}(2008)}]{Vertesi08}
\bibinfo{author}{\bibnamefont{Vert\'esi}, \bibfnamefont{T.}},
  \bibinfo{year}{2008}, \bibinfo{journal}{Phys. Rev.~A}
  \textbf{\bibinfo{volume}{78}}, \bibinfo{pages}{032112}.

\bibitem[{\citenamefont{Vertesi and Bene}(2009)}]{VB09}
\bibinfo{author}{\bibnamefont{Vertesi}, \bibfnamefont{T.}}, and
  \bibinfo{author}{\bibfnamefont{E.}~\bibnamefont{Bene}}, \bibinfo{year}{2009},
  \bibinfo{journal}{Phys. Rev.~A} \textbf{\bibinfo{volume}{80}},
  \bibinfo{pages}{062316}.

\bibitem[{\citenamefont{V\'ertesi and Brunner}(2012)}]{VB12}
\bibinfo{author}{\bibnamefont{V\'ertesi}, \bibfnamefont{T.}}, and
  \bibinfo{author}{\bibfnamefont{N.}~\bibnamefont{Brunner}},
  \bibinfo{year}{2012}, \bibinfo{journal}{Phys. Rev. Lett.}
  \textbf{\bibinfo{volume}{108}}, \bibinfo{pages}{030403}.

\bibitem[{\citenamefont{V\'ertesi and Navascues}(2011)}]{VN11}
\bibinfo{author}{\bibnamefont{V\'ertesi}, \bibfnamefont{T.}}, and
  \bibinfo{author}{\bibfnamefont{M.}~\bibnamefont{Navascues}},
  \bibinfo{year}{2011}, \bibinfo{journal}{Phys. Rev.~A}
  \textbf{\bibinfo{volume}{83}}, \bibinfo{pages}{062112}.

\bibitem[{\citenamefont{V\'ertesi and P\'al}(2009)}]{VP09}
\bibinfo{author}{\bibnamefont{V\'ertesi}, \bibfnamefont{T.}}, and
  \bibinfo{author}{\bibfnamefont{K.}~\bibnamefont{P\'al}},
  \bibinfo{year}{2009}, \bibinfo{journal}{Phys. Rev.~A}
  \textbf{\bibinfo{volume}{79}}, \bibinfo{pages}{042106}.

\bibitem[{\citenamefont{Vertesi and Pal}(2008)}]{VP08}
\bibinfo{author}{\bibnamefont{Vertesi}, \bibfnamefont{T.}}, and
  \bibinfo{author}{\bibfnamefont{K.~F.} \bibnamefont{Pal}},
  \bibinfo{year}{2008}, \bibinfo{journal}{Phys. Rev.~A}
  \textbf{\bibinfo{volume}{77}}, \bibinfo{pages}{042106}.

\bibitem[{\citenamefont{V\'ertesi} \emph{et~al.}(2010)\citenamefont{V\'ertesi,
  Pironio, and Brunner}}]{VPB10}
\bibinfo{author}{\bibnamefont{V\'ertesi}, \bibfnamefont{T.}},
  \bibinfo{author}{\bibfnamefont{S.}~\bibnamefont{Pironio}}, and
  \bibinfo{author}{\bibfnamefont{N.}~\bibnamefont{Brunner}},
  \bibinfo{year}{2010}, \bibinfo{journal}{Phys. Rev. Lett.}
  \textbf{\bibinfo{volume}{104}}, \bibinfo{pages}{060401}.

\bibitem[{\citenamefont{Vidick and Wehner}(2011)}]{VW11}
\bibinfo{author}{\bibnamefont{Vidick}, \bibfnamefont{T.}}, and
  \bibinfo{author}{\bibfnamefont{S.}~\bibnamefont{Wehner}},
  \bibinfo{year}{2011}, \bibinfo{journal}{Phys. Rev.~A}
  \textbf{\bibinfo{volume}{83}}, \bibinfo{pages}{052310}.

\bibitem[{\citenamefont{Vongehr}(2012)}]{vongehr}
\bibinfo{author}{\bibnamefont{Vongehr}, \bibfnamefont{S.}},
  \bibinfo{year}{2012}, \bibinfo{journal}{arXiv:1207.5294} .

\bibitem[{\citenamefont{Walborn} \emph{et~al.}(2011)\citenamefont{Walborn,
  Salles, Gomes, Toscano, and Ribeiro}}]{WSGT+11}
\bibinfo{author}{\bibnamefont{Walborn}, \bibfnamefont{S.~P.}},
  \bibinfo{author}{\bibfnamefont{A.}~\bibnamefont{Salles}},
  \bibinfo{author}{\bibfnamefont{R.~M.} \bibnamefont{Gomes}},
  \bibinfo{author}{\bibfnamefont{F.}~\bibnamefont{Toscano}}, and
  \bibinfo{author}{\bibfnamefont{P.~H.~S.} \bibnamefont{Ribeiro}},
  \bibinfo{year}{2011}, \bibinfo{journal}{Phys. Rev.~A}
  \textbf{\bibinfo{volume}{106}}, \bibinfo{pages}{130402}.

\bibitem[{\citenamefont{Wallman and Bartlett}(2012)}]{WB12}
\bibinfo{author}{\bibnamefont{Wallman}, \bibfnamefont{J.}}, and
  \bibinfo{author}{\bibfnamefont{S.}~\bibnamefont{Bartlett}},
  \bibinfo{year}{2012}, \bibinfo{journal}{Phys. Rev.~A}
  \textbf{\bibinfo{volume}{85}}, \bibinfo{pages}{024101}.

\bibitem[{\citenamefont{Walther} \emph{et~al.}(2005)\citenamefont{Walther,
  Aspelmeyer, Resch, and Zeilinger}}]{WARZ05}
\bibinfo{author}{\bibnamefont{Walther}, \bibfnamefont{P.}},
  \bibinfo{author}{\bibfnamefont{M.}~\bibnamefont{Aspelmeyer}},
  \bibinfo{author}{\bibfnamefont{K.~J.} \bibnamefont{Resch}}, and
  \bibinfo{author}{\bibfnamefont{A.}~\bibnamefont{Zeilinger}},
  \bibinfo{year}{2005}, \bibinfo{journal}{Phys. Rev. Lett.}
  \textbf{\bibinfo{volume}{95}}, \bibinfo{pages}{020403}.

\bibitem[{\citenamefont{Wang and Markham}(2012)}]{WM12}
\bibinfo{author}{\bibnamefont{Wang}, \bibfnamefont{Z.}}, and
  \bibinfo{author}{\bibfnamefont{D.}~\bibnamefont{Markham}},
  \bibinfo{year}{2012}, \bibinfo{journal}{Phys. Rev. Lett.}
  \textbf{\bibinfo{volume}{108}}, \bibinfo{pages}{210407}.

\bibitem[{\citenamefont{Wehner}(2006{\natexlab{a}})}]{wehner:simulate}
\bibinfo{author}{\bibnamefont{Wehner}, \bibfnamefont{S.}},
  \bibinfo{year}{2006}{\natexlab{a}}, in \emph{\bibinfo{booktitle}{Proceedings
  of 23rd STACS}}, volume \bibinfo{volume}{3884} of
  \emph{\bibinfo{series}{LNCS}}, pp. \bibinfo{pages}{162--171}.

\bibitem[{\citenamefont{Wehner}(2006{\natexlab{b}})}]{Wehner06}
\bibinfo{author}{\bibnamefont{Wehner}, \bibfnamefont{S.}},
  \bibinfo{year}{2006}{\natexlab{b}}, \bibinfo{journal}{Phys. Rev. A}
  \textbf{\bibinfo{volume}{73}}, \bibinfo{pages}{022110}.

\bibitem[{\citenamefont{Wehner} \emph{et~al.}(2008)\citenamefont{Wehner,
  Christandl, and Dherty}}]{WCD08}
\bibinfo{author}{\bibnamefont{Wehner}, \bibfnamefont{S.}},
  \bibinfo{author}{\bibfnamefont{M.}~\bibnamefont{Christandl}}, and
  \bibinfo{author}{\bibfnamefont{A.}~\bibnamefont{Dherty}},
  \bibinfo{year}{2008}, \bibinfo{journal}{Phys. Rev.~A}
  \textbf{\bibinfo{volume}{78}}, \bibinfo{pages}{062112}.

\bibitem[{\citenamefont{Weihs} \emph{et~al.}(1998)\citenamefont{Weihs,
  Jennewein, Simon, Weinfurter, and Zeilinger}}]{WJSW+98}
\bibinfo{author}{\bibnamefont{Weihs}, \bibfnamefont{G.}},
  \bibinfo{author}{\bibfnamefont{T.}~\bibnamefont{Jennewein}},
  \bibinfo{author}{\bibfnamefont{C.}~\bibnamefont{Simon}},
  \bibinfo{author}{\bibfnamefont{H.}~\bibnamefont{Weinfurter}}, and
  \bibinfo{author}{\bibfnamefont{A.}~\bibnamefont{Zeilinger}},
  \bibinfo{year}{1998}, \bibinfo{journal}{Phys. Rev. Lett.}
  \textbf{\bibinfo{volume}{81}}, \bibinfo{pages}{5039}.

\bibitem[{\citenamefont{Wenger} \emph{et~al.}(2003)\citenamefont{Wenger,
  Hafezi, Grosshans, Tualle-Brouri, and Grangier}}]{WHGT+03}
\bibinfo{author}{\bibnamefont{Wenger}, \bibfnamefont{J.}},
  \bibinfo{author}{\bibfnamefont{M.}~\bibnamefont{Hafezi}},
  \bibinfo{author}{\bibfnamefont{F.}~\bibnamefont{Grosshans}},
  \bibinfo{author}{\bibfnamefont{R.}~\bibnamefont{Tualle-Brouri}}, and
  \bibinfo{author}{\bibfnamefont{P.}~\bibnamefont{Grangier}},
  \bibinfo{year}{2003}, \bibinfo{journal}{Phys. Rev.~A}
  \textbf{\bibinfo{volume}{67}}, \bibinfo{pages}{012105}.

\bibitem[{\citenamefont{Werner}(1989)}]{Werner89}
\bibinfo{author}{\bibnamefont{Werner}, \bibfnamefont{R.~F.}},
  \bibinfo{year}{1989}, \bibinfo{journal}{Phys. Rev.~A}
  \textbf{\bibinfo{volume}{40}}, \bibinfo{pages}{4277}.

\bibitem[{\citenamefont{Werner and Wolf}(2000)}]{WW00}
\bibinfo{author}{\bibnamefont{Werner}, \bibfnamefont{R.~F.}}, and
  \bibinfo{author}{\bibfnamefont{M.~M.} \bibnamefont{Wolf}},
  \bibinfo{year}{2000}, \bibinfo{journal}{Phys. Rev.~A}
  \textbf{\bibinfo{volume}{61}}, \bibinfo{pages}{062102}.

\bibitem[{\citenamefont{Werner and Wolf}(2001)}]{WW01b}
\bibinfo{author}{\bibnamefont{Werner}, \bibfnamefont{R.~F.}}, and
  \bibinfo{author}{\bibfnamefont{M.~M.} \bibnamefont{Wolf}},
  \bibinfo{year}{2001}, \bibinfo{journal}{Quant. Inf. and Comp.}
  \textbf{\bibinfo{volume}{1}}, \bibinfo{pages}{1}.

\bibitem[{\citenamefont{Werner and Wolf}(2002)}]{WW01}
\bibinfo{author}{\bibnamefont{Werner}, \bibfnamefont{R.~F.}}, and
  \bibinfo{author}{\bibfnamefont{M.~M.} \bibnamefont{Wolf}},
  \bibinfo{year}{2002}, \bibinfo{journal}{Physical Review A}
  \textbf{\bibinfo{volume}{64}}, \bibinfo{pages}{032112}.

\bibitem[{\citenamefont{White} \emph{et~al.}(1999)\citenamefont{White, James,
  Eberhard, and Kwiat}}]{WJEK99}
\bibinfo{author}{\bibnamefont{White}, \bibfnamefont{A.}},
  \bibinfo{author}{\bibfnamefont{D.}~\bibnamefont{James}},
  \bibinfo{author}{\bibfnamefont{P.}~\bibnamefont{Eberhard}}, and
  \bibinfo{author}{\bibfnamefont{P.}~\bibnamefont{Kwiat}},
  \bibinfo{year}{1999}, \bibinfo{journal}{Phys. Rev. Lett.}
  \textbf{\bibinfo{volume}{83}}, \bibinfo{pages}{3103}.

\bibitem[{\citenamefont{Wilms} \emph{et~al.}(2008)\citenamefont{Wilms, Disser,
  Alber, and Percival}}]{WDAP08}
\bibinfo{author}{\bibnamefont{Wilms}, \bibfnamefont{J.}},
  \bibinfo{author}{\bibfnamefont{Y.}~\bibnamefont{Disser}},
  \bibinfo{author}{\bibfnamefont{G.}~\bibnamefont{Alber}}, and
  \bibinfo{author}{\bibfnamefont{I.~C.} \bibnamefont{Percival}},
  \bibinfo{year}{2008}, \bibinfo{journal}{Phys. Rev.~A}
  \textbf{\bibinfo{volume}{78}}, \bibinfo{pages}{032116}.

\bibitem[{\citenamefont{Wiseman} \emph{et~al.}(2007)\citenamefont{Wiseman,
  Jones, and Doherty}}]{WJD07}
\bibinfo{author}{\bibnamefont{Wiseman}, \bibfnamefont{H.}},
  \bibinfo{author}{\bibfnamefont{S.~J.} \bibnamefont{Jones}}, and
  \bibinfo{author}{\bibfnamefont{A.}~\bibnamefont{Doherty}},
  \bibinfo{year}{2007}, \bibinfo{journal}{Phys. Rev. Lett.}
  \textbf{\bibinfo{volume}{98}}, \bibinfo{pages}{140402}.

\bibitem[{\citenamefont{Wittmann} \emph{et~al.}(2012)\citenamefont{Wittmann,
  Ramelow, Steinlechner, Langford, Brunner, Wiseman, Ursin, and
  Zeilinger}}]{WRSL+12}
\bibinfo{author}{\bibnamefont{Wittmann}, \bibfnamefont{B.}},
  \bibinfo{author}{\bibfnamefont{S.}~\bibnamefont{Ramelow}},
  \bibinfo{author}{\bibfnamefont{F.}~\bibnamefont{Steinlechner}},
  \bibinfo{author}{\bibfnamefont{N.}~\bibnamefont{Langford}},
  \bibinfo{author}{\bibfnamefont{N.}~\bibnamefont{Brunner}},
  \bibinfo{author}{\bibfnamefont{H.}~\bibnamefont{Wiseman}},
  \bibinfo{author}{\bibfnamefont{R.}~\bibnamefont{Ursin}}, and
  \bibinfo{author}{\bibfnamefont{A.}~\bibnamefont{Zeilinger}},
  \bibinfo{year}{2012}, \bibinfo{journal}{New J.~Phys}
  \textbf{\bibinfo{volume}{14}}, \bibinfo{pages}{053030}.

\bibitem[{\citenamefont{Wolf and Wullschleger}(2005)}]{WW05}
\bibinfo{author}{\bibnamefont{Wolf}, \bibfnamefont{S.}}, and
  \bibinfo{author}{\bibfnamefont{J.}~\bibnamefont{Wullschleger}},
  \bibinfo{year}{2005}, \bibinfo{journal}{arXiv:quant-ph/0502030v1} .

\bibitem[{\citenamefont{Wood and Spekkens}(2012)}]{WS12}
\bibinfo{author}{\bibnamefont{Wood}, \bibfnamefont{C.}}, and
  \bibinfo{author}{\bibfnamefont{R.}~\bibnamefont{Spekkens}},
  \bibinfo{year}{2012}, \bibinfo{journal}{arXiv:1208.4119} .

\bibitem[{\citenamefont{Yang} \emph{et~al.}(2012)\citenamefont{Yang,
  Cavalcanti, Almeida, Teo, and Scarani}}]{YCAT+12}
\bibinfo{author}{\bibnamefont{Yang}, \bibfnamefont{T.}},
  \bibinfo{author}{\bibfnamefont{D.}~\bibnamefont{Cavalcanti}},
  \bibinfo{author}{\bibfnamefont{M.}~\bibnamefont{Almeida}},
  \bibinfo{author}{\bibfnamefont{C.}~\bibnamefont{Teo}}, and
  \bibinfo{author}{\bibfnamefont{V.}~\bibnamefont{Scarani}},
  \bibinfo{year}{2012}, \bibinfo{journal}{New J. Phys.}
  \textbf{\bibinfo{volume}{14}}, \bibinfo{pages}{013061}.

\bibitem[{\citenamefont{Yang and Navascues}(2013)}]{YN12}
\bibinfo{author}{\bibnamefont{Yang}, \bibfnamefont{T.}}, and
  \bibinfo{author}{\bibfnamefont{M.}~\bibnamefont{Navascues}},
  \bibinfo{year}{2013}, \bibinfo{journal}{Phys. Rev.~A}
  \textbf{\bibinfo{volume}{87}}, \bibinfo{pages}{050102}.

\bibitem[{\citenamefont{Yang} \emph{et~al.}(2011)\citenamefont{Yang,
  Navascu\'es, Sheridan, and Scarani}}]{YNSS11}
\bibinfo{author}{\bibnamefont{Yang}, \bibfnamefont{T.~H.}},
  \bibinfo{author}{\bibfnamefont{M.}~\bibnamefont{Navascu\'es}},
  \bibinfo{author}{\bibfnamefont{L.}~\bibnamefont{Sheridan}}, and
  \bibinfo{author}{\bibfnamefont{V.}~\bibnamefont{Scarani}},
  \bibinfo{year}{2011}, \bibinfo{journal}{Phys. Rev.~A}
  \textbf{\bibinfo{volume}{022105}}, \bibinfo{pages}{83}.

\bibitem[{\citenamefont{Yu and Oh}(2013)}]{YO13}
\bibinfo{author}{\bibnamefont{Yu}, \bibfnamefont{S.}}, and
  \bibinfo{author}{\bibfnamefont{C.}~\bibnamefont{Oh}}, \bibinfo{year}{2013},
  \bibinfo{journal}{arXiv:1306.5330} .

\bibitem[{\citenamefont{Zbinden} \emph{et~al.}(2001)\citenamefont{Zbinden,
  Brendel, Gisin, and Tittel}}]{ZBGT01}
\bibinfo{author}{\bibnamefont{Zbinden}, \bibfnamefont{H.}},
  \bibinfo{author}{\bibfnamefont{J.}~\bibnamefont{Brendel}},
  \bibinfo{author}{\bibfnamefont{N.}~\bibnamefont{Gisin}}, and
  \bibinfo{author}{\bibfnamefont{W.}~\bibnamefont{Tittel}},
  \bibinfo{year}{2001}, \bibinfo{journal}{Phys. Rev.~A}
  \textbf{\bibinfo{volume}{63}}, \bibinfo{pages}{022111}.

\bibitem[{\citenamefont{Zeilinger}(1999)}]{Zeilinger99}
\bibinfo{author}{\bibnamefont{Zeilinger}, \bibfnamefont{A.}},
  \bibinfo{year}{1999}, \bibinfo{journal}{Rev. Mod. Phys.}
  \textbf{\bibinfo{volume}{71}}, \bibinfo{pages}{s288}.

\bibitem[{\citenamefont{Zhang} \emph{et~al.}(2011)\citenamefont{Zhang, Glancy,
  and Knill}}]{ZGK11}
\bibinfo{author}{\bibnamefont{Zhang}, \bibfnamefont{Y.}},
  \bibinfo{author}{\bibfnamefont{S.}~\bibnamefont{Glancy}}, and
  \bibinfo{author}{\bibfnamefont{E.}~\bibnamefont{Knill}},
  \bibinfo{year}{2011}, \bibinfo{journal}{Phys. Rev.~A}
  \textbf{\bibinfo{volume}{84}}, \bibinfo{pages}{062118}.

\bibitem[{\citenamefont{Zhang} \emph{et~al.}(2013)\citenamefont{Zhang, Knill,
  and Glancy}}]{ZKG13}
\bibinfo{author}{\bibnamefont{Zhang}, \bibfnamefont{Y.}},
  \bibinfo{author}{\bibfnamefont{E.}~\bibnamefont{Knill}}, and
  \bibinfo{author}{\bibfnamefont{S.}~\bibnamefont{Glancy}},
  \bibinfo{year}{2013}, \bibinfo{journal}{Phys. Rev. A} 
  \textbf{\bibinfo{volume}{88}}, \bibinfo{pages}{052119}.
  

\bibitem[{\citenamefont{Zhao} \emph{et~al.}(2003)\citenamefont{Zhao, Yang,
  Chen, Zhang, Zukowski, and Pan}}]{ZYCZ+03}
\bibinfo{author}{\bibnamefont{Zhao}, \bibfnamefont{Z.}},
  \bibinfo{author}{\bibfnamefont{T.}~\bibnamefont{Yang}},
  \bibinfo{author}{\bibfnamefont{Y.-A.} \bibnamefont{Chen}},
  \bibinfo{author}{\bibfnamefont{A.-N.} \bibnamefont{Zhang}},
  \bibinfo{author}{\bibfnamefont{M.}~\bibnamefont{Zukowski}}, and
  \bibinfo{author}{\bibfnamefont{J.-W.} \bibnamefont{Pan}},
  \bibinfo{year}{2003}, \bibinfo{journal}{Phys. Rev. Lett.}
  \textbf{\bibinfo{volume}{91}}, \bibinfo{pages}{180401}.

\bibitem[{\citenamefont{Ziegler}(1995)}]{Ziegler95}
\bibinfo{author}{\bibnamefont{Ziegler}, \bibfnamefont{G.~M.}},
  \bibinfo{year}{1995}, \emph{\bibinfo{title}{Lectures on {P}olytopes}}, volume
  \bibinfo{volume}{152} of \emph{\bibinfo{series}{Graduate texts in
  Mathematics}} (\bibinfo{publisher}{Springer-Verlag},
  \bibinfo{address}{New-York}).

\bibitem[{\citenamefont{Zukowski and Brukner}(2002)}]{ZB02}
\bibinfo{author}{\bibnamefont{Zukowski}, \bibfnamefont{M.}}, and
  \bibinfo{author}{\bibfnamefont{C.}~\bibnamefont{Brukner}},
  \bibinfo{year}{2002}, \bibinfo{journal}{Phys. Rev. Lett.}
  \textbf{\bibinfo{volume}{88}}, \bibinfo{pages}{210401}.

\bibitem[{\citenamefont{Zukowski} \emph{et~al.}(2008)\citenamefont{Zukowski,
  Greenberger, Horne, and Zeilinger}}]{ZGHZ08}
\bibinfo{author}{\bibnamefont{Zukowski}, \bibfnamefont{M.}},
  \bibinfo{author}{\bibfnamefont{D.}~\bibnamefont{Greenberger}},
  \bibinfo{author}{\bibfnamefont{M.}~\bibnamefont{Horne}}, and
  \bibinfo{author}{\bibfnamefont{A.}~\bibnamefont{Zeilinger}},
  \bibinfo{year}{2008}, \bibinfo{journal}{Phys. Rev.~A}
  \textbf{\bibinfo{volume}{78}}, \bibinfo{pages}{022111}.

\bibitem[{\citenamefont{Zukowski} \emph{et~al.}(1998)\citenamefont{Zukowski,
  Horodecki, Horodecki, and Horodecki}}]{ZHHH98}
\bibinfo{author}{\bibnamefont{Zukowski}, \bibfnamefont{M.}},
  \bibinfo{author}{\bibfnamefont{R.}~\bibnamefont{Horodecki}},
  \bibinfo{author}{\bibfnamefont{M.}~\bibnamefont{Horodecki}}, and
  \bibinfo{author}{\bibfnamefont{P.}~\bibnamefont{Horodecki}},
  \bibinfo{year}{1998}, \bibinfo{journal}{Phys. Rev.~A}
  \textbf{\bibinfo{volume}{58}}, \bibinfo{pages}{1694}.

\bibitem[{\citenamefont{Zukowski and Kaszlikowski}(1999)}]{ZK99}
\bibinfo{author}{\bibnamefont{Zukowski}, \bibfnamefont{M.}}, and
  \bibinfo{author}{\bibfnamefont{D.}~\bibnamefont{Kaszlikowski}},
  \bibinfo{year}{1999}, \bibinfo{journal}{Phys. Rev.~A}
  \textbf{\bibinfo{volume}{59}}, \bibinfo{pages}{3200}.

\bibitem[{\citenamefont{Zukowski} \emph{et~al.}(1999)\citenamefont{Zukowski,
  Kaszlikowski, Baturo, and Larsson}}]{ZKBL99}
\bibinfo{author}{\bibnamefont{Zukowski}, \bibfnamefont{M.}},
  \bibinfo{author}{\bibfnamefont{D.}~\bibnamefont{Kaszlikowski}},
  \bibinfo{author}{\bibfnamefont{A.}~\bibnamefont{Baturo}}, and
  \bibinfo{author}{\bibfnamefont{J.-A.} \bibnamefont{Larsson}},
  \bibinfo{year}{1999}, \bibinfo{journal}{quant-ph/9910058} .

\bibitem[{\citenamefont{Zukowski} \emph{et~al.}(1993)\citenamefont{Zukowski,
  Zeilinger, Horne, and Ekert}}]{ZZHE93}
\bibinfo{author}{\bibnamefont{Zukowski}, \bibfnamefont{M.}},
  \bibinfo{author}{\bibfnamefont{A.}~\bibnamefont{Zeilinger}},
  \bibinfo{author}{\bibfnamefont{M.}~\bibnamefont{Horne}}, and
  \bibinfo{author}{\bibfnamefont{A.}~\bibnamefont{Ekert}},
  \bibinfo{year}{1993}, \bibinfo{journal}{Phys. Rev. Lett.}
  \textbf{\bibinfo{volume}{71}}, \bibinfo{pages}{4287}.

\end{thebibliography}

\end{document}